\documentclass[twocolumn]{aastex631}
\usepackage{threeparttable}
\usepackage{booktabs}
\usepackage{subfigure}
\usepackage{amsmath}
\usepackage{epstopdf}
\usepackage{color}
\usepackage{enumitem}

\setlist[enumerate]{label=\textbf{\arabic*.}}

\newcommand{\reffig}[1]{Figure \ref{#1}}
\newcommand{\reftable}[1]{Table \ref{#1}}
\newcommand{\refsection}[1]{Section \ref{#1}}
\newcommand{\refsubsection}[1]{Section \ref{#1}}

\newcommand{\refformula}[1]{Equation (\ref{#1})}

\begin{document}
	\title{Estimating Stellar Atmospheric Parameters and [$\alpha$/Fe] for LAMOST O-M type Stars Using a Spectral Emulator}
	\author{Junchao Liang}
	\affil{CAS Key Laboratory of Optical Astronomy, National Astronomical Observatories, Chinese Academy of Sciences, Beijing 100101, People’s Republic of China}
	\affil{School of Astronomy and Space Science, University of Chinese Academy of Sciences, Beijing 100049, People’s Republic of China}
	
	\author{ALi Luo $^\star$}
	\affil{CAS Key Laboratory of Optical Astronomy, National Astronomical Observatories, Chinese Academy of Sciences, Beijing 100101, People’s Republic of China}
	\affil{School of Astronomy and Space Science, University of Chinese Academy of Sciences, Beijing 100049, People’s Republic of China}
	\email{$^\star$ lal@nao.cas.cn}
	
	\author{YinBi Li $^\star$}
	\affil{CAS Key Laboratory of Optical Astronomy, National Astronomical Observatories, Chinese Academy of Sciences, Beijing 100101, People’s Republic of China}
	\email{$^\star$ ybli@bao.ac.cn}
	
	\author{XiaoXiao Ma}
	\affil{CAS Key Laboratory of Optical Astronomy, National Astronomical Observatories, Chinese Academy of Sciences, Beijing 100101, People’s Republic of China}
	\affil{School of Astronomy and Space Science, University of Chinese Academy of Sciences, Beijing 100049, People’s Republic of China}
	
	\author{Shuo Li}
	\affil{CAS Key Laboratory of Optical Astronomy, National Astronomical Observatories, Chinese Academy of Sciences, Beijing 100101, People’s Republic of China}
	\affil{School of Astronomy and Space Science, University of Chinese Academy of Sciences, Beijing 100049, People’s Republic of China}
	
	\author{ShuGuo Ma}
	\affil{CAS Key Laboratory of Optical Astronomy, National Astronomical Observatories, Chinese Academy of Sciences, Beijing 100101, People’s Republic of China}
	
	\author{HaiLing Lu}
	\affil{CAS Key Laboratory of Optical Astronomy, National Astronomical Observatories, Chinese Academy of Sciences, Beijing 100101, People’s Republic of China}
	\affil{School of Astronomy and Space Science, University of Chinese Academy of Sciences, Beijing 100049, People’s Republic of China}
	
	\author{YunJin Zhang}
	\affil{CAS Key Laboratory of Optical Astronomy, National Astronomical Observatories, Chinese Academy of Sciences, Beijing 100101, People’s Republic of China}
	\affil{School of Astronomy and Space Science, University of Chinese Academy of Sciences, Beijing 100049, People’s Republic of China}
	
	\author{Bing Du}
	\affil{CAS Key Laboratory of Optical Astronomy, National Astronomical Observatories, Chinese Academy of Sciences, Beijing 100101, People’s Republic of China}
	
	\author{Xiao Kong}
	\affil{CAS Key Laboratory of Optical Astronomy, National Astronomical Observatories, Chinese Academy of Sciences, Beijing 100101, People’s Republic of China}
	
	\begin{abstract}
		{In this paper, we developed a spectral emulator based on the Mapping Nearby Galaxies at Apache Point Observatory Stellar Library (MaStar) and a grouping optimization strategy to estimate effective temperature ($T_{\text{eff}}$), surface gravity (log $g$), metallicity ([Fe/H]) and the abundance of alpha elements with respect to iron ([$\alpha$/Fe]) for O-M-type stars within the Large Sky Area Multi-Object Fiber Spectroscopic Telescope (LAMOST) low-resolution spectra. The primary aim is to use a rapid spectral-fitting method, specifically the spectral emulator with the grouping optimization strategy, to create a comprehensive catalog for stars of all types within LAMOST, addressing the shortcomings in parameter estimations for both cold and hot stars present in the official LAMOST AFGKM-type catalog. This effort is part of our series of studies dedicated to establishing an empirical spectral library for LAMOST. Experimental results demonstrate that our method is effectively applicable to parameter prediction for LAMOST, with the single-machine processing time within $70$ hr. We observed that the internal error dispersions for $T_{\text{eff}}$, log $g$, [Fe/H], and [$\alpha$/Fe] across different spectral types lie within the ranges of $15-594$ K, $0.03-0.27$ dex, $0.02-0.10$ dex, and $0.01-0.04$ dex, respectively, indicating a good consistency. A comparative analysis with external data highlighted deficiencies in the official LAMOST catalog and issues with MaStar parameters, as well as potential limitations of our method in processing spectra with strong emission lines and bad pixels. The derived atmospheric parameters as a part of this work are available at \url{https://nadc.china-vo.org/res/r101402/}.}
	\end{abstract}
	
	
	\section{Introduction}
	A main task of modern astrophysics is to understand when and how galaxies formed and evolved. Our own galaxy, the Milky Way, offers a unique opportunity to study galaxies in considerable detail by measuring and analyzing the properties of stars \citep{juric2008milky, ivezic2008milky}. The principal properties of stars, such as effective temperature ($T_{\text{eff}}$), surface gravity (log $g$), metallicity ([Fe/H]) and the abundance of alpha elements with respect to iron ([$\alpha$/Fe]), can be measured from spectra. At present, more and more large surveys, such as the Radial Velocity Experiment (\citealt{2006AJ....132.1645S}), the Sloan Extension for Galactic Understanding and Exploration (SEGUE; \citealt{2009AJ....137.4377Y}), the Large Sky Area Multi-object Fiber Spectroscopic Telescope (LAMOST; \citealt{zhao2012lamost}; \citealt{cui2012large}; \citealt{luo2015first}), the Galactic Archaeology with HERMES (\citealt{de2015galah}), the Apache Point Observatory Galactic Evolution Experiment (APOGEE; \citealt{Majewski2017}), the Sloan Digital Sky Survey V (SDSS; \citealt{Almeida_2023}), the Dark Energy Spectroscopic Instrument (DESI; \citeauthor{DESICollaboration2016} \citeyear{DESICollaboration2016}, \citeyear{DESICollaboration2016a}; \citealt{Abareshi2022}), Gaia Radial Velocity Spectrometer (\citealt{GaiaCollaboration2023}), the $4.2$-m William Herschel Telescope Enhanced Area Velocity Explorer (\citealt{Jin2023}), the Multi-Object Optical and Near-infrared Spectrograph (\citealt{Cirasuolo2020}), the upcoming the $4$-metre Multi-Object Spectroscopic Telescope (\citealt{deJong2022}), the Chinese Space Station Telescope (CSST; \citealt{Gong_2019}), provide a large amount of spectra to help us understand the evolution and chemical formation of the Milky Way and even the Universe.
	
	These large surveys have been specially designed to yield full sets of atmospheric parameters for a wide variety of stars, and the methods for constructing spectral emulators \footnote{Spectral Emulator: A method using machine learning to map stellar parameters to spectra \citep{Czekala_2015, tabernero2022steparsyn}.} based on theoretical and empirical spectral libraries are the most widely used spectroscopic techniques for determining stellar atmospheric parameters. Their widespread adoption is evidenced by a variety of publicly available implementations within the community. These methods are foundational to several analytical tools, including the SEGUE atmospheric parameter Pipeline (\citealt{lee2011segue}), My God It’s Full Of Stars (\citealt{sbordone2014mygisfos}), the LAMOST atmospheric parameter Pipeline (LASP; \citealt{luo2015first}), the APOGEE atmospheric parameter and Chemical Abundance Pipeline (\citealt{perez2016aspcap}), Self-consistent ab initio Fitting of Stellar Spectra (The Payne; \citealt{ting2019payne}), Spectrophotometric Modeling of Stars in the Gaia Era (MINESweeper; \citealt{cargile2020minesweeper}), the LAMOST atmospheric parameter Pipeline for M-type stars (LASPM; \citealt{Du_2021}), and a Bayesian code to infer stellar atmospheric parameters using spectral emulators (STEPARSYN; \citealt{tabernero2022steparsyn}).
	
	However, the aforementioned methods have certain limitations. For instance, the spectral emulators for atmospheric parameter determination often involves $\chi^{2}$ optimization, a process that has always been extremely time consuming, especially for millions of spectra. Furthermore, while these methods perform well in predicting parameters for FGK-type stars, catalogues have generally struggled to provide reliable parameters for stars outside of the FGK regime, such as OBA and M stars. For the LAMOST survey, such a vast data set also includes O-M-type stars, but currently, LAMOST has not officially provided atmospheric parameters for O- and B-type stars, and issues still exist in the parameters of A- and M-type stars.
	
	Considering the above problems, we used a spectral emulator based on the Mapping Nearby Galaxies at Apache Point Observatory (MaNGA) Stellar Library (MaStar; \citealt{Yan_2019, abdurro2022seventeenth}) and the new proposed strategy of grouping optimization to estimate $T_{\text{eff}}$, log $g$, [Fe/H], and [$\alpha$/Fe] for O-M-type stars within the LAMOST low-resolution spectra. Our spectral emulator, integrating a principal component analysis (PCA; \citealt{Abdi2010,Jolliffe2016}) and Gaussian process regression (GPR; \citealt{rasmussen2006gaussian}), produces spectra for specified $T_{\text{eff}}$, log $g$, [Fe/H], and [$\alpha$/Fe]. The grouping optimization strategy was employed to achieve a balance between efficiency and accuracy in solving for the minimum $\chi^{2}$. Experimental results show that the combination of the spectral emulator and grouping optimization strategy yields reliable atmospheric parameter predictions while enhancing the efficiency of the spectral-fitting method, making it suitable for large-scale surveys. The measurements of LAMOST low-resolution atmospheric parameters using this approach are viable for the subsequent establishment of an empirical spectral library for LAMOST.
	
	The remainder of this paper is organized as follows. We introduce methodology in \refsection{Methodology}, including motivation and workflow. Then, we describe the data sets used in this work in \refsection{Data introduction}. We discuss the results and compare them to those in previous works in \refsection{Data Experiment}.  Finally, the conclusions are presented in \refsection{Conclusions}. 
	
	
	\section{Methodology} \label{Methodology}
	\subsection{Motivation}  \label{Motivation}
	Our decision to utilize the MaStar and apply the spectral emulator with the grouping optimization strategy for atmospheric parameter prediction was motivated by four considerations:
	\begin{enumerate}
		\item \textbf{Extending atmospheric parameters for M-type and OBA-type stars of the LAMOST survey.} LAMOST officially employs LASP and LASPM to predict atmospheric parameters for AFGK-type and M-type stars, based on the ELODIE \citep{prugniel2007new} and BT-Settl \citep{allard2012models} libraries, respectively. In \refsubsection{M Type} and \refsubsection{AFGK Type}, we found that the parameters predicted by LASPM and LASP for M-type and A-type stars exhibit certain issues. For LASP, the instrumental discrepancies between ELODIE and LAMOST (inconsistent spectral resolution and wavelength coverage) and the sparsity of ELODIE in the domain of hot stars collectively increase the uncertainty in parameter predictions. For LASPM, the issue with certain parameters primarily arises from the fact that synthetic spectra in the optical cannot yet adequately reproduce the spectra of cool stars \citep{Imig_2022}; for example, it is challenging for BT-Settl to reproduce the K- and Na-line pairs sensitive to log $g$ and [Fe/H] \citep{rajpurohit2014high, passegger2016fundamental}. Hence, employing an empirical spectral library that aligns with LAMOST spectra and uniform parameter coverage is essential to enhance the reliability of LAMOST’s atmospheric parameter predictions. In many respects, MaStar spectra and LAMOST spectra are similar, including consistent wavelength coverage and spectral resolution; thus, the issues related to instrumental differences will be reduced when using MaStar to predict LAMOST paramters. Moreover, the MaStar project covers a wide and uniform parameter ranges \citep{Yan_2019}; thus, utilizing it could extend parameters for hot and cool stars, and supplement atmospheric parameters of OB-type stars for LAMOST official parameter catalogs.
		\item \textbf{Testing the reliability of MaStar parameters.} Currently, the parameters of the MaStar are based on parallel predictions by \cite{Hill_2021,Hill_2022}, \cite{Imig_2022}, \cite{lazarz2022sdss}, and Chen et al. (2024, in preparation) \footnote{\url{https://www.sdss4.org/dr17/mastar/mastar-stellar-parameters/}}, with the median of these parameters recommended for use, as referenced by \cite{abdurro2022seventeenth}. However, the validation of the reliability of MaStar's parameters remains an ongoing effort, as indicated by Yan et al. (2024, in preparation). This scenario necessitates a reliability analysis of the parameters when utilizing the MaStar. Given the lack of external data for assessing the reliability of these parameters, particularly for OBA- and M-type stars, we intend to utilize the parameter prediction of LAMOST to compare with external data sets for the verification of MaStar parameters' reliability. This will provide a reference for subsequent calibration work of MaStar parameters.
		\item \textbf{Improving fitting efficiency.} While the spectral emulators are known for their strong interpretability, they tend to be inefficient, with the fitting of a single spectrum being time consuming. This issue is further detailed in the discussions by \cite{ting2016accelerated}, \cite{rui2019analysis}, and \cite{tabernero2022steparsyn}. The challenge escalates when dealing with the tens of millions of spectra from LAMOST. To improve the practical application of the spectral emulator, we have for the first time implemented the grouping optimization strategy, aiming to enhance the efficiency of spectral fitting methods.
		\item \textbf{Deriving homogeneous atmospheric parameters across wide temperature ranges and establishing an empirical spectral library for LAMOST.} Currently, for LAMOST stellar catalogs, the atmospheric parameters for different types of stars are predicted independently by \cite{Du_2021}, \cite{luo2015first} and \cite{xiang2022stellar}. These parameters are not homogenized, making the establishment of an empirical spectral library based on these catalogs unreliable. The existence of the MaStar provides us the opportunity to obtain more homogenized LAMOST parameters. Using these parameters to establish the LAMOST empirical spectral library will enhance its reliability. This work is our first step toward establishing the LAMOST empirical spectral library, and we plan to conduct a series of predictions of atmospheric parameters for LAMOST based on other libraries, such as the BOSZ \citep{bohlin2017new} and MILES \citep{sanchez2006medium, sharma2016new, garcia2021extension}. The recommended catalog will be provided  by combining these prediction parameters, and be further validated within star clusters and Gaia benchmark stars, ultimately being used to establish the LAMOST empirical spectral library.
	\end{enumerate}
	
	\begin{figure*}[!htb]
		\includegraphics[width=6.25in,height=8.5in]{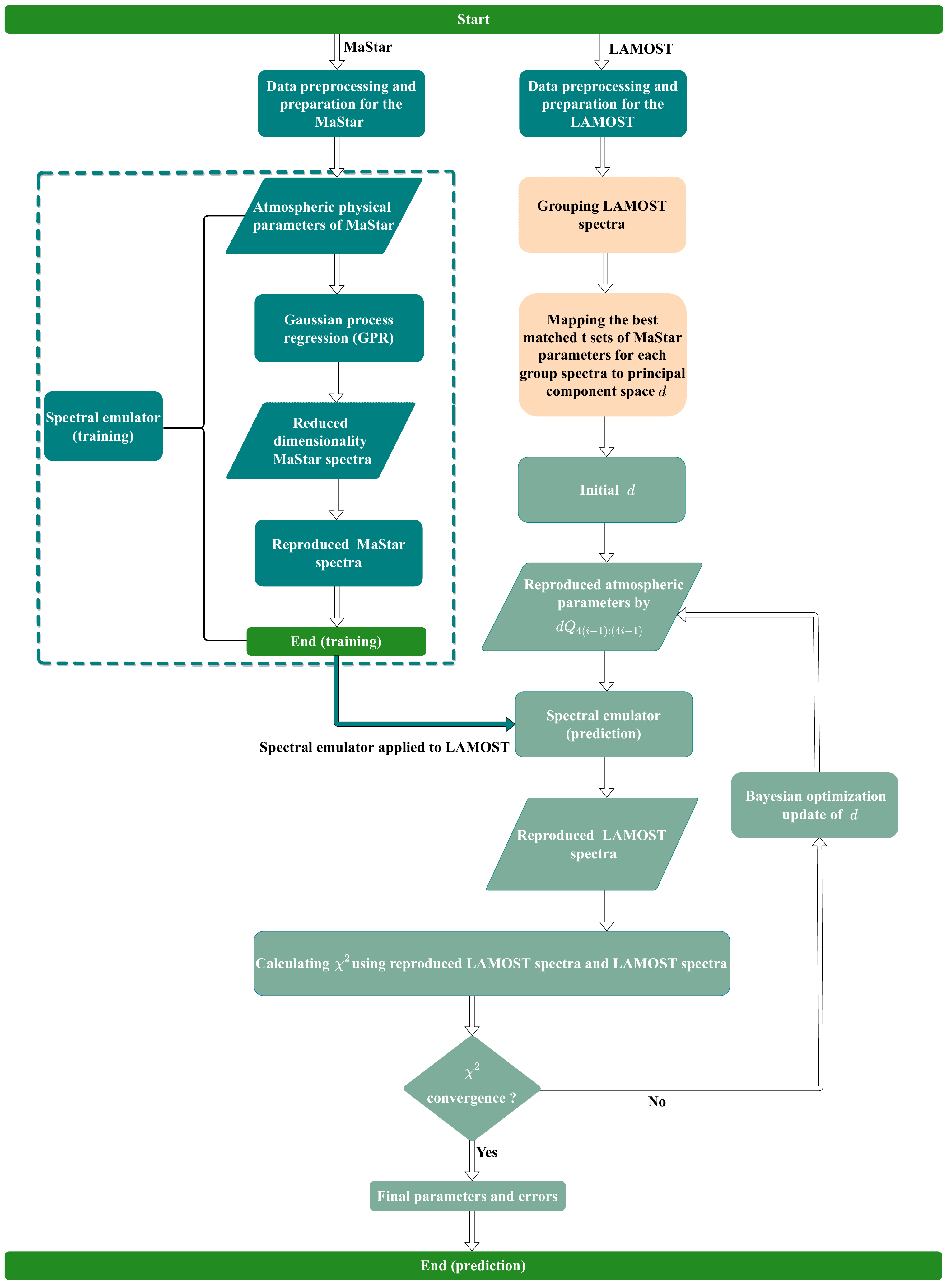}
		\centering
		\caption{Diagram of the workflow for predicting atmospheric parameters. The dashed boxes show the training process of the spectral emulator using the MaStar library. The yellow boxes indicate mapping the stellar atmospheric parameter space to principal component space using PCA within each group of LAMOST spectra. The process from \lq grouping LAMOST spectra\rq \ to \lq $\chi^{2}$ convergence\rq \ is termed the grouping optimization strategy. This strategy uses the trained spectral emulator and Bayesian optimization to find the $\chi^{2}$ minimum in the principal component space, obtaining the final atmospheric physical parameters.}
		\centering
		\label{workflow}
	\end{figure*}
	
	\subsection{workflow}  \label{subsection workflow}
	As shown in \reffig{workflow}, our workflow framework integrates three main parts: training the spectral emulator (dashed boxes), grouping LAMOST spectra and mapping the best matched t sets of MaStar parameters for each group spectra to principal component space (yellow boxes), and using the trained spectral emulator and Bayesian optimization to find the $\chi^{2}$ minimum in the principal component space to obtain the final atmospheric physical parameters (remaining boxes). Notably, the process from \lq grouping LAMOST spectra\rq \ to \lq $\chi^{2}$ convergence\rq \ is collectively referred to as the grouping optimization strategy. Before starting the workflow, we need to preprocess the spectra as mentioned in \refsubsection{Pre-processing for the spectra}, and then, we can proceed with the following steps in the workflow.
	\begin{enumerate}
		\item \textbf{Reducing dimensionality of MaStar spectra.} Before training the spectral emulator, we first performed data preparation on the MaStar library, focusing primarily on the dimensionality reduction of MaStar spectra. We applied PCA to the MaStar spectral library, resulting in the coefficient matrix $B_{nn}$ and the compressed spectra $f$. We discovered that setting PCA to a $95\%$ cumulative variance contribution rate is sufficient to restore the spectra, resulting in the retention of $237$ principal components. This is because the decision to discard $(1-95\%)=5\%$ of spectral information is due to the presence of noise in the spectra. As demonstrated in the work of \citet{bu2015stellar} and \citet{Xiang_2016}, discarding a portion of the information also aids in enhancing the predictive accuracy of atmospheric parameters.
		\item \textbf{Spectral emulator.} We have developed a spectral emulator using GPR that maps any given atmospheric parameters ($x_{*}$) to the best-estimated compressed spectra, denoted as $\bar{f}_{*}$. The final reproduced spectra ($F(x_{*})$) and uncertainties are as follows:
		\begin{equation}
			\begin{cases}
				F(x_{*}) = B_{nn}\bar{f}_{*} \\
				cov(F(x_{*})) = B_{nn}cov(\bar{f}_{*})B_{nn}^{T}.
			\end{cases}
			\label{spectral emulator 1}
		\end{equation}
		We set a combination of the radial basis function and constant kernel \citep{rasmussen2006gaussian} as the kernel function $\phi$ in the spectral emulator. This combination affords the model significant flexibility and adaptability, enabling it to effectively capture the nonlinear characteristics of the data and address noise, while also enhancing the model’s ability to regulate the overall level of data variation. The optimal hyperparameters of $\phi$, including a length scale of the kernel of $1.5$ and an amplitude of $2.1$, were determined through cross validation in MaStar.
		\item \textbf{Grouping LAMOST spectra.} 
		We adjusted the number of spectra used in the $\chi^{2}$ optimization to balance spectral-fitting speed with prediction accuracy for atmospheric parameters. Before grouping LAMOST spectra, we considered the priors of stellar spectral types and atmospheric physical parameters. Considering that the LAMOST spectral analysis pipeline (also called the $1$D Pipeline; \citealt{luo2015first}) has classified the spectra, we initially performed organizing based on spectral types to ensure that similar spectra were grouped together. Additionally, we initialized the LAMOST stellar atmospheric physical parameters using the correlation function interpolation (CFI) method \citep{du2012comparison, luo2015first}, and subsequently sorted these initial values in ascending order to further group spectra with similar parameters, serving as the second prior for spectral grouping. These two priors work together to ensure that each group consists of similar spectra with close atmospheric parameters, indicating that they are correlated within the groups. In \refsubsection{Setting t and N in the workflow}, we will determine that the recommended number of spectra in each group ($N_{\text{group}}$). Thus, for each group of spectra, our initial objective function is:
		\begin{equation}
			\chi^{2}=\frac{1}{N_{\text{group}}}\sum_{i=1}^{N_{\text{group}}}(L_{i}-F(x_{i}))^{2},
			\label{workflow1}
		\end{equation}
		where $L_{i}$ is the $i$th LAMOST spectrum, $x_{i}$ is the unsolved atmospheric parameter of $L_{i}$, $F(x_{i})$ (\refformula{spectral emulator 1}) is the reproduced spectrum using MaStar spectral emulator. For convenience in the following, we refer to the spectra of each group as a \lq concatenated spectrum\rq, denoted as $L=\{L_{1},L_{2},...,L_{N_{\text{group}}}\}$.
		\item \textbf{Mapping the best matched $t$ sets of MaStar parameters for each group spectra to principal component space.} We first obtain $t$ sets of parameters for each LAMOST \lq concatenated spectrum\rq \ from the MaStar parameter space.
		Specifically, for $T_{\text{eff}} < 10,000$ K, we set the parameter prior intervals as [$T_{\text{eff}}-1000$, $T_{\text{eff}}+1000$], [log $g-1$, log $g+1$], [[Fe/H]$-0.5$, [Fe/H]$+0.5$], and [[$\alpha$/Fe]$-0.2$, [$\alpha$/Fe]$+0.2$]; for $T_{\text{eff}} \ge 10000$ K, the temperature interval is adjusted to [$T_{\text{eff}}-5000$, $T_{\text{eff}}+5000$], while the other parameter intervals remain consistent with those for $T_{\text{eff}} < 10,000$ K. We then used the CFI method to match each LAMOST spectrum with the MaStar library, sorting the CFI values in descending order. 
		For the top $t$ sets of sorted MaStar parameters, if they fall within the above prior intervals, we consider these $t$ sets as the matched parameters for each LAMOST spectrum. If only $k < t$ sets fall within the intervals, we supplement with $t-k$ sets of MaStar parameters closest to the first $T_{\text{eff}}$ (i.e., the initial $T_{\text{eff}}$) of the sorted MaStar parameters (starting from the $(t+1)$th parameters), and consider these $t$ sets as the matched parameters. The above set intervals can avoid obviously different parameters among the $t$ samples (such as matching M-type stars to OBA-type star parameters). By repeating the above process, we finally obtained a parameter matrix $\Theta$ with $t$ rows and $4N_{\text{group}}$ columns ($4$ represents the four atmospheric parameters) for each LAMOST \lq concatenated spectrum\rq, and then applied the PCA method (retaining $98\%$ of the cumulative variance contribution) to transform $\Theta$, resulting in the new variable $D$ ($t$ rows and $m$ columns) and the coefficient matrix $Q$ ($m$ rows and $4N_{\text{group}}$ columns). Therefore, the objective function in the stellar atmospheric parameter space, denoted by \refformula{workflow1}, can be transformed into the following form in the principal component space for optimization:
		\begin{equation}
			\chi^{2}=\frac{1}{N_{\text{group}}}\sum_{i=1}^{N_{\text{group}}}(L_{i}-F(dQ_{4(i-1):(4i-1)}))^{2},
			\label{workflow3}
		\end{equation}
		where $d=(d_{1},d_{2},...,d_{m})$ is a new matrix of variables of $D$ with one row and $m$ columns, $Q_{4(i-1):(4i-1)}$ is a coefficient matrix with $m$ rows and four columns composed of the $(4(i-1))$th to $(4i-1)$th columns of $Q$. To ensure that \refformula{workflow3} is optimized within reasonable intervals, we set the optimization boundaries for $d_{i}$ based on the minimum and maximum values of the $i$th column of $D$. The stable optimization boundaries are also a key consideration in determining $t$, as discussed in \refsubsection{Setting t and N in the workflow}.
		\item \textbf{Reproducing LAMOST spectra and calculating $\chi^{2}$.} Once the spectral emulator (\refformula{spectral emulator 1}) was established, it enables the generation of spectra for any given atmospheric parameters. Thus, for any $d$ in step $4$, the spectra corresponding to any $T_{\text{eff}}$, log $g$, [Fe/H] and [$\alpha$/Fe] can be reproduced according to $F(dQ_{4(i-1):(4i-1)})$. This allows for the quantification of the discrepancy between the generated and LAMOST spectra using \refformula{workflow3}.
		\item \textbf{Bayesian optimization.} We have established the boundary values for $d$ in steps $1$ through $5$, corresponding to the boundary values of the $\chi^{2}$ function. Next, we employed the Python package of bayesian-optimization to solve for the minimum point of \refformula{workflow3}. Bayesian optimization estimates the $\chi^{2}$ value by constructing a surrogate model using GPR and balances exploration and exploitation using the upper confidence bound acquisition function. \footnote{\url{https://github.com/fmfn/BayesianOptimization}} This method allows for the integration of prior knowledge and iteratively refines the search space and predictions by updating posterior probabilities, typically outperforming traditional methods in computational efficiency and in finding global optima. Additionally, Bayesian optimization is particularly well suited for problems where sample costs are high, objective function evaluations are expensive, the function form is unknown or noisy, and gradient information is difficult to obtain or nonexistent \citep{Dewancker2015, wang2016bayesian, santoni2024comparison}. Considering the complex functional form of the \refformula{workflow3} and the fact that the boundary values of $d$ can serve as priors to enhance optimization efficiency, using the Bayesian optimization is more reasonable. In the optimization process, we randomly selected $10$ points within the optimization boundaries of $d$ to build the initial surrogate model and used the maximum CFI value point as the first exploitation to refine the surrogate model in the known optimal boundaries of $d$, further optimizing \refformula{workflow3}.
		\item \textbf{$\chi^{2}$ convergence.} We have set convergence criteria for \refformula{workflow3}: if the value of $\chi^{2}$ remains the same for five consecutive iterations, we halt the iteration and output $d$. The final predicted atmospheric parameters are given by $dQ_{4(i-1):(4i-1)}$.
		\item \textbf{Final parameters and errors.} Upon convergence of the Bayesian optimization, the optimal atmospheric parameters for LAMOST spectra are determined. To assess the statistical reliability of these parameters, we employed a statistical model to estimate uncertainties for each parameter, as detailed in \refsubsection{uncertainties}. Finally, we derived the atmospheric parameters along with their corresponding uncertainties (internal errors).
	\end{enumerate}
	
	\section{Data introduction} \label{Data introduction}
	The stellar spectra utilized in this study comprises of two parts: The MaStar \citep{Yan_2019,abdurro2022seventeenth} and LAMOST stellar spectra from the $10$th data release (DR$10$) of low-resolution spectroscopic survey (LRS). The MaStar was used to train the spectral emulator, which is subsequently used for predicting stellar atmospheric parameters of LAMOST DR$10$ LRS.
	\subsection{MaStar} \label{MaStar}
	The MaStar is a project in the SDSS-IV \citep{blanton2017sloan} to build a large, well-calibrated, high-quality empirical library covering the wavelength range $3622-10,354$ \text{\AA} at a resolving power of $R\sim1800$ \citep{Yan_2019}. The survey was designed to collect empirical stellar spectra that can be used to model galaxy observations from MaNGA survey.
	
	The MaStar project, utilizing the $2.5$ m Sloan Foundation Telescope at Apache Point Observatory \citep{Gunn_2006}, has amassed a stellar library of $59,266$ high-quality spectra for $24,290$ stars \citep{abdurro2022seventeenth}. These observations were made possible through coordination with the APOGEE-2N survey, employing MaNGA fiber bundles to capture optical spectra. The fiber bundle method enhances flux calibration accuracy over traditional single-fiber techniques.
	
	The initial version of the MaStar derived its atmospheric parameters from various surveys such as APOGEE, LAMOST, and SEGUE, leading to a diversity in sources and potential systematic differences. Additionally, some stars, selected based on photometry, lacked existing atmospheric parameters. To address this, the MaStar team decided to uniformly derive atmospheric parameters for the entire sample using their spectra, which are characterized by high signal-to-noise ratios (S/N) and excellent flux calibration. Several parallel initiatives were undertaken to determine these parameters from MaStar spectra, as detailed in works by \cite{Hill_2021,Hill_2022}, \cite{Imig_2022}, \cite{lazarz2022sdss}, and Chen et al. (2024, in preparation). In Data Release $17$ (DR$17$) of SDSS, the master set of parameters was derived by taking the median of the parameters of the four methods when they are available and are considered valid. Yan et al. (2024, in preparation) will provide an exhaustive description of this catalog.
	
	Based on the SDSS data release website and introductions to the MaStar by \cite{abdurro2022seventeenth}, we selected spectra with a S/N $>$ $20$, and masked bad pixels. This yielded a library of $18,654$ spectra, covering a wide range in atmospheric parameters space, roughly from $3500$ to $30,000$ K in $T_{\text{eff}}$, from $0$ to $5$ dex in log $g$, from $-2.5$ to $0.5$ dex in [Fe/H], and from $-0.4$ to $0.4$ dex in [$\alpha$/Fe].
	\subsection{The LRS Spectra of LAMOST DR10} \label{Observed Spectra from LAMOST DR10}
	LAMOST, alternatively referred to as the Guoshoujing Telescope, with an effective aperture of $4$ m and size of $6.67$ m for the main mirror, is a Schmidt telescope located at the Xinglong Observatory northeast of Beijing, China \citep{luo2015first}. For LAMOST LRS, the LAMOST can simultaneously collect $4,000$ fiber spectra with $3\arcsec3$ diameter in a wild field ($5^{\circ}$) with a resolving power of $R\sim1800$, and the spectra are separated into a blue channel covering $3700-5900$ \text{\AA} and a red channel covering $5700-9000$ \text{\AA}, with an overlapping region between the two channels at $5700-5900$ \text{\AA} \citep{cui2012large,zhao2012lamost,jin2023large}.
	
	After the commissioning from $2009$ to $2010$, LAMOST began a pilot survey in $2011$ \citep{luo2012data}. Subsequently, it launched the regular LRS from $2012$ September, which has two major parts \citep{zhao2012lamost, yan2022overview}: the LAMOST Extra Galactic Survey, focused on the sciences of galaxies and cosmology; and the LAMOST Experiment for Galactic Understanding and Exploration survey, aimed at investigating the stars and the Milky Way. 
	
	LAMOST DR$10$ includes a total of $11,817,430$ low-resolution spectra, comprising $6,006,056$ unique stars and $2,202,342$ stars with repeat observations. Of these spectra, $11,473,644$ are classified as stars, $263,444$ are classified as galaxies, and $80,342$ are classified as quasars. It also provides AFGKM-type stellar catalogs, including $7,478,650$ AFGK-type and $876,134$ M-type stellar spectra. In this work, we employed the spectral emulator to predict atmospheric parameters for $11,473,644$ stellar spectra.
	
	\subsection{Preprocessing for the spectra} \label{Pre-processing for the spectra}
	In contrast to the ELODIE and MILES libraries, MaStar covers a broader wavelength range. For M-type stars, it includes TiO bands ($7050$ and $8430$ \text{\AA}), K- and Na-line pairs around $7680$ and $8190$ \text{\AA}. For FGK-type stars, it covers Ca~\textsc{ii}~Triplet ($8400-8700$ \text{\AA}) spectral lines. In OBA-type stars, it includes the Paschen series ($8200$ \text{\AA}). Using these absorption lines is essential for more accurate prediction of parameters in both cool and hot stars, as demonstrated in \cite{Du_2021} and \cite{xiang2022stellar}. Considering covering these spectral lines as comprehensively as possible, we used a wavelength range of $3800-8700$ \text{\AA}, and implemented a five-step pre-processing as follows.
	
	\begin{figure*}[htb]
		\centering
		\subfigure{
			\begin{minipage}[t]{0.45\linewidth}
				\centering
				\includegraphics[width=3.2in, height=2.5in]{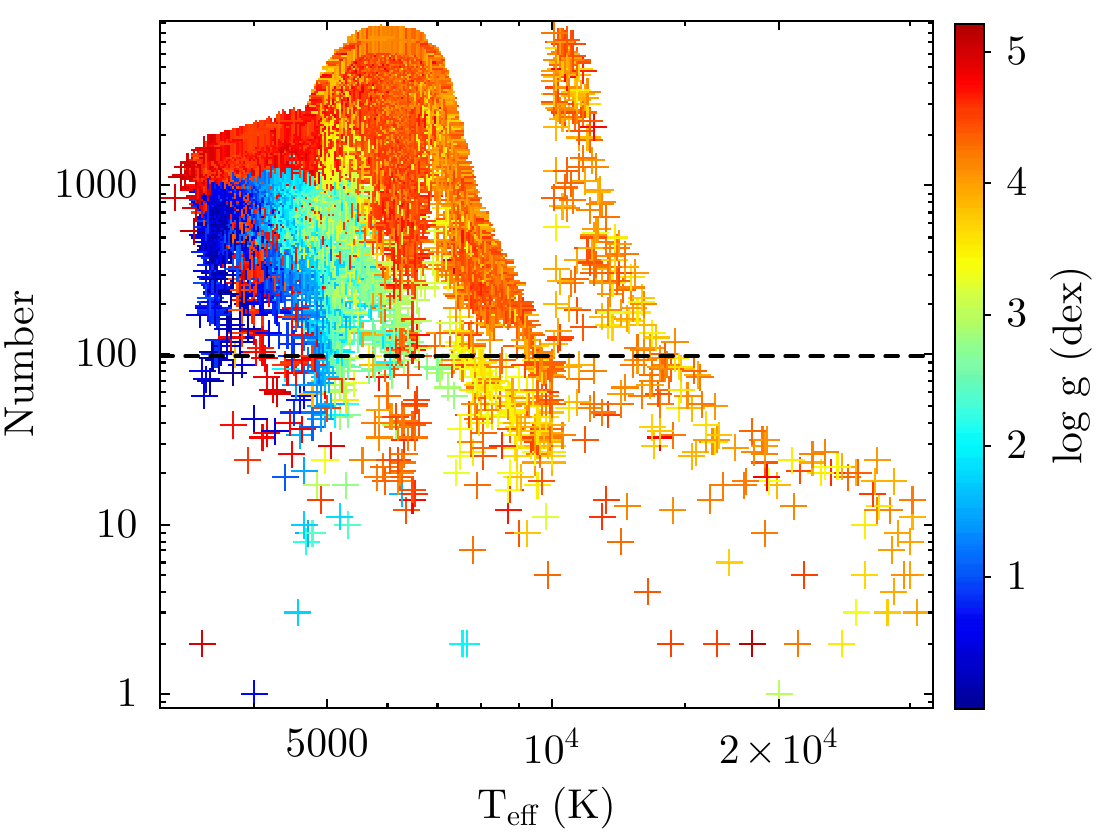}
				\centering
			\end{minipage}
		}
		\centering
		\subfigure{
			\begin{minipage}[t]{0.45\linewidth}
				\centering
				\includegraphics[width=3.2in, height=2.5in]{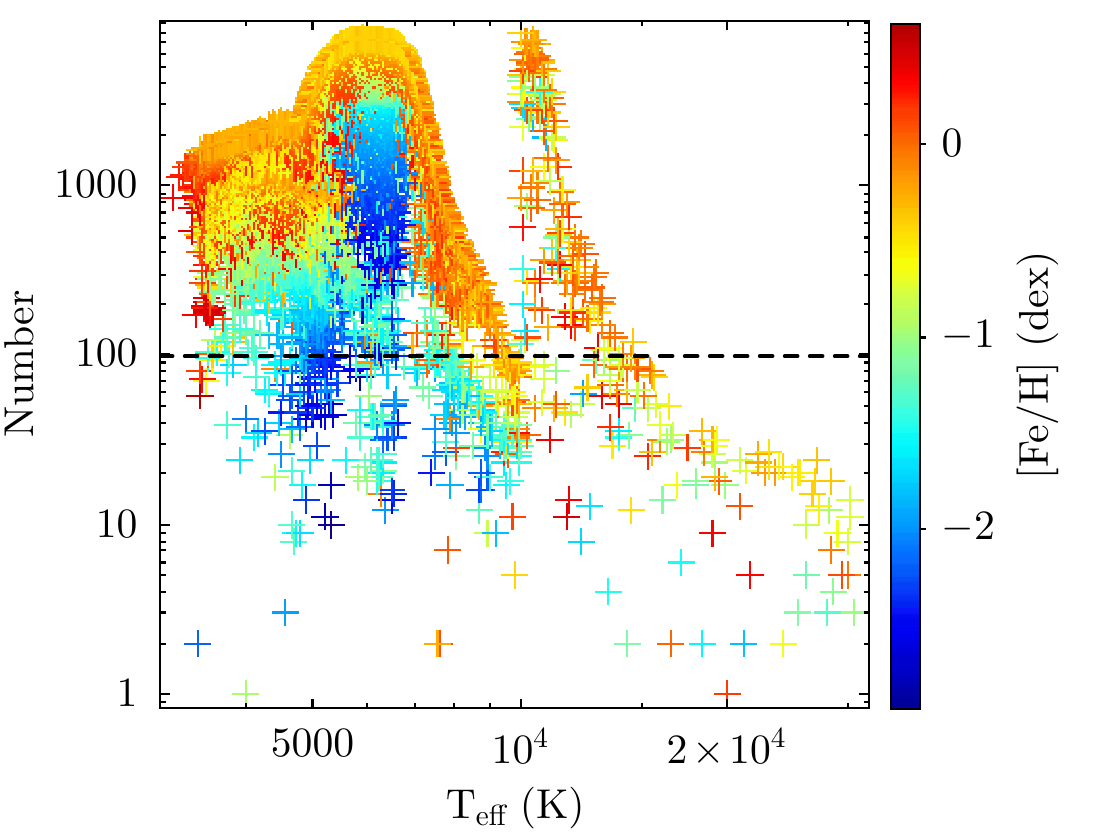}
				\centering
			\end{minipage}
		}
		\caption{MaStar sample numbers as a function of $T_{\text{eff}}$  within prior parameter intervals for parameters of each MaStar spectrum. For $T_{\text{eff}} < 10,000$ K, the prior intervals are set to [$T_{\text{eff}}-1000$, $T_{\text{eff}}+1000$], [log $g-1$, log $g+1$], [[Fe/H]$-0.5$, [Fe/H]$+0.5$], [[$\alpha$/Fe]$-0.2$, [$\alpha$/Fe]$+0.2$]. For $T_{\text{eff}} \ge 10,000$ K, the intervals are [$T_{\text{eff}}-5000$, $T_{\text{eff}}+5000$], with other settings consistent with $T_{\text{eff}} < 10,000$ K.}
		\centering
		\label{count_t}
	\end{figure*}
	
	\begin{enumerate}
		\item \textbf{Masking bad pixels and contaminated absorption lines.} We masked four regions contaminated by the telluric absorption ($6850-6960$, $7210-7350$, $7560-7720$, and $8105-8240$ \text{\AA}), the dichroic region ($5700-6050$ \text{\AA}), and the diffuse interstellar band directory \citep{hobbs2008catalog}. This is crucial as contamination in these regions could result in unreliable spectral flux, and inaccurate parameter predictions.
		\item \textbf{Shifting to rest frame.} To ensure a proper one-to-one correspondence between LAMOST spectra and MaStar spectra dimensions, we need to shift the wavelengths of LAMOST spectra to the rest frame using the \refformula{wl},
		\begin{equation}
			\lambda_{rest}=\frac{\lambda_{obs}}{1+z},
			\label{wl}
		\end{equation}
		where $\lambda_{obs}$ and $z$ respectively represent the observed wavelength and redshift, both provided in the LAMOST fits files \citep{luo2015first}. $\lambda_{rest}$ represents the rest-frame wavelength.
		\item \textbf{Interpolating the spectra.} To ensure consistent wavelength coverage and uniform wavelength intervals for each spectrum, spline interpolation techniques described by \cite{Dierckx1993} were employed within the $[3800, 8700]$ \text{\AA}. Each spectrum from MaStar and LAMOST was resampled within this range with a step size of $1$ \text{\AA}, resulting in spectra comprising $4901$ dimensions.
		\item \textbf{Normalizing the spectra.} To mitigate the impact of imperfect flux calibration and interstellar reddening on MaStar and LAMOST spectra, we applied the normalization technique proposed by \cite{Imig_2022}. This method involves computing the median value within a $400$ pixel window and using it to map the spectral flux onto a consistent range.
		\item \textbf{Rescaling anomalous data in the normalized spectra.} We noted that near-zero median values could inflate the normalized flux, and we assigned a value of $1$ to any normalized flux that exceeds $2$.
	\end{enumerate}
	
	\begin{figure*}[!htb]
		\centering
		\subfigure{
			\begin{minipage}[t]{0.45\linewidth}
				\centering
				\includegraphics[width=3.2in, height=2.4in]{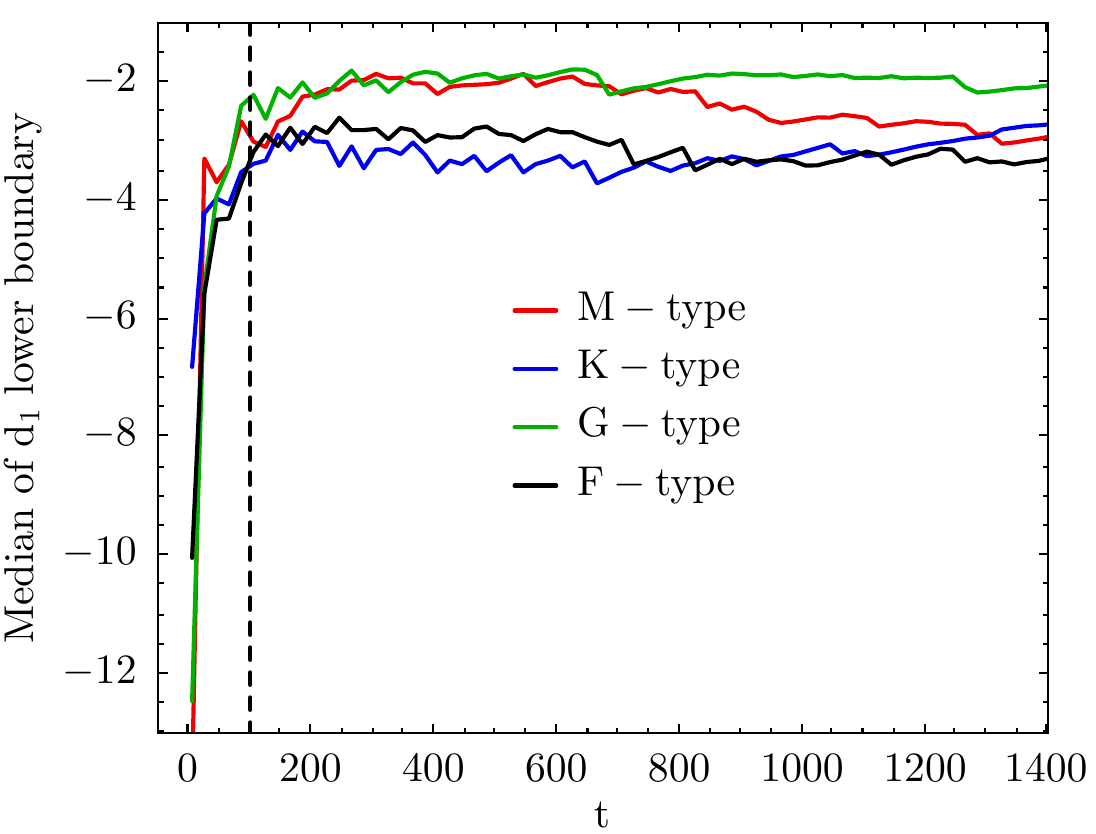}
				\centering
			\end{minipage}
		}
		\centering
		\subfigure{
			\begin{minipage}[t]{0.45\linewidth}
				\centering
				\includegraphics[width=3.2in, height=2.4in]{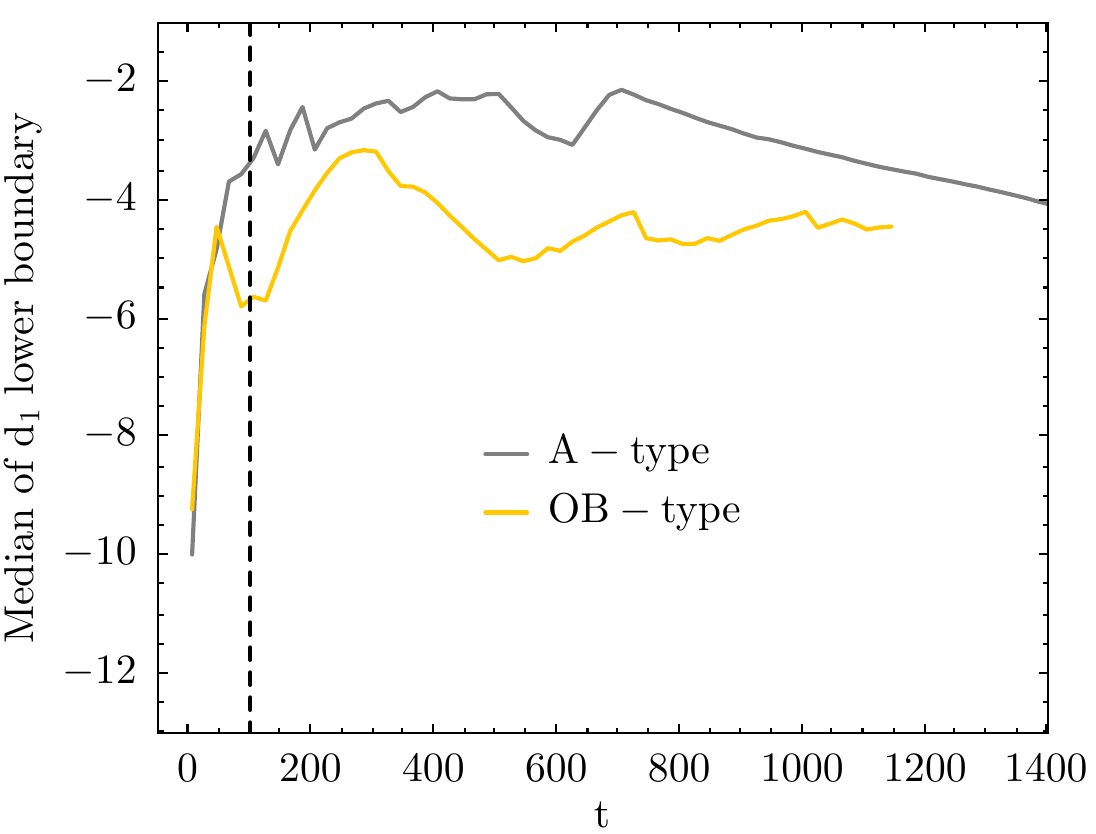}
				\centering
			\end{minipage}
		}
		\centering
		\subfigure{
			\begin{minipage}[t]{0.45\linewidth}
				\centering
				\includegraphics[width=3.2in, height=2.4in]{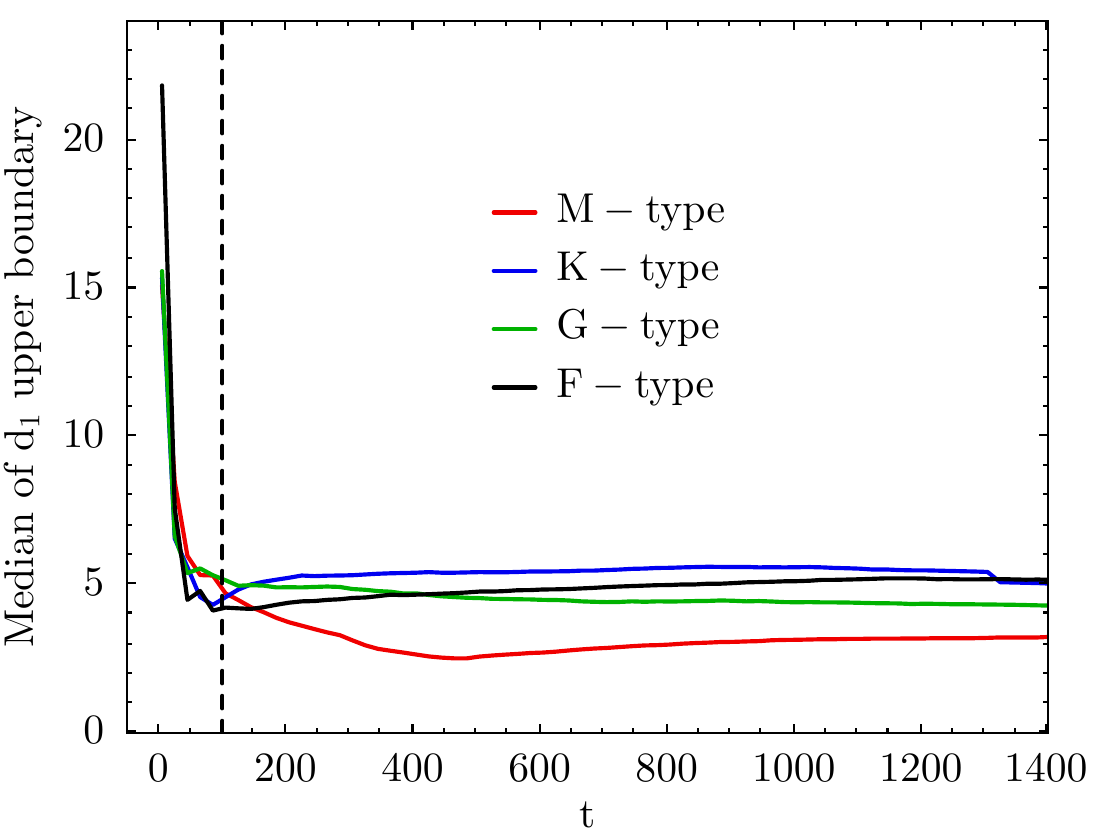}
				\centering
			\end{minipage}
		}
		\centering
		\subfigure{
			\begin{minipage}[t]{0.45\linewidth}
				\centering
				\includegraphics[width=3.2in, height=2.4in]{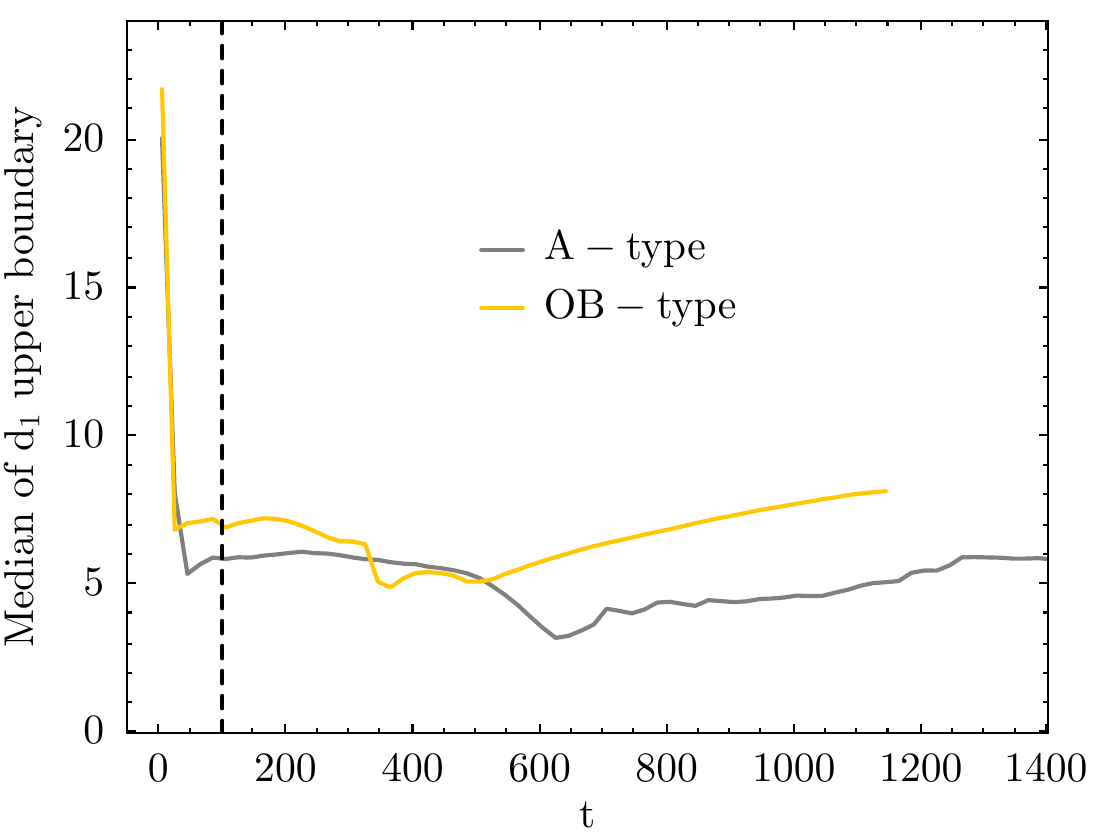}
				\centering
			\end{minipage}
		}
		\caption{The median of the lower and upper boundaries of $d_{1}$ across different $N_{\text{group}}$. The first and second rows represent the lower and upper boundaries of $d_{1}$, respectively, and the different curves represent different spectral types.}
		\centering
		\label{PC1_low_up}
	\end{figure*}
	
	\begin{figure*}[!htb]
		\includegraphics[width=6.1in,height=4.3in]{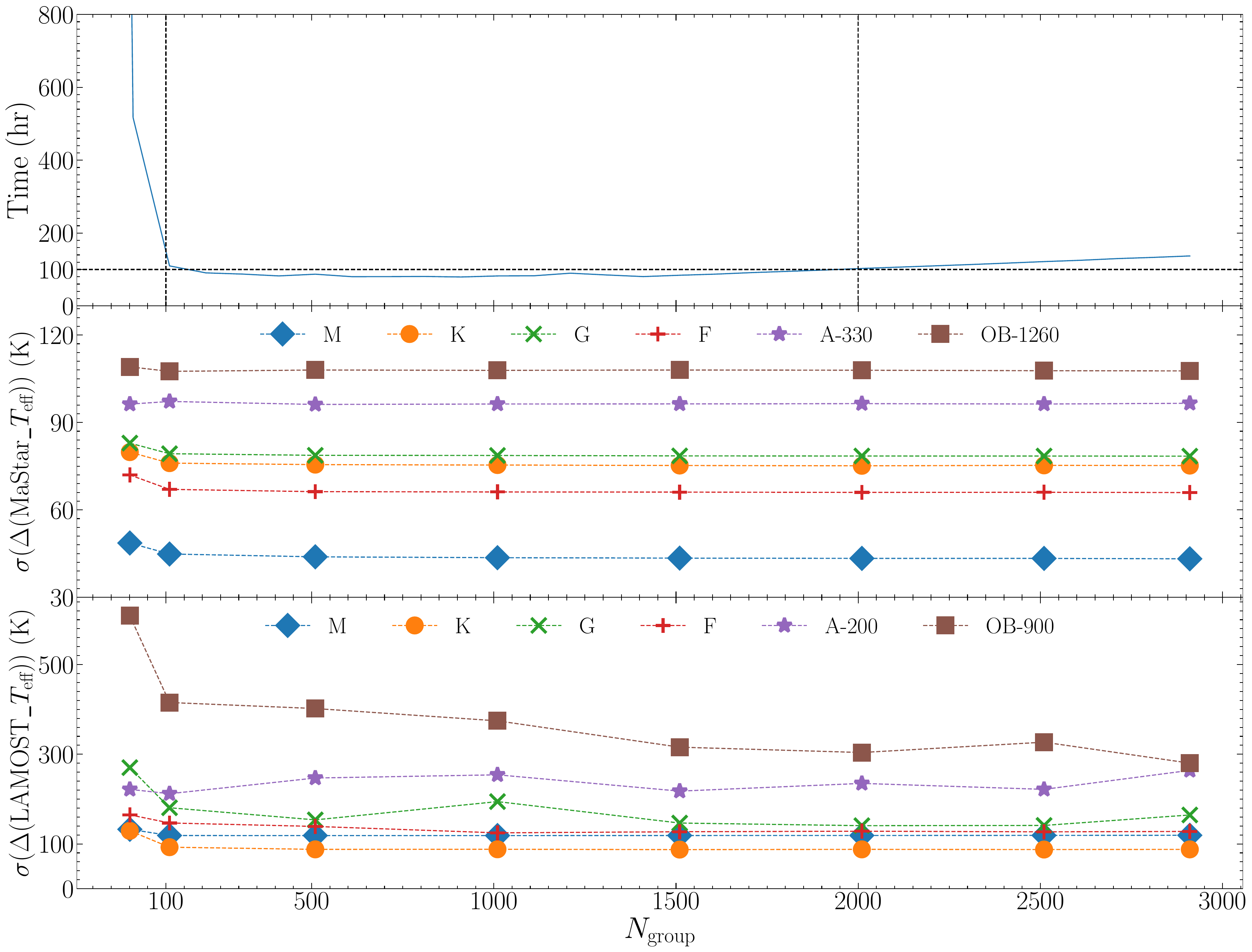}
		\centering
		\caption{Efficiency and accuracy of the grouping optimization strategy. The upper panel shows the time taken to predict atmospheric parameters in LAMOST DR$10$ for different $N_{\text{group}}$. The middle panel displays the dispersion of $T_{\text{eff}}$ errors in the grouping optimization strategy across different spectral types for MaStar, where $\Delta$(MaStar$\texttt{\_}T_{\text{eff}}$) represents the difference between predicted and true $T_{\text{eff}}$. The bottom panel illustrates the dispersion of $T_{\text{eff}}$ errors in the grouping optimization strategy across different spectral types for LAMOST (randomly selected $10,000$ spectra from each spectral type), where $\Delta$(LAMOST$\texttt{\_}T_{\text{eff}}$)  represents the difference between grouped and ungrouped predictions.}
		\centering
		\label{Batch size}
	\end{figure*}
	
	After the above five steps, we can input the MaStar and LAMOST spectra into the workflow outlined in \refsection{subsection workflow}, and predict the stellar atmospheric parameters for low-resolution stellar spectra of LAMOST DR$10$.
	
	\section{Experiment} \label{Data Experiment}
	In this section, we first set up the  parameters of $t$ and $N_{\text{group}}$ in the workflow, and then evaluated the performance of the spectral emulator, the rationality of the prior spectral grouping, and the appropriateness of the $t$ settings. The workflow was subsequently applied to predict atmospheric parameters for LAMOST DR$10$. Based on these predictions, we defined internal parameter errors and established a recommended parameter catalog. Comparing the recommended catalog with external data, we identified and addressed deficiencies in LASPM and LASP for both cool and hot stars, as well as highlighted certain issues with the MaStar. Additionally, we analyzed the challenges encountered with spectral-fitting methods, the MaStar, and the preprocessing for the spectra.
	
	The experiments were carried out in a PyCharm 2019.3.3 environment running on a personal computer with an Intel core i7 2.50 GHz CPU and 16 GB of memory. We connected PyCharm 2019.3.3 to a remote server equipped with an Intel Xeon E5-2620 v4 2.10GHz processor and 64 GB of RAM.
	
	\subsection{Setting $t$ and $N_{\text{group}}$} \label{Setting t and N in the workflow}
	In addition to the hyperparameters of the spectral emulator mentioned in \refsubsection{subsection workflow}, predicting LAMOST spectral parameters also requires specifying $t$ and $N_{\text{group}}$.
	\begin{enumerate}
		\item \textbf{The setting of $t$.} First, we roughly determined the range of $t$. Specifically, for the MaStar library, we set prior intervals (in step $4$ of \refsubsection{subsection workflow}) for the atmospheric parameters of each MaStar spectrum and then counted the number of MaStar samples within each interval. As shown in \reffig{count_t}, $97.64\%$ of the prior parameter intervals include more than $100$ sets of MaStar parameters, but for stars with $T_{\text{eff}} > 10,000$ K, approximately half of the prior intervals contain fewer than $100$ MaStar samples. For high-temperature stars, setting $t$ too high introduces a larger proportion of lower-temperature stars into the $t$ sets of MaStar parameters, resulting in an underestimation of their $T_{\text{eff}}$. Therefore, considering the number of low- and high-temperature stars within the parameter intervals, setting $t$ around $100$ is appropriate. On the other hand, we considered the stable optimization boundaries of $d$ to determine $t$. Specifically, we set $N_{\text{group}}$ to [$5$, $2000$] with a step size of $100$, and the interval of $t$ to [$5$, $1500$] with a step size of $20$. We then calculated the median of the lower and upper boundaries of $d_{1}$ across different $N_{\text{group}}$. As shown in \reffig{PC1_low_up}, we noted significant fluctuations in $d_{1}$ when $t < 100$, increasing the risk of overestimating its lower boundary and underestimating the upper boundary (increasing the likelihood of getting trapped in local minima), whereas $d_{1}$ tends to relatively stabilize when $t > 100$. This means that within the prior parameter intervals, as $t$ increases up to $100$, the atmospheric parameters of LAMOST \lq concatenated spectrum\rq \ are effectively captured by these $t$ sets of MaStar parameters, and further increasing $t$ does not significantly impact this representation. The boundaries of the other $d_{i}$ exhibit similar behavior to $d_{1}$ as $t$ changes. Considering all the above factors, we advise setting $t=100$ to maintain stable principal component boundaries and ensure the prior parameter intervals cover reasonable parameters.
		\item \textbf{The setting of $N_{\text{group}}$.} We used the MaStar library to determine the appropriate value of $N_{\text{group}}$, which need to consider the balance between efficiency and accuracy in optimizing the objective funtion. We adjusted the $N_{\text{group}}$ to repeatly implement the workflow, ranging from $1$ to $3000$ in increments of $10$ ($N_{\text{group}}=1$ represents the scenario without grouping optimization). As shown in the upper and middle subplots of \reffig{Batch size}, we focused on $T_{\text{eff}}$ to illustrate the relationships between the total time \footnote{The total time is determined by multiplying the grouping optimization time in the MaStar library by the ratio of the total number of LAMOST spectra to the total number of MaStar spectra.} and the dispersion of $T_{\text{eff}}$ errors with the $N_{\text{group}}$, respectively, 
		and the figure reveals that the $N_{\text{group}}$ within the $100$-$2000$ are most effective, limiting the total spectral-fitting time to under $100$ hr while also ensuring stable dispersion of $T_{\text{eff}}$ errors (the same is true for other atmospheric parameters). We also found our spectral grouping method does not significantly increase parameter errors and that there is an increase in the total spectral-fitting time for $N_{\text{group}}$ exceeding $2000$. These results validate the feasibility of the grouping optimization strategy, suggesting that the $N_{\text{group}}$ within the range of $100$-$2000$ yields the best performance. In our experiments, the $N_{\text{group}}$ was set to $1000$, and the total time was approximately $70$ hr. Compared to the scenario without grouping optimization, the efficiency was improved by approximately $10$ times (\reffig{Batch size}). It is noteworthy that, under single-machine conditions, our method without grouping optimization achieved an efficiency comparable to LASP. 
		
	\end{enumerate}
	
	\subsection{Performance Assessment} \label{Performance Assessment}
	\subsubsection{Evaluating the performance of the spectral emulator} \label{Evaluating the performance of the spectral emulator}
	
	\begin{figure*}[!htb]
		\centering
		\subfigure{
			\begin{minipage}[t]{0.45\linewidth}
				\centering
				\includegraphics[width=3in, height=1.5in]{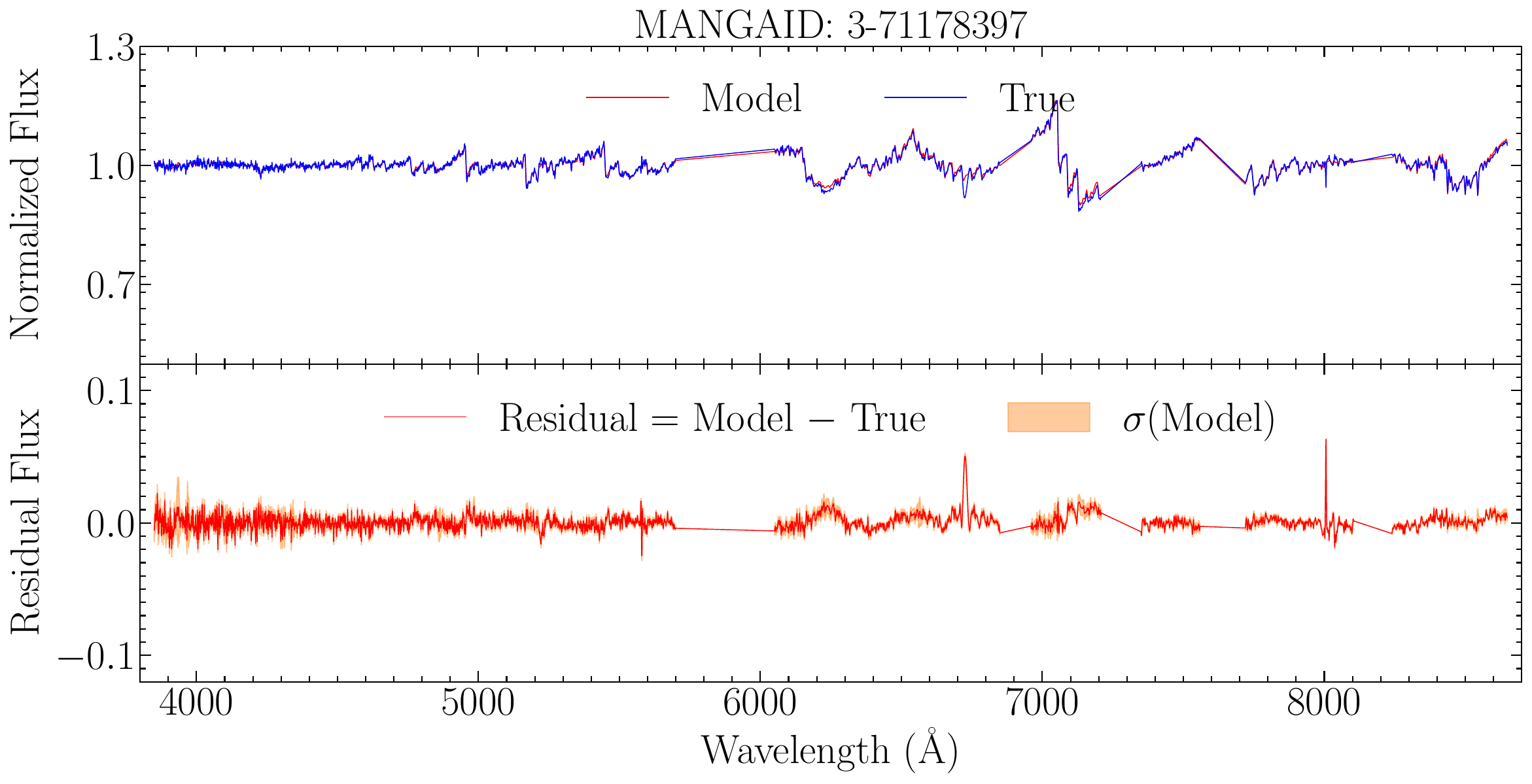}
				\centering
			\end{minipage}
		}
		\centering
		\subfigure{
			\begin{minipage}[t]{0.45\linewidth}
				\centering
				\includegraphics[width=3in, height=1.5in]{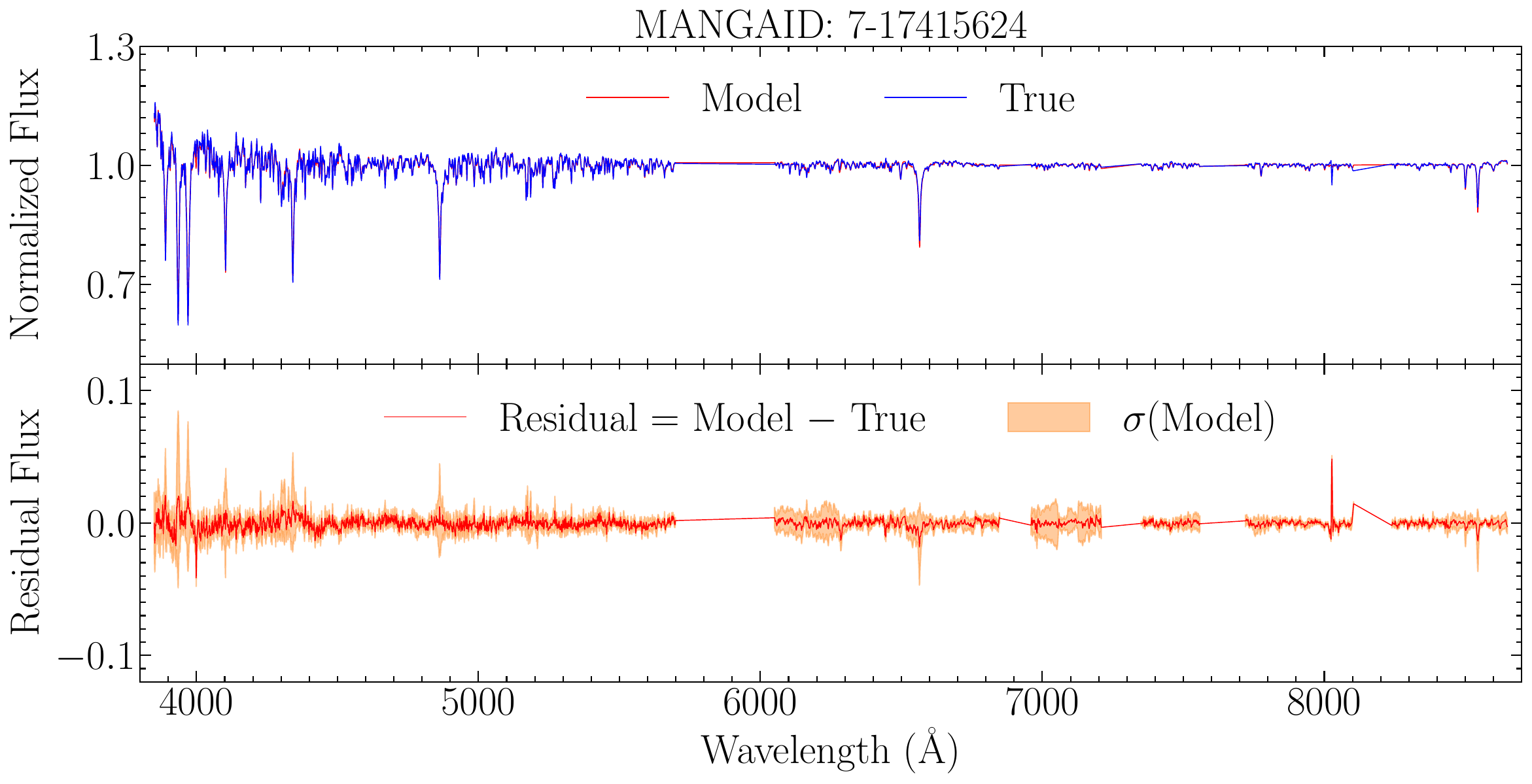}
				\centering
			\end{minipage}
		}
		\centering
		\subfigure{
			\begin{minipage}[t]{0.45\linewidth}
				\centering
				\includegraphics[width=3in, height=1.5in]{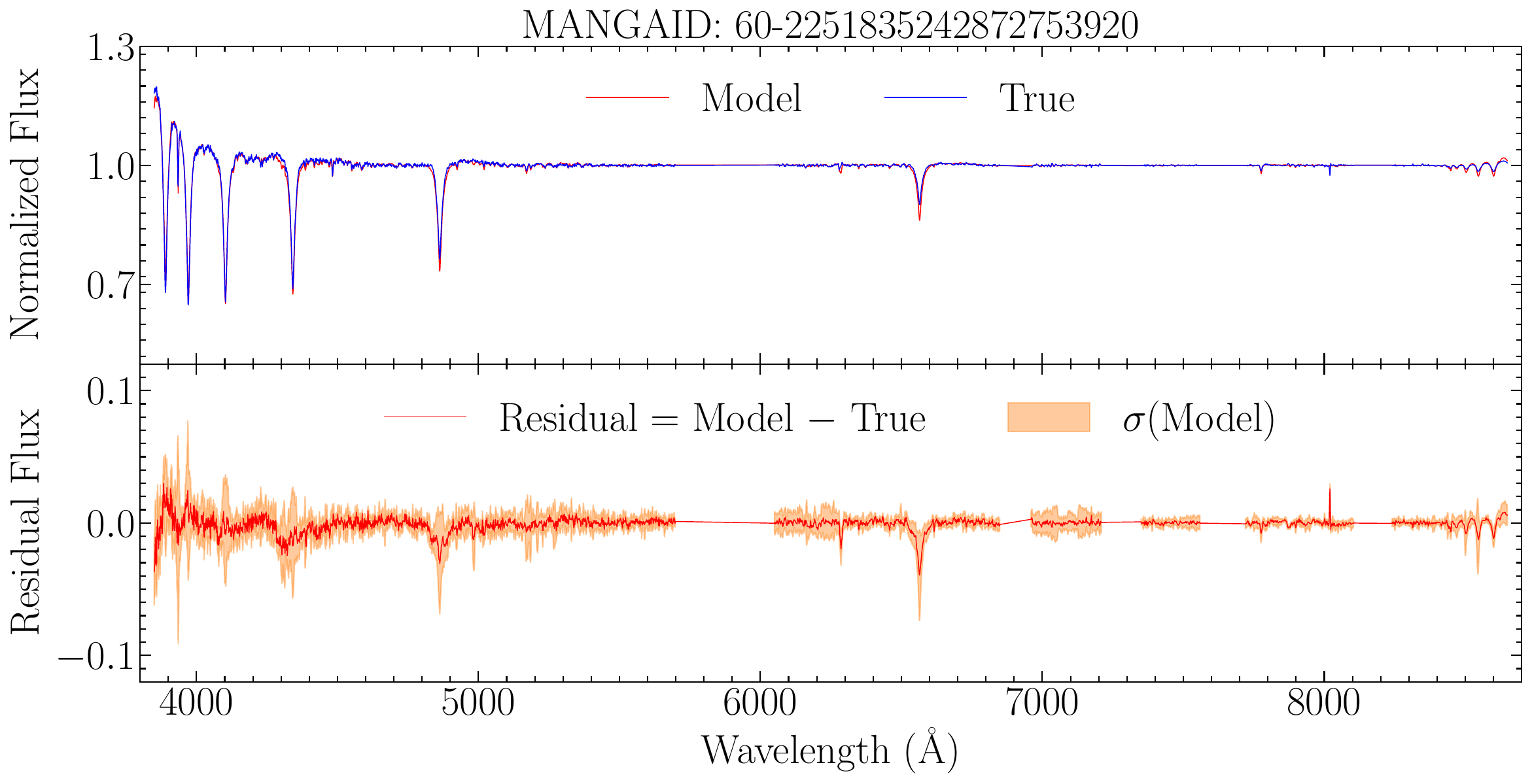}
				\centering
			\end{minipage}
		}
		\centering
		\subfigure{
			\begin{minipage}[t]{0.45\linewidth}
				\centering
				\includegraphics[width=3in, height=1.5in]{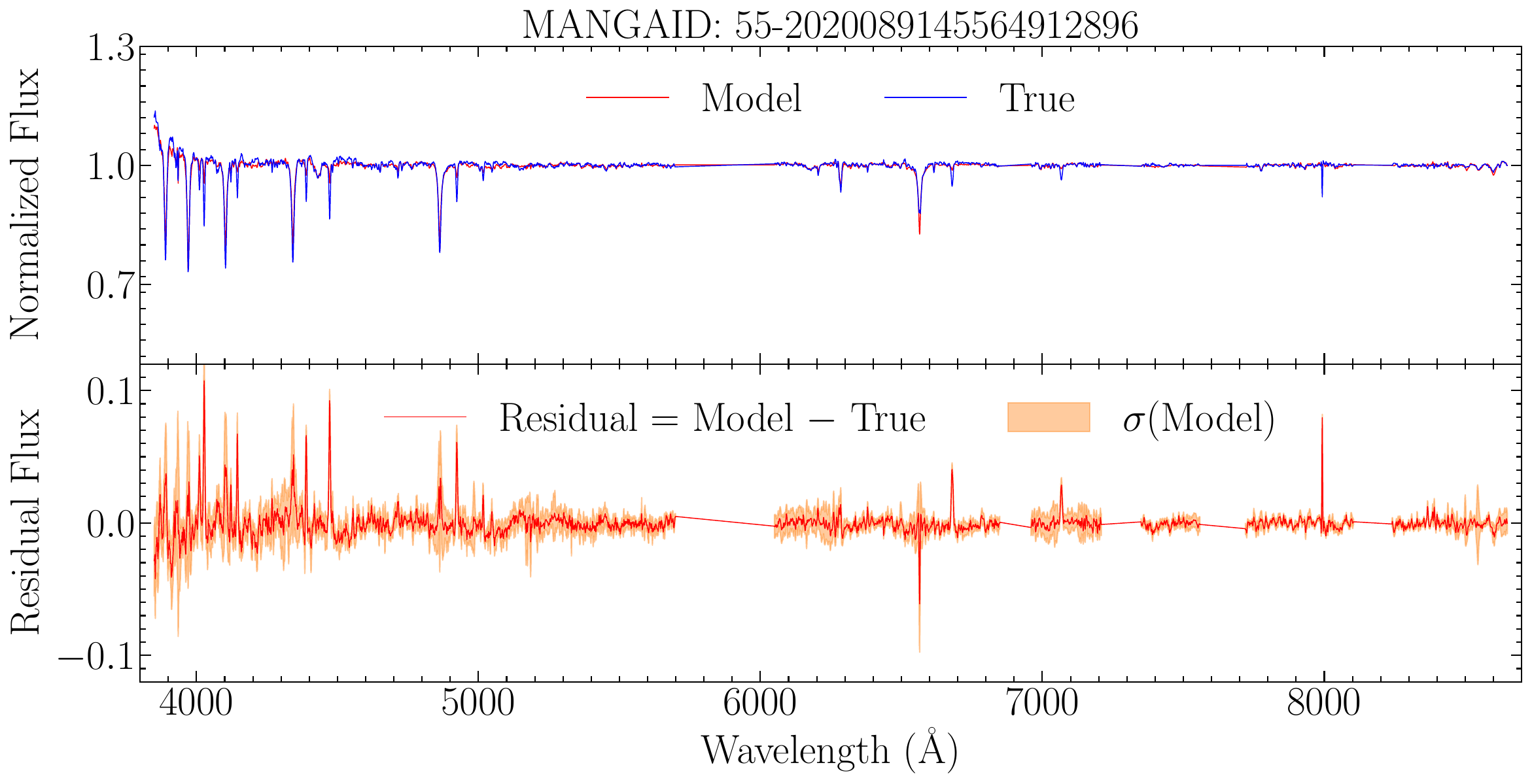}
				\centering
			\end{minipage}
		}
		\caption{Examples of spectra generated by the spectral emulator. The yellow shading represents the uncertainty generated by the spectral emulator (defined in \refformula{spectral emulator 1}).}
		\centering
		\label{GPR 1234}
	\end{figure*}
	
	\begin{figure*}[!htb]
		\centering
		\subfigure{
			\begin{minipage}[t]{0.45\linewidth}
				\centering
				\includegraphics[width=3.2in, height=2.18in]{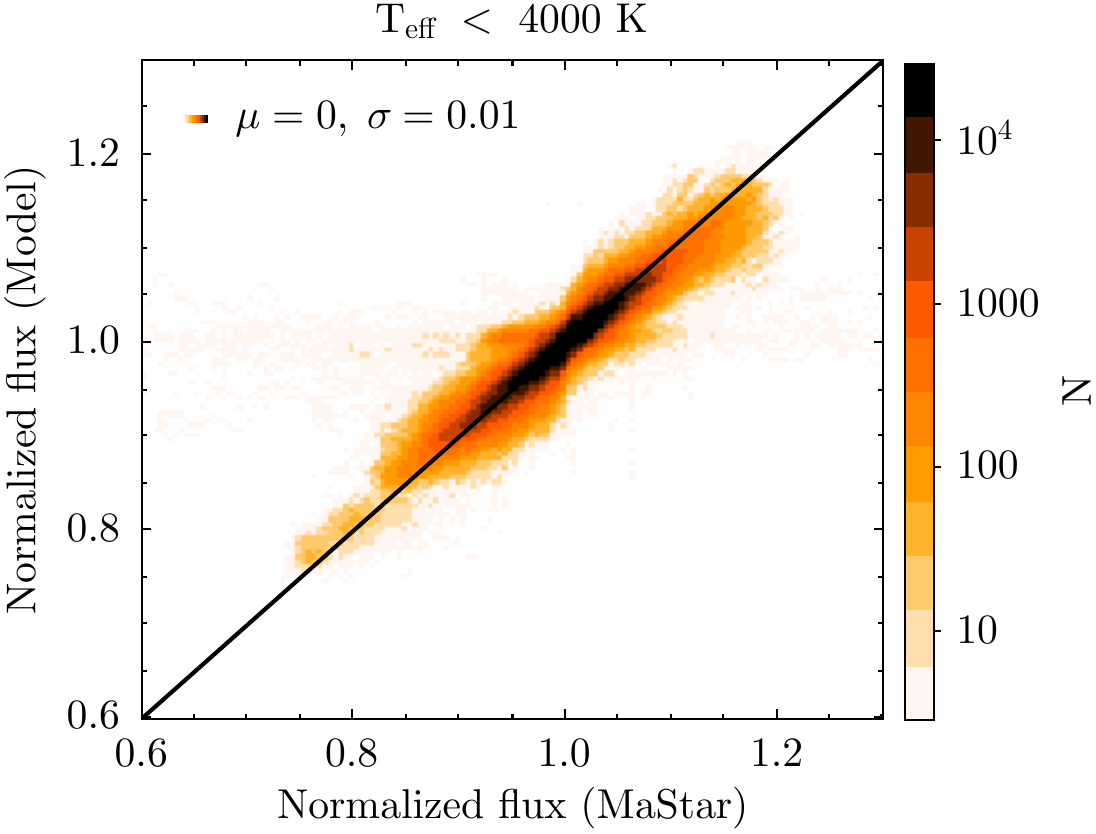}
				\centering
			\end{minipage}
		}
		\centering
		\subfigure{
			\begin{minipage}[t]{0.45\linewidth}
				\centering
				\includegraphics[width=3.2in, height=2.18in]{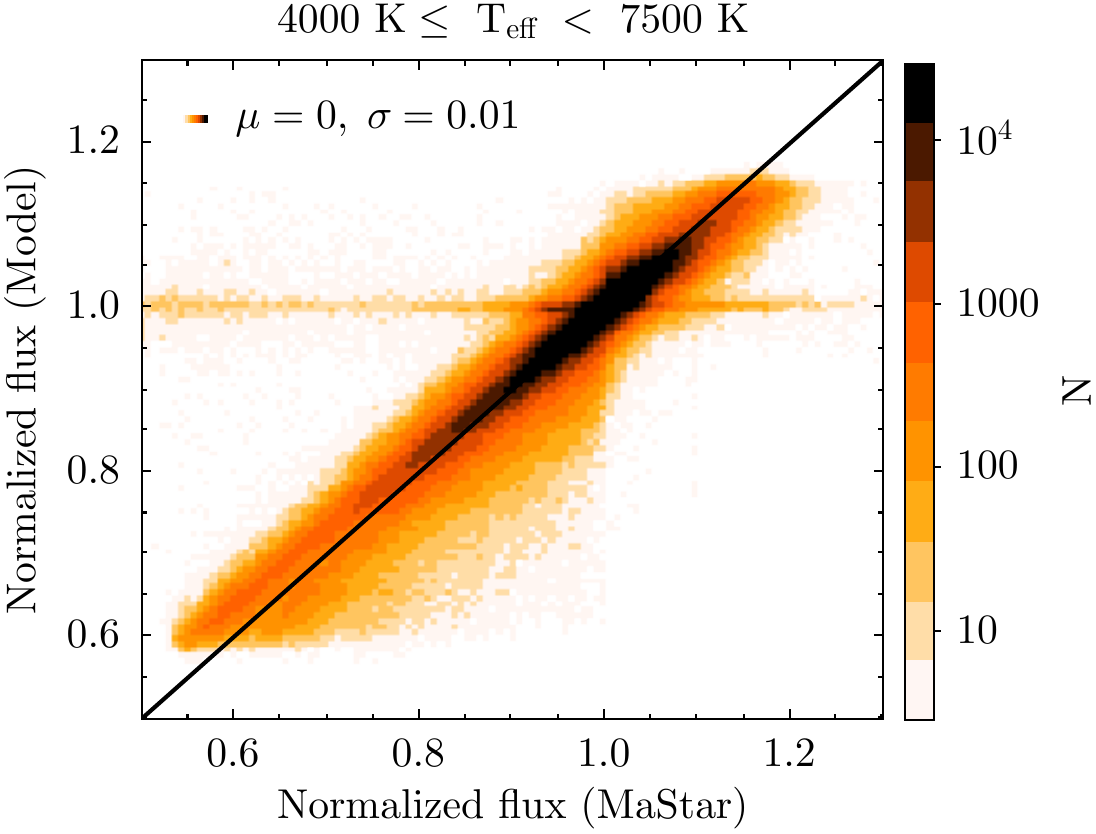}
				\centering
			\end{minipage}
		}
		\centering
		\subfigure{
			\begin{minipage}[t]{0.45\linewidth}
				\centering
				\includegraphics[width=3.32in, height=2.18in]{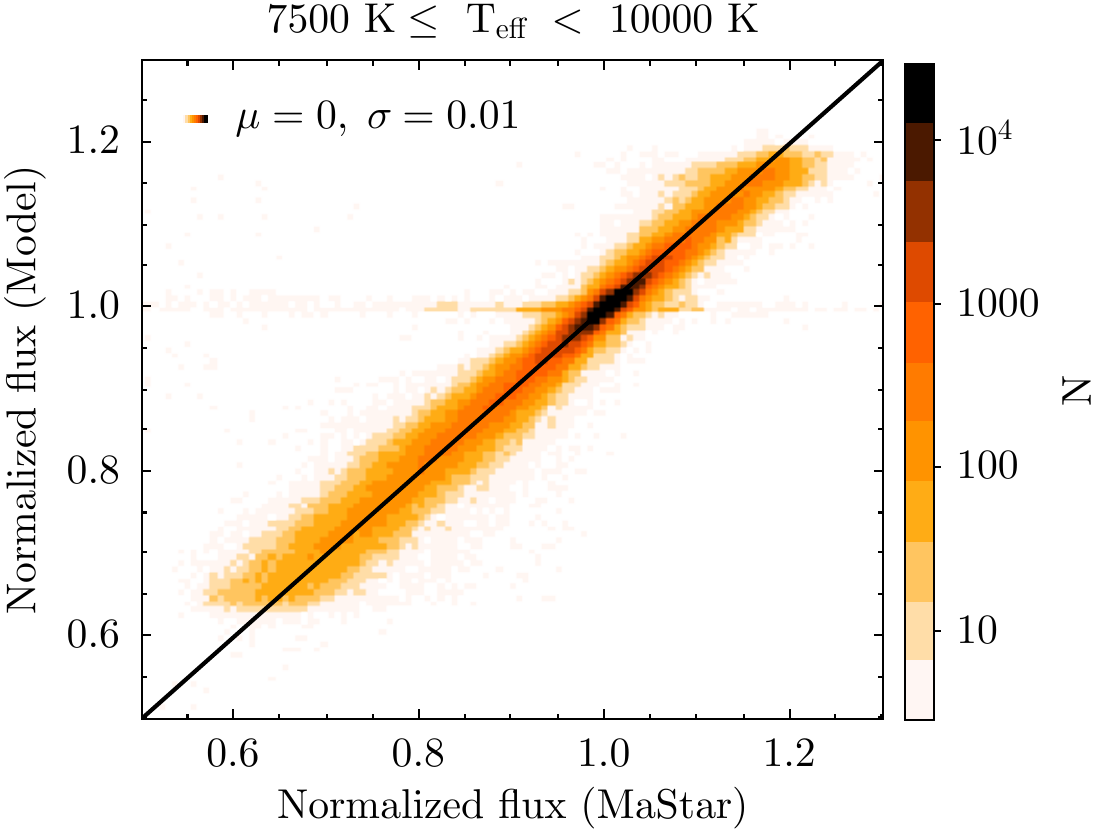}
				\centering
			\end{minipage}
		}
		\centering
		\subfigure{
			\begin{minipage}[t]{0.45\linewidth}
				\centering
				\includegraphics[width=3.32in, height=2.18in]{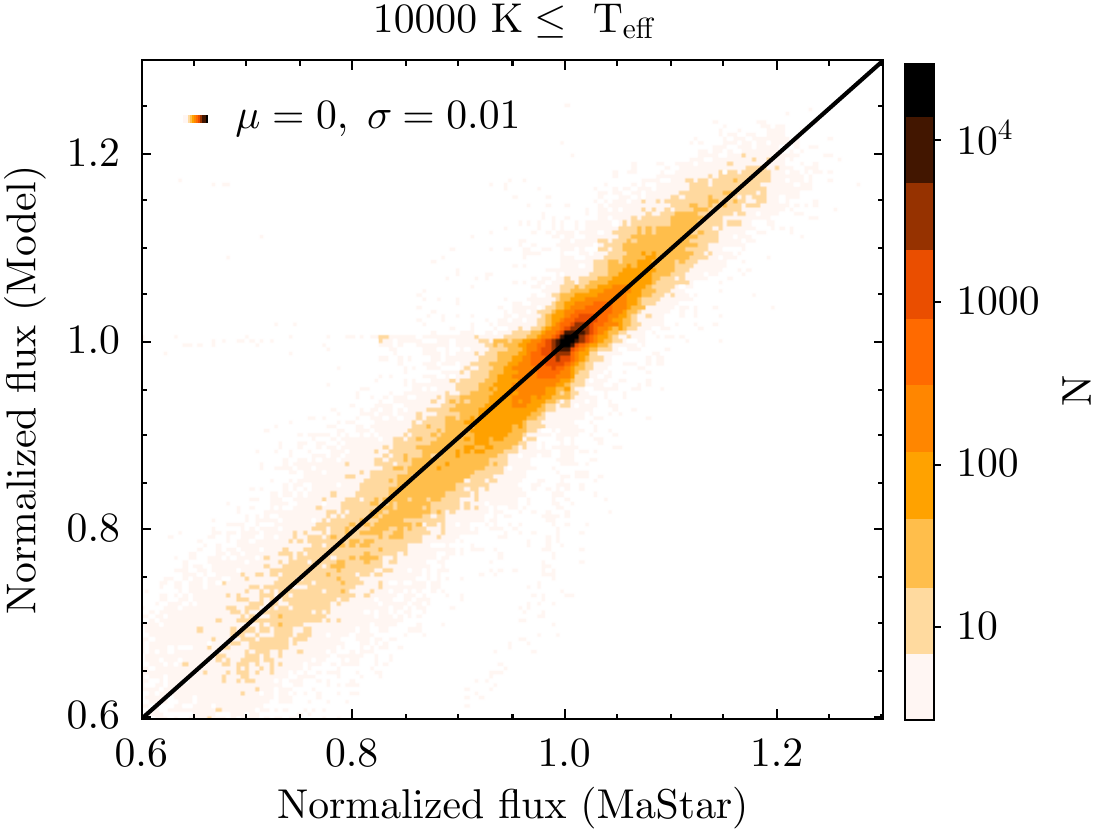}
				\centering
			\end{minipage}
		}
		\caption{Performance of the spectral emulator in different $T_{\text{eff}}$ bins. One-to-one plots compared the normalized flux in MaStar (horizontal axis) to the flux generated by the spectral emulator (vertical axis). $\mu$ and $\sigma$ denote the sample mean and standard deviation, respectively, and apply to all subsequent uses. Here, $\mu$ and $\sigma$ represent the mean and standard deviation of the difference between the predicted and the true fluxes.}
		\centering
		\label{GPR Teff range 1234}
	\end{figure*}

	\begin{figure*}[!htb]
		\includegraphics[width=6.1in,height=4.3in]{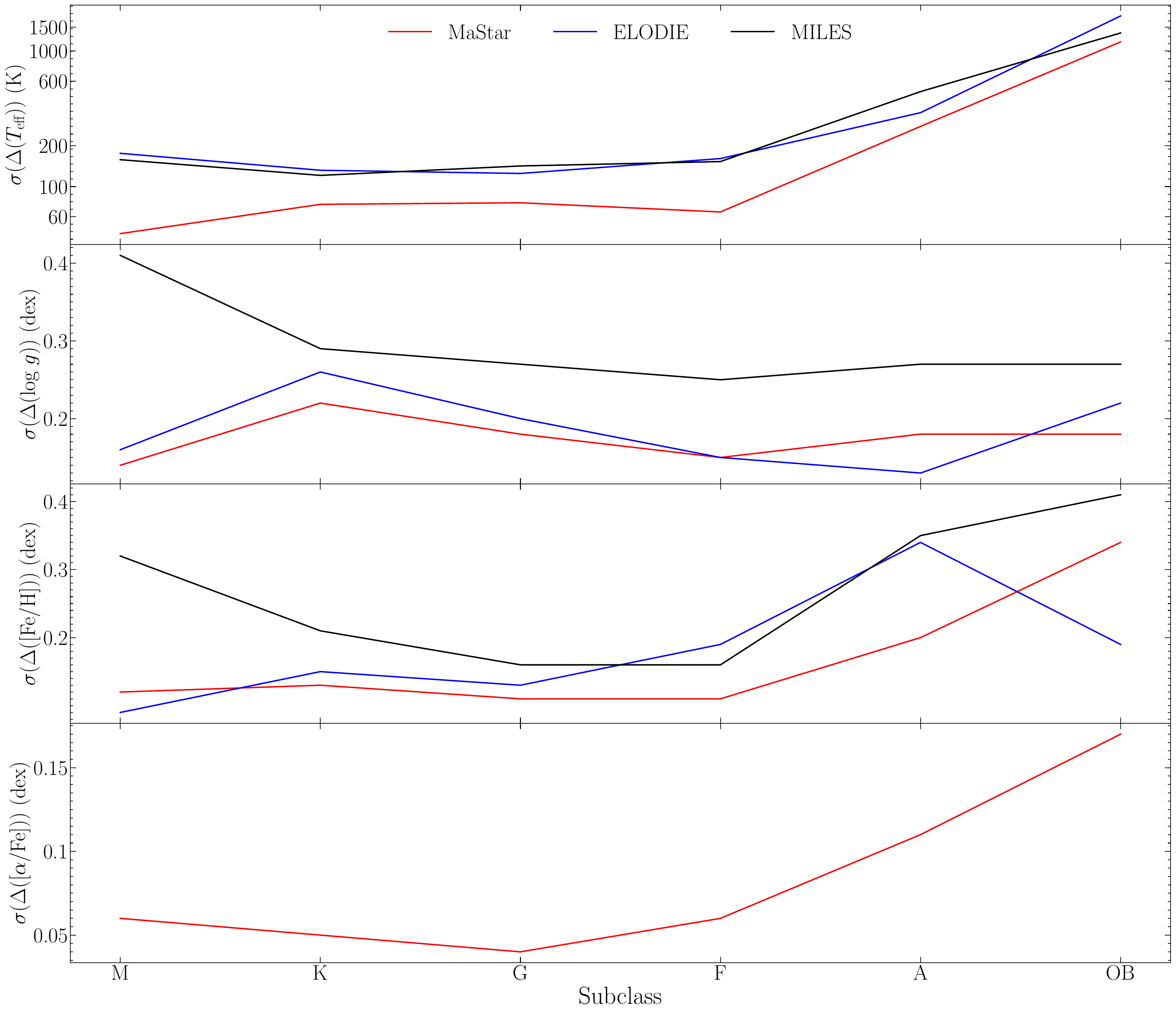}
		\centering
		\caption{Comparison of the performance of spectral emulators constructed by the MaStar, ELODIE, and MILES libraries. The horizontal axis represents spectral subclasses, and the vertical axis shows the dispersion of parameter errors (the predicted values minus true values) for different spectral libraries. The colored curves in each subplot corresponds to different empirical spectral libraries.}
		\centering
		\label{MaStar ELODIE MILES}
	\end{figure*}
	
	\begin{figure*}[!htb]
		\centering
		\subfigure{
			\begin{minipage}[t]{0.45\linewidth}
				\centering
				\includegraphics[width=3.2in, height=2.1in]{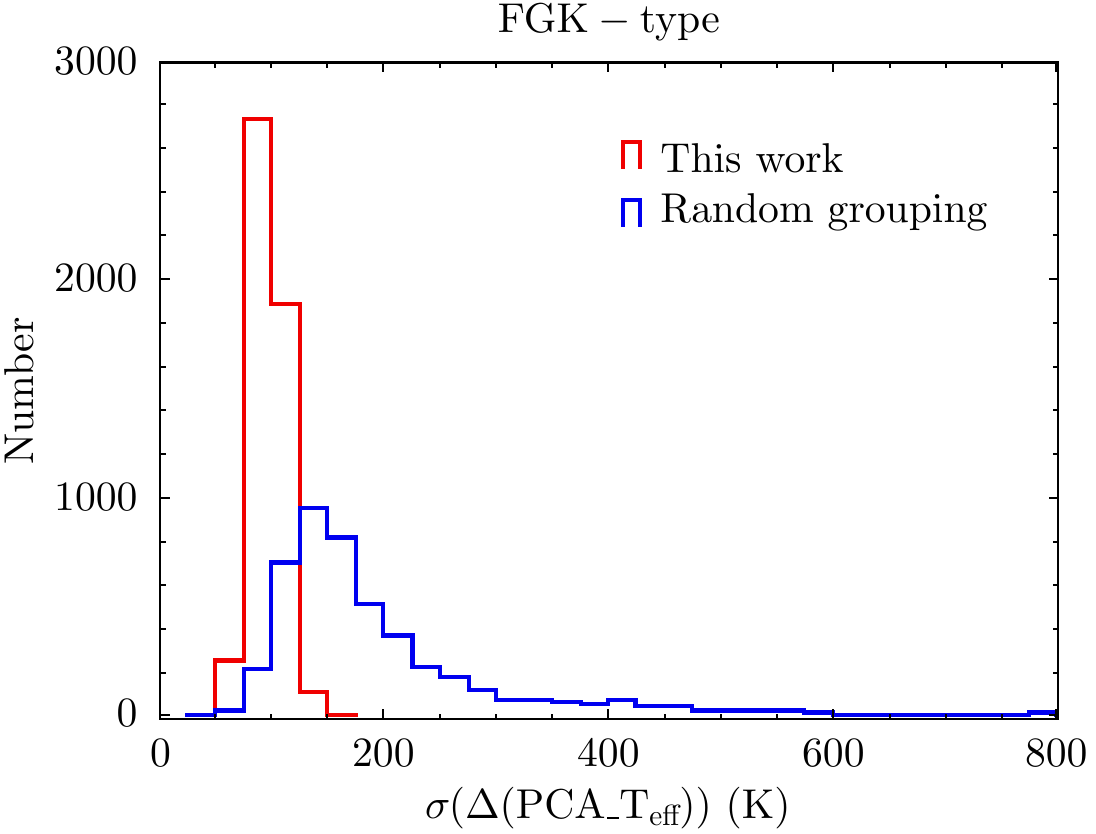}
				\centering
			\end{minipage}
		}
		\centering
		\subfigure{
			\begin{minipage}[t]{0.45\linewidth}
				\centering
				\includegraphics[width=3.2in, height=2.1in]{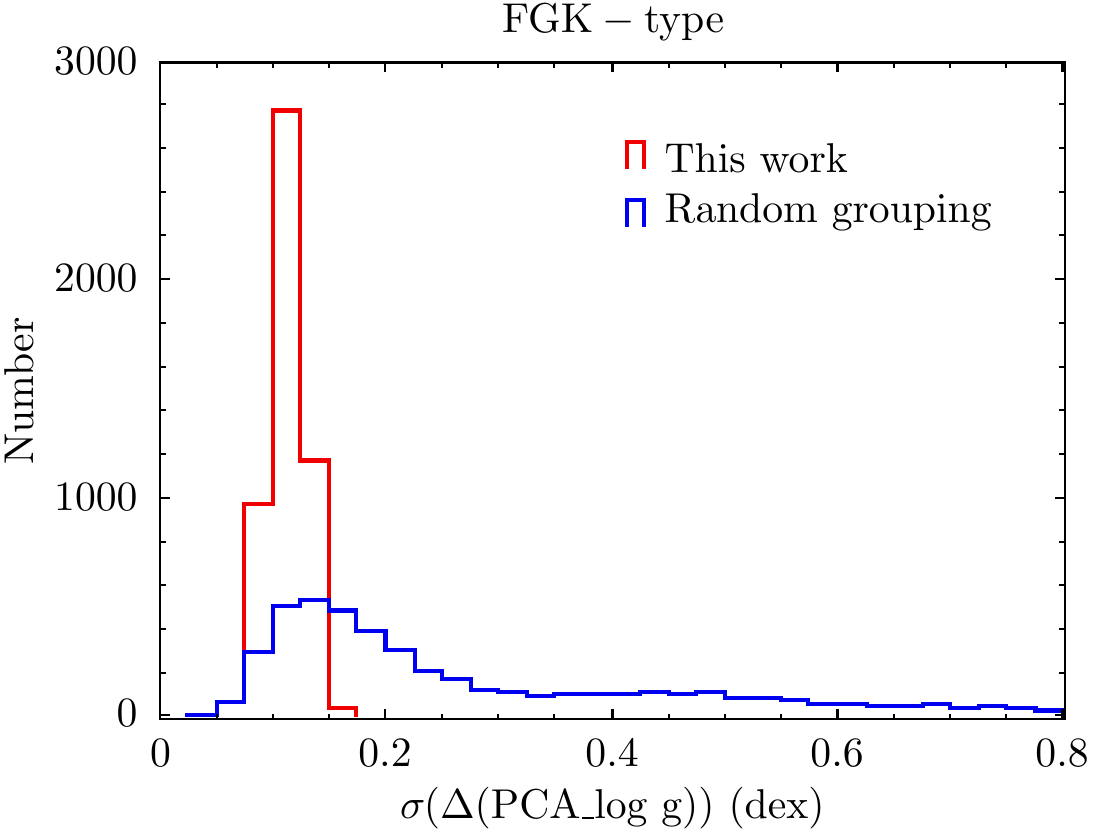}
				\centering
			\end{minipage}
		}
		\centering
		\subfigure{
			\begin{minipage}[t]{0.45\linewidth}
				\centering
				\includegraphics[width=3.2in, height=2.1in]{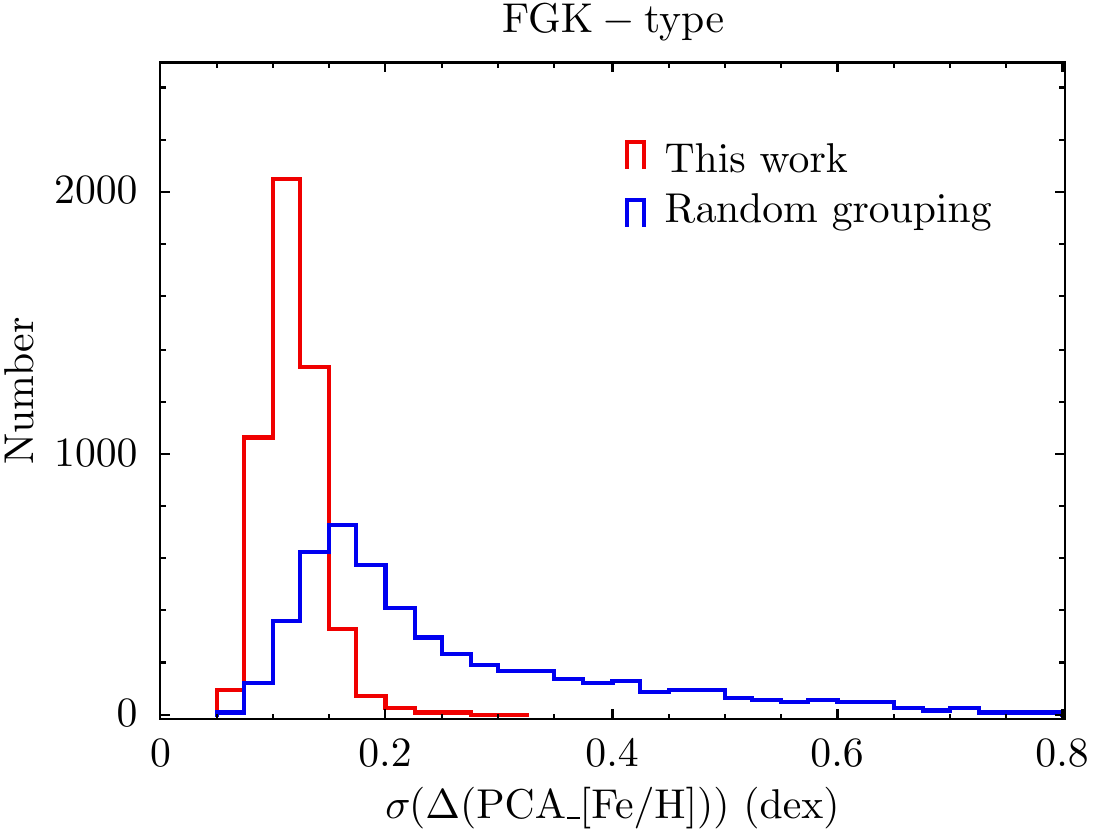}
				\centering
			\end{minipage}
		}
		\centering
		\subfigure{
			\begin{minipage}[t]{0.45\linewidth}
				\centering
				\includegraphics[width=3.2in, height=2.1in]{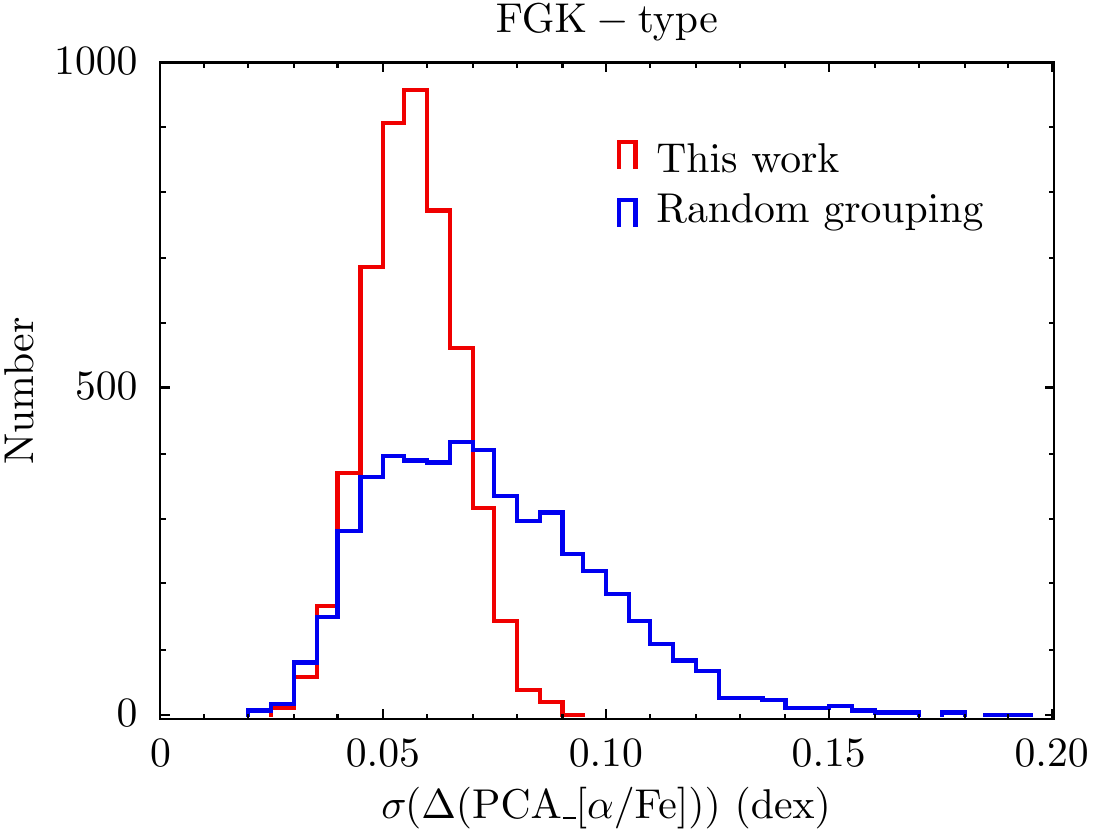}
				\centering
			\end{minipage}
		}
		\caption{The histograms of the standard deviation for each column of $\Delta(\text{PCA}\_T_{\text{eff}})$, $\Delta$(PCA\_log $g$), $\Delta$({PCA}\_[Fe/H]), and $\Delta$(PCA\_[$\alpha$/Fe]) using different grouping methods ($t=100$). The red and blue histograms represent the results using the prior grouping (this work) and random grouping methods, respectively.}
		\centering
		\label{sample_test}
	\end{figure*}
	
	We initially developed a spectral emulator based on MaStar library to generate spectra corresponding to various atmospheric physical parameters. The performance of the spectral emulator can be assessed by its ability to accurately reproduce the MaStar spectra within four $T_{\text{eff}}$ bins ($T_{\text{eff}} < 4000$ K, $4000 \le  T_{\text{eff}} < 6000$ K, $6000 \le  T_{\text{eff}} < 10,000$ K, and $T_{\text{eff}} \ge 10,000$ K). As in \reffig{GPR 1234}, four examples of reproducing results show that certain critical absorption lines were almost entirely preserved, while, simultaneously, noise was effective filtered. Additionally, we also statistically analyzed the one-to-one correspondence of flux at each wavelength for all spectra within the four $T_{\text{eff}}$ bins. As shown in \reffig{GPR Teff range 1234}, we found that the spectral emulator accurately reproduced the MaStar spectra with a systematic offset of $0$ and a dispersion of $0.01$ (the horizontal structures at a reproduced spectral flux of $1$ are due to the settings applied in step $5$ of \refsubsection{Pre-processing for the spectra}). Statistically, this indicates that we can reproduce MaStar normalized spectra across $99.73\%$ of the wavelength coverage area with an error within $-0.03$ to $0.03$.
	
	We also developed the spectral emulator based on ELODIE and MILES libraries to validate the methodology across these diverse spectral data sets. The ELODIE library contains $1959$ spectra obtained with the $1.93$ m Observatoire de Haute-Provence telescope, covering $3900-6800$ \text{\AA} at a resolving power of $42,000$ \citep{prugniel2007new}. The MILES library consists of $985$ stellar spectra obtained with the $2.5$ m Isaac Newton Telescope, covering $3525-7500$ \text{\AA} at a resolution of 2.5 \text{\AA} full width at half maximum \citep{sanchez2006medium}. Considering the sparse parameter distribution of the ELODIE and MILES libraries, we applied the grouping optimization method only to the MaStar library. As shown in \reffig{MaStar ELODIE MILES}, the dispersion of errors in $T_{\text{eff}}$, log $g$, and [Fe/H] for FGK-type stars of the ELODIE and MILES libraries was confined to approximately $150$ K, $0.3$ dex, and $0.2$ dex, respectively. In contrast, the MaStar exhibited lower dispersions, especially for M- and OBA-type. This indicates that the spectral emulator based on the MaStar is more reliable. This can be attributed to two factors.
	\begin{enumerate}
		\item MaStar provides a more uniform coverage in parameter space for both cool and hot stars, enhancing the stability of the spectral emulator and thereby reducing the dispersion in parameter errors. As shown in \citeauthor{xiang2015lamost} (\citeyear{xiang2015lamost}, \citeyear{Xiang_2017}), the parameter space distribution of MILES stars exhibits holes and clusters, which adversely affect the accuracy of parameter predictions.
		\item MaStar exhibits higher consistency in atmospheric parameters. The stellar atmospheric parameters in the MILES library are collected from various literature sources, leading to potential systematic errors \citep{Xiang_2017}. This issue is also presented in the ELODIE and earlier versions of the MaStar. However, the latest version of MaStar has parameters derived from a series of independent measurement conducted by the MaStar team, employing median values as the final parameters, thereby ensuring statistical consistency.
	\end{enumerate}
	Considering all the above aspects, we believe the MaStar holds potential for atmospheric parameter prediction.
	
	\begin{figure*}[!htb]
		\centering
		\subfigure{
			\begin{minipage}[t]{0.45\linewidth}
				\centering
				\includegraphics[width=3.2in, height=2.3in]{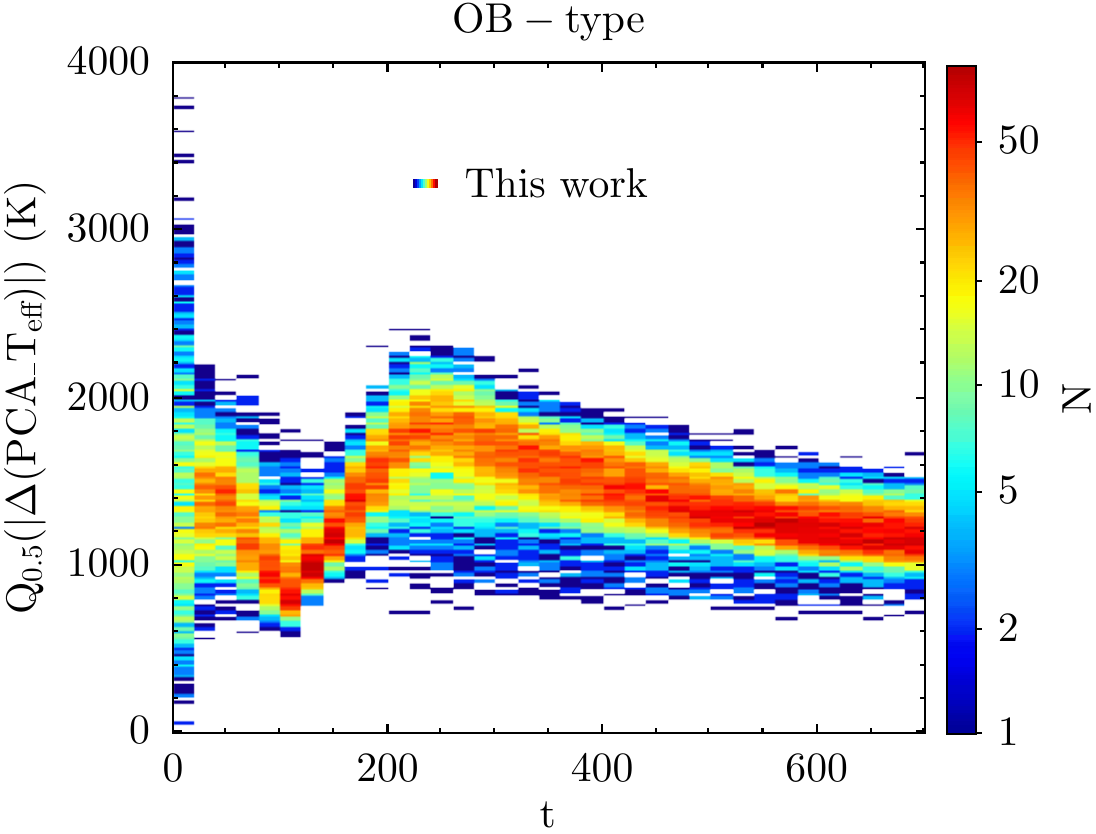}
				\centering
			\end{minipage}
		}
		\centering
		\subfigure{
			\begin{minipage}[t]{0.45\linewidth}
				\centering
				\includegraphics[width=3.2in, 
				height=2.3in]{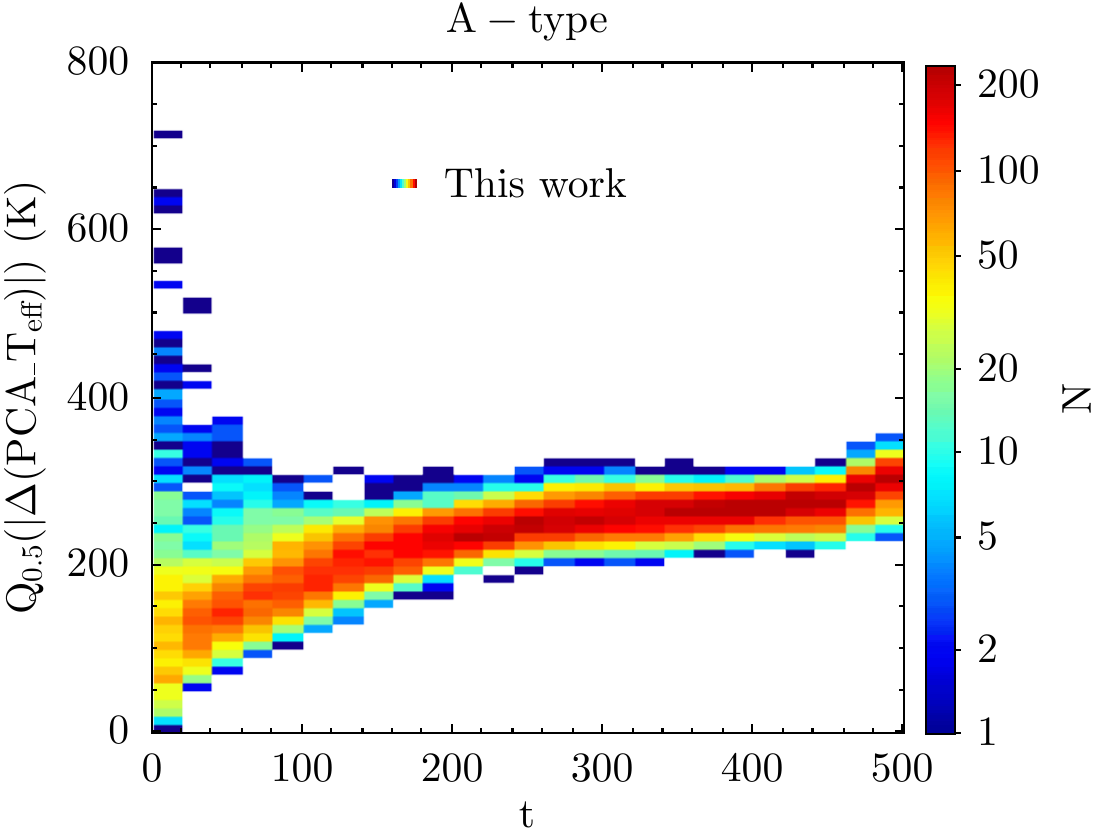}
				\centering
			\end{minipage}
		}
		\centering
		\subfigure{
			\begin{minipage}[t]{0.45\linewidth}
				\centering
				\includegraphics[width=3.2in, height=2.3in]{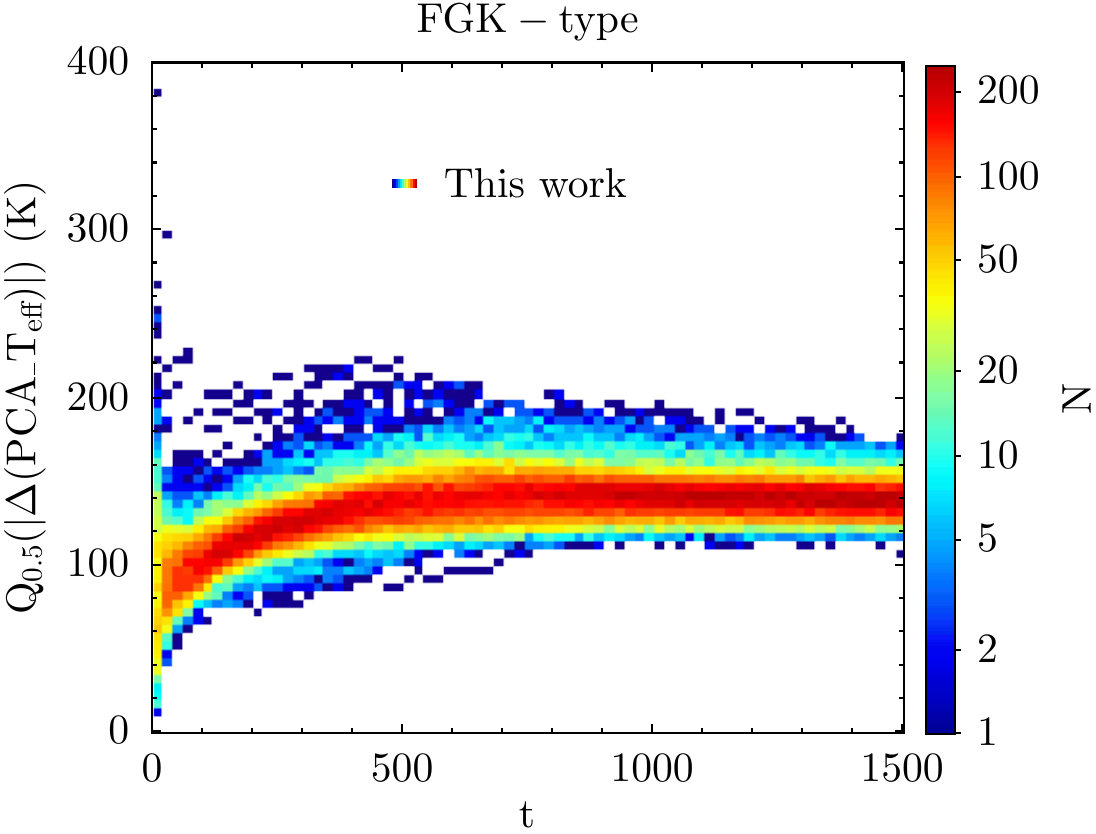}
				\centering
			\end{minipage}
		}
		\centering
		\subfigure{
			\begin{minipage}[t]{0.45\linewidth}
				\centering
				\includegraphics[width=3.2in, 
				height=2.3in]{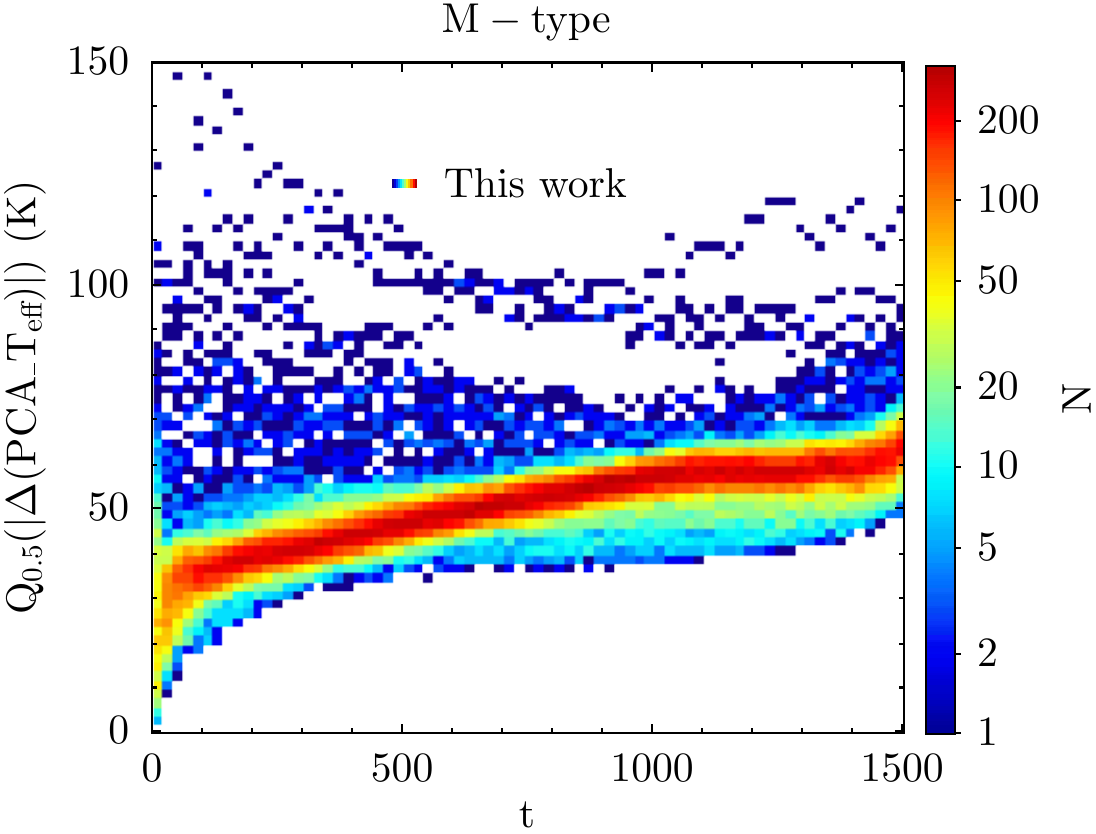}
				\centering
			\end{minipage}
		}
		\caption{The median distribution of the absolute values in each column of $\Delta$(PCA\_$T_{\text{eff}}$) for different $t$ values. Each panel includes $5000$ median values corresponding to the same $t$ value; the colors represent the density of the $5000$ median values at each $t$.}
		\centering
		\label{sample_size_test}
	\end{figure*}
	
	\subsubsection{Evaluating spectral grouping method and the setting of $t=100$} \label{Evaluating spectral grouping method and the setting of the $t$}
	
	For any given LAMOST \lq concatenated spectrum\rq, we obtained the best matched $t$ sets of MaStar parameters and mapped them to the principal component space in order to optimize \refformula{workflow3}. Therefore, before optimizing \refformula{workflow3}, it is necessary to evaluate the rationality of the prior grouping and the setting of $t$. Once the spectral grouping method and $t$ were determined, we also needed to assess the differences in LAMOST atmospheric parameter predictions between using the grouping optimization method and not using it. To do this, we first defined the parameter reproduction error and then used it to evaluate the prior grouping and the appropriateness of setting $t$. This setting was then applied to LAMOST spectra to evaluate its impact on parameter predictions, as detailed below:
	\begin{enumerate}
		\item \textbf{Definition of parameter reproduction error.} We randomly selected a \lq concatenated spectrum\rq \ consisting of $N_{\text{group}}$ LAMOST spectra and obtained its best matched $t$ sets of MaStar parameters. We then calculated the reproduction error of these $t$ sets of MaStar parameters when mapping from the principal component space to the atmospheric physical parameter space. Specifically, we randomly divided these $t$ sets of MaStar parameters into five subdatasets (denoted as $\Theta_{1}, \Theta_{2},..., \Theta_{5}$), each containing $t/5$ samples. Four of these subdata sets were used to map the MaStar parameter space to the principal component space, obtaining the coefficient matrix $Q$. The remaining subdataset ($\Theta_{i}, i=1,2,...,5$) was used to calculate the reproduction errors ($\Theta_{i} Q^{+}  Q-\Theta_{i}$, where $Q^{+}$ is the pseudoinverse of $Q$) when mapping the principal component space back to the atmospheric physical parameter space using $Q$. We sequentially calculated the reproduction errors of the atmospheric physical parameters in these five subdata sets. For each $\Theta_{i}$, the reproduction error is represented as a matrix with $t/5$ rows and $4N_{\text{group}}$ columns ($4$ represents the four atmospheric parameters). To facilitate the statistical analysis of the reproduction errors for $T_{\text{eff}}$, log $g$, [Fe/H], and [$\alpha$/Fe] in each $\Theta_{i}$, we concatenated the $5$ for error matrices by rows into a single matrix with $t/5$ rows and $4N_{\text{group}}\times 5$ columns ($5$ represents the number of subdata sets). We denote the final reproduction error matrices for $T_{\text{eff}}$, log $g$, [Fe/H], and [$\alpha$/Fe] (each matrix with $t/5$ rows and $5N_{\text{group}}$ columns) as $\Delta(\text{PCA}\_T_{\text{eff}})$, $\Delta$(PCA\_log $g$), $\Delta$({PCA}\_[Fe/H]), and $\Delta$(PCA\_[$\alpha$/Fe]), respectively.
		\item \textbf{Evaluating the rationality of prior grouping of spectra.}  We first used FGK-type stars to evaluate the reliability of spectral grouping method by comparing the dispersion in each dimension of reproduction error matrix between prior spectral grouping and random grouping (without considering prior conditions), respectively. As shown in \reffig{sample_test}, the prior grouping method yields smaller dispersion value in the reproduction errors compared to the random grouping method, which indicates that the prior grouping method effectively reproduces the stellar atmospheric physical parameter space. We conducted similar tests on other types of stars and observed that our spectral grouping method is also more effective than random grouping in reproducing parameters.
		\item \textbf{Evaluating the rationality of $t=100$.} For a LAMOST \lq concatenated spectrum\rq, we calculated the reproduction error for the matched $t$ (from $5$ to $1500$ in steps of $20$) sets of MaStar parameters. To capture the changes in each dimension of the reproduction error matrix as $t$ increases, we used the median of absolute reproduction error as the statistic for each dimensional reproduction error. As shown in \reffig{sample_size_test}, taking $T_{\text{eff}}$ as an example, for OB-type stars, the $Q_{0.5}(|\Delta(\text{PCA}\_T_{\text{eff}})|)$ in each $T_{\text{eff}}$ dimension decreases as $t$ increases when $t < 100$. However, when $t > 100$, the $Q_{0.5}(|\Delta(\text{PCA}\_T_{\text{eff}})|)$ initially increases with $t$ until $t = 300$, after which it decreases and gradually stabilizes. This trend is attributed to the limited number of high-temperature star samples within the prior parameter range (\reffig{count_t}). When $100 < t < 300$, these $t$ sets of parameters include a small number of low-temperature stars, which adversely affects the reproduction of high-temperature star parameters, leading to an increase in the reproduction error of $T_{\text{eff}}$. When $t > 300$, these $t$ sets of parameters become dominated by low-temperature stars, causing the $Q_{0.5}(|\Delta(\text{PCA}\_T_{\text{eff}})|)$ to decrease and stabilize, but leading to an underestimation of $T_{\text{eff}}$ for high-temperature stars. For AFGKM-type stars, larger $t$ values result in larger parameter reproduction errors. Considering that, when $t=100$, the $T_{\text{eff}}$ reproduction errors for OB-, A-, FGK-, and M-type stars are approximately $800$, $150$, $100$, and $40$ K, respectively, the error levels are acceptable. Therefore, we believe that the setting of $t=100$ is feasible.
		\item \textbf{Evaluating the impact of prior grouping and setting $t=100$ on LAMOST parameter predictions.} We randomly selected $10,000$ spectra from each spectral type in the LAMOST to evaluate the differences between parameters predicted by the prior grouping method and those predicted without grouping. As shown in the bottom subplot of \reffig{Batch size}, for different types of stars, the difference in parameter predictions gradually decreases as $N_{\text{group}}$ increases and stabilizes when $N_{\text{group}} > 100$. This indicates that, when $N_{\text{group}} > 100$, the optimal solution of objective function obtained using the prior grouping method aligns more closely with the solution obtained without grouping optimization, compared to when $N_{\text{group}} < 100$. Therefore, we believe that it is feasible to use prior grouping with $N_{\text{group}}>100$ and $t=100$.
	\end{enumerate}
	
	Based on the results of these analyses, we conclude that setting $t=100$ and using the prior spectral grouping method is feasible.
	
	\begin{figure*}[htb]
		\centering
		\subfigure{
			\begin{minipage}[t]{0.3\linewidth}
				\centering
				\includegraphics[width=2.2in, height=1.9in]{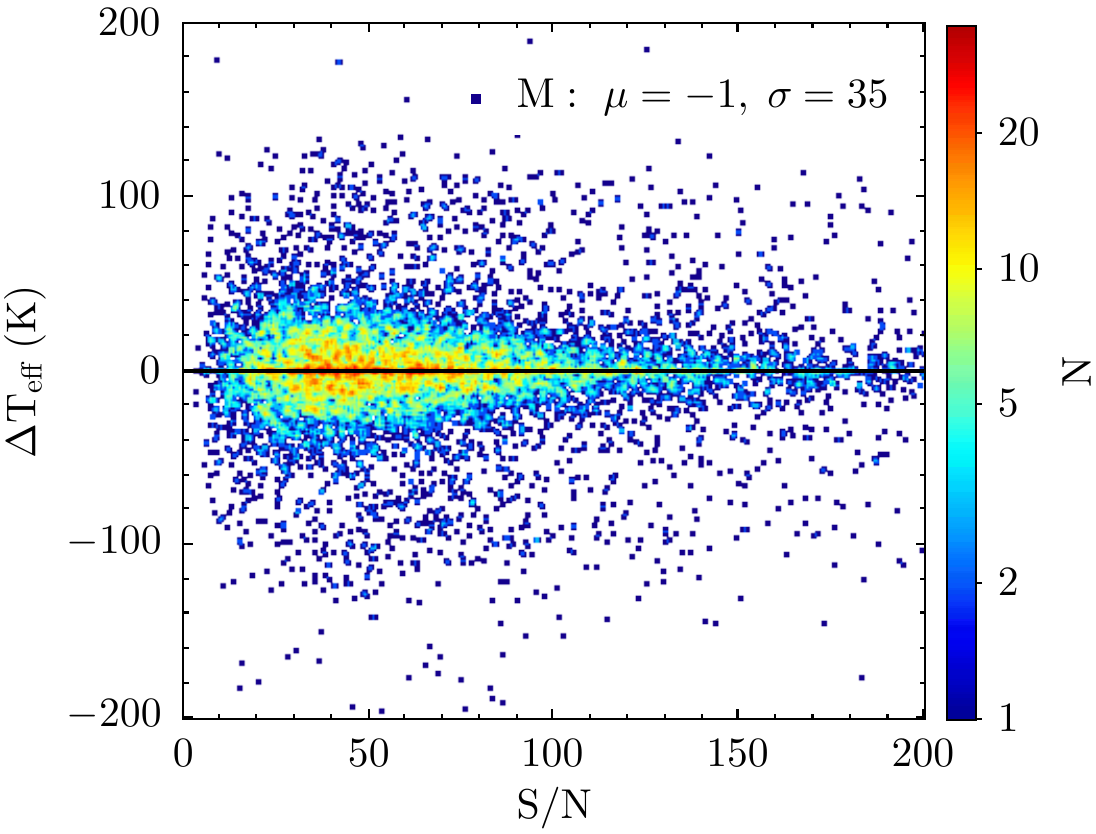}
				\centering
			\end{minipage}
		}
		\centering
		\subfigure{
			\begin{minipage}[t]{0.3\linewidth}
				\centering
				\includegraphics[width=2.2in, height=1.9in]{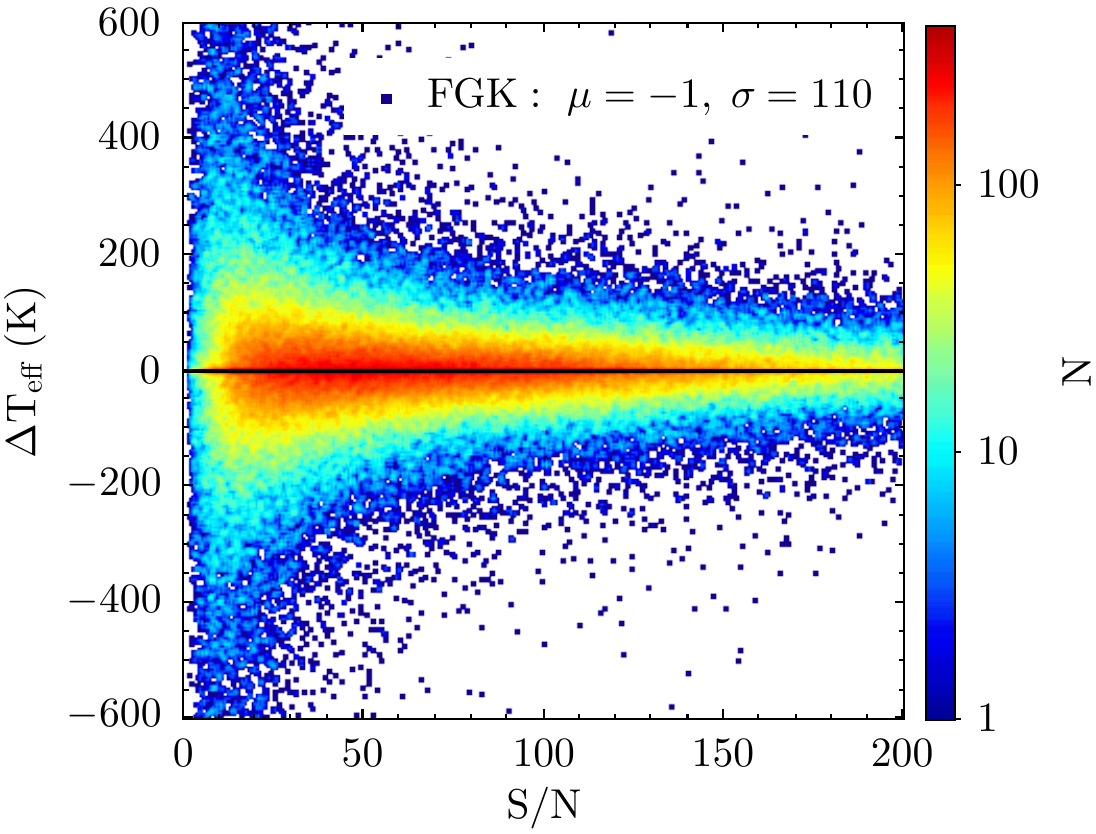}
				\centering
			\end{minipage}
		}
		\centering
		\subfigure{
			\begin{minipage}[t]{0.3\linewidth}
				\centering
				\includegraphics[width=2.2in, height=1.9in]{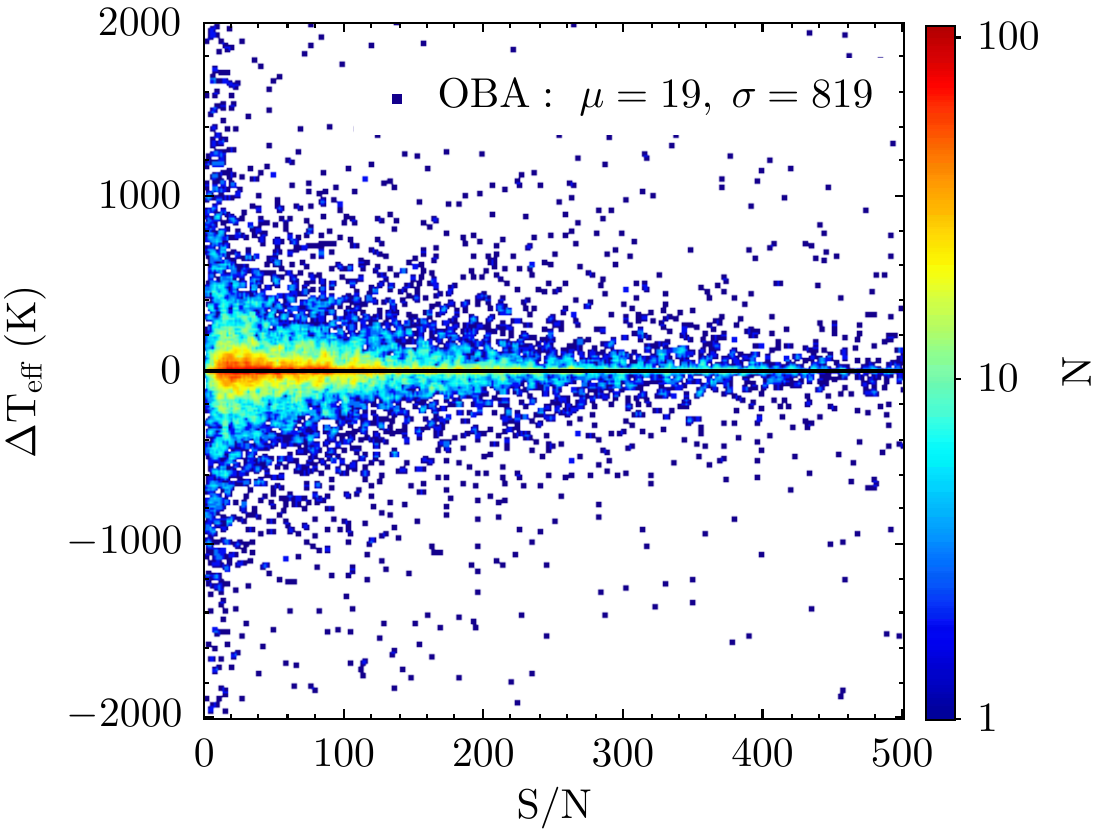}
				\centering
			\end{minipage}
		}
		\centering
		\subfigure{
			\begin{minipage}[t]{0.3\linewidth}
				\centering
				\includegraphics[width=2.2in, height=1.9in]{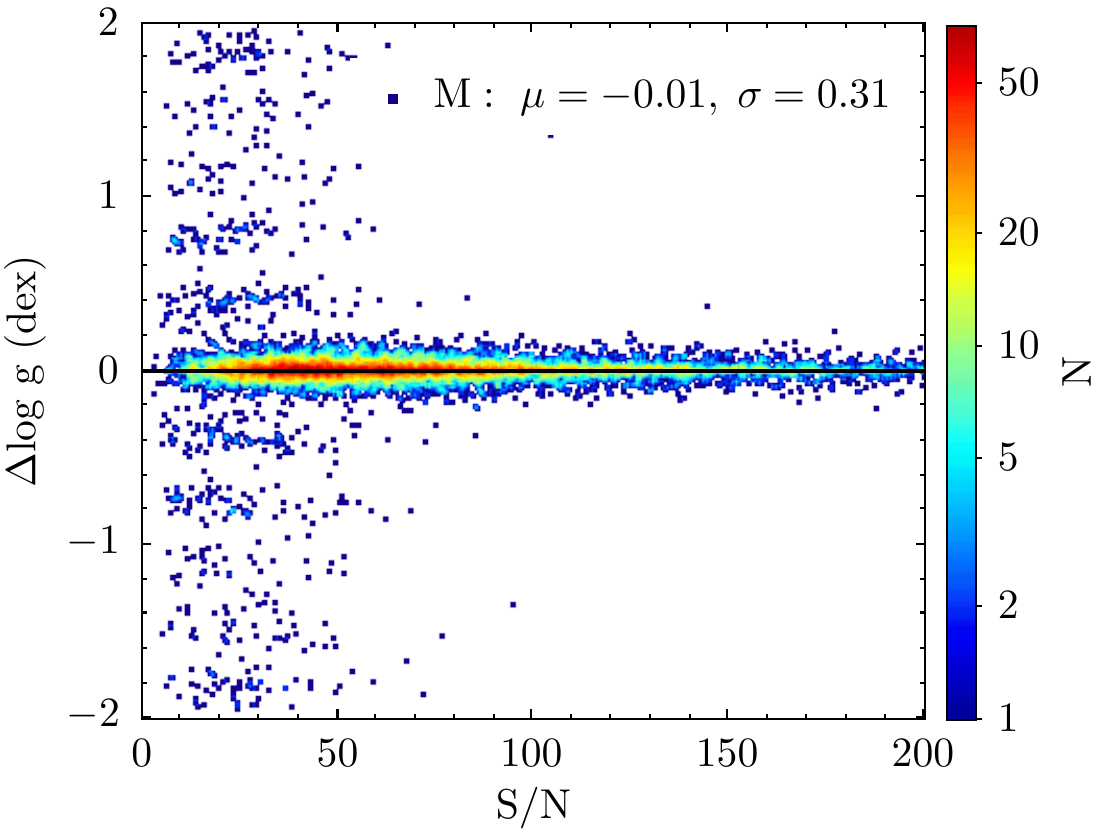}
				\centering
			\end{minipage}
		}
		\centering
		\subfigure{
			\begin{minipage}[t]{0.3\linewidth}
				\centering
				\includegraphics[width=2.2in, height=1.9in]{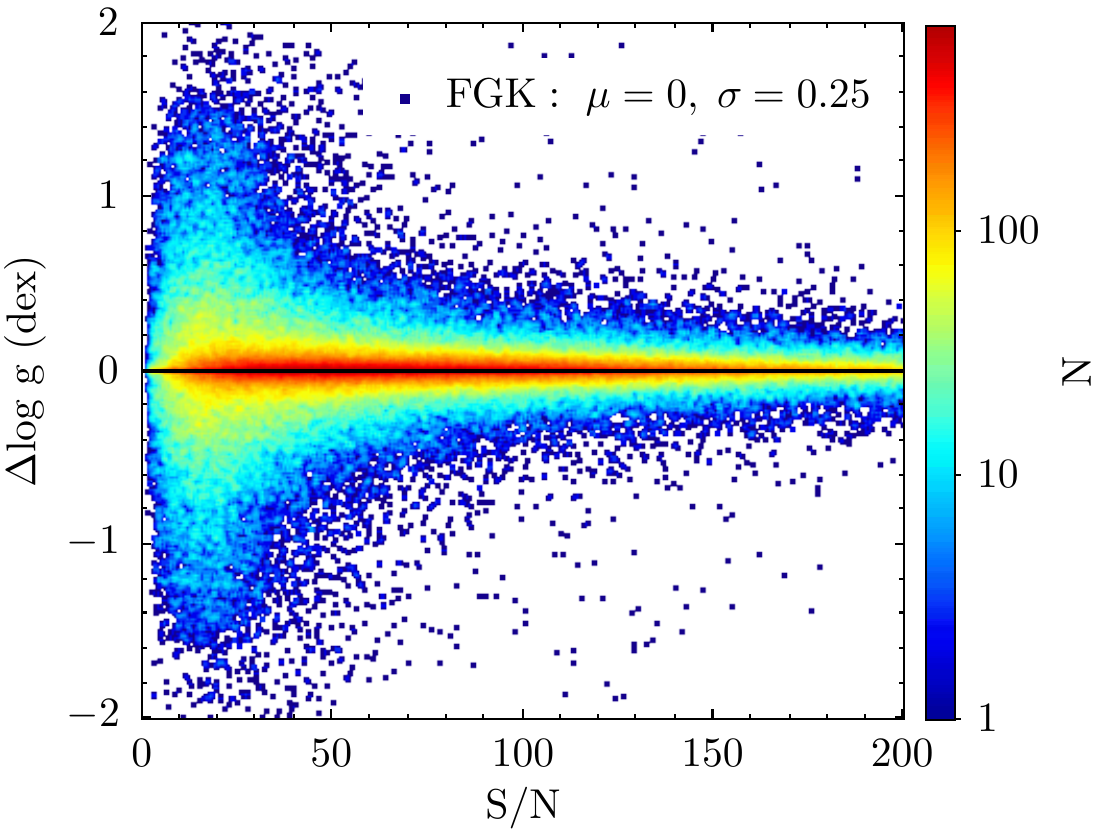}
				\centering
			\end{minipage}
		}
		\centering
		\subfigure{
			\begin{minipage}[t]{0.3\linewidth}
				\centering
				\includegraphics[width=2.2in, height=1.9in]{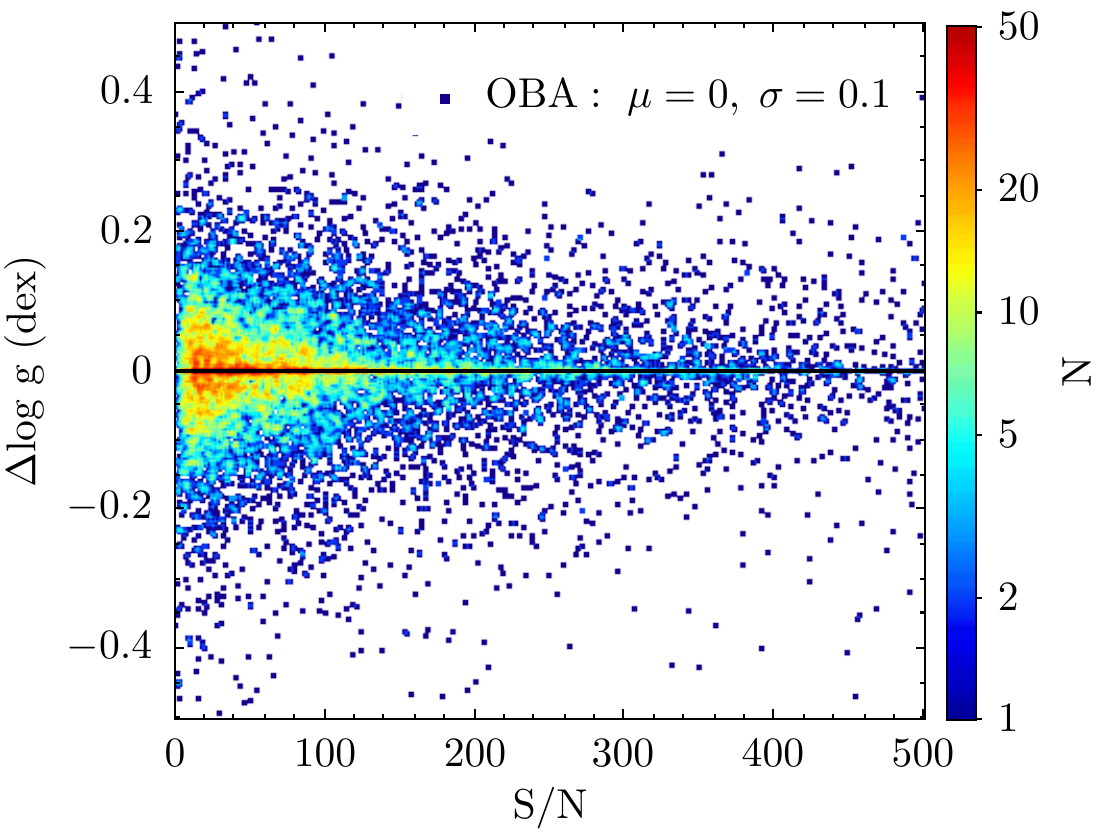}
				\centering
			\end{minipage}
		}
		\centering
		\subfigure{
			\begin{minipage}[t]{0.3\linewidth}
				\centering
				\includegraphics[width=2.2in, height=1.9in]{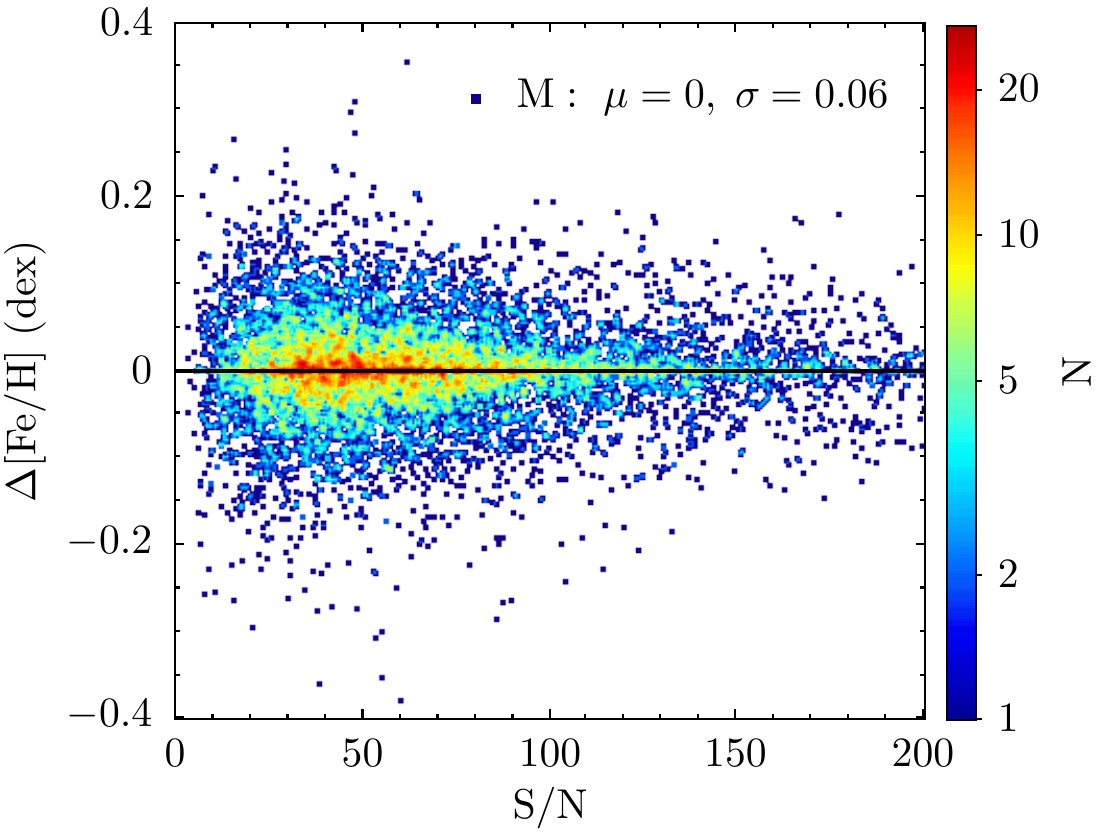}
				\centering
			\end{minipage}
		}
		\centering
		\subfigure{
			\begin{minipage}[t]{0.3\linewidth}
				\centering
				\includegraphics[width=2.2in, height=1.9in]{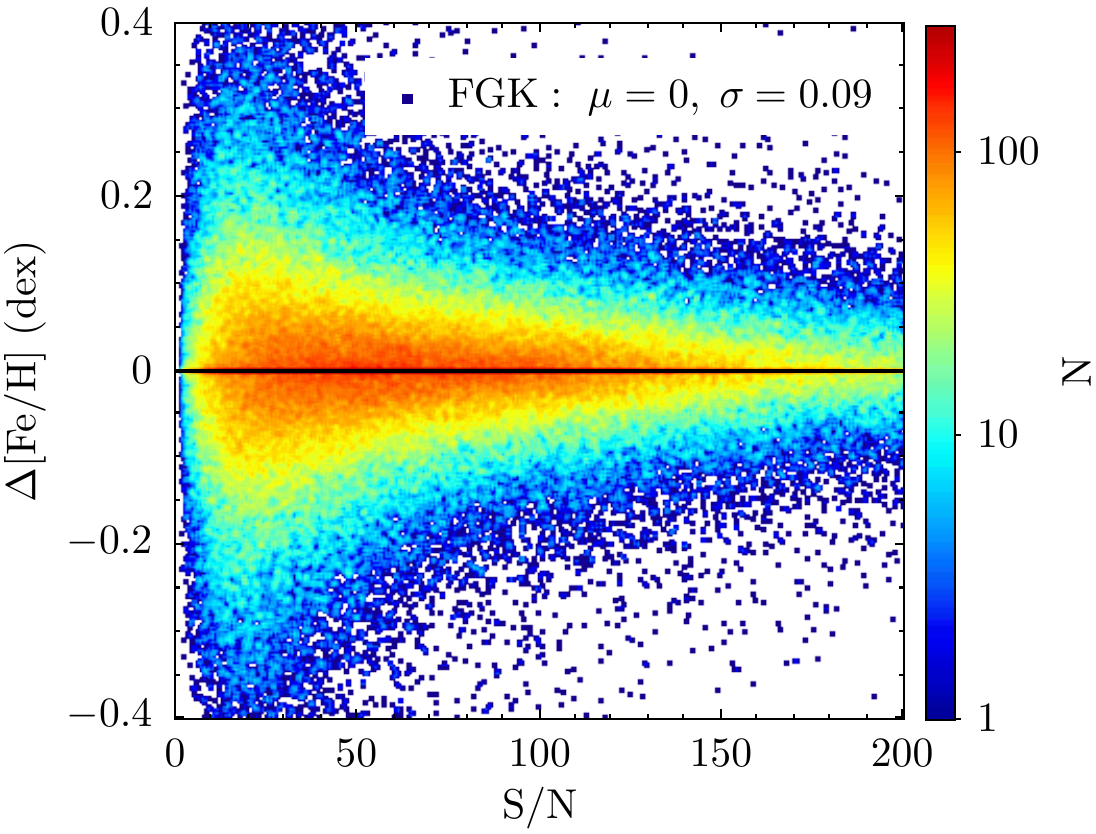}
				\centering
			\end{minipage}
		}
		\centering
		\subfigure{
			\begin{minipage}[t]{0.3\linewidth}
				\centering
				\includegraphics[width=2.2in, height=1.9in]{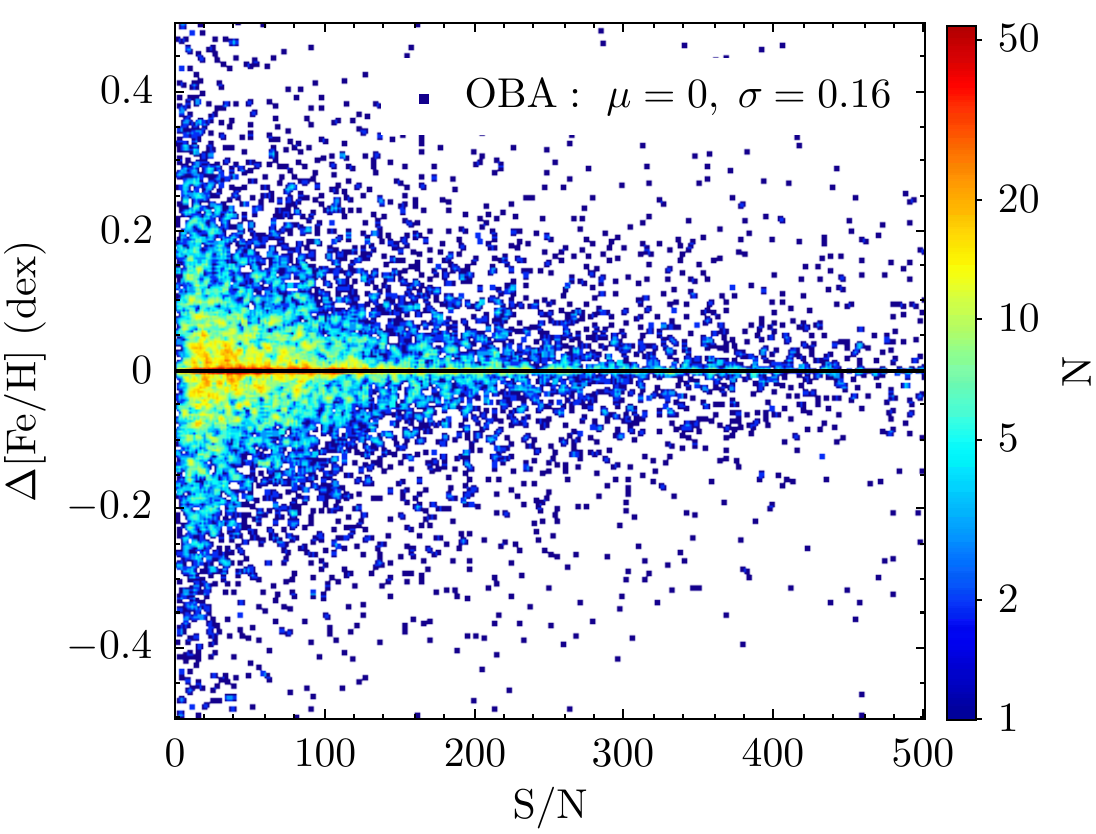}
				\centering
			\end{minipage}
		}
		\centering
		\subfigure{
			\begin{minipage}[t]{0.3\linewidth}
				\centering
				\includegraphics[width=2.2in, height=1.9in]{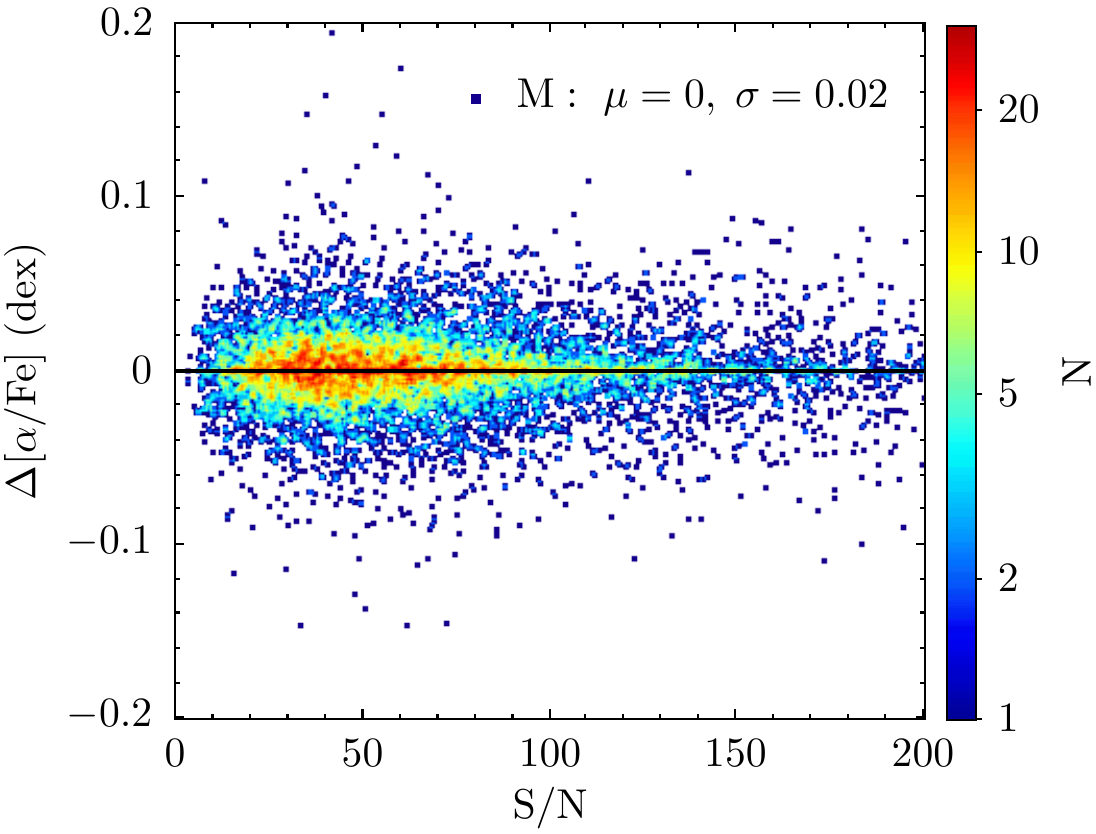}
				\centering
			\end{minipage}
		}
		\centering
		\subfigure{
			\begin{minipage}[t]{0.3\linewidth}
				\centering
				\includegraphics[width=2.2in, height=1.9in]{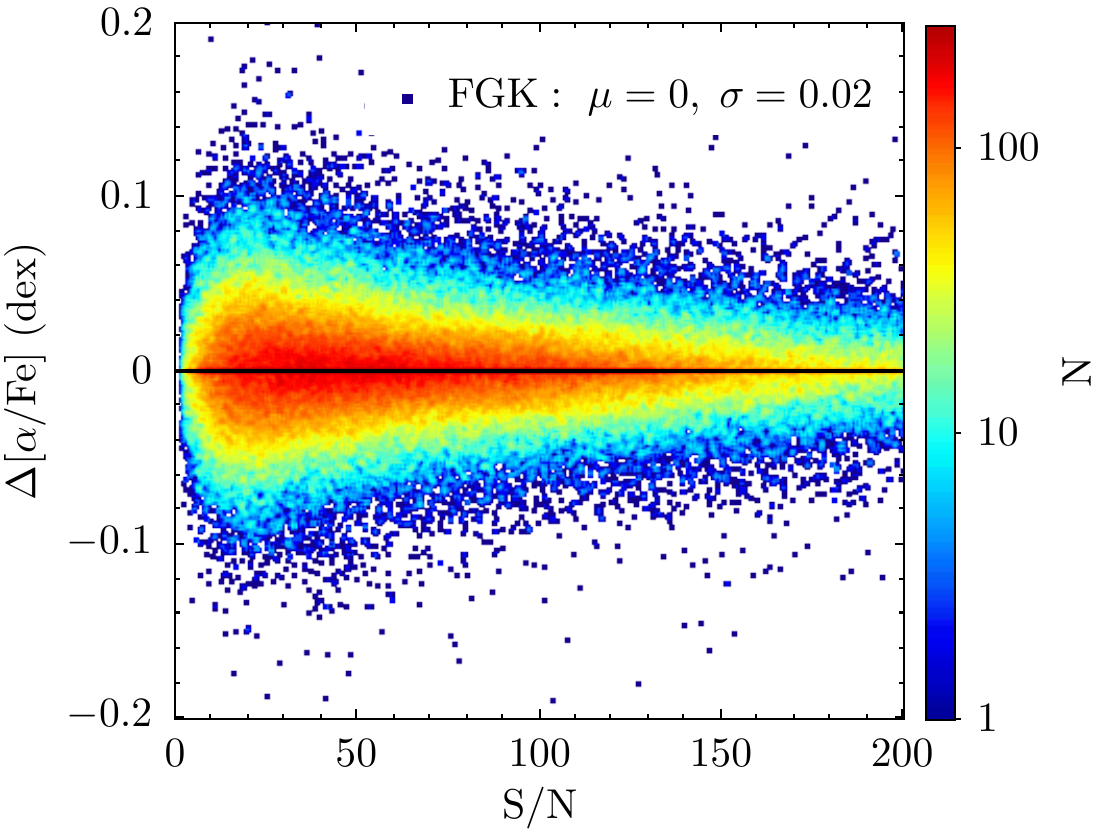}
				\centering
			\end{minipage}
		}
		\centering
		\subfigure{
			\begin{minipage}[t]{0.3\linewidth}
				\centering
				\includegraphics[width=2.2in, height=1.9in]{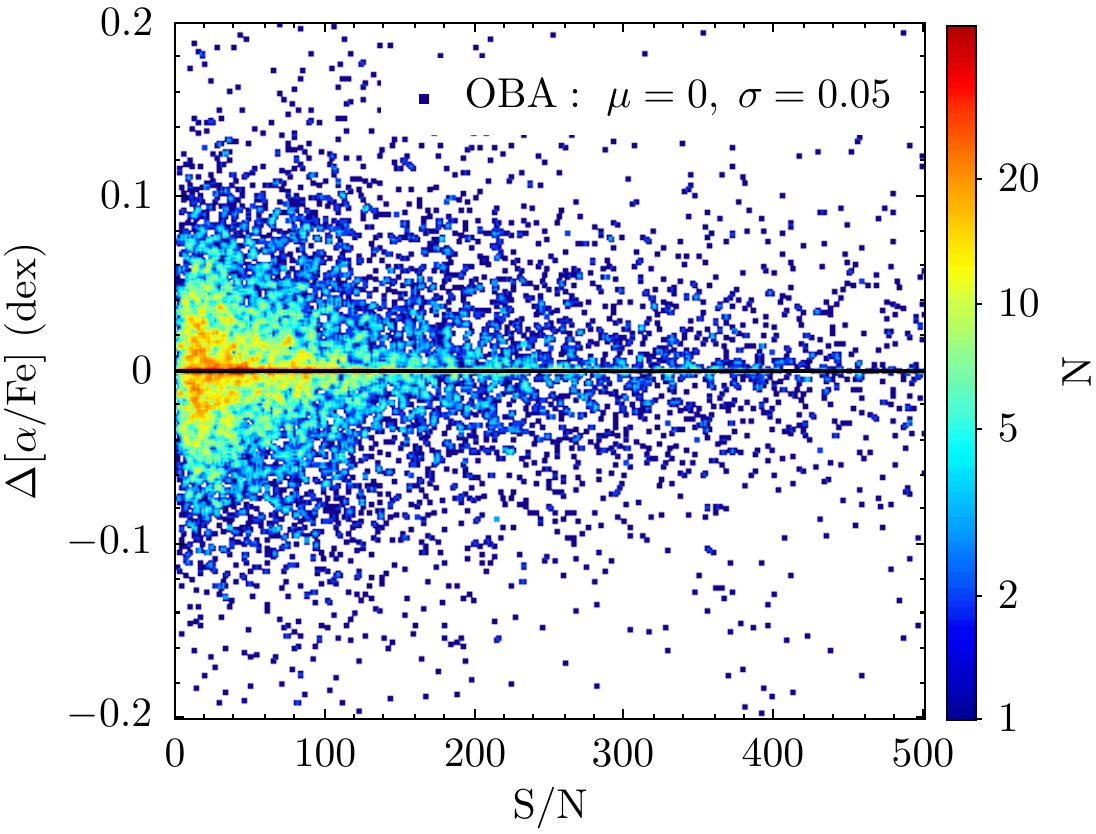}
				\centering
			\end{minipage}
		}
		\caption{Differences in atmospheric parameters from repeat spectral observations across M-type, FGK-type, and OBA-type stars (columns (1)-(3), respectively) as a function of S/N. Only repeat observations carried out on different observation nights and with S/N differences of less than $20\%$ are adopted. The horizontal axis shows the mean S/N of the repeat spectral observations, and the vertical axis shows the differences in atmospheric parameter predictions.}
		\centering
		\label{random err}
	\end{figure*}
	\begin{figure*}[!htb]
		\centering
		\subfigure{
			\begin{minipage}[t]{0.3\linewidth}
				\centering
				\includegraphics[width=2.1in, height=1.8in]{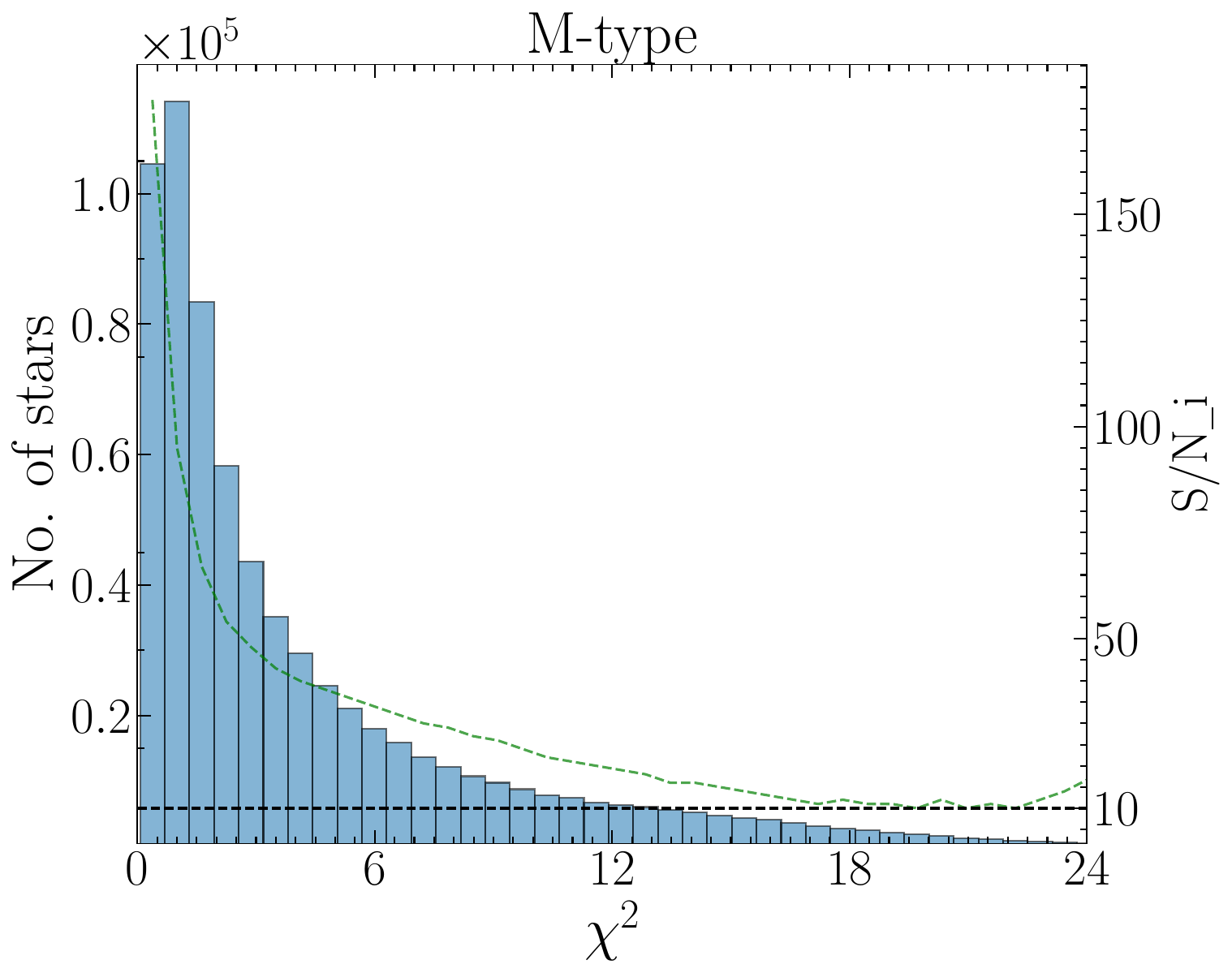}
				\centering
			\end{minipage}
		}
		\centering
		\subfigure{
			\begin{minipage}[t]{0.3\linewidth}
				\centering
				\includegraphics[width=2.1in, height=1.8in]{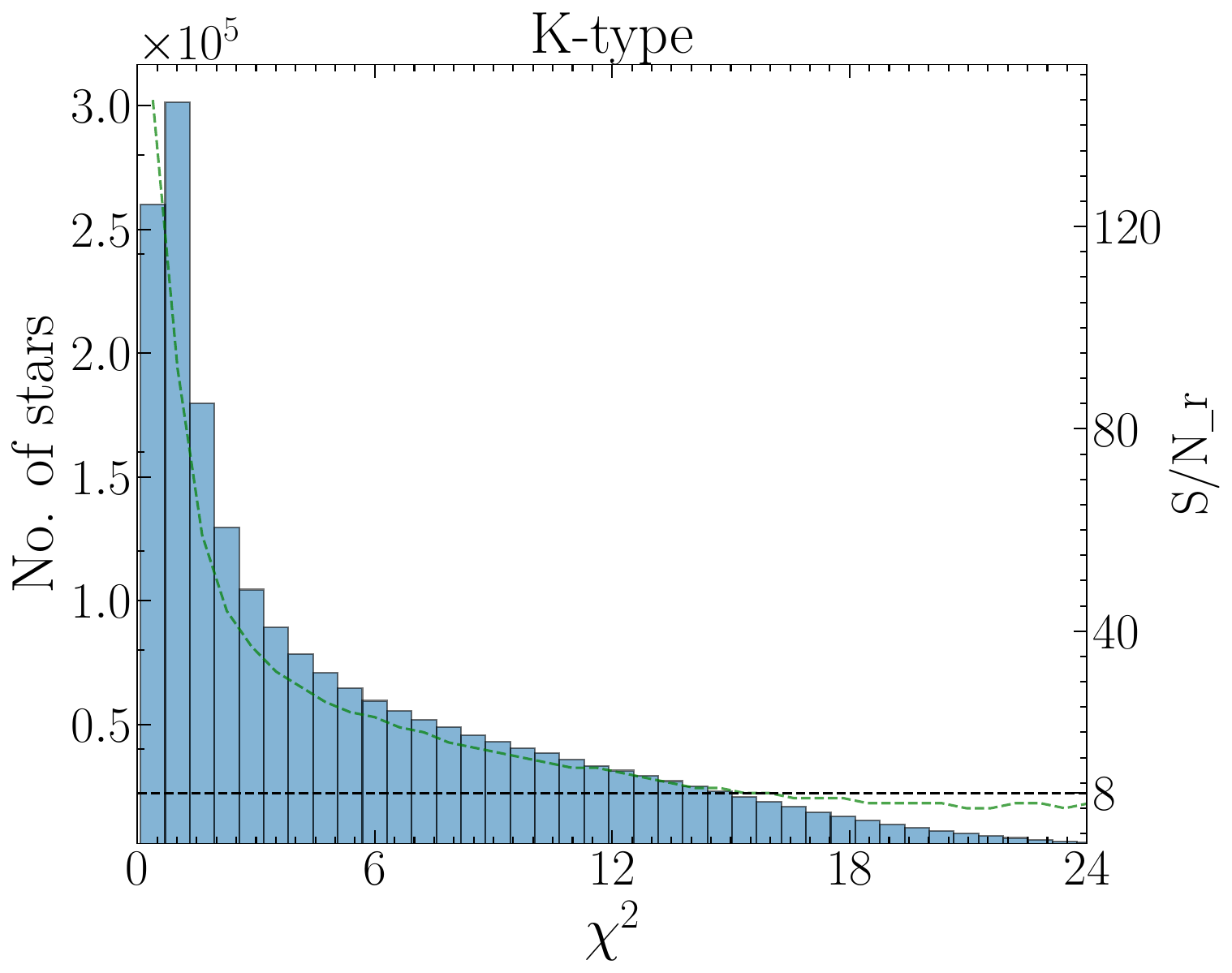}
				\centering
			\end{minipage}
		}
		\centering
		\subfigure{
			\begin{minipage}[t]{0.3\linewidth}
				\centering
				\includegraphics[width=2.1in, height=1.8in]{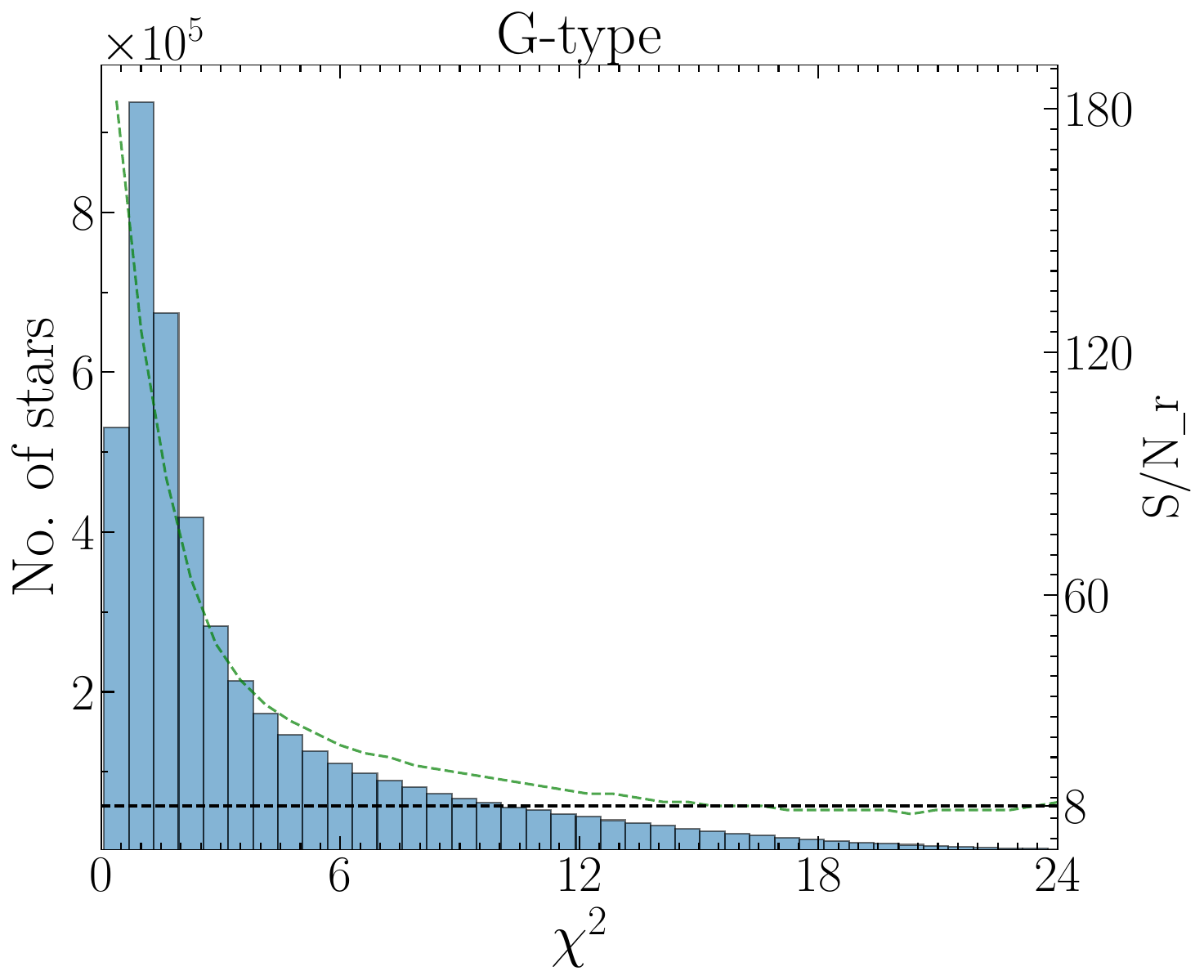}
				\centering
			\end{minipage}
		}
		\centering
		\subfigure{
			\begin{minipage}[t]{0.3\linewidth}
				\centering
				\includegraphics[width=2.1in, height=1.8in]{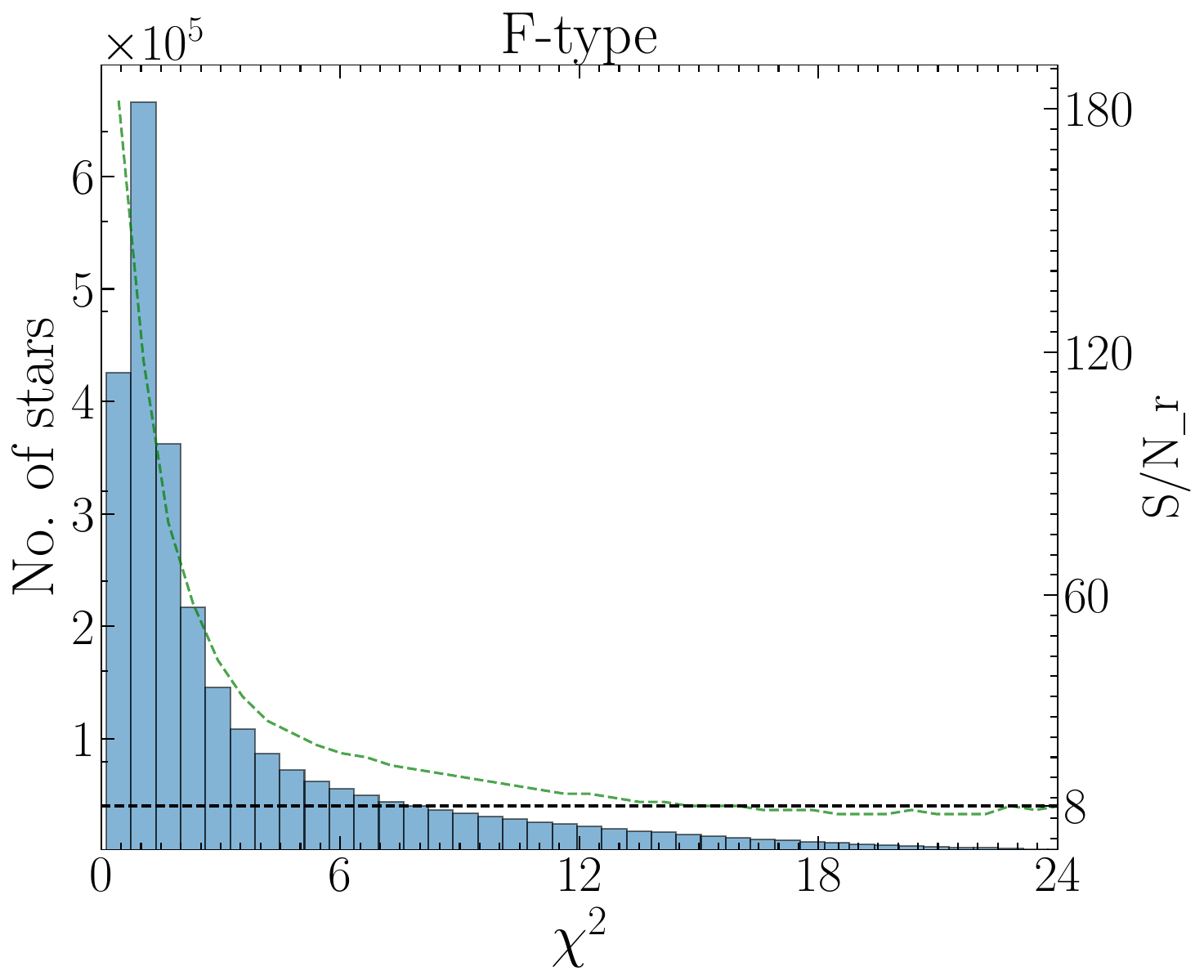}
				\centering
			\end{minipage}
		}
		\centering
		\subfigure{
			\begin{minipage}[t]{0.3\linewidth}
				\centering
				\includegraphics[width=2.1in, height=1.8in]{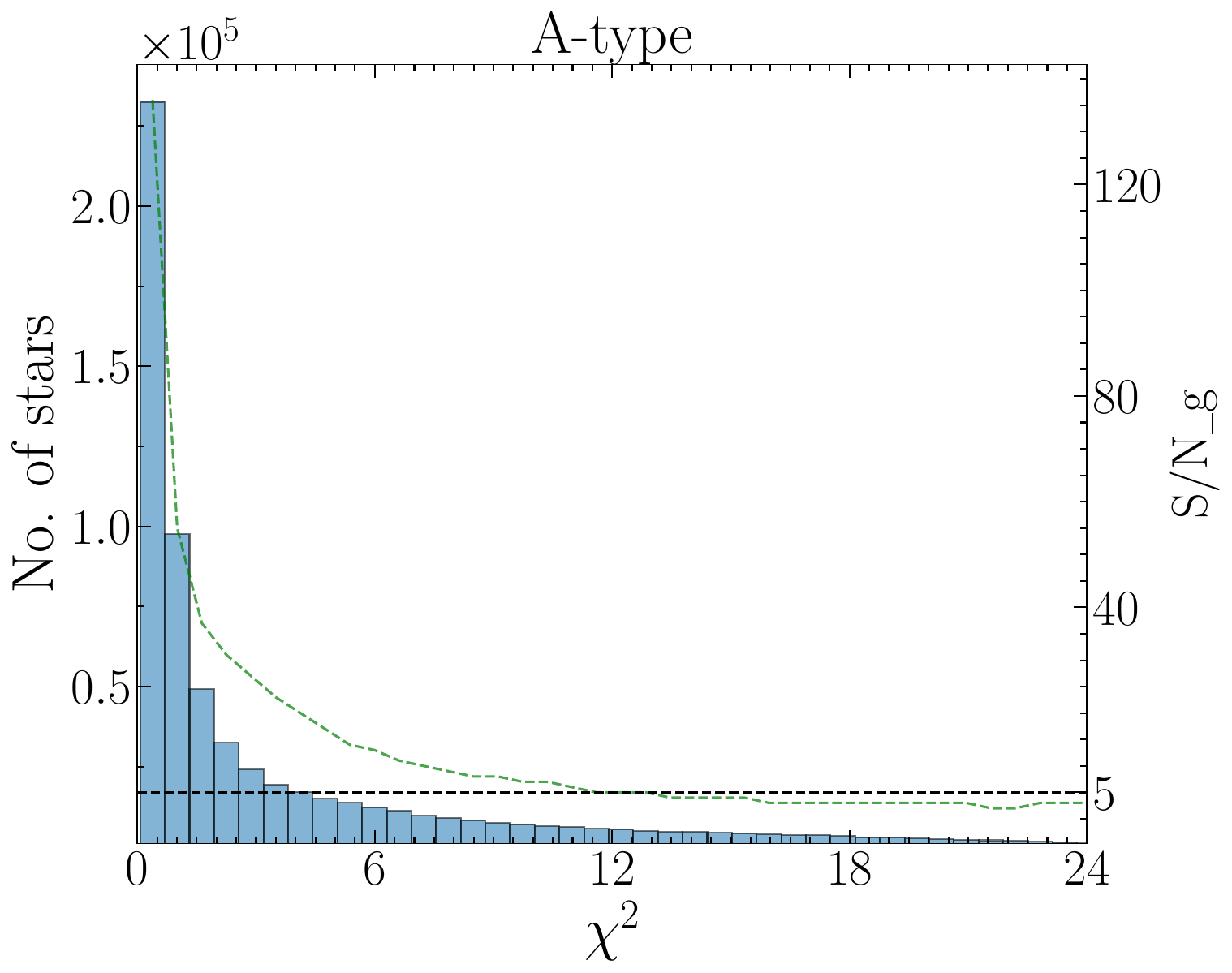}
				\centering
			\end{minipage}
		}
		\centering
		\subfigure{
			\begin{minipage}[t]{0.3\linewidth}
				\centering
				\includegraphics[width=2.1in, height=1.8in]{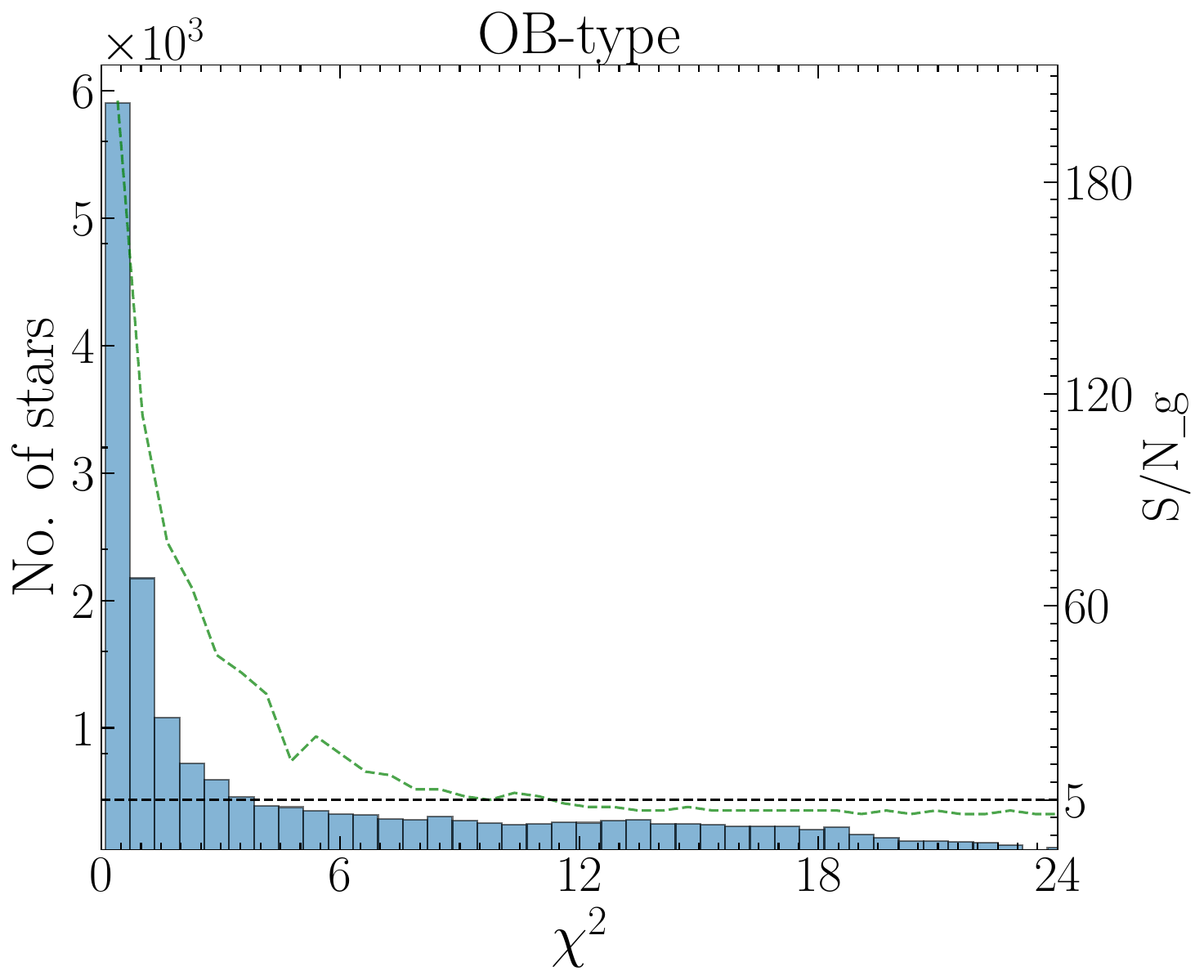}
				\centering
			\end{minipage}
		}
		\caption{Distribution of $\chi^{2}$ values for different spectral types and the variation of S/N with $\chi^{2}$ distribution. The spectral classifications were derived from the LAMOST spectral analysis pipeline \citep{luo2015first}. The horizontal axis represents the $\chi^{2}$ values, the left y-axis of each subplot indicates the number of stars, and the right y-axis represents the S/N. In each subplot, the histogram illustrates the distribution of $\chi^{2}$ values of different spectral types, the green curve denotes the variation of the average S/N with $\chi^{2}$ within each bin, and the black horizontal dashed line indicates the threshold for the S/N. The spectra with S/Ns below this threshold were not included in our recommended catalog.}
		\centering
		\label{loss}
	\end{figure*}
	
	\subsection{The uncertainties of the atmospheric parameters} \label{uncertainties}
	The uncertainties in the derived atmospheric parameters, composed of systematic and random errors, were estimated using statistical methods. They can be estimated as
	\begin{equation}
		\Delta (X)=\sqrt{\Delta _{\text{sys}}^{2}(X)+\Delta _{\text{ran}}^{2}(X)} ,
		\label{error}
	\end{equation}
	where X represents $T_{\text{eff}}$, log $g$, [Fe/H], and [$\alpha$/Fe]. $\Delta _{\text{sys}}$ and $\Delta _{\text{ran}}$ represent the systematic and random errors. 
	
	In previous studies, \citeauthor{xiang2015lamost} (\citeyear{xiang2015lamost}, \citeyear{Xiang_2017}) utilized a quadratic function to establish error models for both systematic and random errors in LAMOST parameters. However, we found that their random error model in relation to $T_{\text{eff}}$ tended to decrease (or increase) monotonically when the S/N $ < 30$ (or S/N $ > 30$). Additionally, we also noted that in comparison to generative models (such as GPR), discriminative models (such as quadratic functions) are unable to capture the uncertainty introduced by error models, which could potentially underestimate the internal errors of the parameters in \citeauthor{xiang2015lamost} (\citeyear{xiang2015lamost}, \citeyear{Xiang_2017}), an aspect that merits careful consideration. Therefore, we opted for GPR to establish the error models (\refformula{sys ran}), aiming to more accurately capture the systematic and random errors of LAMOST parameters.
	\begin{equation}
		\begin{cases}
			\Delta_{\text{sys}}(X)=\text{GPR}_{\text{sys}}(\theta) \\
			\Delta_{\text{ran}}(X)=\text{GPR}_{\text{ran}}(\theta, \text{S/N}),
		\end{cases}
		\label{sys ran}
	\end{equation}
	where X represents $T_{\text{eff}}$, log $g$, [Fe/H], and [$\alpha$/Fe]. $\theta$=\{$T_{\text{eff}}$, log $g$, [Fe/H], [$\alpha$/Fe]\}. $\text{GPR}_{\text{sys}}$ and $\text{GPR}_{\text{ran}}$ represent the GPR models used to establish systematic and random errors, respectively. To derive the uncertainties for all LAMOST atmospheric parameters using \refformula{sys ran}, we first established the following training sets:
	
	\begin{enumerate}
		\item \textbf{Training set for $\text{GPR}_{\text{sys}}$.} The systematic errors are the inherent to the workflow, including contributions from potential issues with the MaStar and errors introduced by the spectral emulator. In alignment with the methodology employed by \cite{xiang2015lamost}, we adopted the prediction errors in atmospheric parameters from the MaStar library as the training set for $\text{GPR}_{\text{sys}}$ (as shown in \reffig{MaStar ELODIE MILES}). In the training set, we noted that the dispersion values of systematic errors for parameters, except for log $g$, increased with increasing $T_{\text{eff}}$. For different type stars, the dispersion values of errors in $T_{\text{eff}}$, log $g$, [Fe/H], and [$\alpha$/Fe] were respectively in the ranges of $40-1400$ K, $0.09-0.22$ dex, $0.08-0.31$ dex, and $0.04-0.16$ dex. This approach is indicative of the theoretical lower bound of error inherent in our workflow when applied to LAMOST.
		\item \textbf{Training set for $\text{GPR}_{\text{ran}}$.} The random errors are induced by the spectral noises, and are functions of the S/N and stellar atmospheric parameters \citep{xiang2015lamost}. Comparing the atmospheric parameters among repeated observations thus gives the training set for $\text{GPR}_{\text{ran}}$. \reffig{random err} illustrates that the random errors of the training set decrease with increasing S/N; the trend appears to be independent of spectral type. For $T_{\text{eff}}$, [Fe/H], and [$\alpha$/Fe], the random error dispersions increase from $35$ to $819$ K, $0.06$ to $0.16$ dex, and $0.02$ to $0.05$ dex respectively with increasing $T_{\text{eff}}$. However, log $g$ random error dispersions are unrelated to spectral type, and aligned with \cite{Hill_2021}. It is notable that the random errors in log $g$ increase for cold giants with S/N $ < 50$. We found that spectra with absolute random errors in log $g$ greater than $0.5$ exhibited strong emission lines or bad pixels that were not recorded in the official LAMOST fits files, and the spectral preprocessing method (\refsubsection{Pre-processing for the spectra}) was not fully effective in removing these anomalies (\reffig{unnormal spectral}), leading to an overestimation of log $g$ (see \refsection{AFGK Type}). Considering that log $g$ of cold giants in MaStar are less than $2$ dex, but our estimates range from $2$ to $4$ (beyond the parameter space of the MaStar), we therefore excluded them from our recommended catalog in \refsubsection{Convergence and quality assessment cuts}. After removing these abnormal spectra, the dispersions of the random errors in log $g$ for M-type and FGK-type decreased from $0.31$ to $0.11$ dex, and $0.25$ to $0.14$ dex, respectively.
	\end{enumerate}
	
	Subsequently, we utilized these training sets to determine the hyperparameters for \refformula{sys ran}, which were then applied to derive the uncertainties for all LAMOST atmospheric parameters. Compared to the above error models using the quadratic functions, we noted that the random errors derived by \refformula{sys ran} decrease with the S/N increasing as shown in \reffig{random err}, and the uncertainties introduced by \refformula{sys ran} for $T_{\text{eff}}$, log $g$, [Fe/H], and [$\alpha$/Fe] are respectively within $60$ K, $0.04$ dex, $0.02$ dex, and $0.01$ dex, which have been incorporated into the internal error.
	
	\begin{table*}[!htb]
		\normalsize
		\centering
		\caption{Descriptions for the stellar label of the recommended catalog}
		\label{stellar label catalog}
		\setlength{\tabcolsep}{3pt}
		\begin{threeparttable}
			\begin{tabular*}{\textwidth}{p{0.25\textwidth} >{\centering\arraybackslash}p{0.5\textwidth} >{\centering\arraybackslash}p{0.25\textwidth}}
				\toprule
				\toprule
				Field & Description & References\\
				\midrule
				obsid & The number ID of this spectrum \\
				uid & Unique source identifier \\
				ra\underline{\hspace{1ex}}obs (deg) & R.A. from the LAMOST DR$10$ catalog (J$2000$) \\
				dec\underline{\hspace{1ex}}obs (deg) & Decl. from the LAMOST DR$10$ catalog (J$2000$) \\
				subclass & The detailed spectral type for each star \\
				snr\underline{\hspace{1ex}}u/g/r/i/z & S/N of $u$, $g$, $r$, $i$, and $z$ bands \\
				z & Redshift from LAMOST fits header \\
				z\underline{\hspace{1ex}}err & Uncertainty in redshift \\
				$T_{\text{eff}}$ (K) & Effective temperature \\
				$T_{\text{eff}}$\underline{\hspace{1ex}}err (K) \tnote{a} & Uncertainty in $T_{\text{eff}}$ \\
				log $g$ (dex) & Surface gravity \\ 
				log $g$\underline{\hspace{1ex}}err (dex) \tnote{a} & Uncertainty in log $g$ \\
				\text{[Fe/H]} (dex) & Metallicity \\ 
				\text{[Fe/H]}\underline{\hspace{1ex}}err (dex) \tnote{a} & Uncertainty in \text{[Fe/H]} \\
				\text{[$\alpha$/Fe]} (dex) & The abundance of alpha elements with respect
				to iron \\
				\text{[$\alpha$/Fe]}\underline{\hspace{1ex}}err (dex) \tnote{a} & Uncertainty in \text{[$\alpha$/Fe]} \\
				log $g$\underline{\hspace{1ex}}Imig (dex) & Replace median log $g$ with see \refsubsection{Future improvements} & \cite{Imig_2022} \\ 
				$\chi^{2}$ & The difference between the observed spectrum and the generated spectrum \\
				sum\underline{\hspace{1ex}}ormask \tnote{b} & Spectral quality flag, $0$ means good quality; other values indicate issues. \\
				FLAG\underline{\hspace{1ex}}S/N & S/N flags for different types of stars, $0$ for recommended, and $1$ is not recommended. \\
				FLAG\underline{\hspace{1ex}}$\chi^{2}$ & $\chi^{2}$ flags for different types of stars, $0$ for recommended, and $1$ is not recommended. \\
				FLAG\underline{\hspace{1ex}}\text{[$\alpha$/Fe]} & \text{[$\alpha$/Fe]} flags for different types of stars, and $0$ for recommended, $1$ is not recommended. \\
				gaia\underline{\hspace{1ex}}id & Gaia DR$3$ Source ID obtained by TOPCAT & \citep{2005ASPC..347...29T,2024arXiv240101156T} \\
				r\underline{\hspace{1ex}}geo & Geometric distance derived from Gaia parallax & \cite{bailer2021estimating} \\
				
				G/BP/RP (mag) & Gaia DR$3$ magnitudes \\
				G\underline{\hspace{1ex}}err/BP\underline{\hspace{1ex}}err/RP\underline{\hspace{1ex}}err (mag) & Uncertainty in Gaia DR$3$ magnitudes \\
				r\underline{\hspace{1ex}}geo\underline{\hspace{1ex}}$16$th & Distance at $16$th percentile of the probability distribution \\
				r\underline{\hspace{1ex}}geo\underline{\hspace{1ex}}$84$th & Distance at $84$th percentile of the probability distribution \\
				ebv\underline{\hspace{1ex}}bayes$19$ & $E(B-V)$ interpolated from the $3$D map using r\underline{\hspace{1ex}}geo & \cite{Green_2019} \\
				
				\bottomrule
			\end{tabular*}
			\begin{tablenotes}
				\item[a] The uncertainty represents the internal error in the predicted values of atmospheric parameters in the recommended catalog, as defined by \refformula{error}.
				\item[b] The ormask is a field in LAMOST fits header, which represents the spectral quality of each wavelength. A value of $0$ signifies high spectral quality, while a nonzero value suggests otherwise \citep{luo2015first}. The sum\underline{\hspace{1ex}}ormask is the sum of ormask at each wavelength, and the value of $0$ and nonzero have the same meanings as the ormask.
			\end{tablenotes}
		\end{threeparttable}
	\end{table*}
	
	\begin{table*}[!htb]
		\normalsize
		\centering
		\caption{Internal Errors and Spectroscopic Statistics in the Recommended Catalog}
		\label{Inner err}
		\begin{tabular*}{\textwidth}{@{\extracolsep{\fill}}cccccccc}
			\toprule
			\toprule
			subclass\tnote{a} & $T_{\text{eff}}$\tnote{b} & log $g$\tnote{b} & [Fe/H]\tnote{2} & [\(\alpha\)/Fe]\tnote{b} & Number & $\chi^{2}$\tnote{b} &  min\underline{\hspace{1ex}}S/N\\
			& (K) & (dex) & (dex) & (dex) & \\
			\midrule
			\multicolumn{8}{c}{log $g$ \( < \) 3.8 dex} \\
			\midrule
			M  & 37 \(\pm\) 17   & 0.41 \(\pm\) 0.27 & 0.07 \(\pm\) 0.04 & 0.03 \(\pm\) 0.01 & 23,843 & 0.70 \(\pm\) 0.56 & 10 \\
			K  & 108 \(\pm\) 82  & 0.35 \(\pm\) 0.20 & 0.04 \(\pm\) 0.02 & 0.02 \(\pm\) 0.01 & 285,490 & 2.27 \(\pm\) 1.75 & 8 \\
			G  & 80 \(\pm\) 57   & 0.23 \(\pm\) 0.14 & 0.05 \(\pm\) 0.04 & 0.02 \(\pm\) 0.01 & 1,195,975 & 1.68 \(\pm\) 1.34 & 8 \\
			F  & 115 \(\pm\) 86  & 0.30 \(\pm\) 0.16 & 0.07 \(\pm\) 0.04 & 0.02 \(\pm\) 0.01 & 88,143 & 2.95 \(\pm\) 1.39 & 8 \\
			A  & 233 \(\pm\) 135 & 0.12 \(\pm\) 0.04 & 0.15 \(\pm\) 0.10 & 0.03 \(\pm\) 0.02 & 34,188 & 0.77 \(\pm\) 0.58 & 5 \\
			OB & 472 \(\pm\) 505 & 0.14 \(\pm\) 0.07 & 0.25 \(\pm\) 0.09 & 0.04 \(\pm\) 0.03 & 783 & 0.53 \(\pm\) 0.32 & 8 \\
			\midrule
			\multicolumn{8}{c}{log $g$ \( \geq \) 3.8 dex} \\
			\midrule
			M  & 15 \(\pm\) 15   & 0.08 \(\pm\) 0.15 & 0.02 \(\pm\) 0.02 & 0.05 \(\pm\) 0.01 & 361,106 & 1.59 \(\pm\) 1.12 & 10 \\
			K  & 33 \(\pm\) 26   & 0.07 \(\pm\) 0.05 & 0.03 \(\pm\) 0.02 & 0.02 \(\pm\) 0.01 & 614,952 & 1.13 \(\pm\) 0.65 & 8 \\
			G  & 25 \(\pm\) 16   & 0.06 \(\pm\) 0.05 & 0.03 \(\pm\) 0.02 & 0.02 \(\pm\) 0.01 & 1,574,741 & 1.41 \(\pm\) 0.66 & 8\\
			F  & 27 \(\pm\) 20  & 0.06 \(\pm\) 0.05 & 0.04 \(\pm\) 0.03 & 0.01 \(\pm\) 0.01 & 1,681,127 & 1.31 \(\pm\) 0.75 & 8\\
			A  & 117 \(\pm\) 94 & 0.05 \(\pm\) 0.03 & 0.09 \(\pm\) 0.07 & 0.03 \(\pm\) 0.02 & 364,815 & 0.84 \(\pm\) 0.71 & 5 \\
			OB & 707 \(\pm\) 594 & 0.08 \(\pm\) 0.05 & 0.21 \(\pm\) 0.08 & 0.05 \(\pm\) 0.04 & 8,387 & 0.81 \(\pm\) 0.69 & 5 \\
			\bottomrule
		\end{tabular*}
	\end{table*}
	\begin{figure*}[htb]
		\centering
		\subfigure{
			\begin{minipage}[t]{0.48\linewidth}
				\centering
				\includegraphics[width=3.2in, height=2.5in]{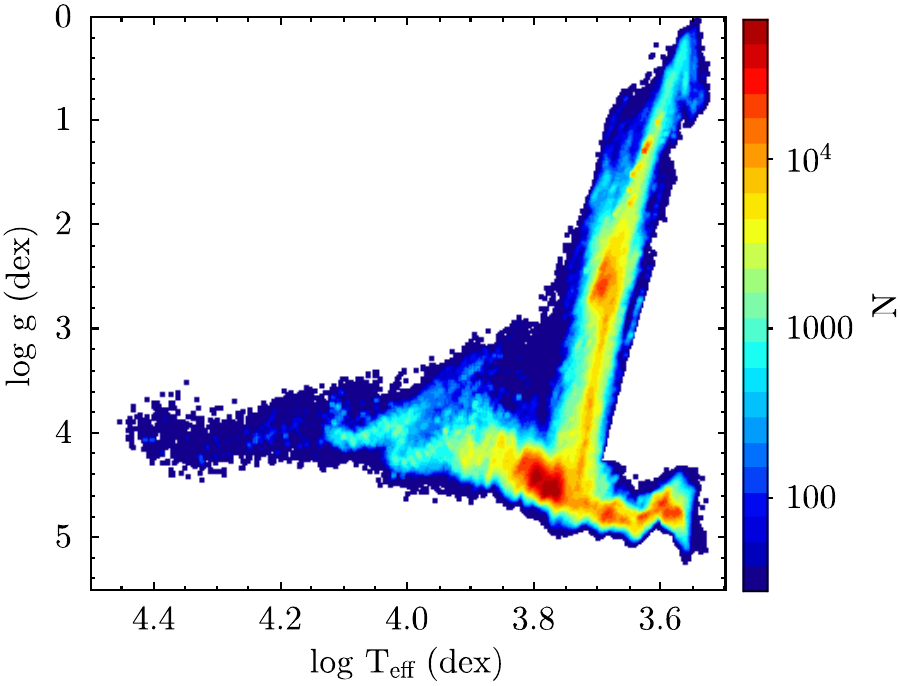}
				\centering
			\end{minipage}
		}
		\centering
		\subfigure{
			\begin{minipage}[t]{0.45\linewidth}
				\centering
				\includegraphics[width=3.2in, height=2.5in]{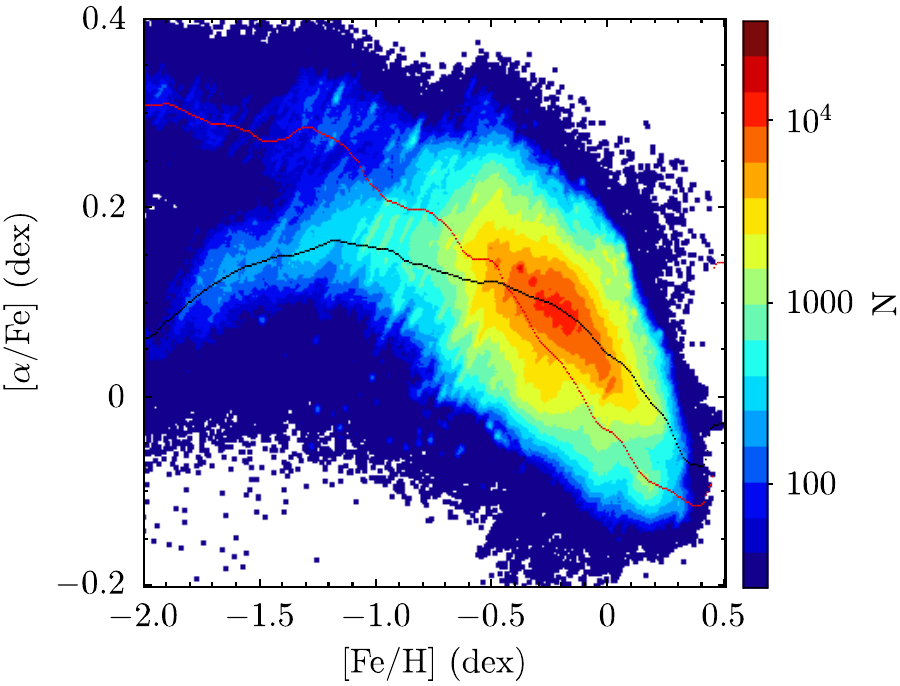}
				\centering
			\end{minipage}
		}
		\caption{The $T_{\text{eff}}$$-$log $g$ (left pannel) and [Fe/H]$-$[$\alpha$/Fe] (right pannel) diagrams for the recommended catalog. Both figures are color-coded by the stellar number density. In the right panel, the red and black curves respectively represent the relationship between the median values of [$\alpha$/Fe] and [Fe/H] for giants and dwarfs. The parameter distribution is similar to that of the MaStar \citep{abdurro2022seventeenth}.}
		\centering
		\label{Teff logg FeH AFe}
	\end{figure*}
	
	\subsection{Convergence and quality assessment cuts} \label{Convergence and quality assessment cuts}
	We provide parameters for $10,344,033$ spectra, or $90\%$ of the full LAMOST DR$10$. The remaining $10\%$ were excluded due to spectral quality issues (S/N $ \le 0 $) or for falling outside the MaStar's atmospheric parameter ranges, where parameter extrapolation is deemed unreliable. Moreover, we proposed a recommended catalog of atmospheric parameters including two stringent quality control indicators.
	\begin{enumerate}
		\item \textbf{FLAG for S/N.} As shown in \reffig{loss}, the distribution of $\chi^{2}$ values for various spectral types is negatively correlated with high S/N. When the S/N is less than a threshold, the $\chi^{2}$ values become unrelated to it. Taking M-type spectra as an example, when S/N $ > 10$, the $\chi^{2}$ values vary with S/N as expected (as the S/N decreases, the $\chi^{2}$ value increases); thus, spectra with S/N $ > 10$ were marked as FLAG\underline{\hspace{1ex}}S/N = $0$. However, when the S/N $ < 10$, the $\chi^{2}$ values become unrelated to it, possibly due to poor spectral quality leading to divergence in our method (unreliable atmospheric parameter predictions); thus, spectra with S/N $ < 10$ were marked as FLAG\underline{\hspace{1ex}}S/N = $1$. As illustrated by the black dashed lines in \reffig{loss}, the same method was applied for O-K-type stars.
		\item \textbf{FLAG for $\chi^{2}$.} When the S/N exceeds the threshold shown in \reffig{loss}, the same atmospheric parameters should correspond to similar spectra, implying similar $\chi^{2}$ values, and a large discrepancy in $\chi^{2}$ values for the same atmospheric parameters indicates inaccurate predictions (possibly due to spectral quality). We established a relationship between $\chi^{2}$ values and atmospheric parameters using a linear regression (LR) model in the parameter space of each spectral type (LR($X$) = $\chi^{2}$). This methodology allows us to apply a more stringent threshold criterion, thereby enhancing the reliability of the recommended  catalog. We set a $\chi^{2}$ threshold as LR($X$), marking data where $\chi^{2}>$ LR($X$) as FLAG\underline{\hspace{1ex}}$\chi^{2}$ = $1$, and the rest as FLAG\underline{\hspace{1ex}}$\chi^{2}$ = $0$.
	\end{enumerate}
	
	In our assessment, we advise against using atmospheric parameter predictions with FLAG\underline{\hspace{1ex}}S/N = $1$ and FLAG\underline{\hspace{1ex}}$\chi^{2}$ = $1$. 
	
	Considering that the spectral features crucial for [$\alpha$/Fe] estimations become less pronounced or vanish, and the absence of high-resolution observational data or recommended catalog for benchmarking [$\alpha$/Fe] of hotter stars, we urge prudence with the [$\alpha$/Fe] for stars with $T_{\text{eff}} > 7000$ K on the upper main sequence. Data in this regime were marked with FLAG\underline{\hspace{1ex}}[$\alpha$/Fe] = $1$, indicating that their use is discouraged, whereas all other data were assigned FLAG\underline{\hspace{1ex}}[$\alpha$/Fe] = $0$. \reftable{stellar label catalog} presents a description of the columns in the recommended catalog \footnote{\url{https://nadc.china-vo.org/res/r101402/}}.
	
	\subsection{The fundamental atmospheric parameters} \label{The fundamental atmospheric parameters}
	
	As illustrated in \reftable{Inner err}, we analyzed the atmospheric parameter uncertainties for giants (log $g < 3.8$) and dwarfs (log $g \ge 3.8$) of different spectral types derived from the the $1$D Pipeline within the recommended catalog. Notably, giants exhibit greater internal errors in log $g$ and [Fe/H] compared to dwarfs, which is likely due to the present limitations in MaStar for giants cooler than $T_{\text{eff}} < 4500$ K. These limitations had also been found in the external comparisons and primarily affect the performance for K and M giants. As the $T_{\text{eff}}$ increase in giants and dwarfs, there is an increase in the uncertainty of $T_{\text{eff}}$, which can be attributed to the sparser features in the spectral energy distribution of hotter stars and to the diminished sensitivity of these features to changes in $T_{\text{eff}}$. The uncertainty in log $g$ appears to be independent of spectral type and seemingly depends on whether a star is a giant or a dwarf. We also see an increase in uncertainty for [Fe/H] and [$\alpha$/Fe] at hotter stars, and this is due to less metal absorption lines which makes it difficult to accurately determine [Fe/H] and [$\alpha$/Fe]. The relationship between the uncertainties in atmospheric parameters in the recommended catalog and spectral types aligns with the results obtained by \cite{Hill_2021,Hill_2022} using Markov Chain Monte Carlo (MCMC), and this consistency suggests that the error model in this work is reasonable.
	
	\reffig{Teff logg FeH AFe} shows the distribution number density in the $T_{\text{eff}}$$-$log $g$ (Kiel diagram) and [Fe/H]$-$[$\alpha$/Fe] (Tinsley$-$Wallerstein diagram) for the recommended catalog. In the left panel, the clarity with which the main sequence, red giant branch, horizontal giant branch, and red clump are discernible within the Kiel diagram effectively illustrates the high precision of the atmospheric parameters we have derived from LAMOST data. In the right panel, the diagram clearly shows a negative correlation between [$\alpha$/Fe] and [Fe/H]  for values of [Fe/H] $\ge -0.5$ dex. For giants (red curve), the tendency for stars of lower [Fe/H] to have higher [$\alpha$/Fe] reflects a natural consequence of Galactic chemical evolution (\citeauthor{Xiang_2017} \citeyear{Xiang_2017}, \citeyear{xiang2019abundance}). Moreover, in addition to the dominant negative correlation sequence in the Tinsley$-$Wallerstein diagram, giants also display an additional sequence of thick disk stars with [$\alpha$/Fe] of approximately $0.3$ dex, consistent with \cite{adibekyan2011new}. Notably, the thick disk sequence is not seen in dwarfs (black curve) when [Fe/H] $ < -1.5$ dex; instead, the diagram reveals a positive correlation between [$\alpha$/Fe] and [Fe/H]. For dwarfs with [Fe/H] $ < -1.5$ dex, we identified that $99\%$ of the stars were classified as AF-type ($6300 < T_{\text{eff}} < 7000$ K), exhibiting spectra with strong emission lines or bad pixels not recorded in the official LAMOST fits files. The spectral preprocessing method (\refsubsection{Pre-processing for the spectra}) was not fully effective in removing these anomalies (\reffig{unnormal spectral}), resulting in an overestimation of [$\alpha$/Fe] by $\sim0.1$ dex.
	
	\subsection{External Comparison} \label{External Comparison}
	We compared the atmospheric parameters in the recommended catalog with external data based on spectral types from $1$D Pipeline. This also addressed issues present in LASPM and LASP, providing a reliable catalog for establishing the empirical spectral library of LAMOST. Additionally, we highlighted the shortcomings of the recommended catalog, which are inherited from the MaStar, offering insights for future corrections in MaStar's parameters.
	\subsubsection{M-type Stars: Recommended Catalog versus APOGEE and Compensating LASPM} \label{M Type}
	\begin{figure*}[htb]
		\centering
		\subfigure{
			\begin{minipage}[t]{0.3\linewidth}
				\centering
				\includegraphics[width=2.1in, height=1.8in]{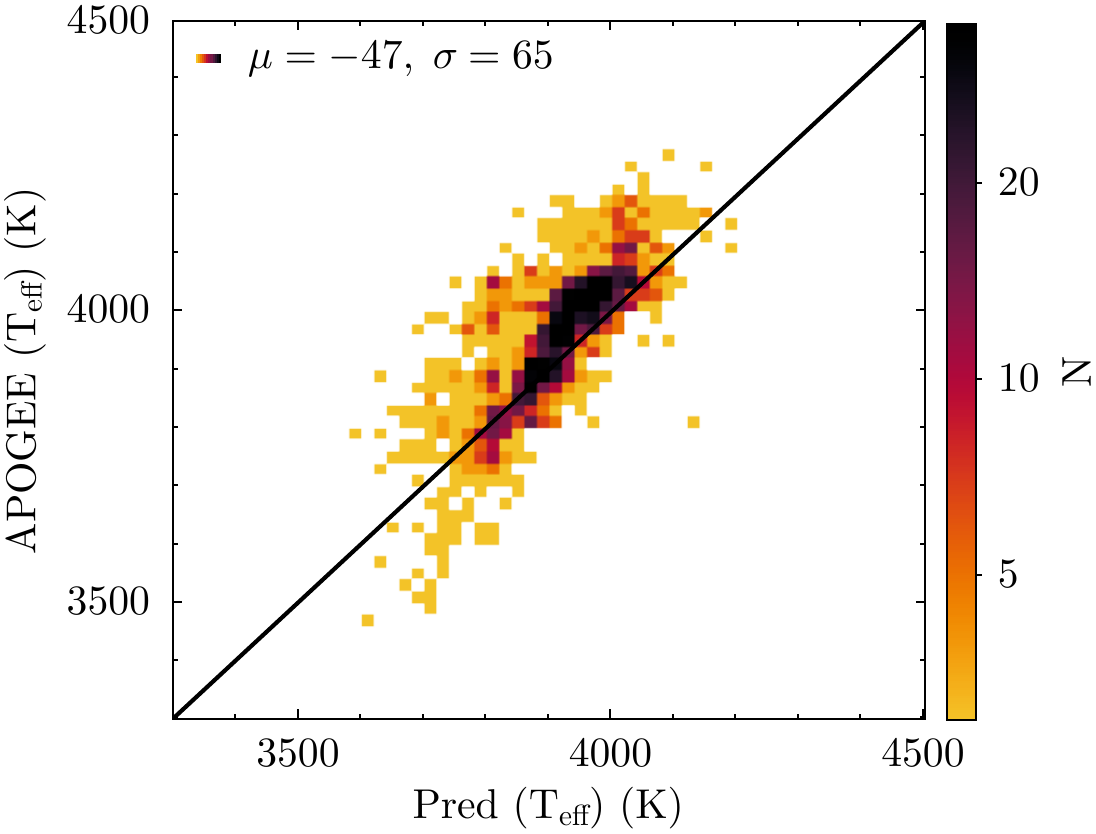}
				\centering
			\end{minipage}
		}
		\centering
		\subfigure{
			\begin{minipage}[t]{0.3\linewidth}
				\centering
				\includegraphics[width=2.1in, height=1.8in]{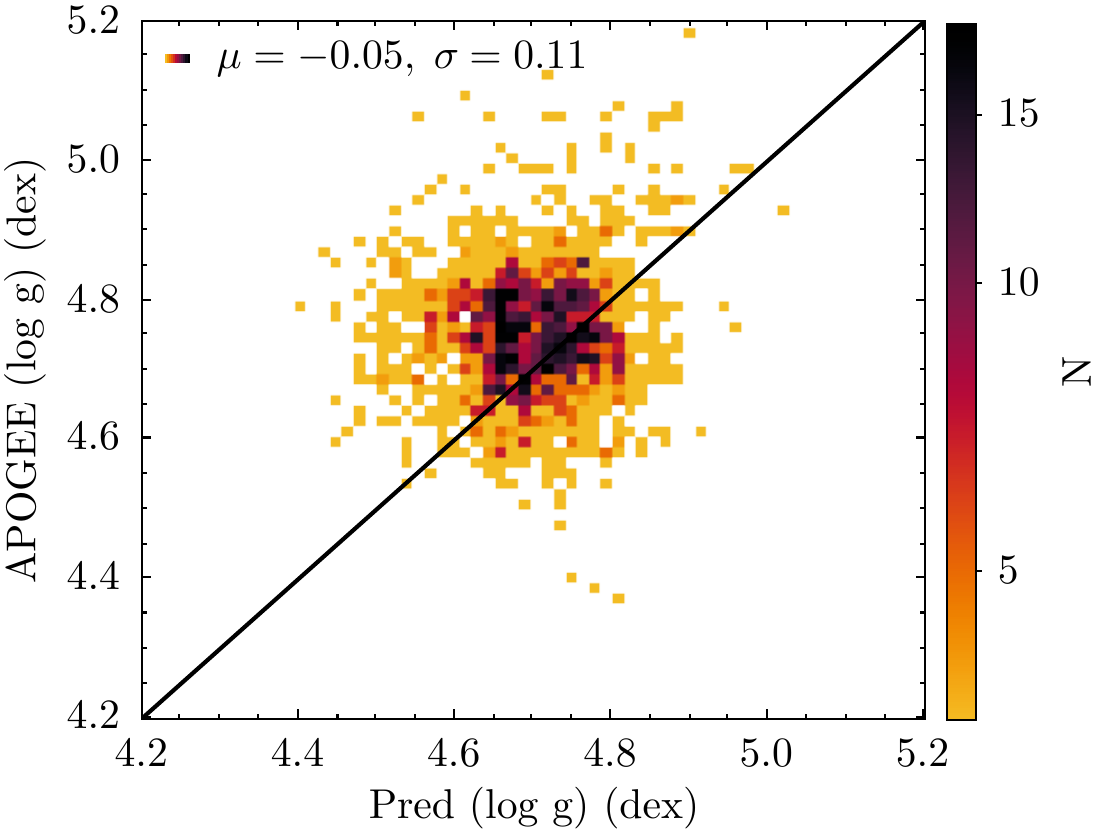}
				\centering
			\end{minipage}
		}
		\centering
		\subfigure{
			\begin{minipage}[t]{0.3\linewidth}
				\centering
				\includegraphics[width=2.1in, height=1.8in]{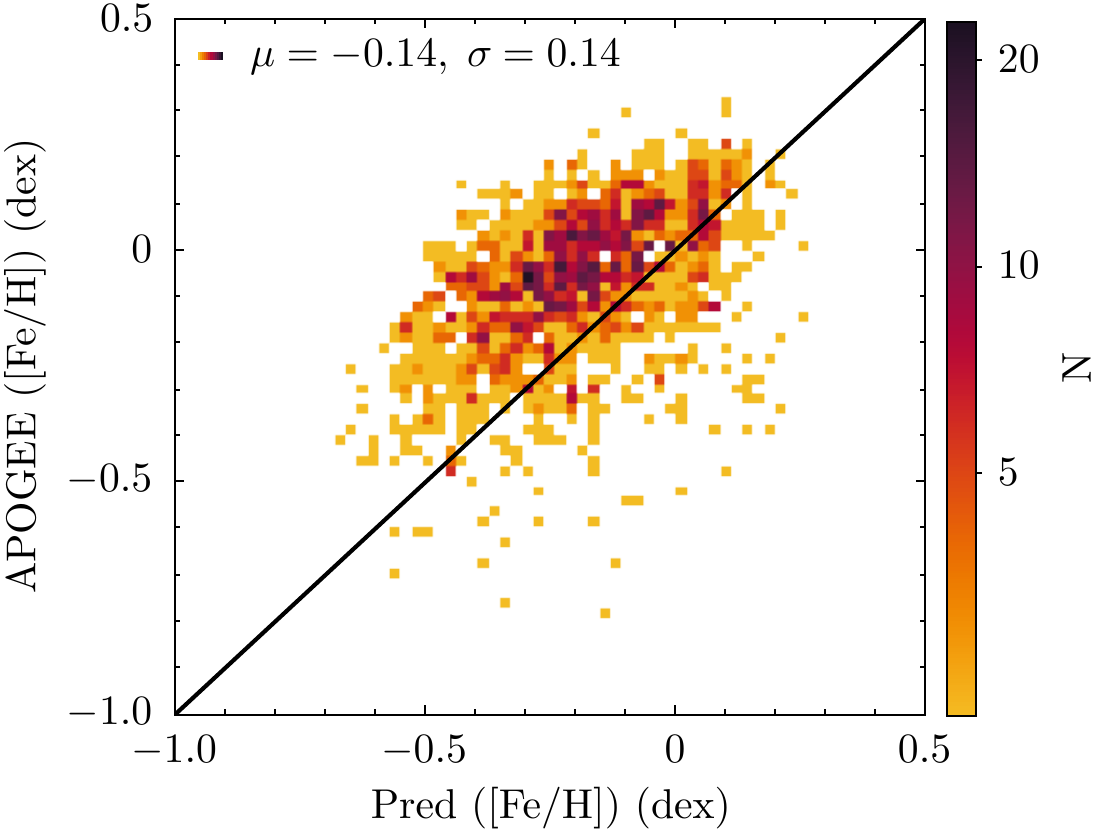}
				\centering
			\end{minipage}
		}
		\caption{Comparison of the dM parameters in the recommended catalog with APOGEE DR$16$. Mean and standard deviation of the parameter differences between the recommended catalog and APOGEE are also shown in the plot.}
		\centering
		\label{dM Teff logg FeH}
	\end{figure*}
	\begin{figure*}[htb]
		\centering
		\subfigure{
			\begin{minipage}[t]{0.3\linewidth}
				\centering
				\includegraphics[width=2.1in, height=1.8in]{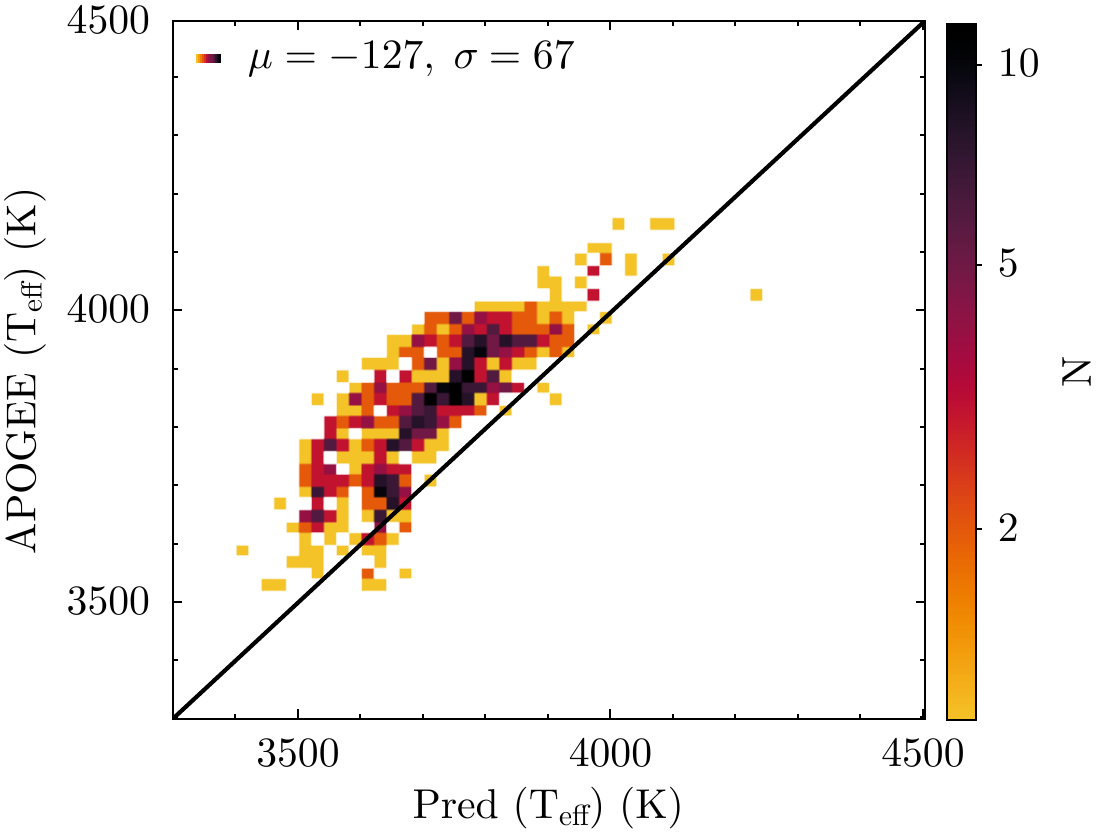}
				\centering
			\end{minipage}
		}
		\centering
		\subfigure{
			\begin{minipage}[t]{0.3\linewidth}
				\centering
				\includegraphics[width=2.1in, height=1.8in]{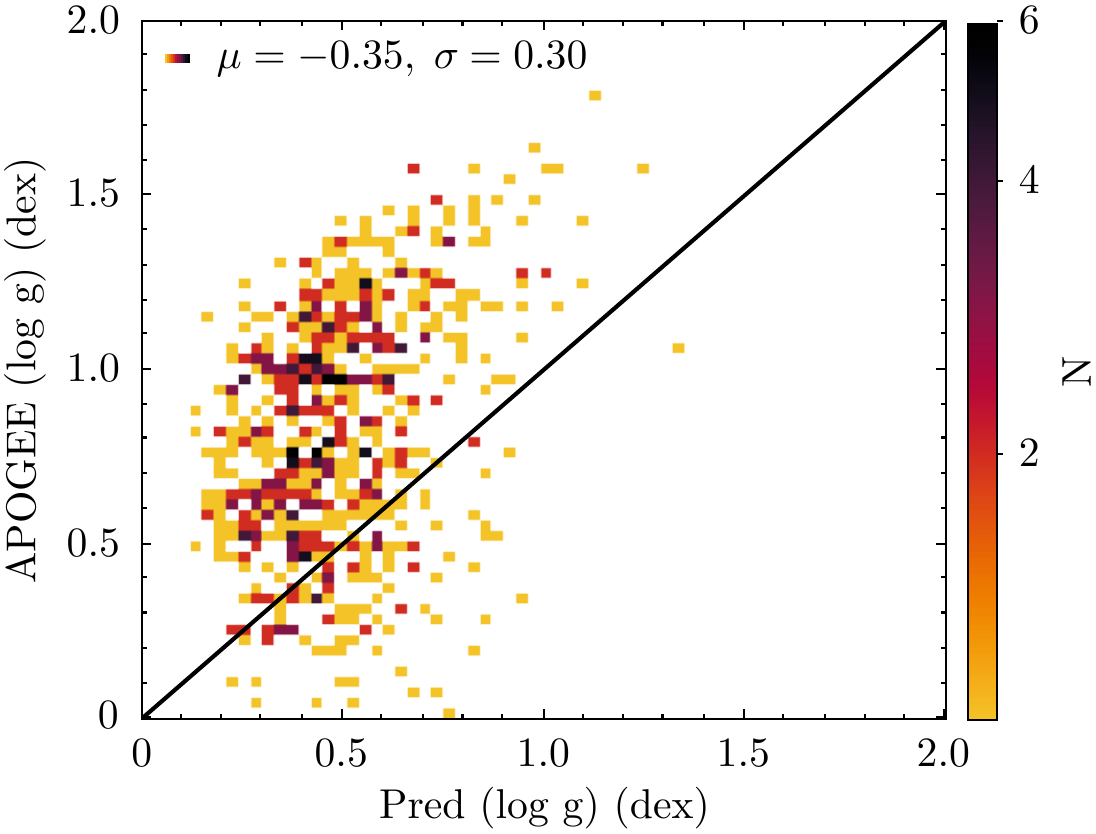}
				\centering
			\end{minipage}
		}
		\centering
		\subfigure{
			\begin{minipage}[t]{0.3\linewidth}
				\centering
				\includegraphics[width=2.1in, height=1.8in]{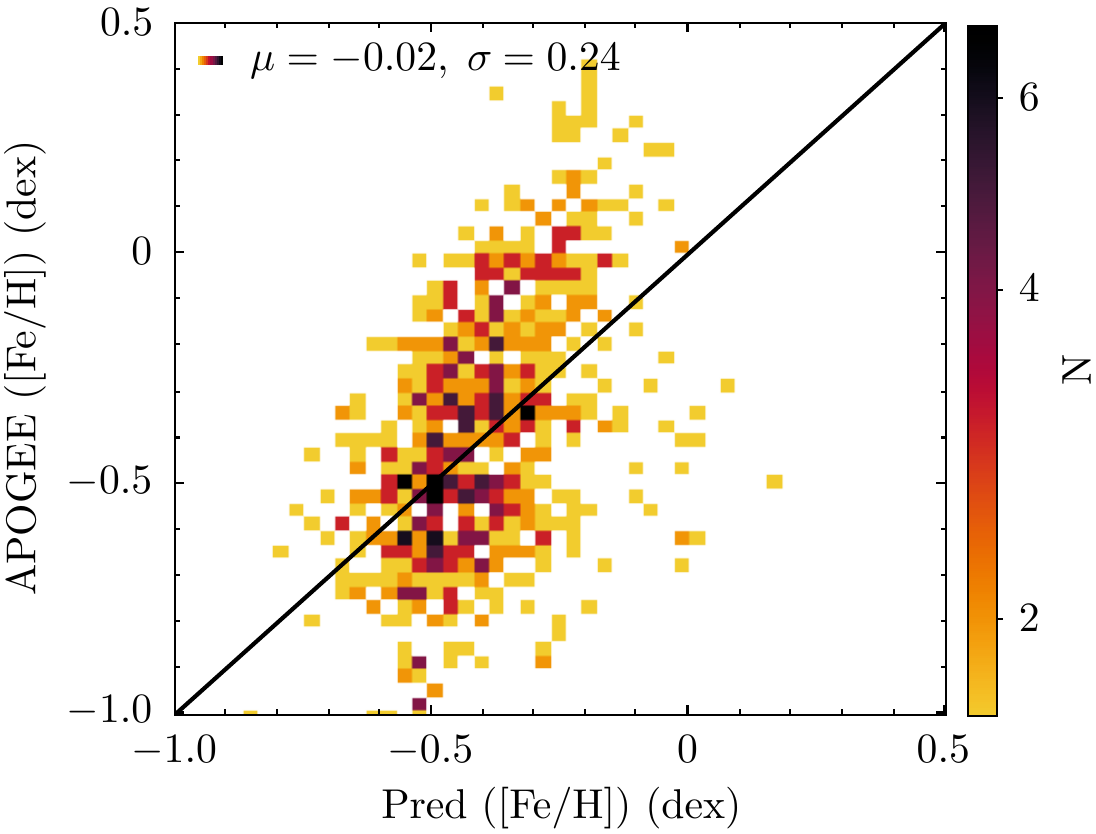}
				\centering
			\end{minipage}
		}
		\caption{Comparison of the gM parameters in the recommended catalog with APOGEE DR16. Mean and standard deviation of the parameter differences between the recommended catalog and APOGEE are also shown in the plot.}
		\centering
		\label{gM Teff logg FeH}
	\end{figure*}
	\begin{figure*}[htb]
		\centering
		\subfigure{
			\begin{minipage}[t]{0.3\linewidth}
				\centering
				\includegraphics[width=2.1in, height=1.8in]{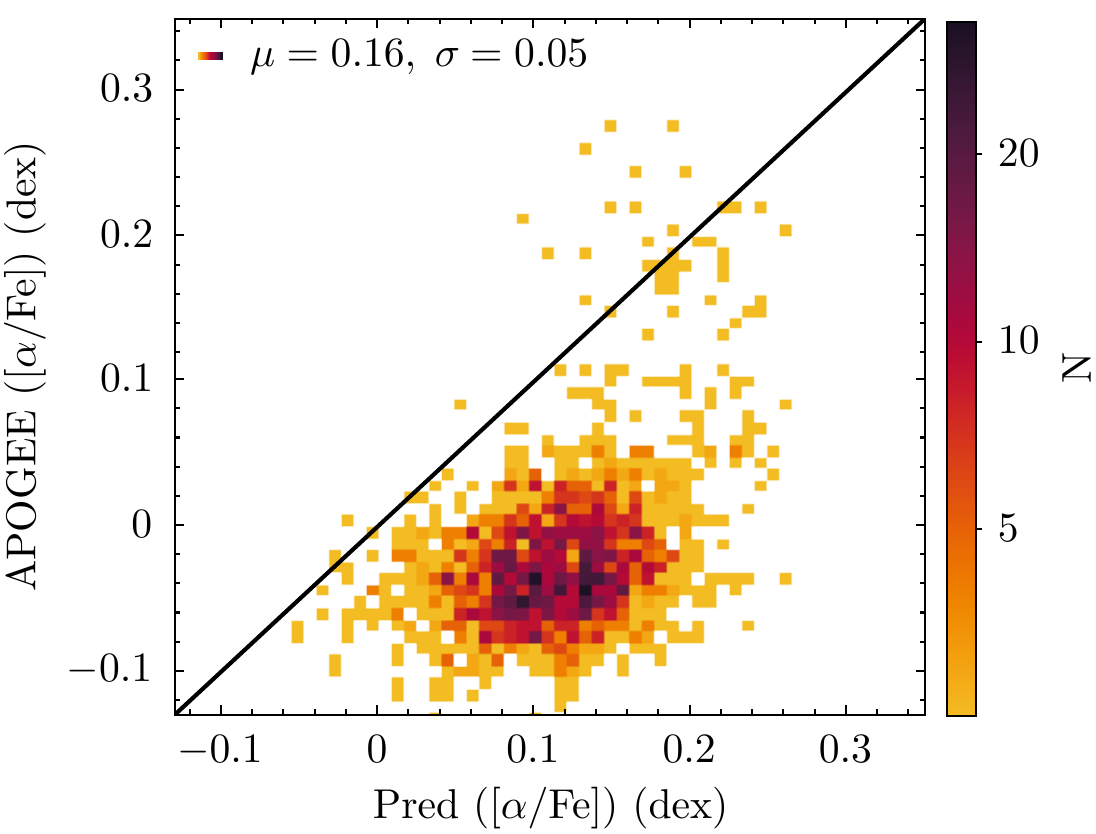}
				\centering
			\end{minipage}
		}
		\centering
		\subfigure{
			\begin{minipage}[t]{0.3\linewidth}
				\centering
				\includegraphics[width=2.1in, height=1.8in]{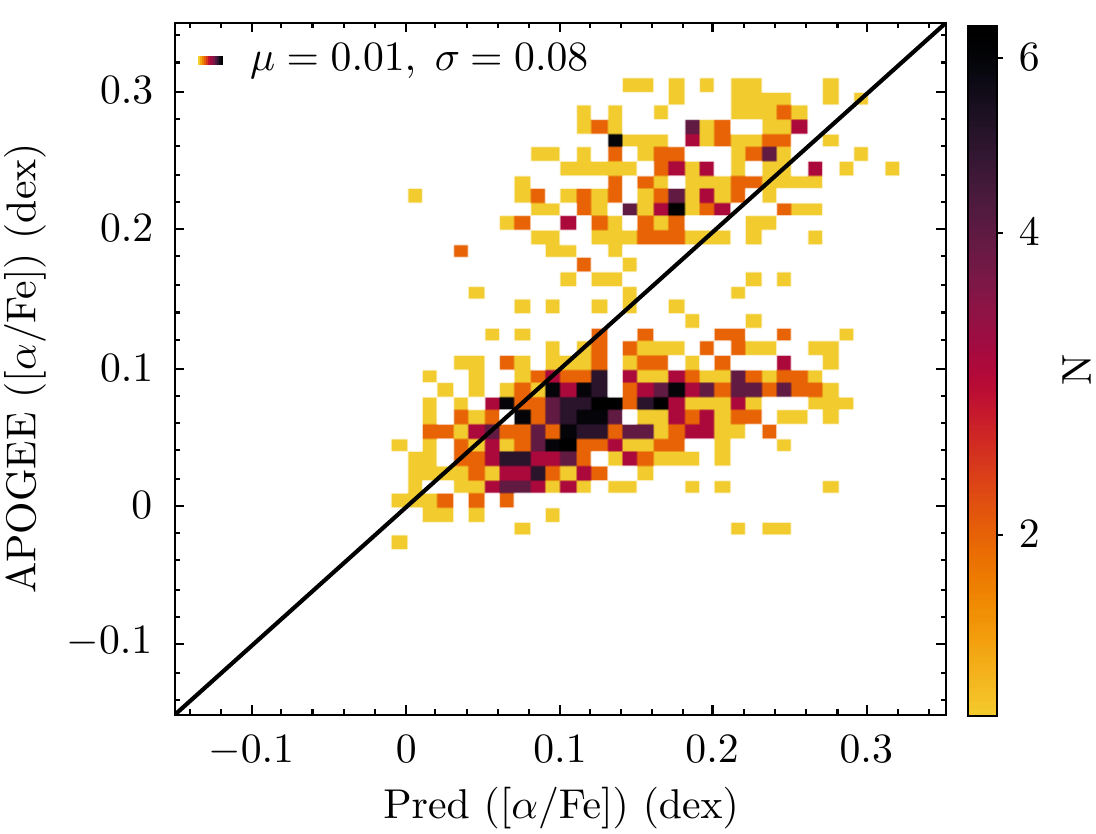}
				\centering
			\end{minipage}
		}
		\centering
		\subfigure{
			\begin{minipage}[t]{0.3\linewidth}
				\centering
				\includegraphics[width=2.1in, height=1.8in]{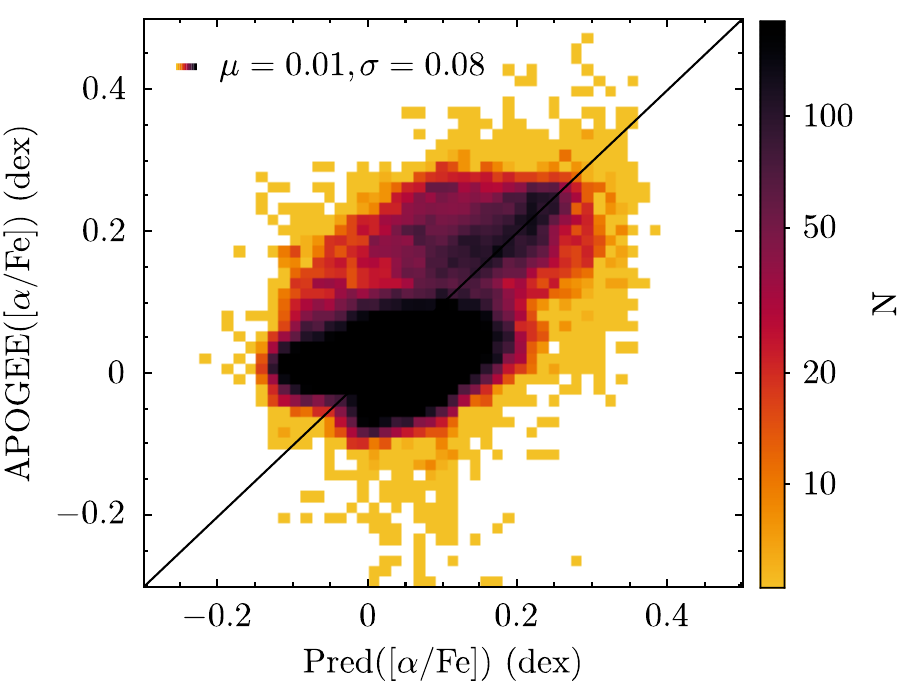}
				\centering
			\end{minipage}
		}
		\caption{Comparison of the [$\alpha$/Fe] in the recommended catalog with APOGEE DR16 in dM (left panel), gM (middle panel) and FGK-type (right panel) stars. Mean and standard deviation of the parameter differences between the recommended catalog and APOGEE are also shown in the plot.}
		\centering
		\label{dM gM FGK AFe}
	\end{figure*}
	\begin{figure*}[htb]
		\centering
		\subfigure{
			\begin{minipage}[t]{0.3\linewidth}
				\centering
				\includegraphics[width=2.1in, height=1.8in]{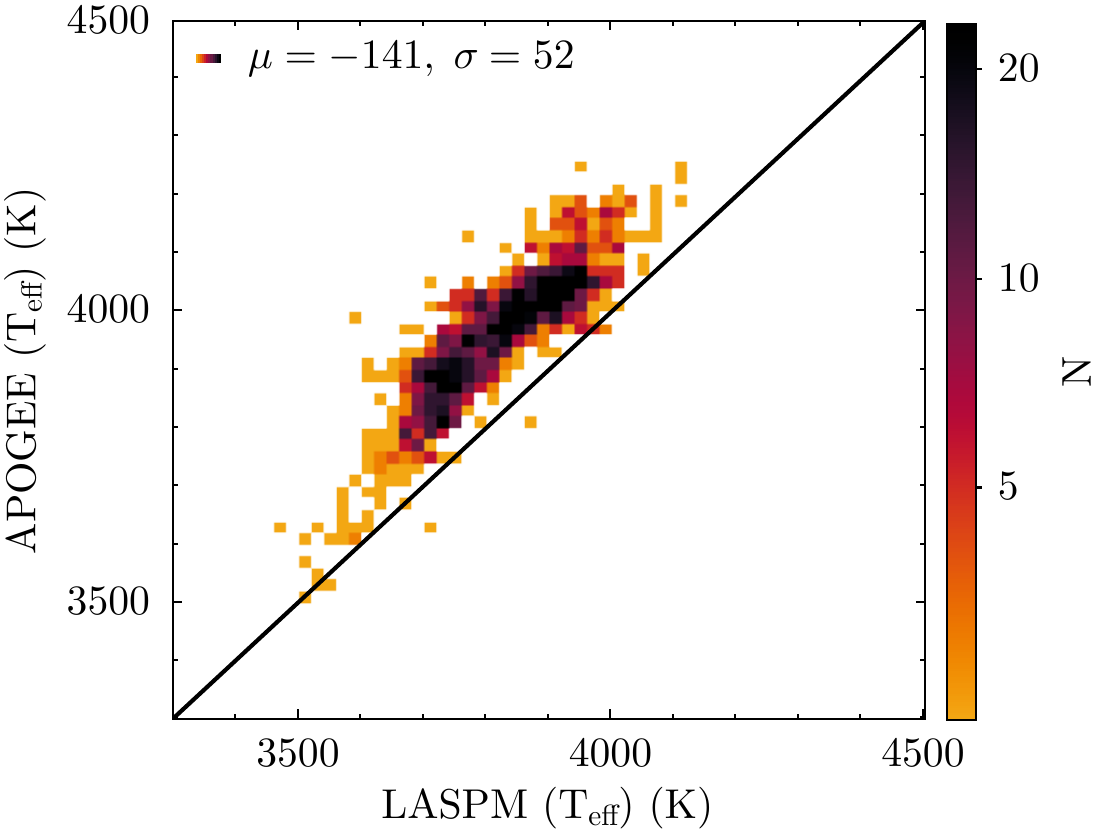}
				\centering
			\end{minipage}
		}
		\centering
		\subfigure{
			\begin{minipage}[t]{0.3\linewidth}
				\centering
				\includegraphics[width=2.1in, height=1.8in]{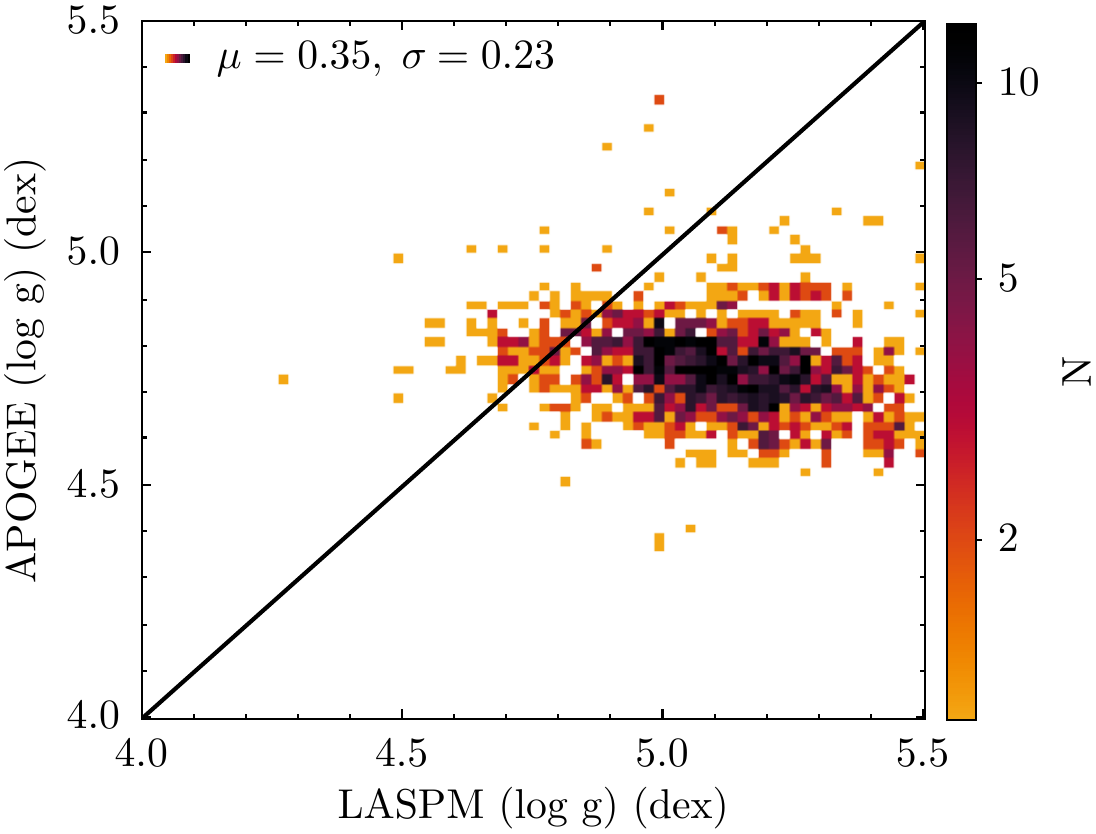}
				\centering
			\end{minipage}
		}
		\centering
		\subfigure{
			\begin{minipage}[t]{0.3\linewidth}
				\centering
				\includegraphics[width=2.1in, height=1.8in]{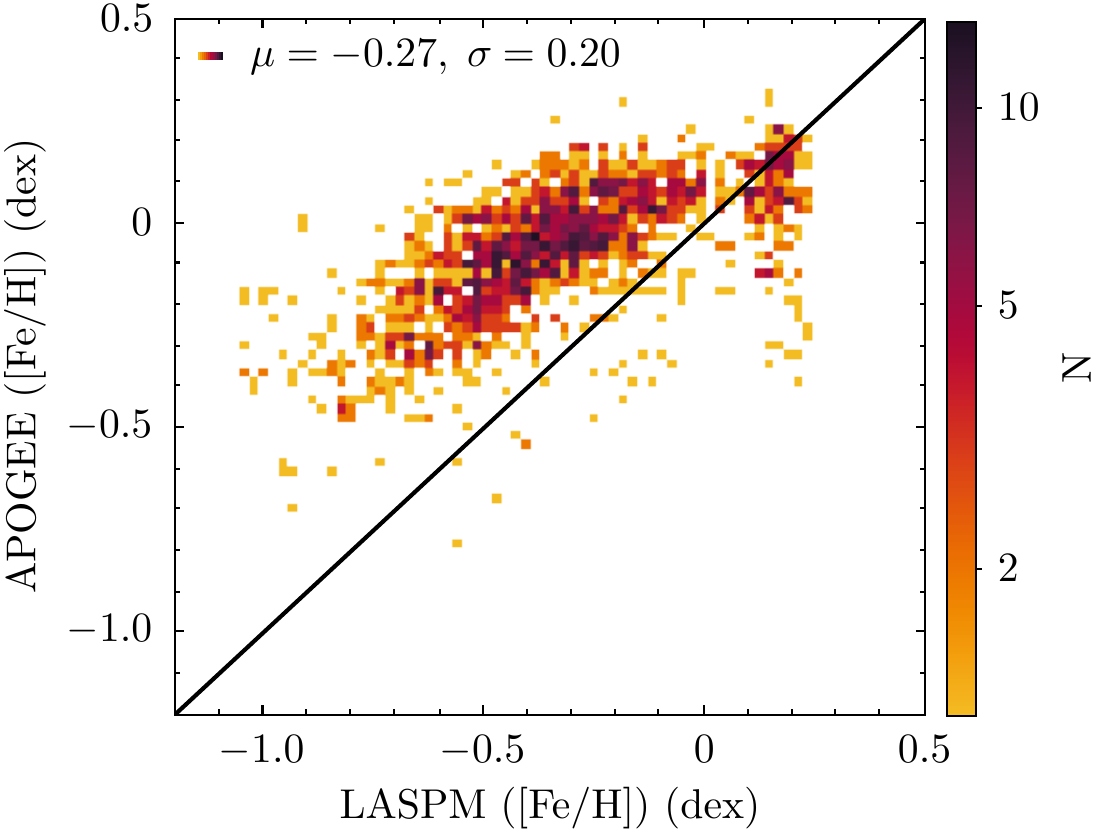}
				\centering
			\end{minipage}
		}
		\caption{Comparison of the dM parameters in LASPM with APOGEE DR$16$. Mean and standard deviation of the parameter differences between LASPM and APOGEE are also shown in the plot.}
		\centering
		\label{dM LASPM Teff logg FeH}
	\end{figure*}
	\begin{figure*}[htb]
		\centering
		\subfigure{
			\begin{minipage}[t]{0.3\linewidth}
				\centering
				\includegraphics[width=2.1in, height=1.8in]{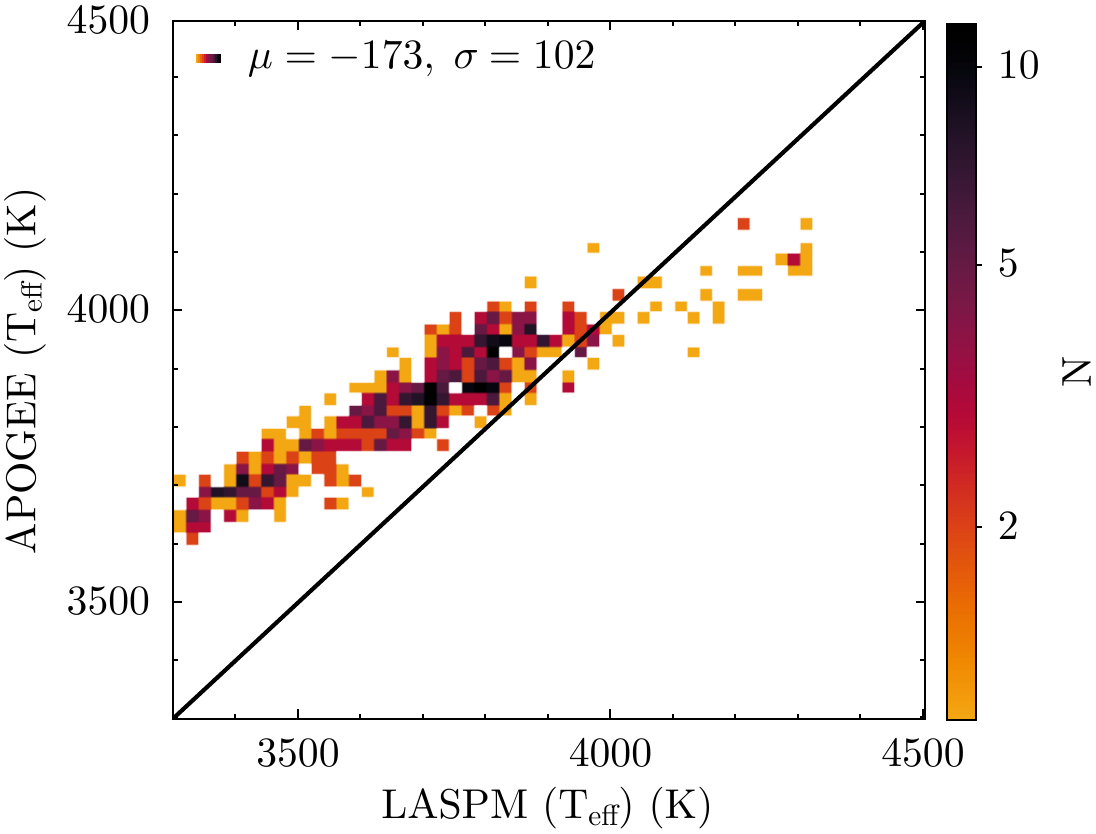}
				\centering
			\end{minipage}
		}
		\centering
		\subfigure{
			\begin{minipage}[t]{0.3\linewidth}
				\centering
				\includegraphics[width=2.1in, height=1.8in]{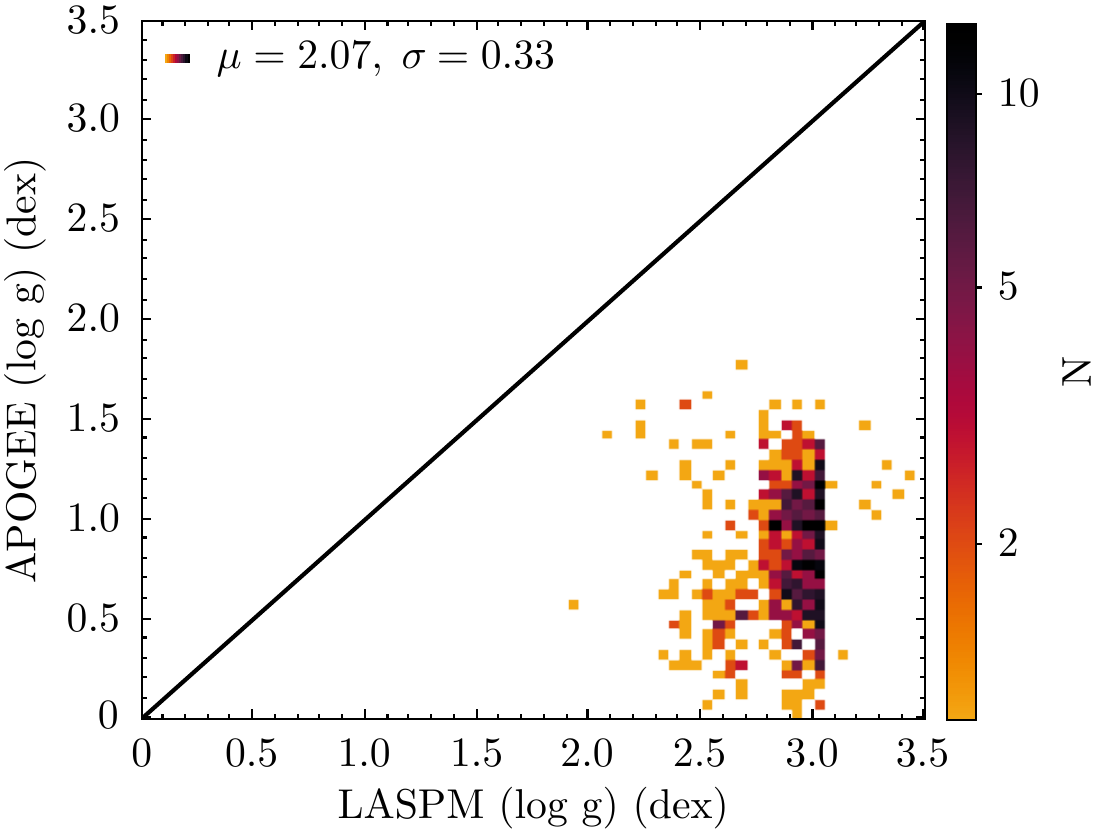}
				\centering
			\end{minipage}
		}
		\centering
		\subfigure{
			\begin{minipage}[t]{0.3\linewidth}
				\centering
				\includegraphics[width=2.1in, height=1.8in]{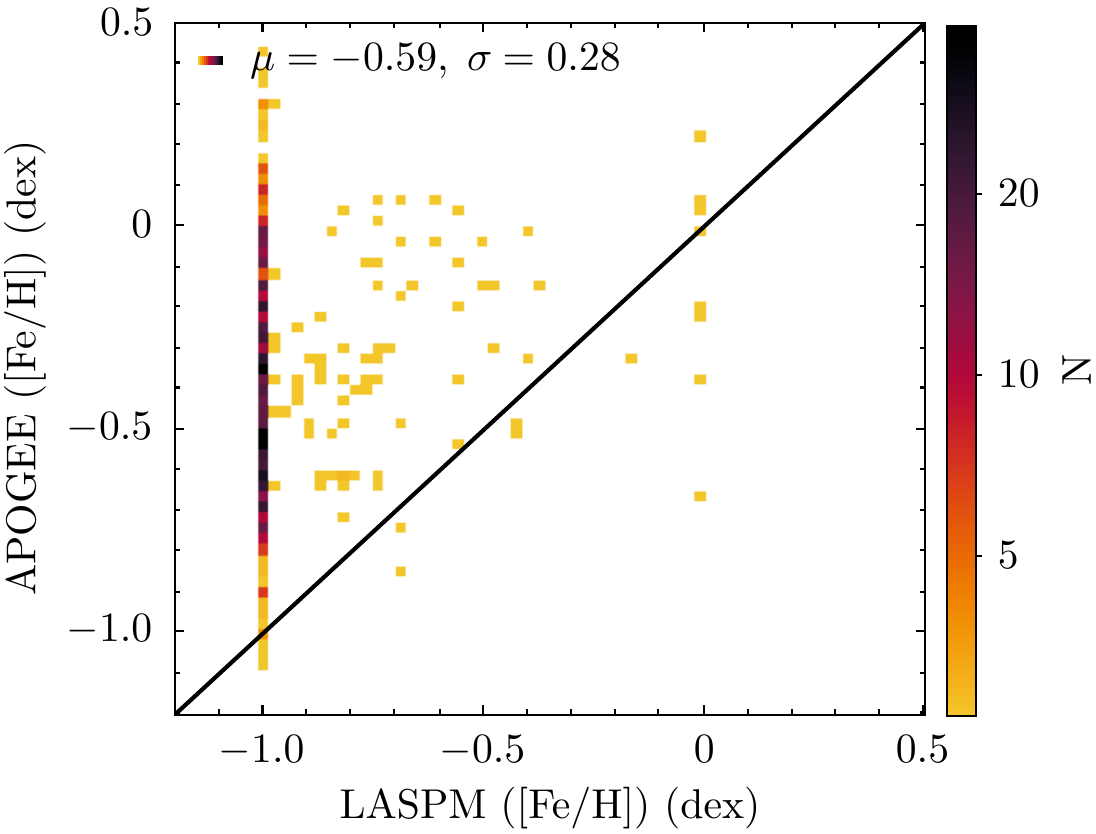}
				\centering
			\end{minipage}
		}
		\caption{Comparison of the gM parameters in LASPM with APOGEE DR$16$. Mean and standard deviation of the parameter differences between LASPM and APOGEE are also shown in the plot.}
		\centering
		\label{gM LASPM Teff logg FeH}
	\end{figure*}
	APOGEE is one of the programs in SDSS $\text{III}$ (\citealt{Majewski2017}) and SDSS $\text{IV}$ (\citealt{blanton2017sloan}), and it is a high-resolution ($R \sim 22,500$), near-infrared (NIR; $\lambda=1.51 - 1.70 \ \mu m$), multiepoch, large spectroscopic survey of the Milky Way (\citealt{santana2021final}). The survey expected to exceed $\sim 700,000$ stars, mainly consisting of Galactic red giants from all stellar populations, as well as the Magellanic Clouds and other nearby dwarf galaxy red giants, and numerous cool (FGKM) dwarf stars (\citealt{zasowski2013target, zasowski2017target}; \citealt{smith2021apogee}). APOGEE Data Release $16$ (APOGEE DR$16$) contains approximately $430,000$ stars covering both the northern and southern sky, from which radial velocities, atmospheric parameters, and chemical abundances of up to $26$ species were determined \citep{2020AJ....160..120J}. We crossmatched our M-type stars with APOGEE DR$16$ allStar-file, following the criteria with ASPCAPFLAG = $0$ and STARFLAG = $0$. We obtained a total of $2420$ M-type dwarfs (dM) and $747$ M-type giants (gM). 
	
	In our analysis (as shown in \reffig{dM Teff logg FeH}, \reffig{gM Teff logg FeH} and \reffig{dM gM FGK AFe}), we noted that the parameters of $T_{\text{eff}}$, log $g$ and [Fe/H] show good agreements with APOGEE data, although there are some systematic offsets. Specifically, for dM, we reported offsets and dispersions of $-47\pm65$ K, $-0.05\pm0.11$ dex, and $-0.14\pm0.14$ dex, respectively. In contrast, gM exhibit offsets and dispersions of $-127\pm67$ K, $-0.35\pm0.30$ dex, and $-0.02\pm0.24$ dex. Notably, [$\alpha$/Fe] predictions for gM stars are more accurate, showing a minor offset of $0.01$ dex and a dispersion of $0.08$ dex relative to APOGEE, compared to a $0.16$ dex offset and $0.05$ dex dispersion for dM stars. We attribute the systematic underestimation of log $g$ predictions by an average of $0.35$ dex relative to APOGEE to potential unreliability of the median log $g$ values in the MaStar, as described in \refsubsection{Future improvements}. The systematic discrepancies in other parameters are likely attributed to the intrinsic differences between NIR and optical spectroscopy. Similar systematic discrepancies were observed in \cite{Li_2021}, who employed the BT-Stell interpolator for $T_{\text{eff}}$ prediction, and in \cite{Ding_2022}, who used the MILES interpolator for estimating three fundamental parameters, when compared with APOGEE data. Our results demonstrated a closer alignment of dM parameters with APOGEE, especially in log $g$, than those reported by \cite{Ding_2022}. Additionally, our gM parameter predictions attain a level of consistency with APOGEE comparable to \cite{Ding_2022}, suggesting that MaStar's M-type parameters within the current empirical spectral library are reliable and match the consistency level of the MILES interpolator with APOGEE.
	
	The LASPM extends the capability of the LASP dealing with cool stars \citep{Du_2021}. Currently, the gM atmospheric parameters provided by this method exhibit certain issues \citep{Qiu_2023}. Validating whether the atmospheric parameters of M-type stars in our recommended catalog can improve the reliability of LASPM is crucial for the subsequent establishment of the LAMOST empirical spectral library. 
	
	We crossmatched these $2420$ dM and $747$ gM with LASPM, and obtained $2046$ dM and $746$ gM common stars with S/N of $i$ band greater than $30$. As illustrated in \reffig{dM LASPM Teff logg FeH} and \reffig{gM LASPM Teff logg FeH}, a comparative analysis between the LASPM and APOGEE data for dM and gM stars reveals certain issues that need to be addressed in LASPM's estimation of log $g$ and [Fe/H]. There, systematic offsets and dispersions for dM are $0.35\pm0.23$ dex,
	and $-0.27\pm0.20$ dex, respectively, and even more pronounced for gM, reaching $2.07\pm0.33$ dex and $-0.59\pm0.28$ dex. These differences are likely attributable to two primary factors:
	\begin{enumerate}
		\item \textbf{Discrepancies between theoretical and observed spectra in M-type stars.} The current theoretical models are inadequate in adequately reproducing the spectra of dM \citep{Kurucz_1979,Dupree_2016}. For example, BT-Settl synthetic spectra encounters difficulties in reproducing the K- and Na-line pairs sensitive to log $g$ and [Fe/H] \citep{rajpurohit2014high, passegger2016fundamental}, while MARCS synthetic spectra \citep{gustafsson2008grid} consistently fail to reproduce the NaD line wings, TiO band head fluxes, several broad absorption features (e.g. $\sim4400-4500$ and $\sim5500-5600$ \text{\AA}), and the wavelength region caused by a spectral depression centred on the $\sim4227$ \text{\AA} neutral Ca resonance line in cool star spectra \citep{Rains_2024}. The atmospheres of these cold stars are rich in molecules like water, methane, and titanium oxide, with transitions not adequately understood or included in many contemporary line lists \citep{Imig_2022}. Furthermore, LASPM exhibits larger discrepancies when using BT-Settl to reproduce gM spectra compared to dM spectra, which may contribute to the lower reliability of gM parameters in LASPM. Du et al. (2024, in preparation) proposes combining the strengths of different theoretical spectral libraries to mitigate the impact of discrepancies with observed spectra.
		\item \textbf{The inherent limitations of linear interpolation.} \cite{du2012comparison} compared the capabilities of nonlinear interpolators with linear interpolators for atmospheric parameter prediction, particularly for stars outside the $4000$-$8000$ K range, and found nonlinear interpolators to be more reliable. The efficacy of linear interpolators is correlated with the grid step size and uniformity. In LASPM, the grid step size in log $g$ and [Fe/H] is $0.5$ dex, which also increases the possibility of linear interpolators falling into local optima of $\chi^{2}$.
	\end{enumerate}
	
	Combining the outcomes from \reffig{dM Teff logg FeH} and \reffig{gM Teff logg FeH} (the recommended catalog versus APOGEE) and \reffig{dM LASPM Teff logg FeH} and \reffig{gM LASPM Teff logg FeH} (LASPM versus APOGEE), our predictions align more consistency with APOGEE, especially regarding  log $g$ and [Fe/H]. Relative to the comparison of LASPM versus APOGEE in log $g$ and [Fe/H] (as shown in \reffig{dM LASPM Teff logg FeH} and \reffig{gM LASPM Teff logg FeH}), our recommended catalog versus APOGEE for dM stars demonstrates a reduction in the absolute values of offsets by $0.3$ and $0.13$ dex, respectively, and a decrease in dispersion values by $0.12$ and $0.06$ dex. For gM stars, the reductions are $1.72$ and $0.57$ dex in offsets, and $0.03$ and $0.04$ dex in dispersion values. Consequently, the recommended catalog effectively improves certain issues present in LASPM.
	\subsubsection{AFGK-type Stars: Recommended Catalog versus APOGEE, PASTEL, LAMOST and Compensating A-type Parameters in LASP} \label{AFGK Type}
	\begin{figure*}[htb]
		\centering
		\subfigure{
			\begin{minipage}[t]{0.3\linewidth}
				\centering
				\includegraphics[width=2.1in, height=1.8in]{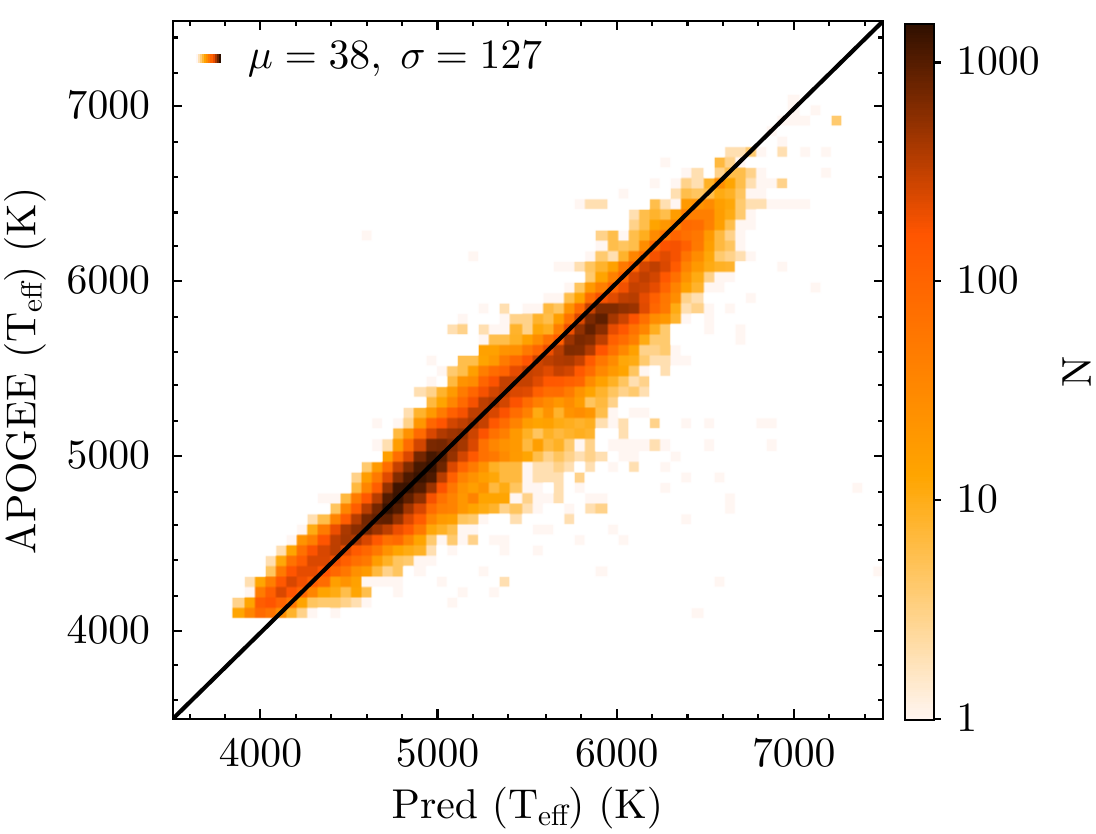}
				\centering
			\end{minipage}
		}
		\centering
		\subfigure{
			\begin{minipage}[t]{0.3\linewidth}
				\centering
				\includegraphics[width=2.1in, height=1.8in]{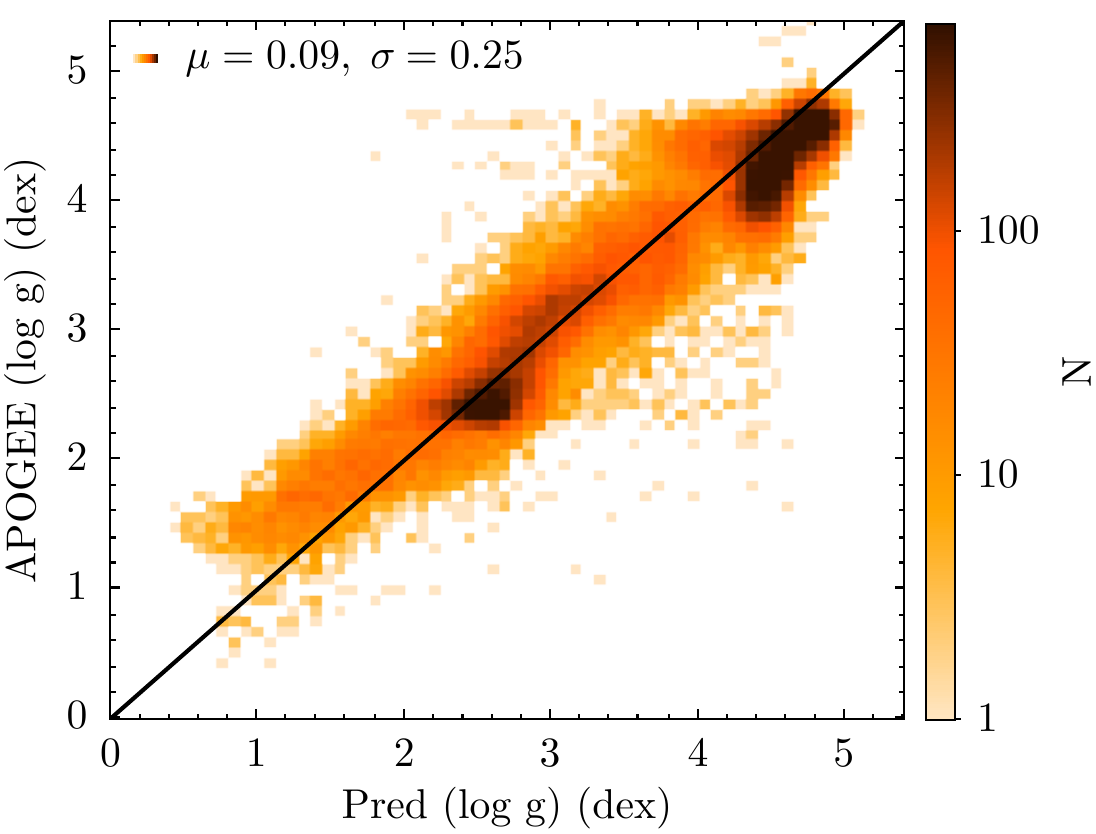}
				\centering
			\end{minipage}
		}
		\centering
		\subfigure{
			\begin{minipage}[t]{0.3\linewidth}
				\centering
				\includegraphics[width=2.1in, height=1.8in]{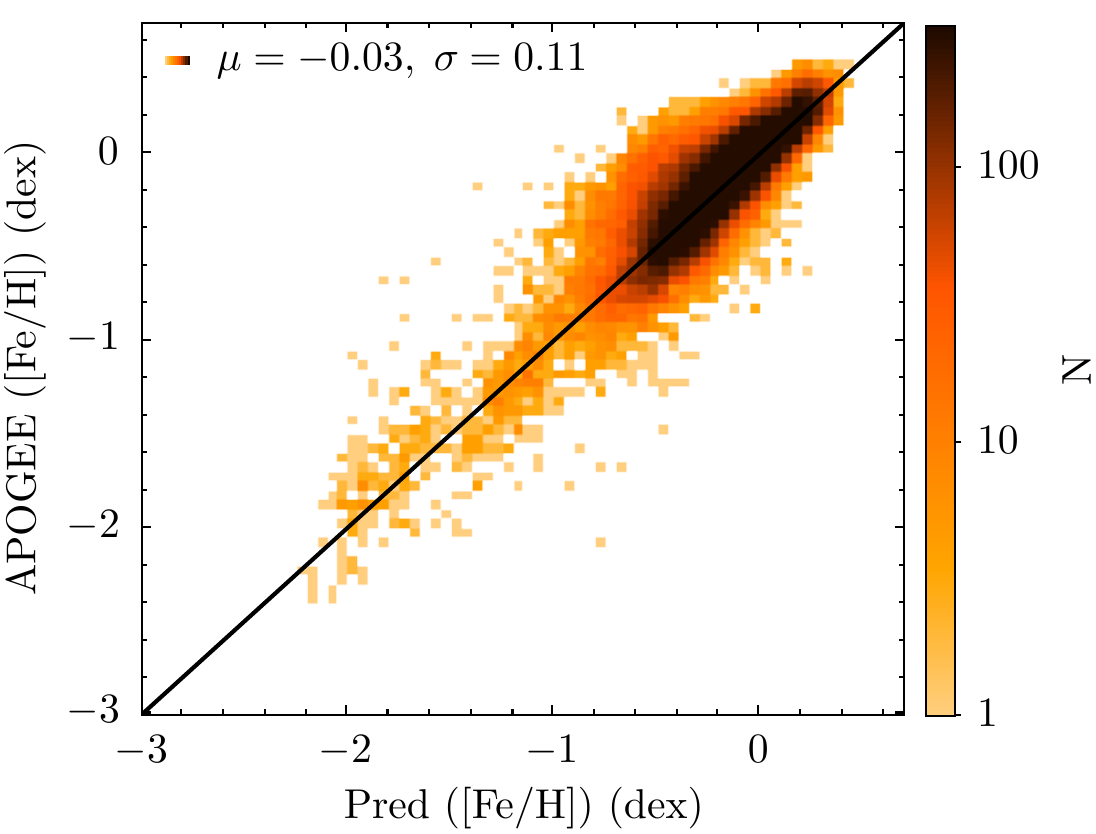}
				\centering
			\end{minipage}
		}
		\caption{Comparison of the FGK-type stars' parameters in the recommended catalog with APOGEE DR16. Mean and standard deviation of the parameter differences between the recommended catalog and APOGEE are also shown in the plot.}
		\centering
		\label{APOGEE FGK Teff logg FeH}
	\end{figure*}
	\begin{figure*}[htb]
		\centering
		\subfigure{
			\begin{minipage}[t]{0.3\linewidth}
				\centering
				\includegraphics[width=2.1in, height=1.8in]{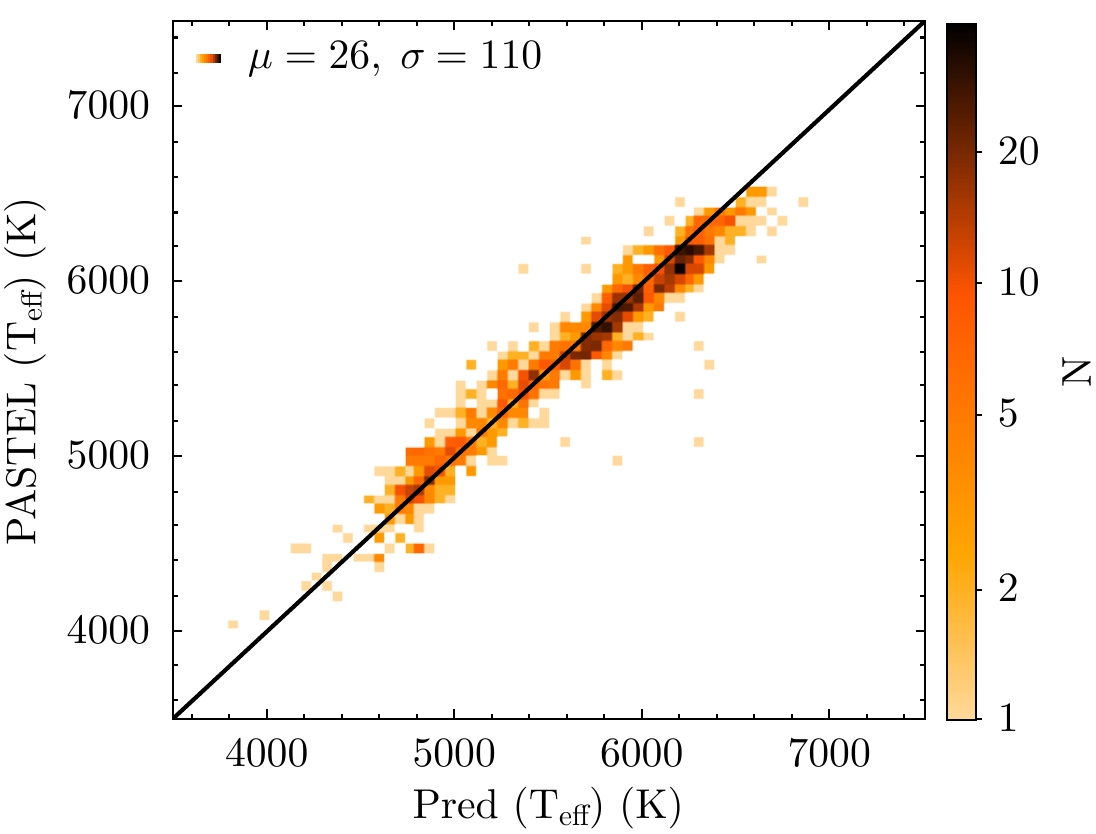}
				\centering
			\end{minipage}
		}
		\centering
		\subfigure{
			\begin{minipage}[t]{0.3\linewidth}
				\centering
				\includegraphics[width=2.1in, height=1.8in]{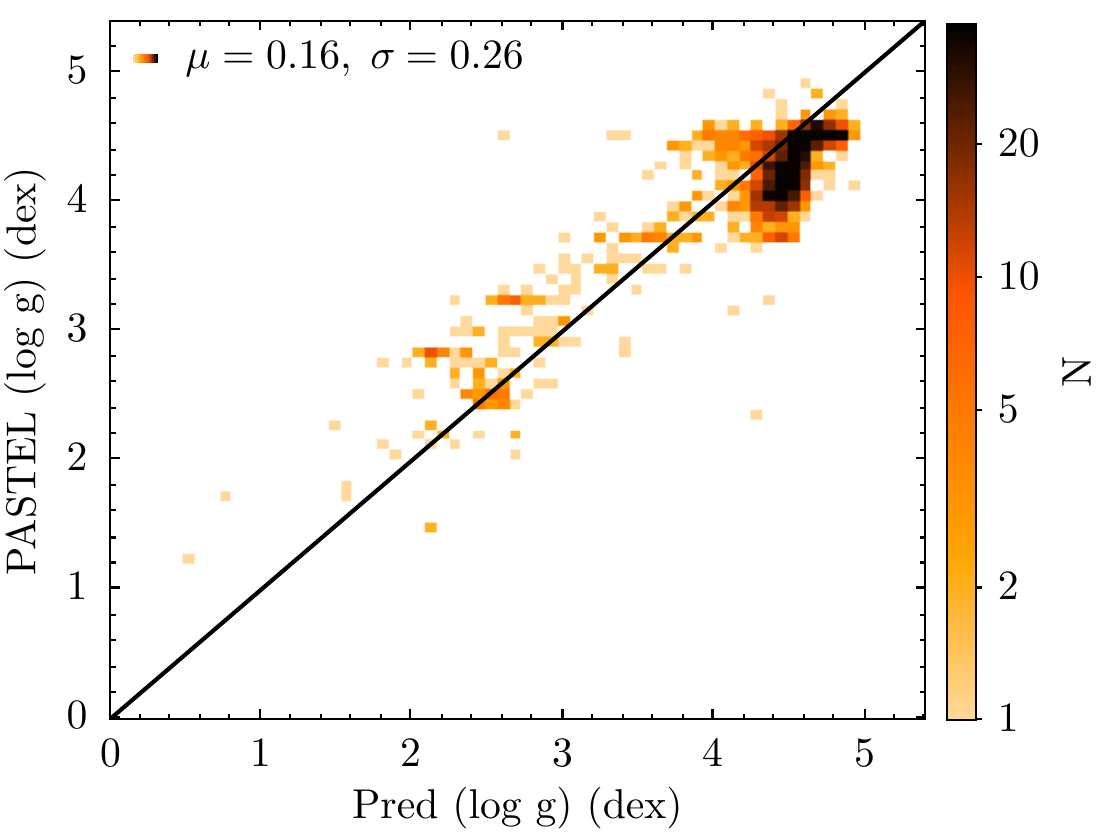}
				\centering
			\end{minipage}
		}
		\centering
		\subfigure{
			\begin{minipage}[t]{0.3\linewidth}
				\centering
				\includegraphics[width=2.1in, height=1.8in]{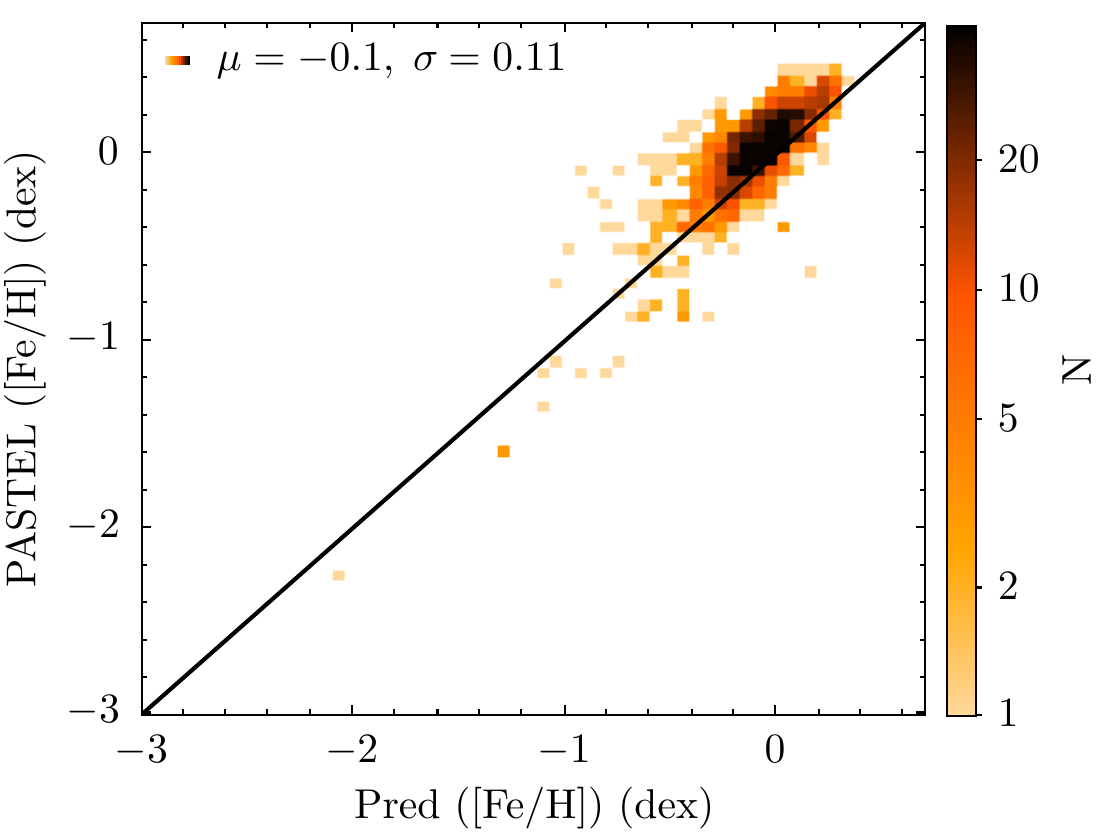}
				\centering
			\end{minipage}
		}
		\caption{Comparison of the FGK-type stars' parameters in the recommended catalog with PASTEL. Mean and standard deviation of the parameter differences between the recommended catalog and PASTEL are also shown in the plot.}
		\centering
		\label{PASTEL FGK Teff logg FeH}
	\end{figure*}
	\begin{figure*}[htb]
		\centering
		\subfigure{
			\begin{minipage}[t]{0.3\linewidth}
				\centering
				\includegraphics[width=2.1in, height=1.8in]{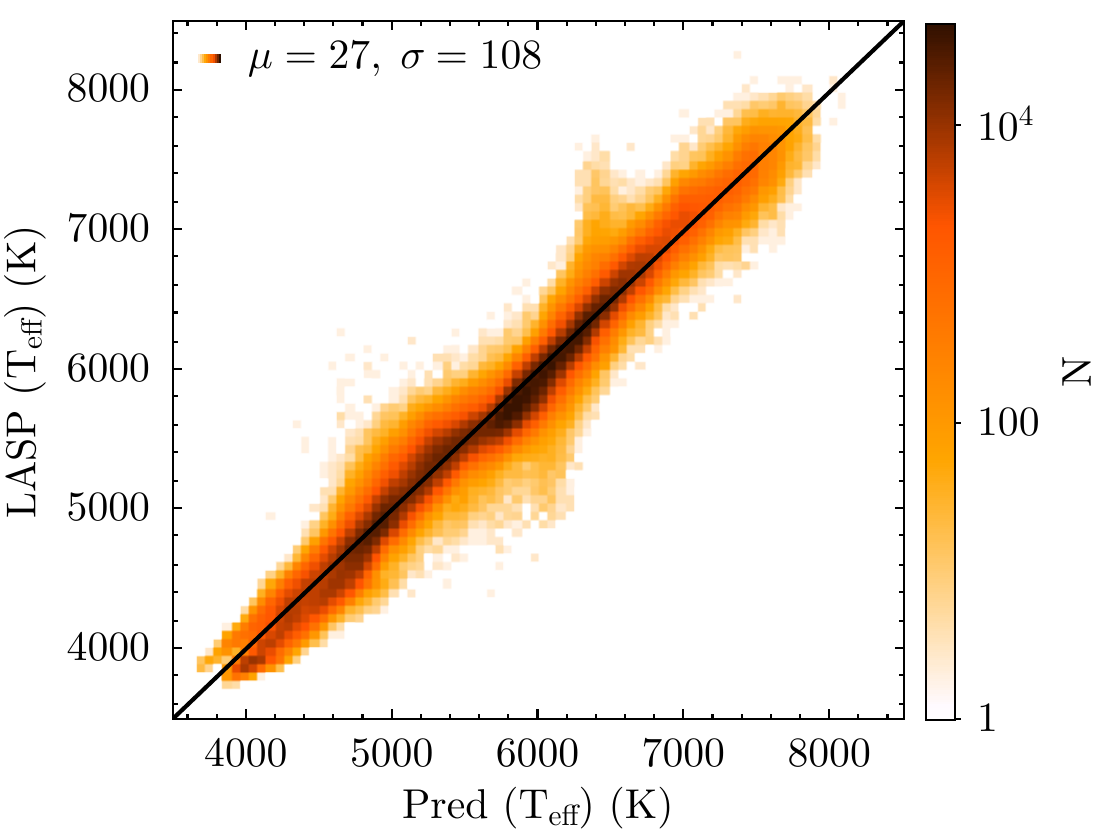}
				\centering
			\end{minipage}
		}
		\centering
		\subfigure{
			\begin{minipage}[t]{0.3\linewidth}
				\centering
				\includegraphics[width=2.1in, height=1.8in]{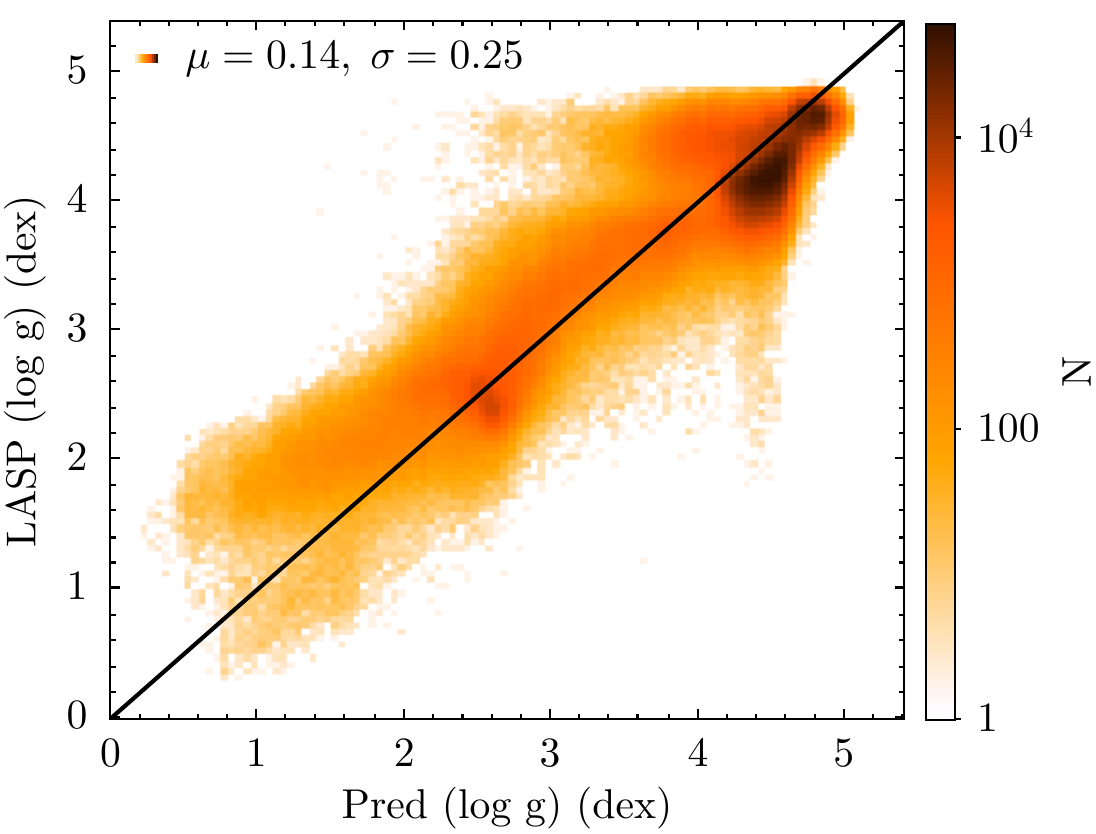}
				\centering
			\end{minipage}
		}
		\centering
		\subfigure{
			\begin{minipage}[t]{0.3\linewidth}
				\centering
				\includegraphics[width=2.1in, height=1.8in]{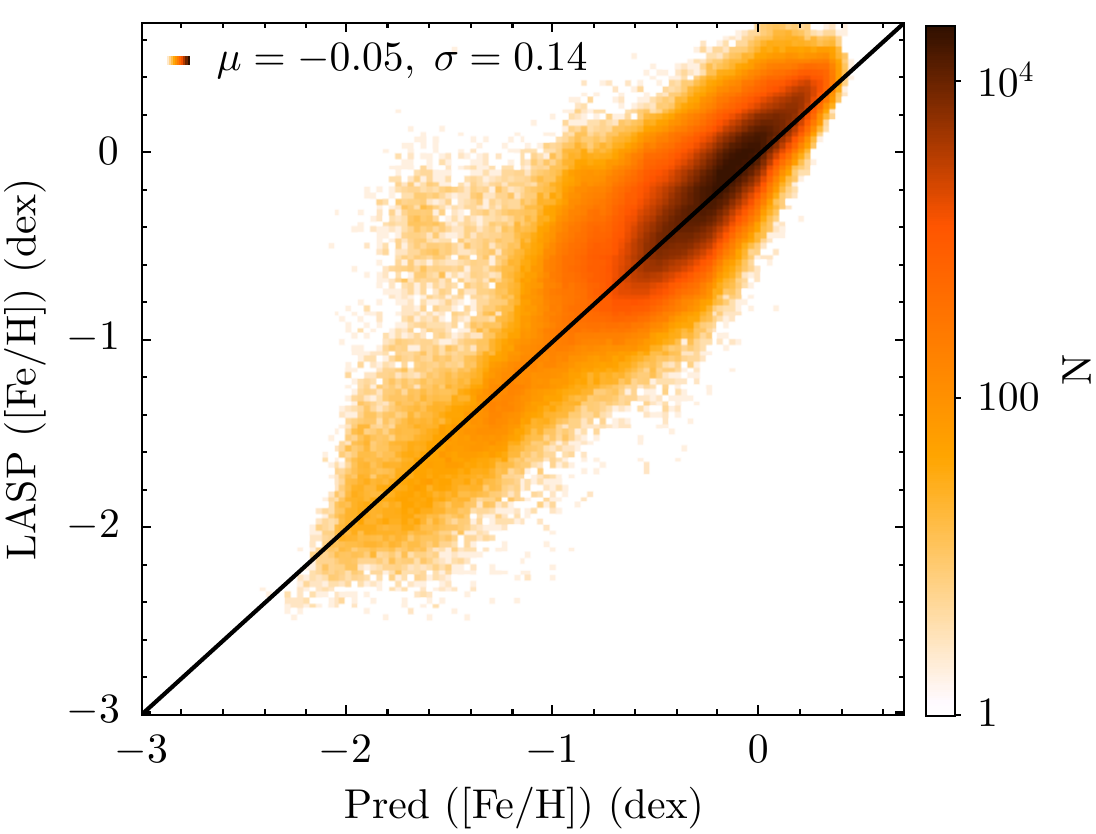}
				\centering
			\end{minipage}
		}
		\caption{Comparison of the FGK-type stars' parameters in the recommended catalog with LAMOST. Mean and standard deviation of the parameter differences between the recommended catalog and LAMOST are also shown in the plot.}
		\centering
		\label{LAMOST FGK Teff logg FeH}
	\end{figure*}
	\begin{figure*}[htb]
		\centering
		\subfigure{
			\begin{minipage}[t]{0.45\linewidth}
				\centering
				\includegraphics[width=3in, height=1.5in]{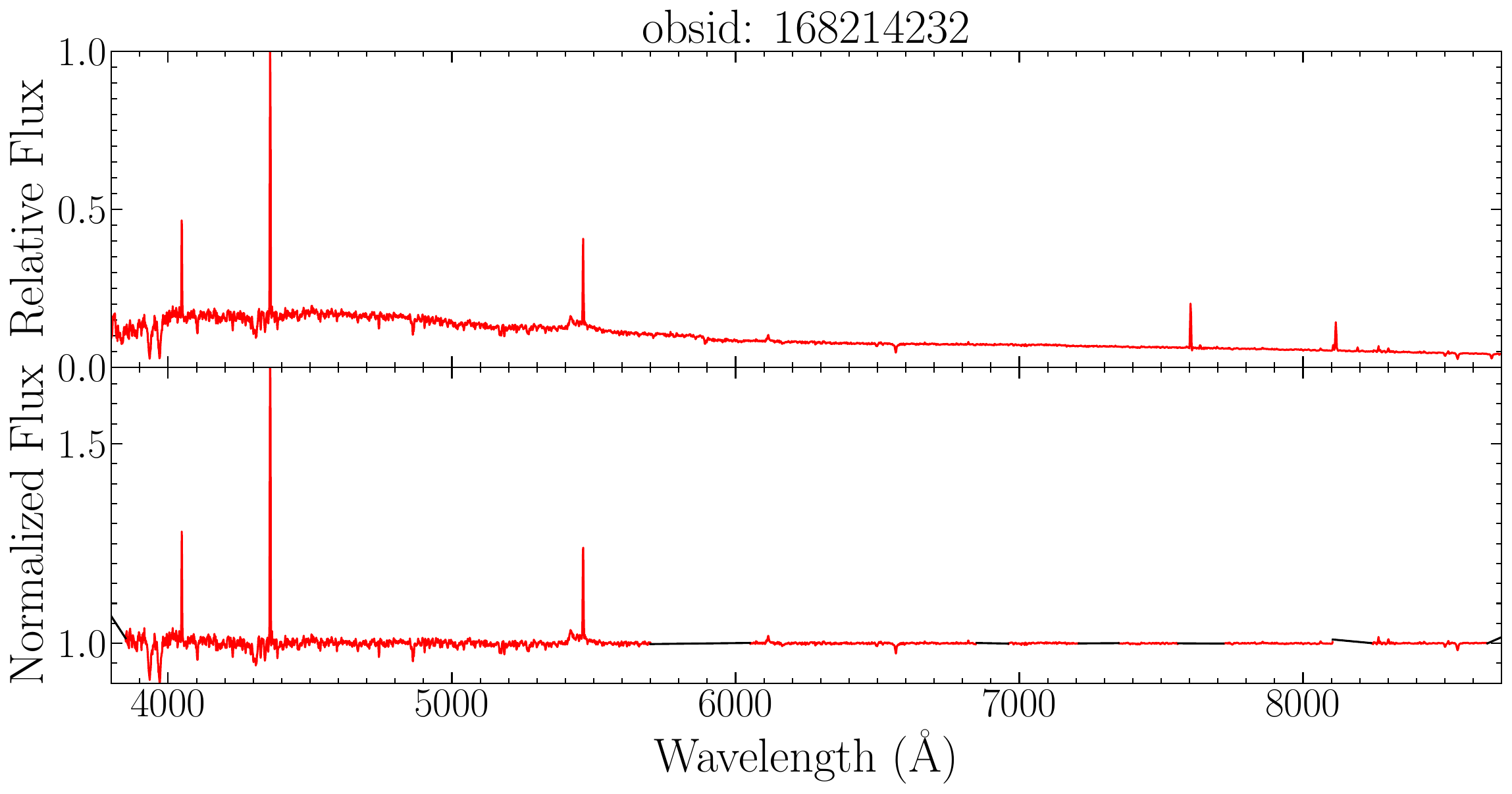}
				\centering
			\end{minipage}
		}
		\centering
		\subfigure{
			\begin{minipage}[t]{0.45\linewidth}
				\centering
				\includegraphics[width=3in, height=1.5in]{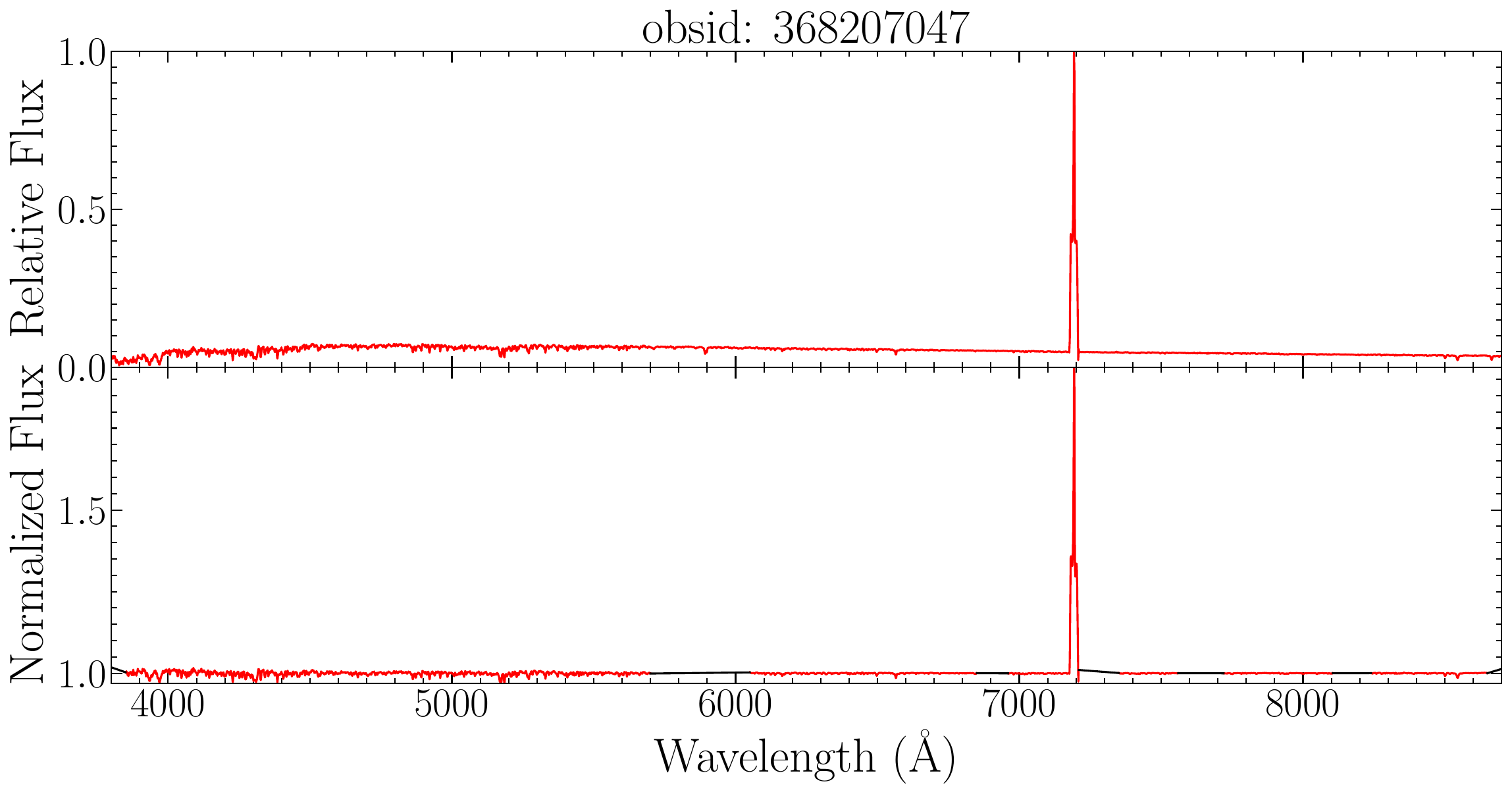}
				\centering
			\end{minipage}
		}
		\centering
		\subfigure{
			\begin{minipage}[t]{0.45\linewidth}
				\centering
				\includegraphics[width=3in, height=1.5in]{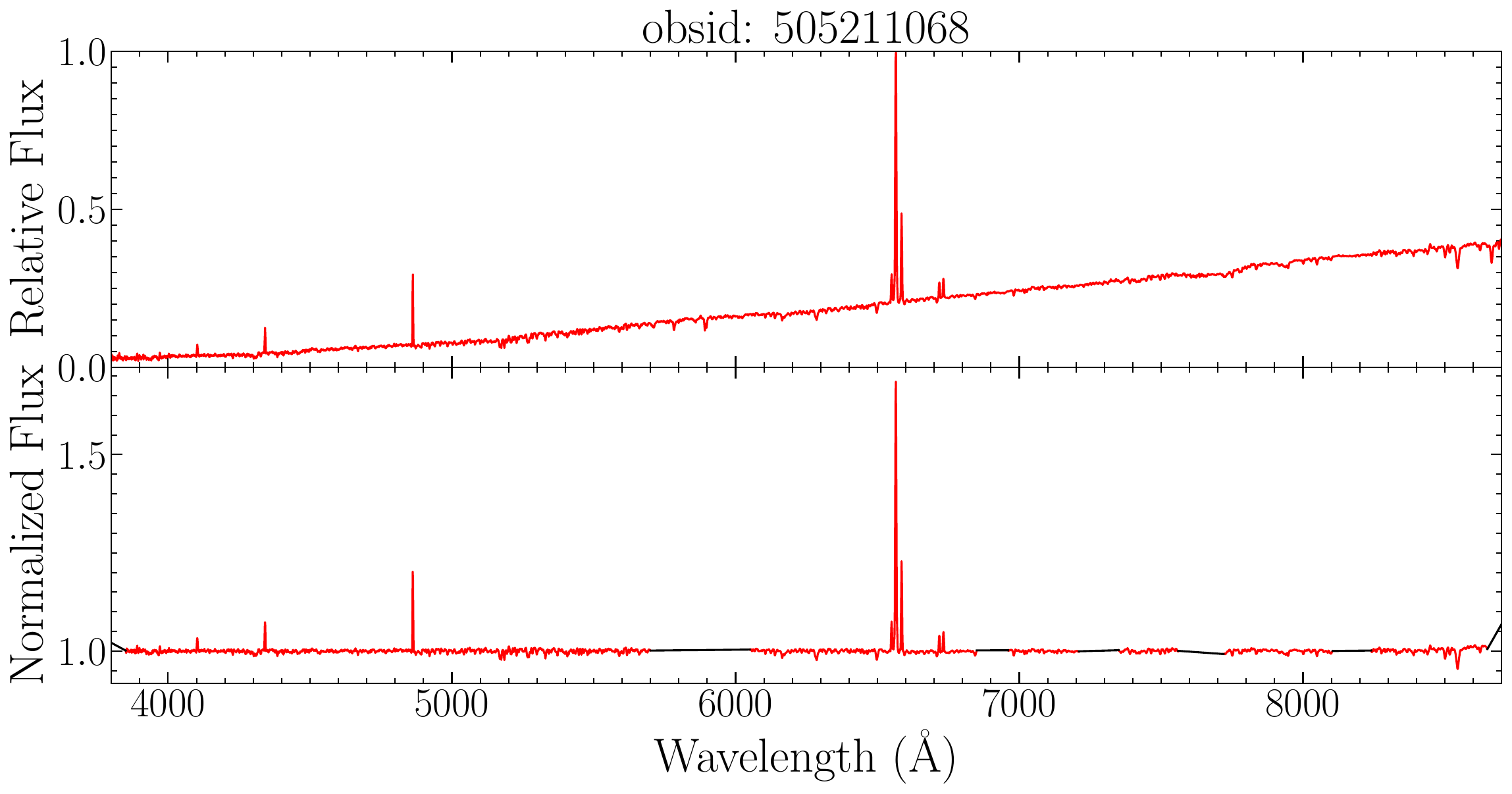}
				\centering
			\end{minipage}
		}
		\centering
		\subfigure{
			\begin{minipage}[t]{0.45\linewidth}
				\centering
				\includegraphics[width=3in, height=1.5in]{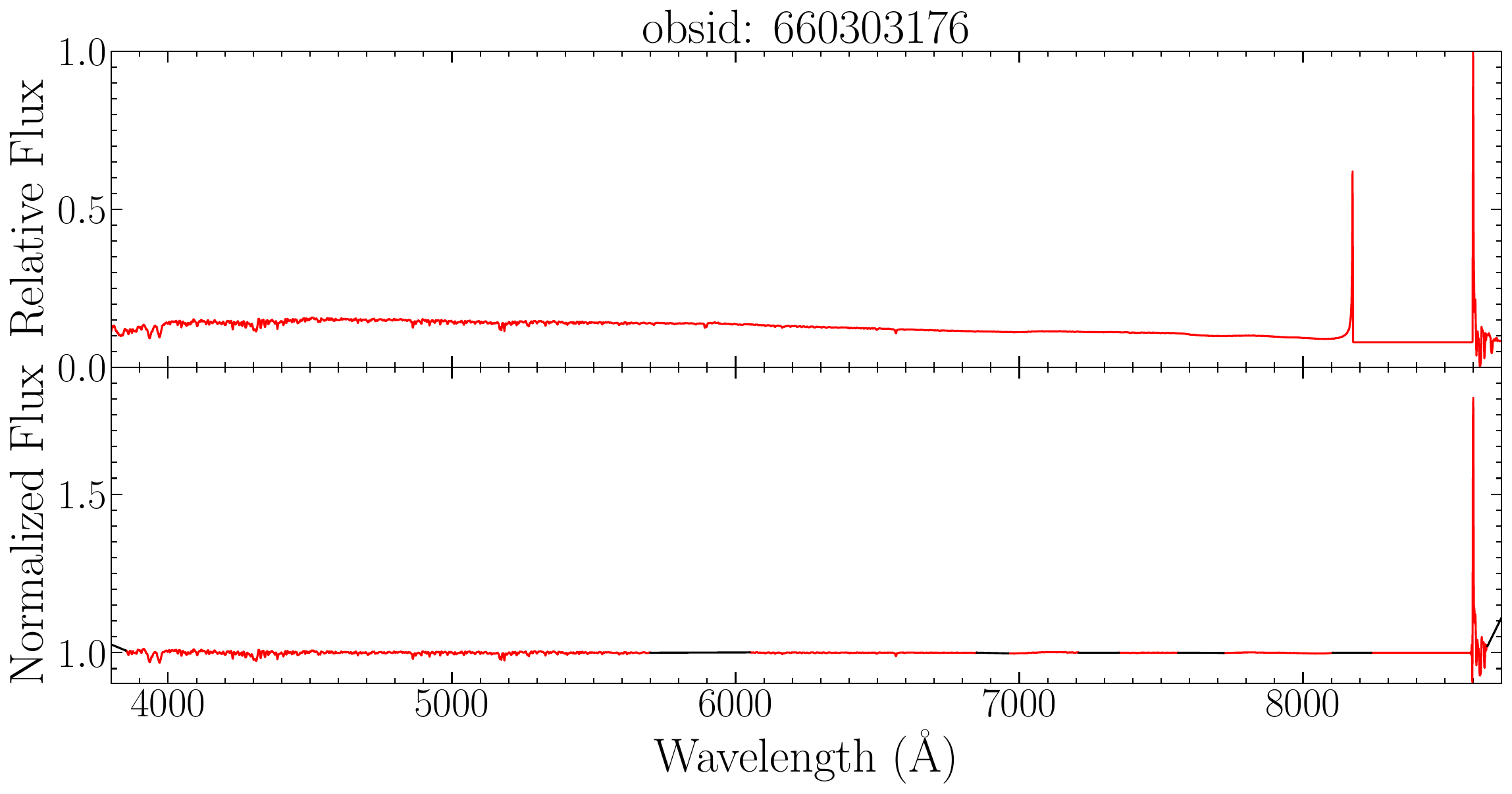}
				\centering
			\end{minipage}
		}
		\caption{Examples of anomalous spectra. The horizontal axis represents the spectral wavelength, while the vertical axis of each subplot displays relative flux of the original spectra scaled to the range of $0$-$1$ and the normalized flux. The black lines indicate the masked wavelength ranges as mentioned in the first preprocessing step of \refsubsection{Pre-processing for the spectra}.}
		\centering
		\label{unnormal spectral}
	\end{figure*}
	\begin{figure*}[htb]
		\centering
		\subfigure{
			\begin{minipage}[t]{0.3\linewidth}
				\centering
				\includegraphics[width=2.1in, height=1.8in]{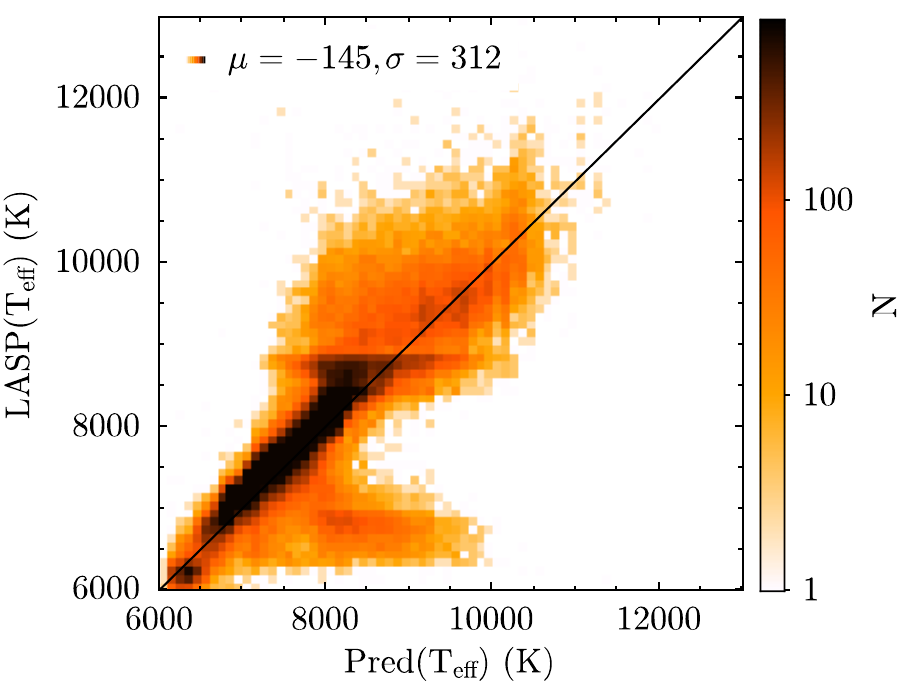}
				\centering
			\end{minipage}
		}
		\centering
		\subfigure{
			\begin{minipage}[t]{0.3\linewidth}
				\centering
				\includegraphics[width=2.1in, height=1.8in]{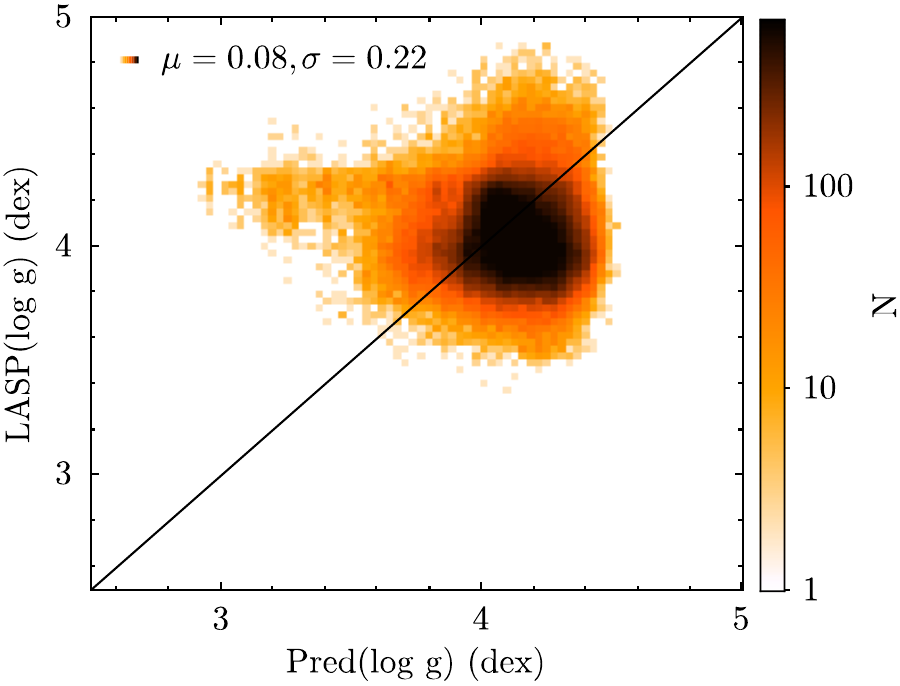}
				\centering
			\end{minipage}
		}
		\centering
		\subfigure{
			\begin{minipage}[t]{0.3\linewidth}
				\centering
				\includegraphics[width=2.1in, height=1.8in]{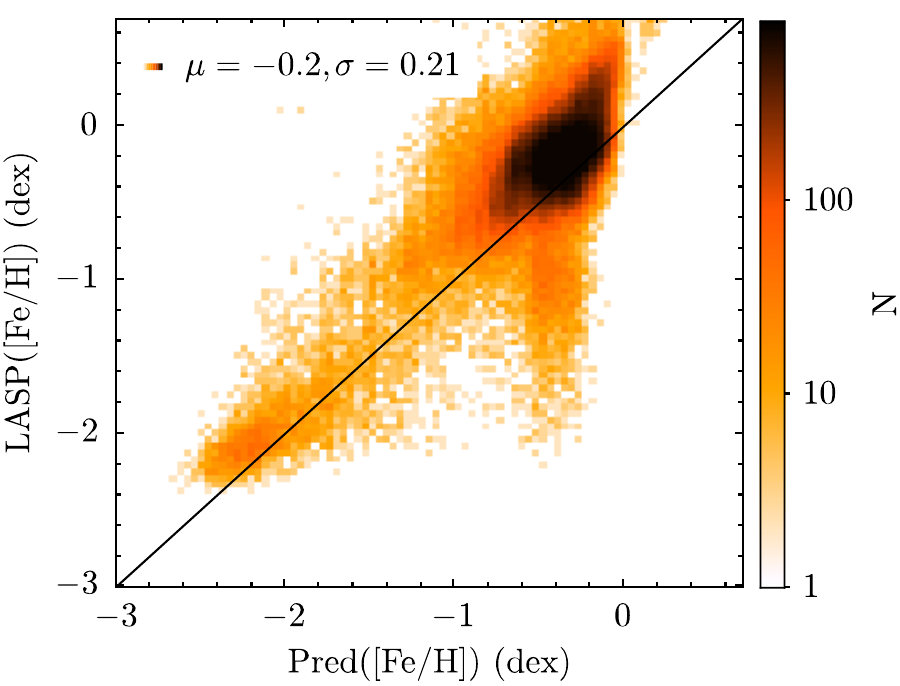}
				\centering
			\end{minipage}
		}
		\caption{Comparison of the A-type stars' parameters in the recommended catalog with LAMOST. Mean and standard deviation of the parameter differences between the recommended catalog and LAMOST are also shown in the plot.}
		\centering
		\label{LAMOST A Teff logg FeH}
	\end{figure*}
	
	In this subsection, we conducted a comparative analysis of the atmospheric parameters for AFGK-type stars in the recommended catalog with those derived from the APOGEE DR16 \citep{2020AJ....160..120J}, PASTEL \citep{soubiran2020vizier}, and LAMOST DR10 \citep{luo2015first}. We further examined the deficiencies in the atmospheric parameters of A-type stars within LASP and propose using the parameters from the recommended catalog as a viable supplement.
	
	After processing the APOGEE data in \refsubsection{M Type}, we crossmatched the FGK-type stars from the recommended catalog with APOGEE DR16, obtaining $88,004$ spectra. As illustrated in \reffig{APOGEE FGK Teff logg FeH} and \reffig{dM gM FGK AFe} (right panel), the offsets and scatters for $T_{\text{eff}}$, log $g$, [Fe/H], and [$\alpha$/Fe] are $38\pm127$ K, $0.09\pm0.25$ dex, $-0.03\pm0.11$ dex, and $0.01\pm0.08$ dex, respectively, indicating good consistency for these parameters, especially for [Fe/H] and [$\alpha$/Fe]. We noted that, for $T_{\text{eff}}$ greater than $6000$ K, $T_{\text{eff}}$ and log $g$ appear to be overestimated by approximately $169$ K and $0.29$ dex, respectively. \cite{xiang2015lamost} also observed a similar trend when predicting the atmospheric parameters of low-resolution LAMOST spectra using the MILES library. This trend may be attributed to differences between empirical and theoretical templates, as well as inconsistencies between optical and NIR spectra. Additionally, we observed a systematic discrepancy in log $g$ for cold giants ($T_{\text{eff}}$ $<$ $4500$ K and log $g$ $<$ $3.8$ dex) compared to APOGEE. As described in \refsubsection{Future improvements}, we confirmed that this discrepancy was due to a systematic bias in the median log $g$ provided by MaStar. We suggest that the parameter log $g$ of MaStar for cold giants require further calibration, as indicated by ongoing work referenced on the SDSS official website.
	
	PASTEL (\citeauthor{soubiran2010pastel} \citeyear{soubiran2010pastel}, \citeyear{soubiran2020vizier}) is a regularly updated bibliographical catalogue compiling determinations of $T_{\text{eff}}$, log $g$, and [Fe/H], which
	were obtained from detailed analyses of high-resolution, high S/N spectra, carried out with the
	help of model atmospheres. The techniques employed in predicting stellar atmospheric parameters continue to advance
	and develop. According to the description provided in \cite{soubiran2016pastel}, one of the recent innovations
	introduced by PASTEL is the inclusion of $T_{\text{eff}}$ measurements obtained through the fundamental method,
	utilizing angular diameters and total flux, as well as log $g$ determined via asteroseismology.
	Furthermore, the $T_{\text{eff}}$, log $g$ and [Fe/H] are now more frequently determined using automated
	methods applied to large samples comprising hundreds of stars, leading to the substantial increase in the content
	of PASTEL. \cite{soubiran2020vizier} updated the PASTEL catalogue, which includes $81,362$ records and
	supersedes the two previous versions of the [Fe/H] catalogue (\citealt{soubiran1997catalogue,soubiran2001catalogue}).
	Overall, the stellar atmospheric parameters provided by the PASTEL catalog are quite reliable.
	
	For the comparison, we obtained $1779$ data by crossmatching our recommended catalog with the PASTEL. As shown in \reffig{PASTEL FGK Teff logg FeH}, the offsets and scatters for $T_{\text{eff}}$, log $g$, and [Fe/H] are $26\pm110$ K, $0.16\pm0.26$ dex, and $-0.1\pm0.11$ dex, respectively, indicating good consistency for $T_{\text{eff}}$ and [Fe/H]. The [Fe/H] values show a relatively high systematic error of $0.1$ dex in comparison with PASTEL, a trend that aligns with the comparison of [Fe/H] for medium-resolution LAMOST spectra predicted by \cite{Wang_2023} with PASTEL. 
	The cause of this bias remains unclear and warrants further investigation. Similar to \reffig{APOGEE FGK Teff logg FeH}, we found that the log $g$ are systematically $0.2$ dex higher than the PASTEL values between $4$ and $5$ dex, which may be attributed to offsets arising from the median log $g$ in MaStar, and we analyzed this situation in \refsubsection{Future improvements}.
	
	\begin{figure*}[htb]
		\centering
		\subfigure{
			\begin{minipage}[t]{0.45\linewidth}
				\centering
				\includegraphics[width=3.2in, height=2.2in]{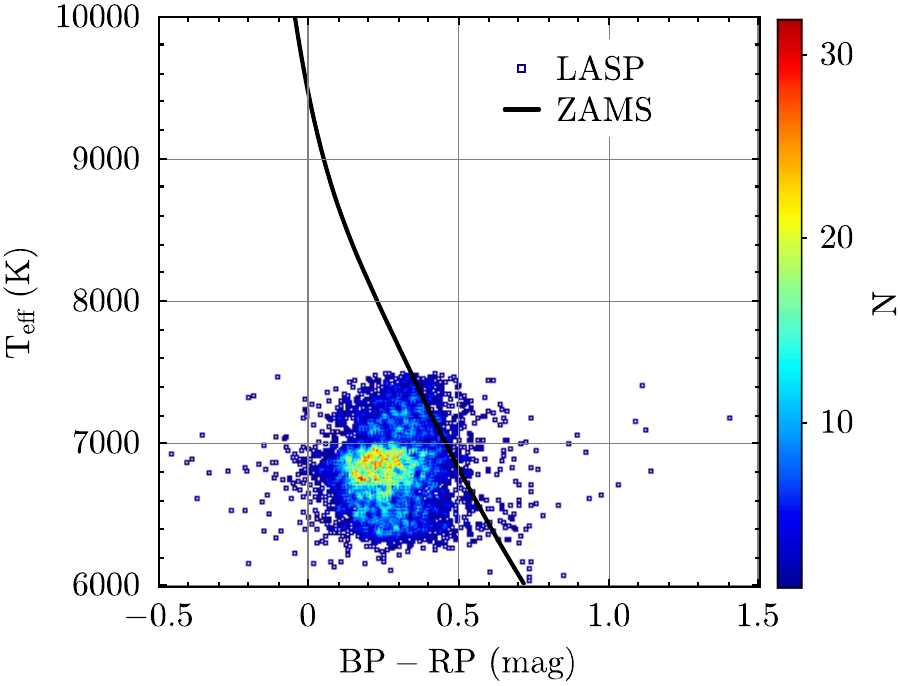}
				\centering
			\end{minipage}
		}
		\centering
		\subfigure{
			\begin{minipage}[t]{0.45\linewidth}
				\centering
				\includegraphics[width=3.2in, height=2.2in]{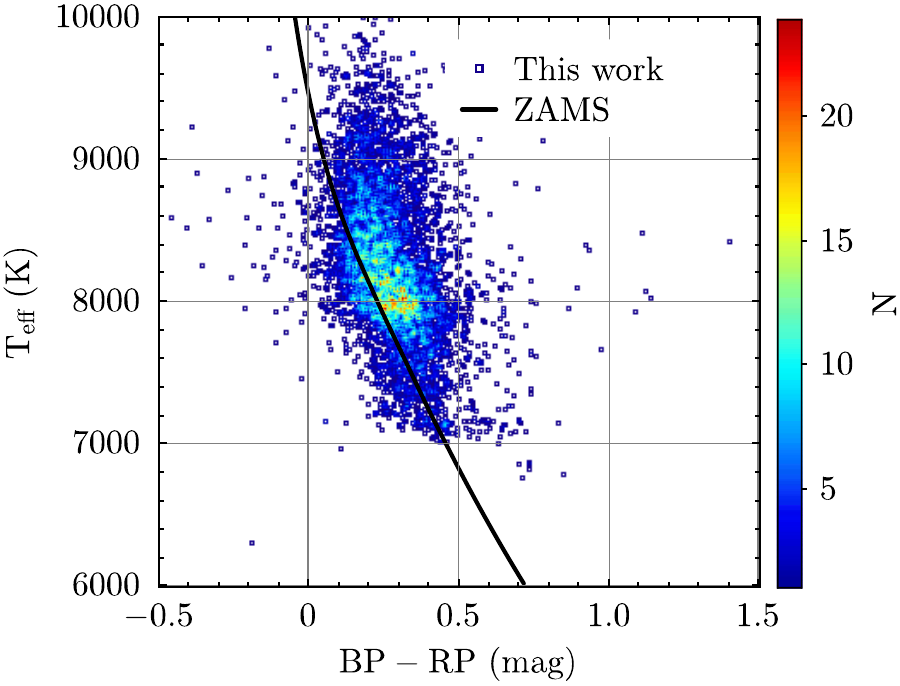}
				\centering
			\end{minipage}
		}
		\centering
		\subfigure{
			\begin{minipage}[t]{0.45\linewidth}
				\centering
				\includegraphics[width=3.2in, height=2.2in]{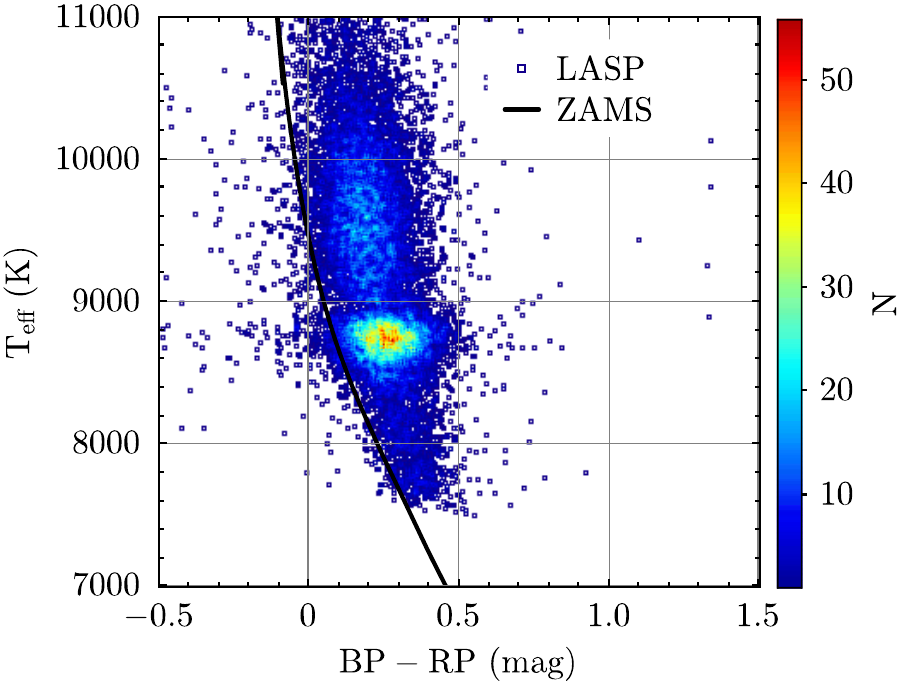}
				\centering
			\end{minipage}
		}
		\centering
		\subfigure{
			\begin{minipage}[t]{0.45\linewidth}
				\centering
				\includegraphics[width=3.2in, height=2.2in]{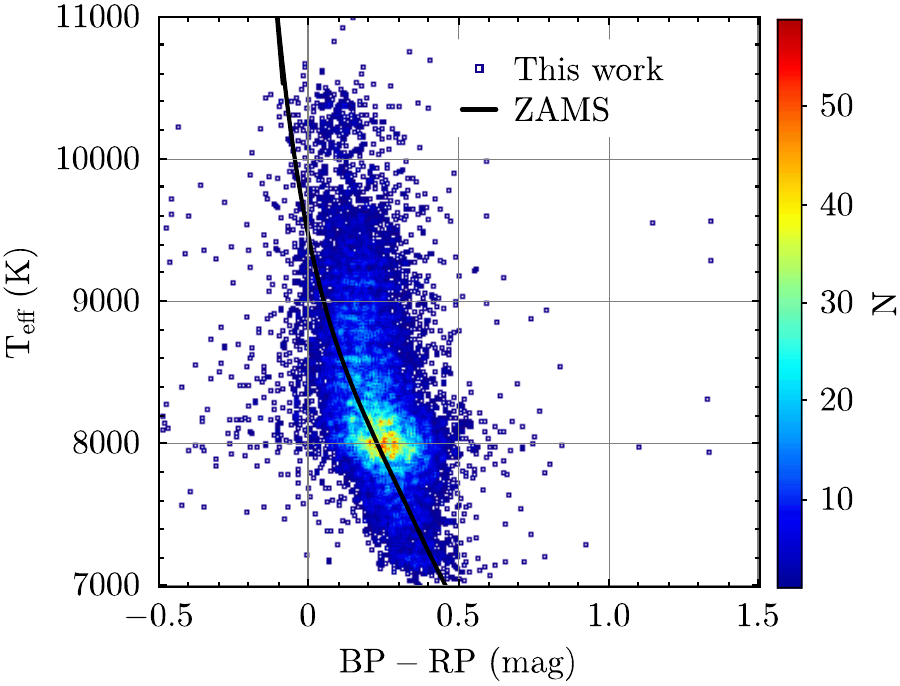}
				\centering
			\end{minipage}
		}
		\caption{$T_{\text{eff}}$ as a function of dereddened (BP-RP) color. The top left and top right panels, respectively, show the relationship between LASP's $T_{\text{eff}}$ (for $T_{\text{eff}} < 7500$ K), this work's $T_{\text{eff}}$, and the dereddened (BP-RP) color. The bottom left and bottom right panels respectively show the relationship when LASP's $T_{\text{eff}} \ge 7500$ K, between LASP's $T_{\text{eff}}$, this work's $T_{\text{eff}}$, and the dereddened (BP-RP) color. The black curve delineates the MIST isochrone's $T_{\text{eff}}$ as a function of (BP-RP) color for ZAMS with initial [Fe/H] = $0$ dex.}
		\centering
		\label{LAMOST A 7500} 
	\end{figure*}
	
	The LAMOST official stellar atmospheric parameter pipeline, LASP, is derived through a two-step process: the CFI and refined optimization with University of Lyon spectroscopic analysis software (\citealt{wu2011automatic}). For comprehensive details on LAMOST, refer to \refsection{Observed Spectra from LAMOST DR10}. To ensure high internal reliability of LASP, we selected FGK-type stars with an S/N greater than $20$ and internal errors in $T_{\text{eff}}$, log $g$, and [Fe/H] less than $200$ K, $0.2$ dex, and $0.1$ dex, respectively. For A-type stars, the criteria were the S/N in $g$ band greater than $20$ and internal errors in $T_{\text{eff}}$, log $g$, and [Fe/H] less than $500$ K, $0.2$ dex, and $0.1$ dex. Matching records via the observation ID (obsid) field with LASP yielded $4,135,169$ spectra for FGK-type and $183,524$ for A-type stars.
	
	As shown in \reffig{LAMOST FGK Teff logg FeH}, for FGK-type stars, the offsets and scatter for $T_{\text{eff}}$, log $g$, and [Fe/H] are respectively $27\pm108$ K, $0.14\pm0.25$ dex, and $-0.05\pm0.14$ dex, indicating good consistency for $T_{\text{eff}}$ and [Fe/H]. For log $g$, we identified two significant differences attributable to LAMOST spectral quality, and one significant difference was likely caused by the MaStar. We noted $3834$ spectra (the first difference) classified as giants (log $g < 3.5$ dex) in LASP with the recommended catalog log $g > 4$ dex, and $4770$ specatra (the second difference) classified as dwarfs (log $g > 4$ dex) in LASP with the recommended catalog log $g < 3.5$ dex. The two significant difference also correspond to regions where LASP's $T_{\text{eff}}$ is around $7000$ K (the recommended catalog $\sim6500$ K), and LASP's $T_{\text{eff}}$ around $5000$ K (the recommended catalog $\sim6000$ K), with the first difference also in regions with LASP's [Fe/H] from $-1$ to $0$ dex and the recommended catalog [Fe/H] from $-2$ to $-1$ dex. The significant differences in the atmospheric parameters for these sources are primarily due to the spectral quality of LAMOST. Among these spectra, $2584$ and $2942$ respectively have high S/N and no MASK records in the LAMOST official fits files. However, these spectra are indeed mostly contaminated by strong emission lines and bad pixels, leading to abnormal flux values. As shown in \reffig{unnormal spectral}, we showed spectral examples for obsid = $168214232$, $368207047$, $505211068$, and $660303176$. We noted that the preprocessing in \refsection{Pre-processing for the spectra} cannot effectively remove these abnormal flux values, thus result in significant discrepancies in parameter prediction, especially in log $g$. We plan to incorporate the outlier-detection method by \cite{wei2013mining} and \cite{lu2021study} for parameter prediction in the future to identify abnormal spectra and improve the internal reliability of the recommended catalog. The third difference was likely caused by the MaStar. It is consistent with \reffig{APOGEE FGK Teff logg FeH}, where we found that log $g$ were underestimated by $0.5$ dex in the recommended catalog for cold giants. The recommended catalog in this region will be updated following the release of MaStar team's parameter corrections.
	
	\begin{figure*}[htb]
		\centering
		\subfigure{
			\begin{minipage}[t]{0.3\linewidth}
				\centering
				\includegraphics[width=2.1in, height=1.8in]{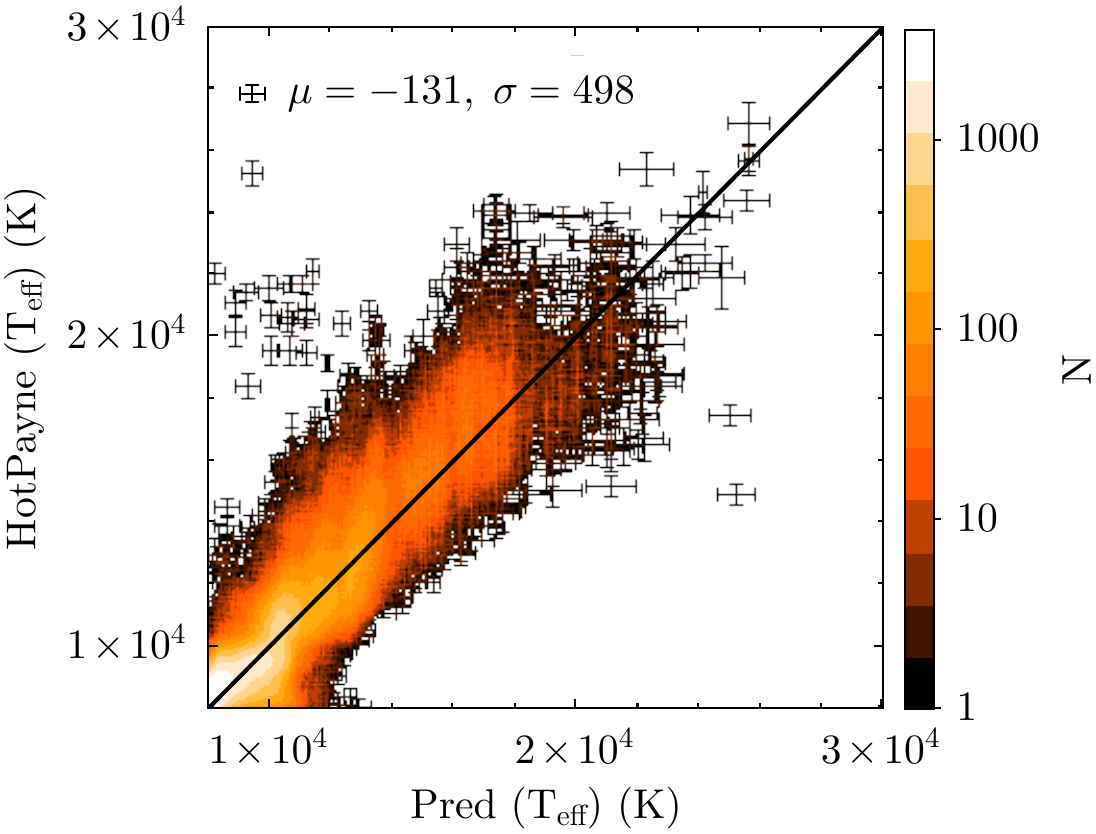}
				\centering
			\end{minipage}
		}
		\centering
		\subfigure{
			\begin{minipage}[t]{0.3\linewidth}
				\centering
				\includegraphics[width=2.1in, height=1.8in]{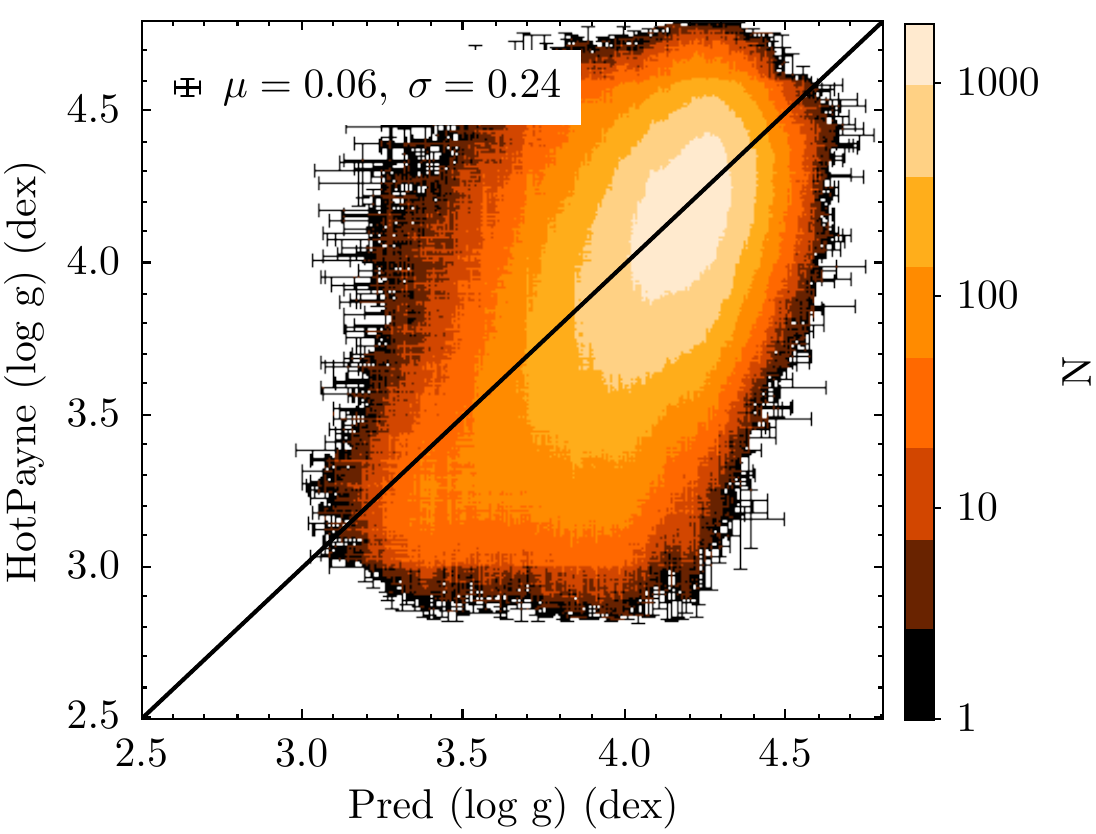}
				\centering
			\end{minipage}
		}
		\centering
		\subfigure{
			\begin{minipage}[t]{0.3\linewidth}
				\centering
				\includegraphics[width=2.1in, height=1.8in]{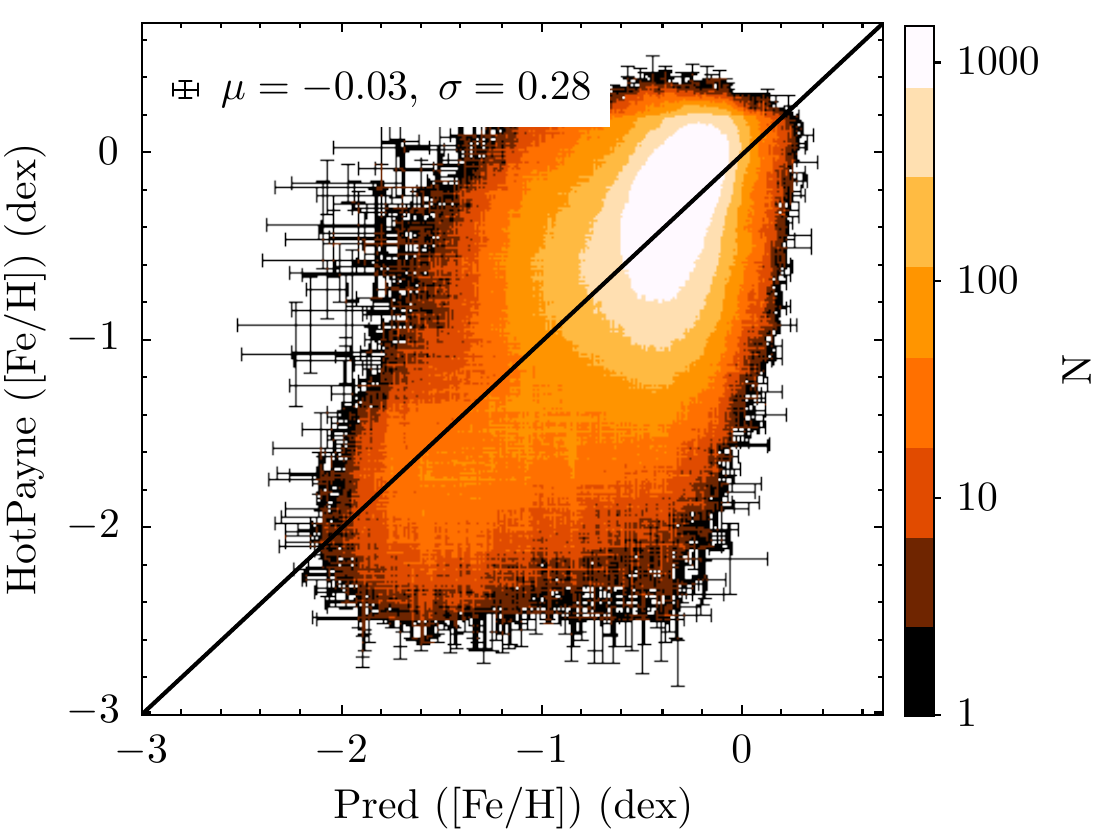}
				\centering
			\end{minipage}
		}
		\caption{Comparison of the OBA-type stars' parameters in the recommended catalog with HotPayne. Mean and standard deviation of the parameter differences between the recommended catalog and HotPayne are also shown in the plot. The error bars represent the uncertainty of each parameter.}
		\centering
		\label{HotPayne OBA Teff logg FeH}
	\end{figure*}
	\begin{figure*}[htb]
		\centering
		\subfigure{
			\begin{minipage}[t]{0.3\linewidth}
				\centering
				\includegraphics[width=2.1in, height=1.8in]{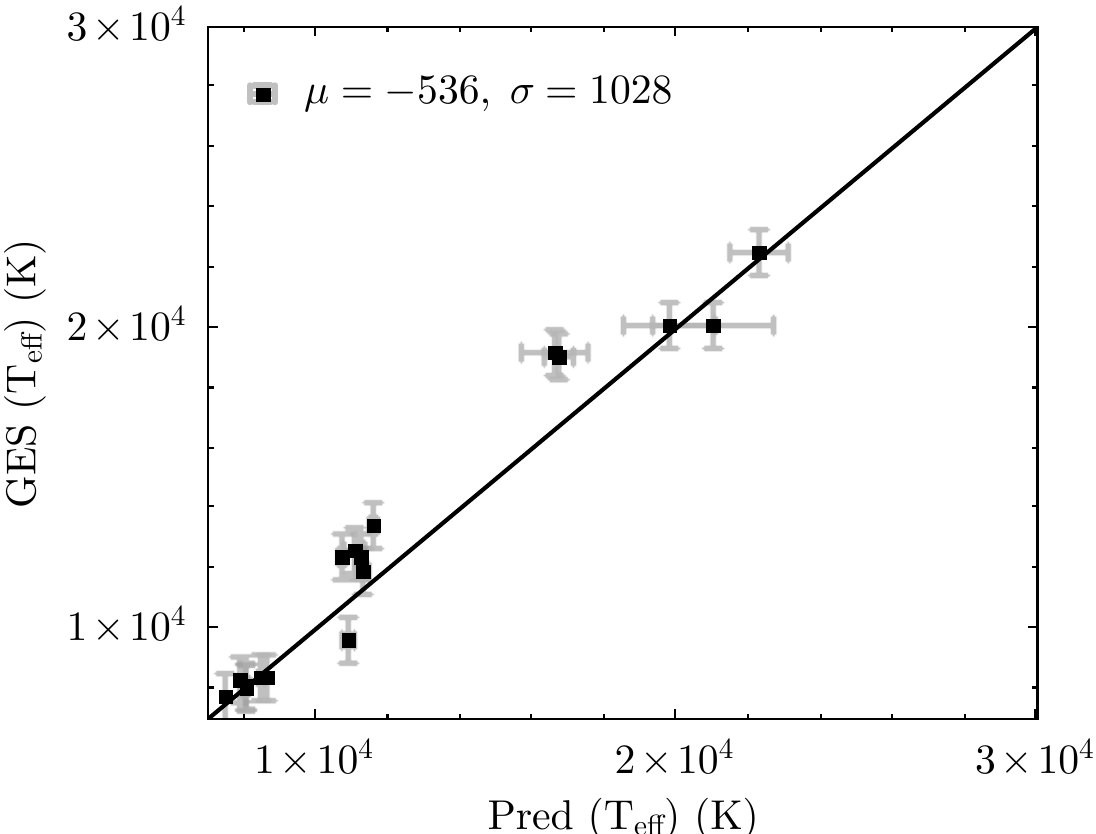}
				\centering
			\end{minipage}
		}
		\centering
		\subfigure{
			\begin{minipage}[t]{0.3\linewidth}
				\centering
				\includegraphics[width=2.1in, height=1.8in]{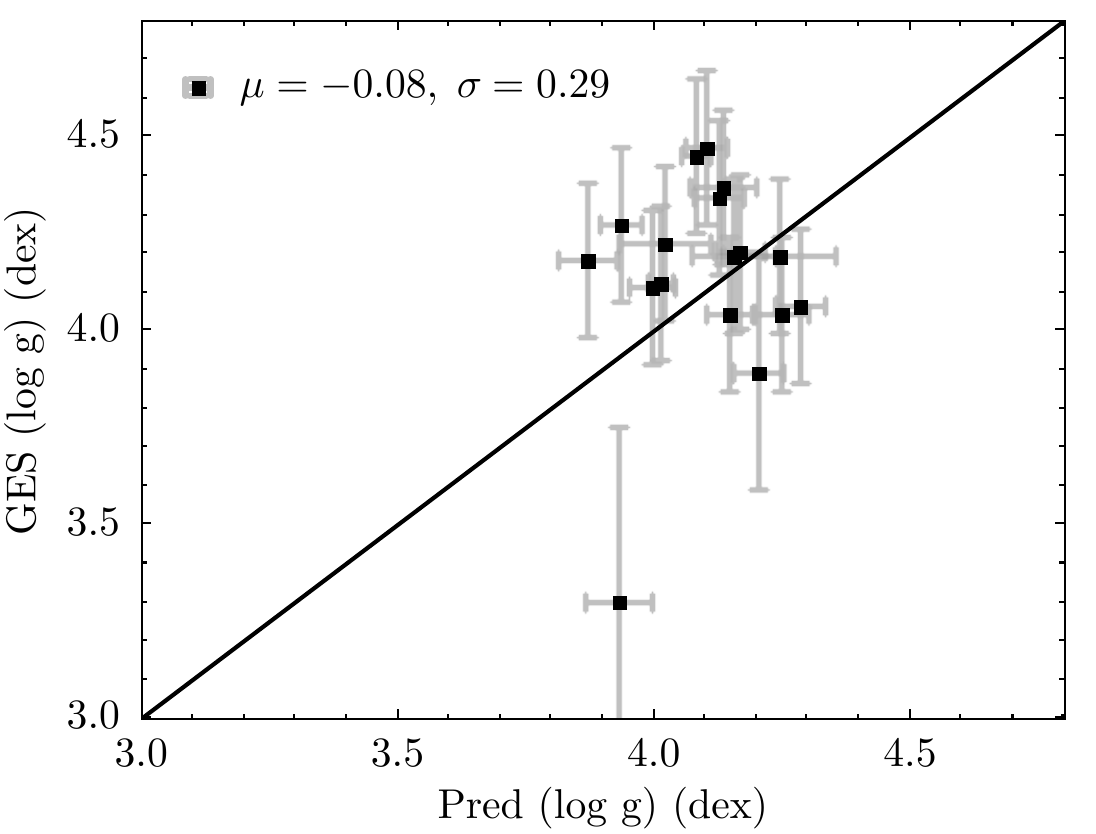}
				\centering
			\end{minipage}
		}
		\centering
		\subfigure{
			\begin{minipage}[t]{0.3\linewidth}
				\centering
				\includegraphics[width=2.1in, height=1.8in]{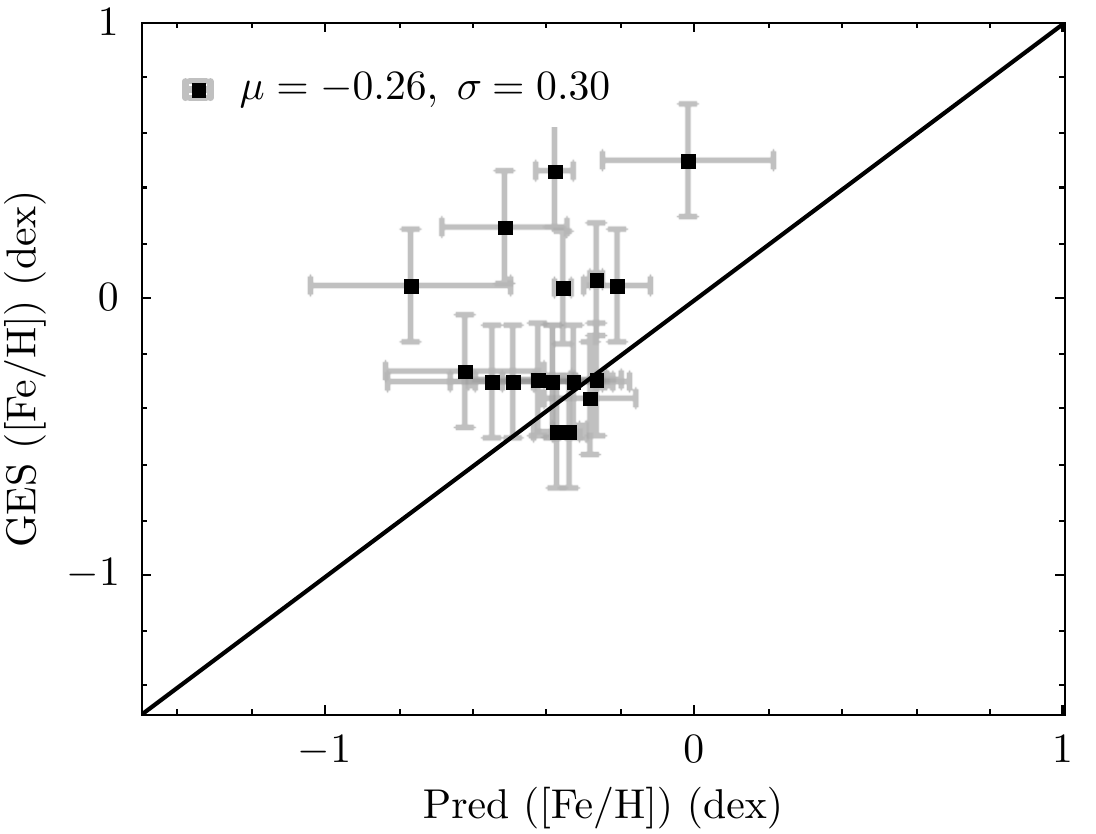}
				\centering
			\end{minipage}
		}
		\caption{Comparison of the OBA-type stars' parameters in the recommended catalog with Gaia-ESO. Mean and standard deviation of the parameter differences between the recommended catalog and Gaia-ESO are also shown in the plot. The error bars represent the uncertainty of each parameter.}
		\centering
		\label{GES OBA Teff logg FeH}
	\end{figure*}
	
	We compared the parameters of A-type stars with LASP, observing offsets and scatters in $T_{\text{eff}}$, log $g$, and [Fe/H] as $-145\pm312$ K, $0.08\pm0.22$ dex, and $-0.2\pm0.21$ dex, respectively, as shown in \reffig{LAMOST A Teff logg FeH}. Notably, two discrepancies in $T_{\text{eff}}$ were identified: our recommended catalog tends to overestimate $T_{\text{eff}}$ for LASP's $T_{\text{eff}} < 7500$ K ($11\%$ of cases) and underestimate when LASP's $T_{\text{eff}} \ge 7500$ K ($13\%$ of cases). The overestimated $T_{\text{eff}}$ region corresponds to LASP's [Fe/H] underestimated to $-2$ to $-1$ dex (the recommended catalog, $-1$ to $0$ dex). To assess reliability, we displayed $T_{\text{eff}}$ as a function of the BP-RP color of Gaia DR3, correcting the color for interstellar reddening estimated using the 3D reddening map of \cite{Green_2019} and extinction coefficients ($R_{\text{BP}} = 3.24$, $R_{\text{RP}} = 1.91$) from \cite{Chen_2018}. As shown in \reffig{LAMOST A 7500}, after dereddening, our recommended catalog's $T_{\text{eff}}$ aligns with zero-age main sequence stars' (ZAMS) MIST isochrones \citep{2016ApJS..222....8D,Choi_2016}, unlike LASP. Therefore, we believed these discrepancies may be attributed to the unreliability of the A-type parameters in LASP. This unreliability likely originates from the sparsity of the ELODIE library in hot star regions and the inadequate spectral wavelength range selected by the LASP for predicting the atmospheric parameters of A-type stars.	Given that LASP sequentially predicts $T_{\text{eff}}$, [Fe/H], and log $g$, the unreliability of $T_{\text{eff}}$ can lead to unreliability in [Fe/H] and log $g$. Therefore, we recommend cautious use of LASP parameters in this region. We have reported this issue to the LAMOST team, and anticipate further corrections to these parameters in a future data release.
	
	\subsubsection{Comparing OBA-type atmospheric parameters with HotPayne, Gaia-ESO Survey} \label{OBA Type}
	HotPayne, developed by \cite{xiang2022stellar}, is a spectral analysis tool developed for determining stellar labels from low-resolution spectra of hot stars, particularly those with $T_{\text{eff}} > 7500$ K. This method is based on the PAYNE and uses Kurucz's ATLAS12/SYNTHE local thermodynamic equilibrium (LTE) spectra as underlying models. It has enabled the estimation of various stellar labels, including $T_{\text{eff}}$, log $g$, and [Fe/H] for approximately $330,000$ hot stars in LAMOST DR$6$. 
	
	Following the guidelines set by \cite{xiang2022stellar}, we selected spectra with S/N $> 30$ and chi2ratio $ < 10$. Additionally, for A-type stars, we utilized spectra with internal errors of less than $500$ K, $0.2$ dex, and $0.4$ dex for $T_{\text{eff}}$, log $g$, and [Fe/H], respectively. For OB-type stars, we used spectra with internal errors less than $1000$ K, $0.2$ dex, and $0.4$ dex. By matching with the obsid field, we acquired a total of $130,145$ spectra with parameters of the recommended catalog and HotPayne. As illustrated in \reffig{HotPayne OBA Teff logg FeH}, the offsets and scatters for $T_{\text{eff}}$, log $g$, and [Fe/H] are $-131\pm498$ K, $0.06\pm0.24$ dex, and $-0.03\pm0.28$ dex, respectively, indicating good consistency for these parameters.
	
	\begin{figure*}[!htb]
		\includegraphics[width=7.1in,height=2.3in]{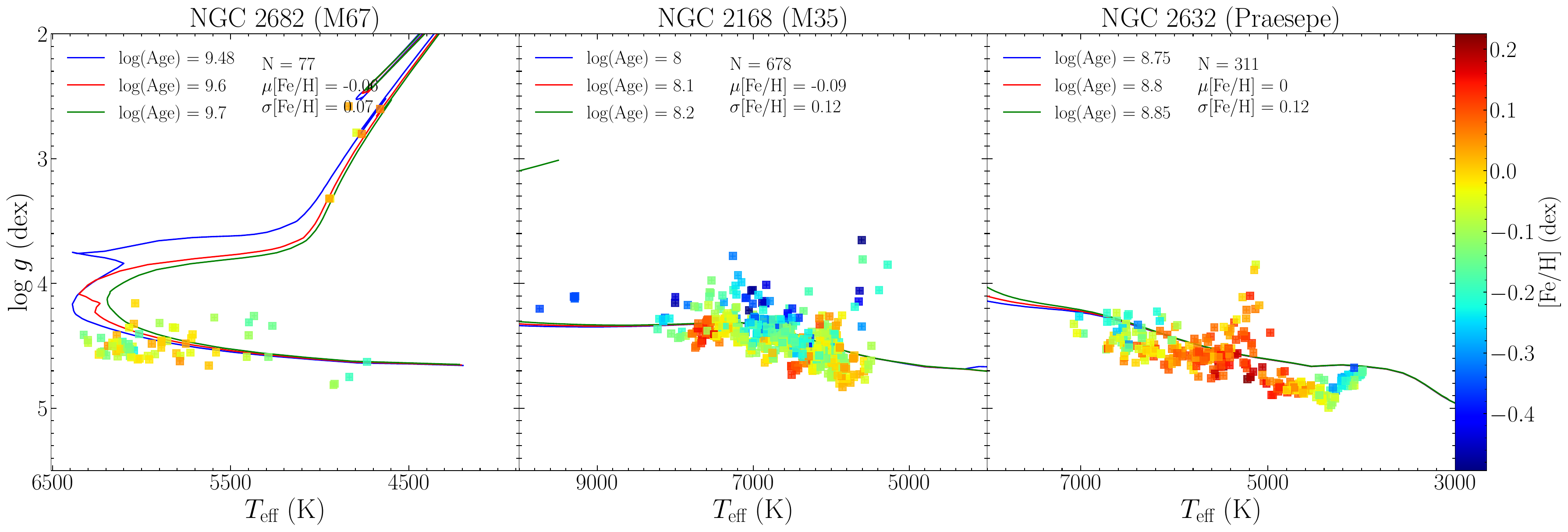}
		\centering
		\caption{The distributions of $T_{\text{eff}}$, log $g$, and [Fe/H] in the recommended catalog for M$67$ (left panel), M$35$ (middle panel), and Praesepe (right panel). The different colored curves represent PARSEC V$2.0$ isochrones of various ages. The [Fe/H] for the OCs were taken from \cite{fu2022lamost}, while the ages were derived from \cite{carrera2019extended}, \cite{sung1999ubvi}, and \cite{salaris2004age} respectively.}
		\centering
		\label{our cluster}
	\end{figure*}
	\begin{figure*}[!htb]
		\includegraphics[width=7.1in,height=2.3in]{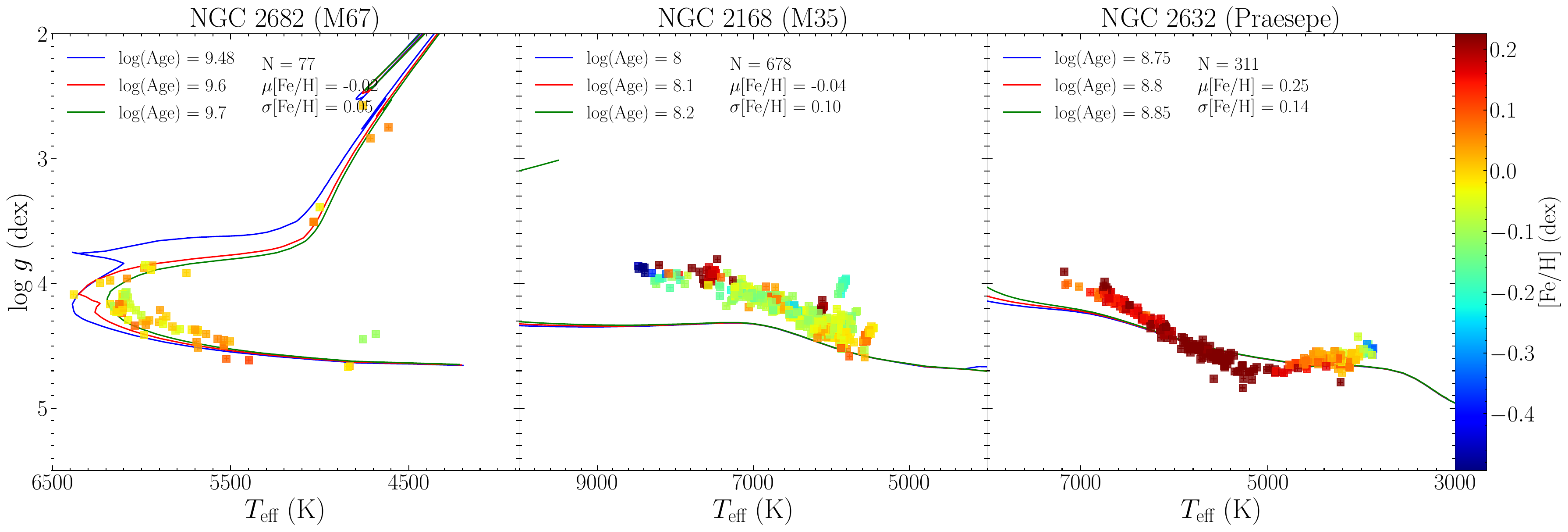}
		\centering
		\caption{The distributions of $T_{\text{eff}}$, log $g$, and [Fe/H] in \cite{fu2022lamost} for M$67$ (left panel), M$35$ (middle panel), and Praesepe (right panel). The different colored curves represent PARSEC V$2.0$ isochrones of various ages. The [Fe/H] for the OCs were taken from \cite{fu2022lamost}, while the ages were derived from \cite{carrera2019extended}, \cite{sung1999ubvi}, and \cite{salaris2004age} respectively.}
		\centering
		\label{LASP cluster}
	\end{figure*}
	
	The Gaia-ESO Survey (GES) is a public spectroscopic survey that has targeted more than $10^5$ stars, encompassing all major components of the Milky Way, from the end of $2011$ through $2018$, with its final public data release occurring in $2022$ May \citep{hourihane2023gaia}. In total, $115,614$ unique stars were observed using either the medium-resolution (R $\sim$ $20,000$) GIRAFFE setups or the high-resolution (R $\sim$ $47,000$) UVES instrument. The survey's spectral analysis is coordinated by several working groups, each specialized in analyzing one or more of the diverse stellar samples. Working Group $13$ (WG$13$) has been designated the responsibility of the spectral analysis of the hottest stars (O, B, and A types, $T_{\text{eff}} > 7000$ K) observed within the GES framework \citep{blomme2022gaia}. The parameters of OBA-type stars were derived by weighting non-local thermodynamic equilibrium (NLTE) and LTE, ensuring more accurate modeling of the complex atmospheric processes of these hot stars.
	
	We crossmatched the OBA-type stars in the recommended catalog with the GES, identifying a total of $18$ spectra. As shown in \reffig{GES OBA Teff logg FeH}, the offsets and scatters for $T_{\text{eff}}$, log $g$, and [Fe/H] are $-536\pm1028$ K, $-0.08\pm0.29$ dex, and $-0.26\pm0.3$ dex, respectively. Significantly, the comparision results reveal that, compared to the GES, the recommended catalog systematically underestimates $T_{\text{eff}}$, log $g$, and [Fe/H] by $1260$ K, $0.15$ dex, and $0.36$ dex, respectively, for the range of $10,000 < T_{\text{eff}} < 20,000$ K. However, the recommended catalog aligned well with GES parameters for $T_{\text{eff}} > 20,000$ K, with systematic differences in $T_{\text{eff}}$, log $g$, and [Fe/H] being $135$ K, $-0.06$ dex, and $-0.1$ dex, respectively. This pattern of systematic discrepancies is also present in the comparison of parameters between HotPayne and GES. Given that the atmospheric parameters predicted by both MaStar and HotPayne are based on LTE assumptions, we believe that these differences are likely attributable to the inherent discrepancies between LTE and NLTE models. Although repeated observations confirm consistency in the recommended catalog’s atmospheric parameters (\reffig{random err}), comparisons with the GES highlight NLTE effects' systematic impact on the parameters of hot stars. Consequently, we are actively extending our parameter prediction method to incorporate NLTE models and plan to publish a comprehensive catalog of NLTE atmospheric parameters in our forthcoming study.
	
	\begin{figure*}[htb]
		\centering
		\subfigure{
			\begin{minipage}[t]{0.45\linewidth}
				\centering
				\includegraphics[width=3.2in, height=2.4in]{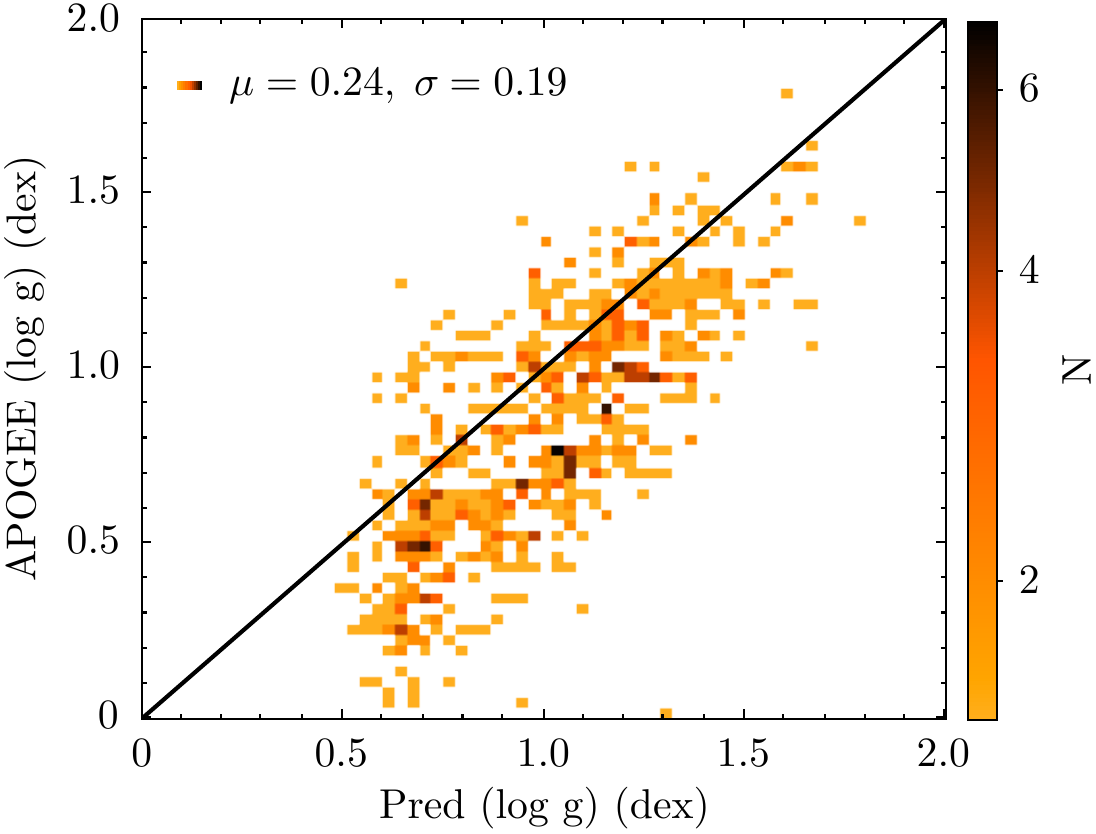}
				\centering
			\end{minipage}
		}
		\centering
		\subfigure{
			\begin{minipage}[t]{0.45\linewidth}
				\centering
				\includegraphics[width=3.2in, height=2.4in]{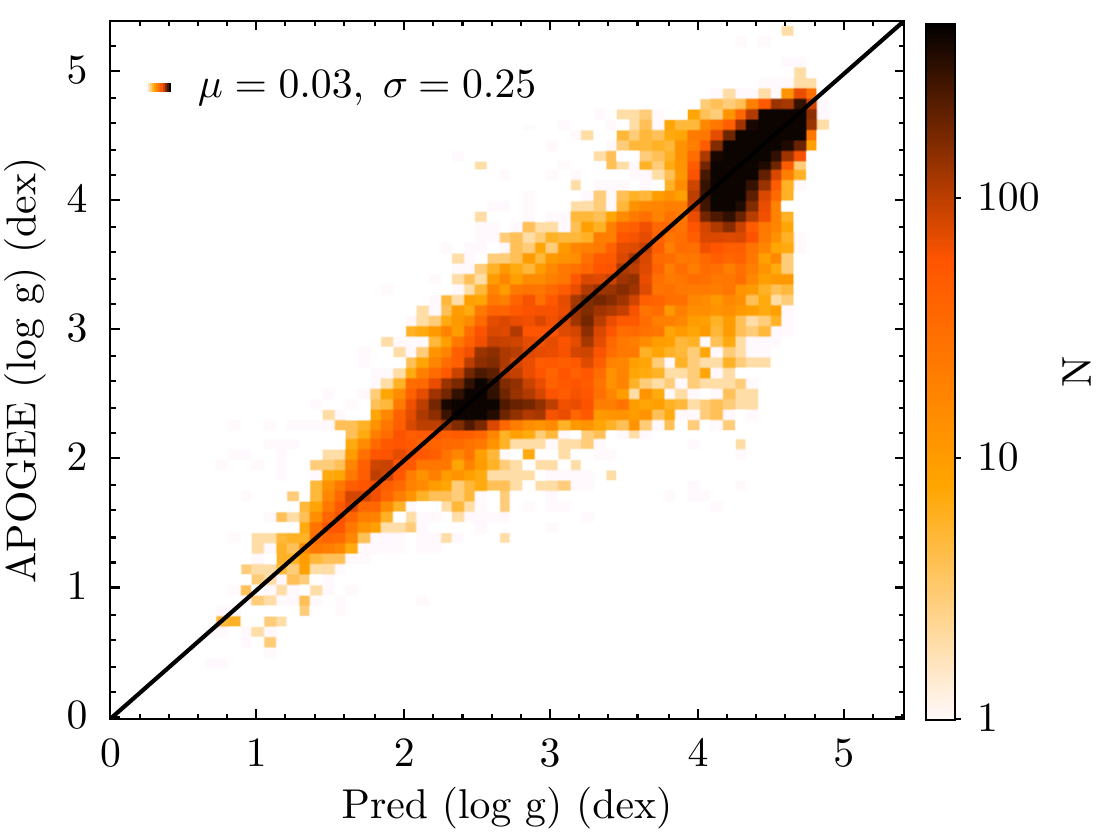}
				\centering
			\end{minipage}
		}
		\centering
		\subfigure{
			\begin{minipage}[t]{0.45\linewidth}
				\centering
				\includegraphics[width=3.2in, height=2.4in]{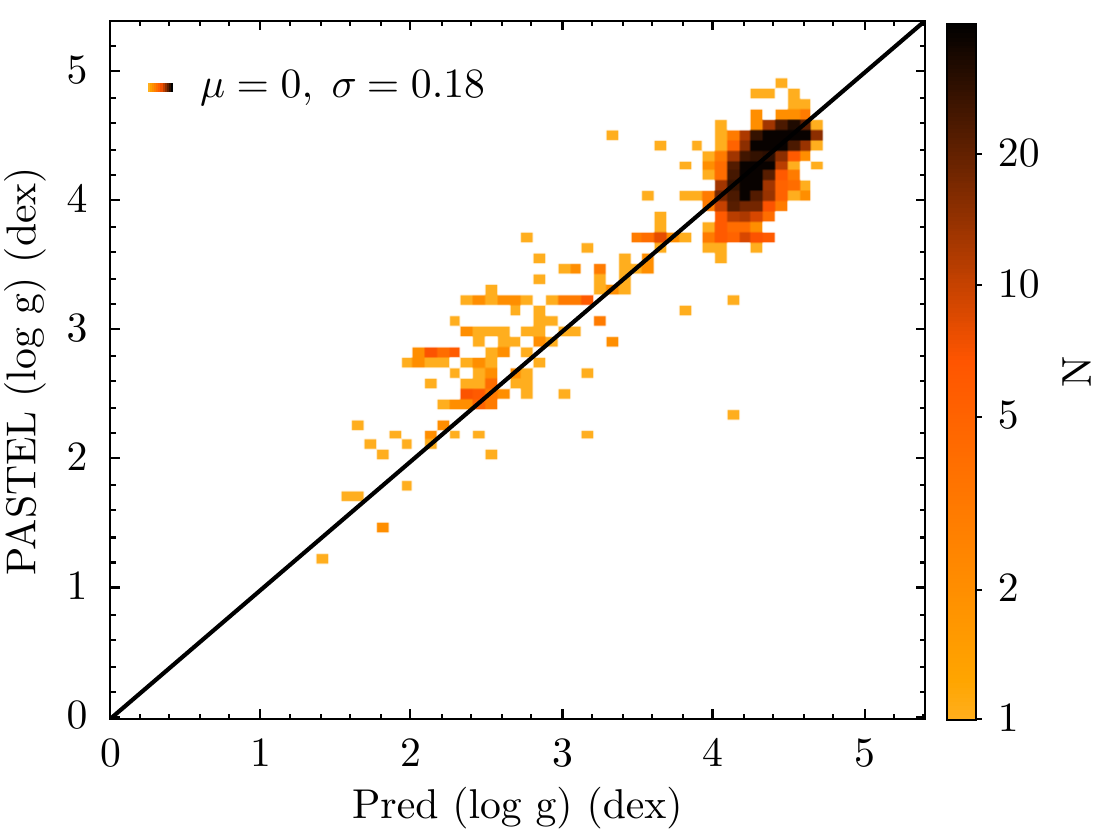}
				\centering
			\end{minipage}
		}
		\centering
		\subfigure{
			\begin{minipage}[t]{0.45\linewidth}
				\centering
				\includegraphics[width=3.2in, height=2.4in]{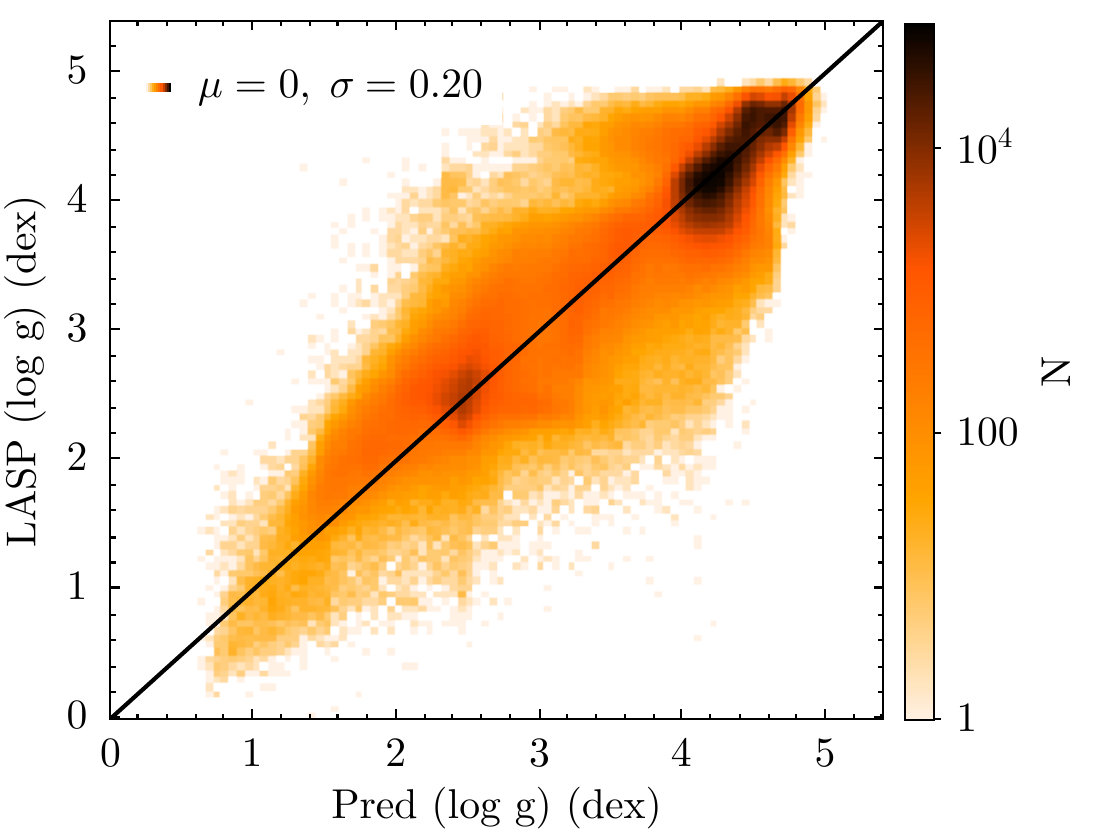}
				\centering
			\end{minipage}
		}
		\caption{Comparison of the log $g$ in the recommended catalog (log $g$ from \citealt{Imig_2022} replaces the MaStar's median log $g$ for predicting LAMOST parameters) with external data. The top left, top right, bottom left, and bottom right panels respectively represent the comparison of log $g$ with APOGEE's gM, APOGEE's FGK-type, PASTEL's FGK-type, and LASP's FGK-type stars. Mean and standard deviation of log $g$ differences between the recommended catalog and external data are also shown in each plot.}
		\centering
		\label{IJ logg}
	\end{figure*}
	
	\subsubsection{Applying the recommended catalog to open clusters} \label{clusters}
	Stars within open clusters (OCs) are considered to have formed almost simultaneously from a single gas cloud, exhibiting almost identical metallicity. Therefore, OCs offer a good test bed to check the consistency of metallicity determinations.
	
	We conducted a consistency test for the [Fe/H] among the M$67$, M$35$, and Praesepe clusters by crossmatching the recommended catalog with the member stars of these clusters as provided by \cite{fu2022lamost}. \reffig{our cluster} displays log $g$ as a function of $T_{\text{eff}}$ for member stars of these clusters. The PARSEC V$2.0$ isochrones also \citep{nguyen2022parsec} were overplotted for the three clusters. As shown in \reffig{our cluster}, the parameters of the recommended catalog generally are consistent with the isochrones in the $T_{\text{eff}}$$-$log $g$ plane. The [Fe/H] within the recommended catalog for the clusters M$67$, M$35$, and Praesepe shows little variation, with low dispersions of $0.07$, $0.12$, and $0.12$ dex, and their average [Fe/H] of $-0.06$, $-0.09$, and $0$ dex, respectively. These values closely match the [Fe/H] provided by \cite{fu2022lamost}, which are $-0.02$ dex for M$67$, $-0.04$ dex for M$35$, and $0.25$ dex for Praesepe, respectively. Notably, as \reffig{LASP cluster} illustrates, we noted a higher consistency with the isochrones for M$67$ for atmospheric parameters measured using LASP in \cite{fu2022lamost}, yet for M$35$ and Praesepe, log $g$ was systematically underestimated by $0.3$ dex for $T_{\text{eff}} > 6500$ K, highlighting the advantages of the recommended catalog in hot stars.
	
	\subsection{Future improvements} \label{Future improvements}
	This is our first attempt to predict stellar atmospheric parameters ($3500 < T_{\text{eff}} < 30,000$ K) for LAMOST low-resolution spectra based on the MaStar and to improve the efficiency of $\chi^{2}$ optimization using the grouping optimization strategy. Although we have demonstrated the feasibility of the grouping optimization strategy and the reliability of the recommended catalog, this work contains a few imperfections that need to be improved in future work.
	
	We plan to extend the grouping optimization strategy to error models involving $\chi^{2}$ function optimization. The parameter errors obtained using \refformula{error} assume that the systematic and random errors are independent. However, achieving this independence in practice is challenging. In such cases, Monte Carlo (MC) or MCMC \citep{andrieu2008tutorial} are typically used to derive the posterior distribution of parameters (the mean as the parameter prediction value and the standard deviation as the parameter error). To our knowledge, applying MC and MCMC to parameter prediction is currently very time consuming. For example, STEPARSYN \citep{tabernero2022steparsyn} requires $0.67$ hr to predict the atmospheric parameters of a high-resolution stellar spectrum using MCMC. Even with a $16$ computer SPARK cluster, generative spectrum networks (GSN; \citealt{rui2019analysis}) still needs $40.5$ hr to complete parameter predictions for $5.3$ million LAMOST low-resolution stellar spectra. If the grouping optimization strategy can be successfully applied to MC and MCMC, we will improve the efficiency of methods such as STEPARSYN and GSN and obtain more reasonable internal parameter errors in parameter predictions.
	
	We will update the log $g$ in the recommended catalog for subsequent work. As discussed in \refsubsection{M Type} and \refsubsection{AFGK Type}, we found significant differences of log $g$ between the recommended catalog and the external data for gM and FGK-type stars. Therefore, it is necessary to clarify the cause of these differences and make corrections to improve the reliability of the recommended catalog. We used the log $g$ from \cite{Imig_2022} to replace the median log $g$ from the MaStar and applied them to the LAMOST parameter predictions. As shown in \reffig{IJ logg}, we found that the predicted values of log $g$ are more consistent with external data. This indicates that the workflow (\refsubsection{subsection workflow}) is reasonable, and the difference in log $g$ is likely due to certain issues caused by the median log $g$ adopted by the MaStar. Further corrections to the MaStar by Yan et al. (2024, in preparation) will reduce the systematic errors present in our recommended catalog. This effort will be part of the next update of the recommended catalog.
	
	We will refine preprocessing of the spectra. To our knowledge, more than $10\%$ of hot star spectra contain emission lines \citep{xiang2022stellar}, and approximately $0.2\%$ of FGK-type stars are expected to have strong spectral emission lines and bad pixels (\refsubsection{AFGK Type}). In the current work, the spectral preprocessing method (\refsubsection{Pre-processing for the spectra}) was not fully effective in removing these anomalies (\reffig{unnormal spectral}). Therefore, the parameters might be problematic in unknown ways, adding uncertainty to these stars. Furthermore, Li et al. (2024, in preparation) found that the currently provided radial velocity by LAMOST have certain issues in M-type stars, which could affect the reliability of parameter predictions for M-type stars. In future work, we will use the outlier-detection method by \cite{wei2013mining} and \cite{lu2021study} to mask flux outliers and apply the radial velocities provided by Li et al. (2024, in preparation) to refine the preprocessing of the spectra.
	\section{Conclusions} \label{Conclusions}
	In this study, we have successfully demonstrated, for the first time, the application of the spectral emulator to derive the atmospheric parameters of O-M type stars, including $T_{\text{eff}}$, log $g$, [Fe/H], and [$\alpha$/Fe], from the low-resolution spectra of LAMOST. This method provides homogeneous parameters for a wider range of stellar types compared to other techniques currently available for LAMOST spectra.
	
	We constructed the spectral emulator based on the MaStar and employed the grouping optimization strategy to achieve a balance between efficiency and accuracy in our spectral-fitting method. The experimental results indicate that the workflow (\refsubsection{subsection workflow}) is feasible, capable of predicting parameters for $11,473,644$ spectra within $70$ hr while ensuring internal errors of $T_{\text{eff}}$, log $g$, [Fe/H], and [$\alpha$/Fe] within the ranges of $15-594$ K, $0.03-0.27$ dex, $0.02-0.10$ dex, and $0.01-0.04$ dex, respectively. To our knowledge, under single-machine operation, the grouping optimization strategy can improve efficiency by about $10$ times compared to spectral emulator methods such as LASP.
	
	Through a comparison of the recommended catalog with external data, we achieved four main results.
	\begin{enumerate}
		\item We identified and addressed certain issues in LASPM and LASP for M-type and A-type stars, respectively. We attribute the issues in M-type stars primarily to the limitations of the BT-Settl library and linear interpolators, while the issues in A-type stars are likely due to the inadequate spectral wavelength range selected by the LASP for reliably measuring parameters and the sparsity of ELODIE for hot stars. 
		\item We supplemented relatively reliable atmospheric parameters for M-type and OBA-type stars of LAMOST, providing data support for our subsequent development of an empirical spectral library for LAMOST.
		\item We highlighted the potential unreliability of the median log $g$ in the MaStar. Subsequently, we repredicted and supplemented the recommended catalog with log $g$ from \cite{Imig_2022}.
		\item We discovered limitations in the spectral preprocessing when addressing strong emission lines and bad pixels, resulting in approximately $0.2\%$ of FGK-type stellar atmospheric parameters being unreliable. Furthermore, by analyzing the distribution of cluster member stars in the $T_{\text{eff}}$$-$log $g$ panel, we noted that LASP may provide more reliable atmospheric parameters for stars with $T_{\text{eff}} < 6500$ K, likely because our full-spectrum fitting is more sensitive to outliers.
	\end{enumerate}
	
	Moreover, our work provides the first recommended catalog for the subsequent establishment of a LAMOST empirical spectral library covering a wide range of parameter space. Our efforts also offer a viable methodology for other large-scale surveys, such as the upcoming CSST optical survey.
	
	\section*{Acknowledgments}
	
	We would like to thank Mao-Sheng Xiang for providing us with useful comments.
	
	This work is supported by the National Key R\&D Program of China No. 2019YFA0405502, the National Natural Science Foundation of China (12273075, 12273078, 12261141689), and the science research grants from the China Manned Space Project with NO.CMS-CSST-2021-A10, CMS-CSST-2021-B05.
	
	Guoshoujing Telescope (the Large Sky Area Multi-Object Fiber Spectroscopic Telescope LAMOST) is a National Major Scientific Project built by the Chinese Academy of Sciences. Funding for the project has been provided by the National Development and Reform Commission. LAMOST is operated and managed by the National Astronomical Observatories, Chinese Academy of Sciences.
	
	Funding for the Sloan Digital Sky Survey IV has been provided by the Alfred P. Sloan Foundation, the U.S. Department of Energy Office of Science, and the Participating Institutions. SDSS acknowledges support and resources from the Center for High Performance Computing  at the University of Utah. The SDSS website is \url{www.sdss4.org}.
	
	SDSS is managed by the Astrophysical Research Consortium for the Participating Institutions of the SDSS Collaboration including the Brazilian Participation Group, the Carnegie Institution for Science, Carnegie Mellon University, Center for Astrophysics-Harvard \& Smithsonian, the Chilean Participation Group, the French Participation Group, Instituto de Astrof\'isica de Canarias, The Johns Hopkins 
	University, Kavli Institute for the Physics and Mathematics of the Universe (IPMU) / University of 
	Tokyo, the Korean Participation Group, Lawrence Berkeley National Laboratory, Leibniz Institut f\"ur Astrophysik 
	Potsdam (AIP),  Max-Planck-Institut f\"ur Astronomie (MPIA Heidelberg), Max-Planck-Institut f\"ur Astrophysik (MPA Garching), Max-Planck-Institut f\"ur Extraterrestrische Physik (MPE), National Astronomical Observatories of China, New Mexico State University, New York University, University of Notre Dame, Observat\'ario Nacional / MCTI, The Ohio State University, Pennsylvania State University, Shanghai Astronomical Observatory, United Kingdom Participation Group, Universidad Nacional Aut\'onoma de M\'exico, University of Arizona, University of Colorado Boulder, University of Oxford, University of Portsmouth, University of Utah, University of Virginia, University of Washington, University of Wisconsin, Vanderbilt University, and Yale University.
	
	This work has made use of data from the European Space Agency (ESA) mission {\it Gaia} (\url{https://www.cosmos.esa.int/gaia}), processed by the {\it Gaia} Data Processing and Analysis Consortium (DPAC,
	\url{https://www.cosmos.esa.int/web/gaia/dpac/consortium}). Funding for the DPAC has been provided by national institutions, in particular the institutions participating in the {\it Gaia} Multilateral Agreement.
	
	$Software$: Astropy \citep{2022ApJ...935..167A}, Bayesian Optimization (\citealt{Stander2002}; \url{https://github.com/fmfn/BayesianOptimization};  \citealt{pmlrv32gardner14}), Joblib (\url{https://joblib.readthedocs.io/en/stable/}), Matplotlib \citep{Hunter:2007}, NumPy \citep{harris2020array}, Pandas \citep{reback2020pandas}, SciPy \citep{Virtanen_2020}, Scikit-learn \citep{abraham2014machine}, TOPCAT \citep{2005ASPC..347...29T,2024arXiv240101156T}.
	
	\bibliographystyle{aasjournal}  
	\footnotesize
	\bibliography{refs}

\begin{thebibliography}{}
\expandafter\ifx\csname natexlab\endcsname\relax\def\natexlab#1{#1}\fi
\providecommand{\url}[1]{\href{#1}{#1}}
\providecommand{\dodoi}[1]{doi:~\href{http://doi.org/#1}{\nolinkurl{#1}}}
\providecommand{\doeprint}[1]{\href{http://ascl.net/#1}{\nolinkurl{http://ascl.net/#1}}}
\providecommand{\doarXiv}[1]{\href{https://arxiv.org/abs/#1}{\nolinkurl{https://arxiv.org/abs/#1}}}

\bibitem[{Abareshi {et~al.}(2022)Abareshi, Aguilar, Ahlen, Alam, Alexander,
  Alfarsy, Allen, Prieto, Alves, Ameel, Armengaud, Asorey, Aviles, Bailey,
  Balaguera-Antolínez, Ballester, Baltay, Bault, Beltran, Benavides, BenZvi,
  Berti, Besuner, Beutler, Bianchi, Blake, Blanc, Blum, Bolton, Bose, Bramall,
  Brieden, Brodzeller, Brooks, Brownewell, Buckley-Geer, Cahn, Cai, Canning,
  Capasso, Rosell, Carton, Casas, Castander, Cervantes-Cota, Chabanier,
  Chaussidon, Chuang, Circosta, Cole, Cooper, Costa, Cousinou, Cuceu, Davis,
  Dawson, Cruz-Noriega, Macorra, Mattia, Costa, Demmer, Derwent, Dey, Dey,
  Dhungana, Ding, Dobson, Doel, Donald-McCann, Donaldson, Douglass, Duan,
  Dunlop, Edelstein, Eftekharzadeh, Eisenstein, Enriquez-Vargas, Escoffier,
  Evatt, Fagrelius, Fan, Fanning, Fawcett, Ferraro, Ereza, Flaugher,
  Font-Ribera, Forero-Romero, Frenk, Fromenteau, Gänsicke, Garcia-Quintero,
  Garrison, Gaztañaga, Gerardi, Gil-Marín, A~Gontcho, Gonzalez-Morales,
  Gonzalez-de Rivera, Gonzalez-Perez, Gordon, Graur, Green, Grove, Gruen,
  Gutierrez, Guy, Hahn, Harris, Herrera, Herrera-Alcantar, Honscheid, Howlett,
  Huterer, Iršič, Ishak, Jelinsky, Jiang, Jimenez, Jing, Joyce, Jullo,
  Juneau, Karaçaylı, Karamanis, Karcher, Karim, Kehoe, Kent, Kirkby, Kisner,
  Kitaura, Koposov, Kovács, Kremin, Krolewski, L’Huillier, Lahav, Lambert,
  Lamman, Lan, Landriau, Lane, Lang, Lange, Lasker, Guillou, Leauthaud,
  Van~Suu, Levi, Li, Magneville, Manera, Manser, Marshall, Martini, McCollam,
  McDonald, Meisner, Mena-Fernández, Meneses-Rizo, Mezcua, Miller, Miquel,
  Montero-Camacho, Moon, Moustakas, Mueller, Muñoz-Gutiérrez, Myers,
  Nadathur, Najita, Napolitano, Neilsen, Newman, Nie, Ning, Niz, Norberg,
  Noriega, O’Brien, Obuljen, Palanque-Delabrouille, Palmese, Zhiwei,
  Pappalardo, PENG, Percival, Perruchot, Pogge, Poppett, Porredon, Prada,
  Prochaska, Pucha, Pérez-Fernández, Pérez-Ràfols, Rabinowitz, Raichoor,
  Ramirez-Solano, Ramírez-Pérez, Ravoux, Reil, Rezaie, Rocher, Rockosi, Roe,
  Roodman, Ross, Rossi, Ruggeri, Ruhlmann-Kleider, Sabiu, Safonova, Said,
  Saintonge, Catonga, Samushia, Sanchez, Saulder, Schaan, Schlafly, Schlegel,
  Schmoll, Scholte, Schubnell, Secroun, Seo, Serrano, Sharples, Sholl, Silber,
  Silva, Sirk, Siudek, Smith, Sprayberry, Staten, Stupak, Tan, Tarlé, Tie,
  Tojeiro, Ureña-López, Valdes, Valenzuela, Valluri, Vargas-Magaña, Verde,
  Walther, Wang, Wang, Weaver, Weaverdyck, Wechsler, Wilson, Yang, Yu, Yuan,
  Yèche, Zhang, Zhang, Zhao, Zhou, Zhou, Zou, Zou, Zou, \& Zu}]{Abareshi2022}
Abareshi, B., Aguilar, J., Ahlen, S., {et~al.} 2022, The Astronomical Journal,
  164, 207, \dodoi{10.3847/1538-3881/ac882b}

\bibitem[{Abdi \& Williams(2010)}]{Abdi2010}
Abdi, H., \& Williams, L.~J. 2010, Wiley interdisciplinary reviews.
  Computational statistics, 2, 433, \dodoi{10.1002/wics.101}

\bibitem[{{Abdurro'uf} {et~al.}(2022){Abdurro'uf}, {Accetta}, {Aerts}, {Silva
  Aguirre}, {Ahumada}, {Ajgaonkar}, {Filiz Ak}, {Alam}, {Allende Prieto},
  {Almeida}, {Anders}, {Anderson}, {Andrews}, {Anguiano}, {Aquino-Ort{\'\i}z},
  {Arag{\'o}n-Salamanca}, {Argudo-Fern{\'a}ndez}, {Ata}, {Aubert},
  {Avila-Reese}, {Badenes}, {Barb{\'a}}, {Barger}, {Barrera-Ballesteros},
  {Beaton}, {Beers}, {Belfiore}, {Bender}, {Bernardi}, {Bershady}, {Beutler},
  {Bidin}, {Bird}, {Bizyaev}, {Blanc}, {Blanton}, {Boardman}, {Bolton},
  {Boquien}, {Borissova}, {Bovy}, {Brandt}, {Brown}, {Brownstein}, {Brusa},
  {Buchner}, {Bundy}, {Burchett}, {Bureau}, {Burgasser}, {Cabang}, {Campbell},
  {Cappellari}, {Carlberg}, {Wanderley}, {Carrera}, {Cash}, {Chen}, {Chen},
  {Cherinka}, {Chiappini}, {Choi}, {Chojnowski}, {Chung}, {Clerc}, {Cohen},
  {Comerford}, {Comparat}, {da Costa}, {Covey}, {Crane}, {Cruz-Gonzalez},
  {Culhane}, {Cunha}, {Dai}, {Damke}, {Darling}, {Davidson}, {Davies},
  {Dawson}, {De Lee}, {Diamond-Stanic}, {Cano-D{\'\i}az}, {S{\'a}nchez},
  {Donor}, {Duckworth}, {Dwelly}, {Eisenstein}, {Elsworth}, {Emsellem},
  {Eracleous}, {Escoffier}, {Fan}, {Farr}, {Feng}, {Fern{\'a}ndez-Trincado},
  {Feuillet}, {Filipp}, {Fillingham}, {Frinchaboy}, {Fromenteau}, {Galbany},
  {Garc{\'\i}a}, {Garc{\'\i}a-Hern{\'a}ndez}, {Ge}, {Geisler}, {Gelfand},
  {G{\'e}ron}, {Gibson}, {Goddy}, {Godoy-Rivera}, {Grabowski}, {Green},
  {Greener}, {Grier}, {Griffith}, {Guo}, {Guy}, {Hadjara}, {Harding},
  {Hasselquist}, {Hayes}, {Hearty}, {Hern{\'a}ndez}, {Hill}, {Hogg},
  {Holtzman}, {Horta}, {Hsieh}, {Hsu}, {Hsu}, {Huber}, {Huertas-Company},
  {Hutchinson}, {Hwang}, {Ibarra-Medel}, {Chitham}, {Ilha}, {Imig}, {Jaekle},
  {Jayasinghe}, {Ji}, {Johnson}, {Jones}, {J{\"o}nsson}, {Katkov}, {Khalatyan},
  {Kinemuchi}, {Kisku}, {Knapen}, {Kneib}, {Kollmeier}, {Kong}, {Kounkel},
  {Kreckel}, {Krishnarao}, {Lacerna}, {Lane}, {Langgin}, {Lavender}, {Law},
  {Lazarz}, {Leung}, {Leung}, {Lewis}, {Li}, {Li}, {Lian}, {Liang}, {Lin},
  {Lin}, {Lin}, {Lintott}, {Long}, {Longa-Pe{\~n}a}, {L{\'o}pez-Cob{\'a}},
  {Lu}, {Lundgren}, {Luo}, {Mackereth}, {de la Macorra}, {Mahadevan},
  {Majewski}, {Manchado}, {Mandeville}, {Maraston}, {Margalef-Bentabol},
  {Masseron}, {Masters}, {Mathur}, {McDermid}, {Mckay}, {Merloni},
  {Merrifield}, {Meszaros}, {Miglio}, {Di Mille}, {Minniti}, {Minsley},
  {Monachesi}, {Moon}, {Mosser}, {Mulchaey}, {Muna}, {Mu{\~n}oz}, {Myers},
  {Myers}, {Nadathur}, {Nair}, {Nandra}, {Neumann}, {Newman}, {Nidever},
  {Nikakhtar}, {Nitschelm}, {O'Connell}, {Garma-Oehmichen}, {Luan Souza de
  Oliveira}, {Olney}, {Oravetz}, {Ortigoza-Urdaneta}, {Osorio}, {Otter},
  {Pace}, {Padilla}, {Pan}, {Pan}, {Parikh}, {Parker}, {Peirani}, {Pe{\~n}a
  Ram{\'\i}rez}, {Penny}, {Percival}, {Perez-Fournon}, {Pinsonneault},
  {Poidevin}, {Poovelil}, {Price-Whelan}, {B{\'a}rbara de Andrade Queiroz},
  {Raddick}, {Ray}, {Rembold}, {Riddle}, {Riffel}, {Riffel}, {Rix}, {Robin},
  {Rodr{\'\i}guez-Puebla}, {Roman-Lopes}, {Rom{\'a}n-Z{\'u}{\~n}iga}, {Rose},
  {Ross}, {Rossi}, {Rubin}, {Salvato}, {S{\'a}nchez}, {S{\'a}nchez-Gallego},
  {Sanderson}, {Santana Rojas}, {Sarceno}, {Sarmiento}, {Sayres}, {Sazonova},
  {Schaefer}, {Schiavon}, {Schlegel}, {Schneider}, {Schultheis}, {Schwope},
  {Serenelli}, {Serna}, {Shao}, {Shapiro}, {Sharma}, {Shen}, {Shetrone}, {Shu},
  {Simon}, {Skrutskie}, {Smethurst}, {Smith}, {Sobeck}, {Spoo}, {Sprague},
  {Stark}, {Stassun}, {Steinmetz}, {Stello}, {Stone-Martinez},
  {Storchi-Bergmann}, {Stringfellow}, {Stutz}, {Su}, {Taghizadeh-Popp},
  {Talbot}, {Tayar}, {Telles}, {Teske}, {Thakar}, {Theissen}, {Tkachenko},
  {Thomas}, {Tojeiro}, {Hernandez Toledo}, {Troup}, {Trump}, {Trussler},
  {Turner}, {Tuttle}, {Unda-Sanzana}, {V{\'a}zquez-Mata}, {Valentini},
  {Valenzuela}, {Vargas-Gonz{\'a}lez}, {Vargas-Maga{\~n}a}, {Alfaro},
  {Villanova}, {Vincenzo}, {Wake}, {Warfield}, {Washington}, {Weaver},
  {Weijmans}, {Weinberg}, {Weiss}, {Westfall}, {Wild}, {Wilde}, {Wilson},
  {Wilson}, {Wilson}, {Wolf}, {Wood-Vasey}, {Yan}, {Zamora}, {Zasowski},
  {Zhang}, {Zhao}, {Zheng}, {Zheng}, \& {Zhu}}]{abdurro2022seventeenth}
{Abdurro'uf}, {Accetta}, K., {Aerts}, C., {et~al.} 2022, \apjs, 259, 35,
  \dodoi{10.3847/1538-4365/ac4414}

\bibitem[{Abraham {et~al.}(2014)Abraham, Pedregosa, Eickenberg, Gervais,
  Mueller, Kossaifi, Gramfort, Thirion, \& Varoquaux}]{abraham2014machine}
Abraham, A., Pedregosa, F., Eickenberg, M., {et~al.} 2014, Frontiers in
  neuroinformatics, 8, 14, \dodoi{10.3389/fninf.2014.00014}

\bibitem[{{Adibekyan} {et~al.}(2011){Adibekyan}, {Santos}, {Sousa}, \&
  {Israelian}}]{adibekyan2011new}
{Adibekyan}, V.~Z., {Santos}, N.~C., {Sousa}, S.~G., \& {Israelian}, G. 2011,
  \aap, 535, L11, \dodoi{10.1051/0004-6361/201118240}

\bibitem[{Allard {et~al.}(2012)Allard, Homeier, \& Freytag}]{allard2012models}
Allard, F., Homeier, D., \& Freytag, B. 2012, Philosophical Transactions of the
  Royal Society A: Mathematical, Physical and Engineering Sciences, 370, 2765,
  \dodoi{10.1098/rsta.2011.0269}

\bibitem[{Almeida {et~al.}(2023)Almeida, Anderson, Argudo-Fernández, Badenes,
  Barger, Barrera-Ballesteros, Bender, Benitez, Besser, Bird, Bizyaev, Blanton,
  Bochanski, Bovy, Brandt, Brownstein, Buchner, Bulbul, Burchett, Díaz,
  Carlberg, Casey, Chandra, Cherinka, Chiappini, Coker, Comparat, Conroy,
  Contardo, Cortes, Covey, Crane, Cunha, Dabbieri, Davidson, Davis,
  de~Andrade~Queiroz, De~Lee, Méndez~Delgado, Demasi, Di~Mille, Donor, Dow,
  Dwelly, Eracleous, Eriksen, Fan, Farr, Frederick, Fries, Frinchaboy,
  Gänsicke, Ge, González~Ávila, Grabowski, Grier, Guiglion, Gupta, Hall,
  Hawkins, Hayes, Hermes, Hernández-García, Hogg, Holtzman, Ibarra-Medel, Ji,
  Jofre, Johnson, Jones, Kinemuchi, Kluge, Koekemoer, Kollmeier, Kounkel,
  Krishnarao, Krumpe, Lacerna, Lago, Laporte, Liu, Liu, Liu, Lopes,
  Macktoobian, Majewski, Malanushenko, Maoz, Masseron, Masters, Matijevic,
  McBride, Medan, Merloni, Morrison, Myers, Mészáros, Negrete, Nidever,
  Nitschelm, Oravetz, Oravetz, Pan, Peng, Pinsonneault, Pogge, Qiu, Ramirez,
  Rix, Rosso, Runnoe, Salvato, Sanchez, Santana, Saydjari, Sayres, Schlaufman,
  Schneider, Schwope, Serna, Shen, Sobeck, Song, Souto, Spoo, Stassun,
  Steinmetz, Straumit, Stringfellow, Sánchez-Gallego, Taghizadeh-Popp, Tayar,
  Thakar, Tissera, Tkachenko, Toledo, Trakhtenbrot, Fernández-Trincado, Troup,
  Trump, Tuttle, Ulloa, Vazquez-Mata, Alfaro, Villanova, Wachter, Weijmans,
  Wheeler, Wilson, Wojno, Wolf, Xue, Ybarra, Zari, \& Zasowski}]{Almeida_2023}
Almeida, A., Anderson, S.~F., Argudo-Fernández, M., {et~al.} 2023, The
  Astrophysical Journal Supplement Series, 267, 44,
  \dodoi{10.3847/1538-4365/acda98}

\bibitem[{Andrieu \& Thoms(2008)}]{andrieu2008tutorial}
Andrieu, C., \& Thoms, J. 2008, Statistics and computing, 18, 343,
  \dodoi{10.1007/s11222-008-9110-y}

\bibitem[{{Astropy Collaboration} {et~al.}(2022){Astropy Collaboration},
  {Price-Whelan}, {Lim}, {Earl}, {Starkman}, {Bradley}, {Shupe}, {Patil},
  {Corrales}, {Brasseur}, {N{\"o}the}, {Donath}, {Tollerud}, {Morris},
  {Ginsburg}, {Vaher}, {Weaver}, {Tocknell}, {Jamieson}, {van Kerkwijk},
  {Robitaille}, {Merry}, {Bachetti}, {G{\"u}nther}, {Aldcroft},
  {Alvarado-Montes}, {Archibald}, {B{\'o}di}, {Bapat}, {Barentsen},
  {Baz{\'a}n}, {Biswas}, {Boquien}, {Burke}, {Cara}, {Cara}, {Conroy},
  {Conseil}, {Craig}, {Cross}, {Cruz}, {D'Eugenio}, {Dencheva}, {Devillepoix},
  {Dietrich}, {Eigenbrot}, {Erben}, {Ferreira}, {Foreman-Mackey}, {Fox},
  {Freij}, {Garg}, {Geda}, {Glattly}, {Gondhalekar}, {Gordon}, {Grant},
  {Greenfield}, {Groener}, {Guest}, {Gurovich}, {Handberg}, {Hart},
  {Hatfield-Dodds}, {Homeier}, {Hosseinzadeh}, {Jenness}, {Jones}, {Joseph},
  {Kalmbach}, {Karamehmetoglu}, {Ka{\l}uszy{\'n}ski}, {Kelley}, {Kern},
  {Kerzendorf}, {Koch}, {Kulumani}, {Lee}, {Ly}, {Ma}, {MacBride}, {Maljaars},
  {Muna}, {Murphy}, {Norman}, {O'Steen}, {Oman}, {Pacifici}, {Pascual},
  {Pascual-Granado}, {Patil}, {Perren}, {Pickering}, {Rastogi}, {Roulston},
  {Ryan}, {Rykoff}, {Sabater}, {Sakurikar}, {Salgado}, {Sanghi}, {Saunders},
  {Savchenko}, {Schwardt}, {Seifert-Eckert}, {Shih}, {Jain}, {Shukla}, {Sick},
  {Simpson}, {Singanamalla}, {Singer}, {Singhal}, {Sinha}, {Sip{\H{o}}cz},
  {Spitler}, {Stansby}, {Streicher}, {{\v{S}}umak}, {Swinbank}, {Taranu},
  {Tewary}, {Tremblay}, {de Val-Borro}, {Van Kooten}, {Vasovi{\'c}}, {Verma},
  {de Miranda Cardoso}, {Williams}, {Wilson}, {Winkel}, {Wood-Vasey}, {Xue},
  {Yoachim}, {Zhang}, {Zonca}, \& {Astropy Project
  Contributors}}]{2022ApJ...935..167A}
{Astropy Collaboration}, {Price-Whelan}, A.~M., {Lim}, P.~L., {et~al.} 2022,
  \apj, 935, 167, \dodoi{10.3847/1538-4357/ac7c74}

\bibitem[{Bailer-Jones {et~al.}(2021)Bailer-Jones, Rybizki, Fouesneau,
  Demleitner, \& Andrae}]{bailer2021estimating}
Bailer-Jones, C., Rybizki, J., Fouesneau, M., Demleitner, M., \& Andrae, R.
  2021, The Astronomical Journal, 161, 147, \dodoi{10.3847/1538-3881/abd806}

\bibitem[{Blanton {et~al.}(2017)Blanton, Bershady, Abolfathi, Albareti, Prieto,
  Almeida, Alonso-García, Anders, Anderson, Andrews, Aquino-Ortíz,
  Aragón-Salamanca, Argudo-Fernández, Armengaud, Aubourg, Avila-Reese,
  Badenes, Bailey, Barger, Barrera-Ballesteros, Bartosz, Bates, Baumgarten,
  Bautista, Beaton, Beers, Belfiore, Bender, Berlind, Bernardi, Beutler, Bird,
  Bizyaev, Blanc, Blomqvist, Bolton, Boquien, Borissova, van~den Bosch, Bovy,
  Brandt, Brinkmann, Brownstein, Bundy, Burgasser, Burtin, Busca, Cappellari,
  Carigi, Carlberg, Rosell, Carrera, Chanover, Cherinka, Cheung, Chew,
  Chiappini, Choi, Chojnowski, Chuang, Chung, Cirolini, Clerc, Cohen, Comparat,
  da~Costa, Cousinou, Covey, Crane, Croft, Cruz-Gonzalez, Cuadra, Cunha, Damke,
  Darling, Davies, Dawson, de~la Macorra, Dell’Agli, Lee, Delubac, Mille,
  Diamond-Stanic, Cano-Díaz, Donor, Downes, Drory, du~Mas~des Bourboux,
  Duckworth, Dwelly, Dyer, Ebelke, Eigenbrot, Eisenstein, Emsellem, Eracleous,
  Escoffier, Evans, Fan, Fernández-Alvar, Fernandez-Trincado, Feuillet,
  Finoguenov, Fleming, Font-Ribera, Fredrickson, Freischlad, Frinchaboy,
  Fuentes, Galbany, Garcia-Dias, García-Hernández, Gaulme, Geisler, Gelfand,
  Gil-Marín, Gillespie, Goddard, Gonzalez-Perez, Grabowski, Green, Grier,
  Gunn, Guo, Guy, Hagen, Hahn, Hall, Harding, Hasselquist, Hawley, Hearty,
  Hernández, Ho, Hogg, Holley-Bockelmann, Holtzman, Holzer, Huehnerhoff,
  Hutchinson, Hwang, Ibarra-Medel, da~Silva~Ilha, Ivans, Ivory, Jackson,
  Jensen, Johnson, Jones, Jönsson, Jullo, Kamble, Kinemuchi, Kirkby, Kitaura,
  Klaene, Knapp, Kneib, Kollmeier, Lacerna, Lane, Lang, Law, Lazarz, Lee, Goff,
  Liang, Li, Li, Lian, Lima, Lin, Lin, de~Lis, Liu, de~Icaza~Lizaola, Long,
  Lucatello, Lundgren, MacDonald, Machado, MacLeod, Mahadevan, Maia, Maiolino,
  Majewski, Malanushenko, Malanushenko, Manchado, Mao, Maraston,
  Marques-Chaves, Masseron, Masters, McBride, McDermid, McGrath, McGreer,
  Peña, Melendez, Merloni, Merrifield, Meszaros, Meza, Minchev, Minniti,
  Miyaji, More, Mulchaey, Müller-Sánchez, Muna, Munoz, Myers, Nair, Nandra,
  do~Nascimento, Negrete, Ness, Newman, Nichol, Nidever, Nitschelm, Ntelis,
  O’Connell, Oelkers, Oravetz, Oravetz, Pace, Padilla, Palanque-Delabrouille,
  Palicio, Pan, Parejko, Parikh, Pâris, Park, Patten, Peirani,
  Pellejero-Ibanez, Penny, Percival, Perez-Fournon, Petitjean, Pieri,
  Pinsonneault, Pisani, Poleski, Prada, Prakash, de~Andrade~Queiroz, Raddick,
  Raichoor, Rembold, Richstein, Riffel, Riffel, Rix, Robin, Rockosi,
  Rodríguez-Torres, Roman-Lopes, Román-Zúñiga, Rosado, Ross, Rossi, Ruan,
  Ruggeri, Rykoff, Salazar-Albornoz, Salvato, Sánchez, Aguado,
  Sánchez-Gallego, Santana, Santiago, Sayres, Schiavon, da~Silva~Schimoia,
  Schlafly, Schlegel, Schneider, Schultheis, Schuster, Schwope, Seo, Shao,
  Shen, Shetrone, Shull, Simon, Skinner, Skrutskie, Slosar, Smith, Sobeck,
  Sobreira, Somers, Souto, Stark, Stassun, Stauffer, Steinmetz,
  Storchi-Bergmann, Streblyanska, Stringfellow, Suárez, Sun, Suzuki, Szigeti,
  Taghizadeh-Popp, Tang, Tao, Tayar, Tembe, Teske, Thakar, Thomas, Thompson,
  Tinker, Tissera, Tojeiro, Toledo, de~la Torre, Tremonti, Troup, Valenzuela,
  Valpuesta, Vargas-González, Vargas-Magaña, Vazquez, Villanova, Vivek, Vogt,
  Wake, Walterbos, Wang, Weaver, Weijmans, Weinberg, Westfall, Whelan, Wild,
  Wilson, Wood-Vasey, Wylezalek, Xiao, Yan, Yang, Ybarra, Yèche, Zakamska,
  Zamora, Zarrouk, Zasowski, Zhang, Zhao, Zheng, Zheng, Zhou, Zhou, Zhu,
  Zoccali, \& Zou}]{blanton2017sloan}
Blanton, M.~R., Bershady, M.~A., Abolfathi, B., {et~al.} 2017, The Astronomical
  Journal, 154, 28, \dodoi{10.3847/1538-3881/aa7567}

\bibitem[{{Blomme} {et~al.}(2022){Blomme}, {Daflon}, {Gebran}, {Herrero},
  {Lobel}, {Mahy}, {Martins}, {Morel}, {Berlanas}, {Blaz{\`e}re}, {Fr{\'e}mat},
  {Gosset}, {Ma{\'\i}z Apell{\'a}niz}, {Santos}, {Semaan},
  {Sim{\'o}n-D{\'\i}az}, {Volpi}, {Holgado}, {Jim{\'e}nez-Esteban}, {Nieva},
  {Przybilla}, {Gilmore}, {Randich}, {Negueruela}, {Prusti}, {Vallenari},
  {Alfaro}, {Bensby}, {Bragaglia}, {Flaccomio}, {Francois}, {Korn},
  {Lanzafame}, {Pancino}, {Smiljanic}, {Bergemann}, {Carraro}, {Franciosini},
  {Gonneau}, {Heiter}, {Hourihane}, {Jofr{\'e}}, {Magrini}, {Morbidelli},
  {Sacco}, {Worley}, \& {Zaggia}}]{blomme2022gaia}
{Blomme}, R., {Daflon}, S., {Gebran}, M., {et~al.} 2022, \aap, 661, A120,
  \dodoi{10.1051/0004-6361/202142349}

\bibitem[{Bohlin {et~al.}(2017)Bohlin, M{\'e}sz{\'a}ros, Fleming, Gordon,
  Koekemoer, \& Kov{\'a}cs}]{bohlin2017new}
Bohlin, R.~C., M{\'e}sz{\'a}ros, S., Fleming, S.~W., {et~al.} 2017, The
  Astronomical Journal, 153, 234, \dodoi{10.3847/1538-3881/aa6ba9}

\bibitem[{{Bu} \& {Pan}(2015)}]{bu2015stellar}
{Bu}, Y., \& {Pan}, J. 2015, \mnras, 447, 256, \dodoi{10.1093/mnras/stu2063}

\bibitem[{Cargile {et~al.}(2020)Cargile, Conroy, Johnson, Ting, Bonaca, Dotter,
  \& Speagle}]{cargile2020minesweeper}
Cargile, P.~A., Conroy, C., Johnson, B.~D., {et~al.} 2020, The Astrophysical
  Journal, 900, 28, \dodoi{10.3847/1538-4357/aba43b}

\bibitem[{{Carrera} {et~al.}(2019){Carrera}, {Pasquato}, {Vallenari},
  {Balaguer-N{\'u}{\~n}ez}, {Cantat-Gaudin}, {Mapelli}, {Bragaglia}, {Bossini},
  {Jordi}, {Galad{\'\i}-Enr{\'\i}quez}, \& {Solano}}]{carrera2019extended}
{Carrera}, R., {Pasquato}, M., {Vallenari}, A., {et~al.} 2019, \aap, 627, A119,
  \dodoi{10.1051/0004-6361/201935599}

\bibitem[{Chen {et~al.}(2018)Chen, Huang, Yuan, Wang, Fan, Xiang, Zhang, Tian,
  \& Liu}]{Chen_2018}
Chen, B.-Q., Huang, Y., Yuan, H.-B., {et~al.} 2018, Monthly Notices of the
  Royal Astronomical Society, 483, 4277, \dodoi{10.1093/mnras/sty3341}

\bibitem[{Choi {et~al.}(2016)Choi, Dotter, Conroy, Cantiello, Paxton, \&
  Johnson}]{Choi_2016}
Choi, J., Dotter, A., Conroy, C., {et~al.} 2016, The Astrophysical Journal,
  823, 102, \dodoi{10.3847/0004-637x/823/2/102}

\bibitem[{Cirasuolo {et~al.}(2020)Cirasuolo, Fairley, Rees, Gonzalez, Taylor,
  Maiolino, Afonso, Evans, Flores, Lilly, Oliva, Paltani, Vanzi, Abreu,
  Accardo, Adams, Álvarez Méndez, Amans, Amarantidis, Atek, Atkinson,
  Banerji, Barrett, Barrientos, Bauer, Beard, Béchet, Belfiore, Bellazzini,
  Benoist, Best, Biazzo, Black, Boettger, Bonifacio, Bowler, Bragaglia,
  Brierley, Brinchmann, Brinkmann, Buat, Buitrago, Burgarella, Burningham,
  Buscher, Cabral, Caffau, Cardoso, Carnall, Carollo, Castillo, Castignani,
  Catelan, Cicone, Cimatti, Cioni, Clementini, Cochrane, Coelho, Colling,
  Contini, Contreras, Conzelmann, Cresci, Cropper, Cucciati, Cullen, Cumani,
  Curti, Da~Silva, Daddi, Dalessandro, Dalessio, Dauvin, Davidson, De~Laverny,
  Delplancke-Ströbele, De~Lucia, Del~Vecchio, Dessauges-Zavadsky, Di~Matteo,
  Dole, Drass, Dunlop, Dünner, Eales, Ellis, Enriques, Fasola, Ferguson,
  Ferruzzi, Fisher, Flores, Fontana, Forchi, Francois, Franzetti, Gargiulo,
  Garilli, Gaudemard, Gieles, Gilmore, Ginolfi, Gomes, Guinouard, Gutierrez,
  Haigron, Hammer, Hammersley, Haniff, Harrison, Haywood, Hill, Hubin,
  Humphrey, Ibata, Infante, Ives, Ivison, Iwert, Jablonka, Jakob, Jarvis, King,
  Kneib, Laporte, Lawrence, Lee, Li~Causi, Lorenzoni, Lucatello, Luco, Macleod,
  Magliocchetti, Magrini, Mainieri, Maire, Mannucci, Martin, Matute,
  Maurogordato, McGee, Mcleod, McLure, McMahon, Melse, Messias, Mucciarelli,
  Nisini, Nix, Norberg, Oesch, Oliveira, Origlia, Padilla, Palsa, Pancino,
  Papaderos, Pappalardo, Parry, Pasquini, Peacock, Pedichini, Pello, Peng,
  Pentericci, Pfuhl, Piazzesi, Popovic, Pozzetti, Puech, Puzia, Raichoor,
  Randich, Recio-Blanco, Reis, Reix, Renzini, Rodrigues, Rojas,
  Rojas-Arriagada, Rota, Royer, Sacco, Sanchez-Janssen, Sanna, Santos, Sarzi,
  Schaerer, Schiavon, Schnell, Schultheis, Scodeggio, Serjeant, Shen, Simmonds,
  Smoker, Sobral, Sordet, Spérone, Strachan, Sun, Swinbank, Tait, Tereno,
  Tojeiro, Torres, Tosi, Tozzi, Tresiter, Valenti, Valenzuela~Navarro,
  Vanzella, Vergani, Verhamme, Vernet, Vignali, Vinther, Von~Dran, Waring,
  Watson, Wild, Willesme, Woodward, Wuyts, Yang, Zamorani, Zoccali, Bluck, \&
  Trussler}]{Cirasuolo2020}
Cirasuolo, M., Fairley, A., Rees, P., {et~al.} 2020, Published in The Messenger
  vol. 180, pp. 10-17, June 2020., \dodoi{10.18727/0722-6691/5195}

\bibitem[{Cui {et~al.}(2012)Cui, Zhao, Chu, Li, Li, Zhang, Su, Yao, Wang, Xing,
  {et~al.}}]{cui2012large}
Cui, X.-Q., Zhao, Y.-H., Chu, Y.-Q., {et~al.} 2012, Research in Astronomy and
  Astrophysics, 12, 1197, \dodoi{10.1088/1674-4527/12/9/003}

\bibitem[{Czekala {et~al.}(2015)Czekala, Andrews, Mandel, Hogg, \&
  Green}]{Czekala_2015}
Czekala, I., Andrews, S.~M., Mandel, K.~S., Hogg, D.~W., \& Green, G.~M. 2015,
  The Astrophysical Journal, 812, 128, \dodoi{10.1088/0004-637x/812/2/128}

\bibitem[{de~Jong {et~al.}(2022)de~Jong, Bellido-Tirado, Brynnel, Amini, Frey,
  F{\"u}{\ss}lein, G{\"a}bler, Giannone, Johl, Kuba, Lemke, Micheva, Saviauk,
  Steinmetz, Walcher, Winkler, Lind, Loveday, Feltzing, McMahon, Mainieri,
  Pirard, Bensby, Bergemann, Chiappini, Christlieb, Cioni, Comparat, Driver,
  Hook, Irwin, Kneib, Liske, Merloni, Minchev, Richard, Starkenburg, Sullivan,
  Worley, Gaessler, Laurent, Pragt, Remillieux, Rothmaier, Smedley, Stilz,
  Walton, Alexander, Church, Croom, Davies, Heneka, Kacharov, Knoche,
  Kordopatis, Krumpe, Martell, Norberg, Pelisoli, Sharma, Storm, \&
  Tempel}]{deJong2022}
de~Jong, R.~S., Bellido-Tirado, O., Brynnel, J.~G., {et~al.} 2022, in
  Ground-based and Airborne Instrumentation for Astronomy IX, ed. C.~J. Evans,
  J.~J. Bryant, \& K.~Motohara, Vol. 12184, International Society for Optics
  and Photonics (SPIE), 1218414, \dodoi{10.1117/12.2628965}

\bibitem[{{De Silva} {et~al.}(2015){De Silva}, {Freeman}, {Bland-Hawthorn},
  {Martell}, {de Boer}, {Asplund}, {Keller}, {Sharma}, {Zucker}, {Zwitter},
  {Anguiano}, {Bacigalupo}, {Bayliss}, {Beavis}, {Bergemann}, {Campbell},
  {Cannon}, {Carollo}, {Casagrande}, {Casey}, {Da Costa}, {D'Orazi}, {Dotter},
  {Duong}, {Heger}, {Ireland}, {Kafle}, {Kos}, {Lattanzio}, {Lewis}, {Lin},
  {Lind}, {Munari}, {Nataf}, {O'Toole}, {Parker}, {Reid}, {Schlesinger},
  {Sheinis}, {Simpson}, {Stello}, {Ting}, {Traven}, {Watson}, {Wittenmyer},
  {Yong}, \& {{\v{Z}}erjal}}]{de2015galah}
{De Silva}, G.~M., {Freeman}, K.~C., {Bland-Hawthorn}, J., {et~al.} 2015,
  \mnras, 449, 2604, \dodoi{10.1093/mnras/stv327}

\bibitem[{{DESI Collaboration} {et~al.}(2016{\natexlab{a}}){DESI
  Collaboration}, Aghamousa, Aguilar, Ahlen, Alam, Allen, Prieto, Annis,
  Bailey, Balland, Ballester, Baltay, Beaufore, Bebek, Beers, Bell, Bernal,
  Besuner, Beutler, Blake, Bleuler, Blomqvist, Blum, Bolton, Briceno, Brooks,
  Brownstein, Buckley-Geer, Burden, Burtin, Busca, Cahn, Cai, Cardiel-Sas,
  Carlberg, Carton, Casas, Castander, Cervantes-Cota, Claybaugh, Close, Coker,
  Cole, Comparat, Cooper, Cousinou, Crocce, Cuby, Cunningham, Davis, Dawson,
  de~la Macorra, De~Vicente, Delubac, Derwent, Dey, Dhungana, Ding, Doel, Duan,
  Ealet, Edelstein, Eftekharzadeh, Eisenstein, Elliott, Escoffier, Evatt,
  Fagrelius, Fan, Fanning, Farahi, Farihi, Favole, Feng, Fernandez, Findlay,
  Finkbeiner, Fitzpatrick, Flaugher, Flender, Font-Ribera, Forero-Romero,
  Fosalba, Frenk, Fumagalli, Gaensicke, Gallo, Garcia-Bellido, Gaztanaga,
  Fusillo, Gerard, Gershkovich, Giannantonio, Gillet, Gonzalez-de Rivera,
  Gonzalez-Perez, Gott, Graur, Gutierrez, Guy, Habib, Heetderks, Heetderks,
  Heitmann, Hellwing, Herrera, Ho, Holland, Honscheid, Huff, Hutchinson,
  Huterer, Hwang, Laguna, Ishikawa, Jacobs, Jeffrey, Jelinsky, Jennings, Jiang,
  Jimenez, Johnson, Joyce, Jullo, Juneau, Kama, Karcher, Karkar, Kehoe,
  Kennamer, Kent, Kilbinger, Kim, Kirkby, Kisner, Kitanidis, Kneib, Koposov,
  Kovacs, Koyama, Kremin, Kron, Kronig, Kueter-Young, Lacey, Lafever, Lahav,
  Lambert, Lampton, Landriau, Lang, Lauer, Goff, Guillou, Van~Suu, Lee, Lee,
  Leitner, Lesser, Levi, L'Huillier, Li, Liang, Lin, Linder, Loebman, Lukić,
  Ma, MacCrann, Magneville, Makarem, Manera, Manser, Marshall, Martini, Massey,
  Matheson, McCauley, McDonald, McGreer, Meisner, Metcalfe, Miller, Miquel,
  Moustakas, Myers, Naik, Newman, Nichol, Nicola, da~Costa, Nie, Niz, Norberg,
  Nord, Norman, Nugent, O'Brien, Oh, Olsen, Padilla, Padmanabhan, Padmanabhan,
  Palanque-Delabrouille, Palmese, Pappalardo, Pâris, Park, Patej, Peacock,
  Peiris, Peng, Percival, Perruchot, Pieri, Pogge, Pollack, Poppett, Prada,
  Prakash, Probst, Rabinowitz, Raichoor, Ree, Refregier, Regal, Reid, Reil,
  Rezaie, Rockosi, Roe, Ronayette, Roodman, Ross, Ross, Rossi, Rozo,
  Ruhlmann-Kleider, Rykoff, Sabiu, Samushia, Sanchez, Sanchez, Schlegel,
  Schneider, Schubnell, Secroun, Seljak, Seo, Serrano, Shafieloo, Shan,
  Sharples, Sholl, Shourt, Silber, Silva, Sirk, Slosar, Smith, Smoot, Som,
  Song, Sprayberry, Staten, Stefanik, Tarle, Tie, Tinker, Tojeiro, Valdes,
  Valenzuela, Valluri, Vargas-Magana, Verde, Walker, Wang, Wang, Weaver,
  Weaverdyck, Wechsler, Weinberg, White, Yang, Yeche, Zhang, Zhao, Zheng, Zhou,
  Zhou, Zhu, Zou, \& Zu}]{DESICollaboration2016}
{DESI Collaboration}, Aghamousa, A., Aguilar, J., {et~al.} 2016{\natexlab{a}},
  The DESI Experiment Part I: Science,Targeting, and Survey Design,  arXiv,
  \dodoi{10.48550/ARXIV.1611.00036}

\bibitem[{{DESI Collaboration} {et~al.}(2016{\natexlab{b}}){DESI
  Collaboration}, {Aghamousa}, {Aguilar}, {Ahlen}, {Alam}, {Allen}, {Allende
  Prieto}, {Annis}, {Bailey}, {Balland}, {Ballester}, {Baltay}, {Beaufore},
  {Bebek}, {Beers}, {Bell}, {Bernal}, {Besuner}, {Beutler}, {Blake}, {Bleuler},
  {Blomqvist}, {Blum}, {Bolton}, {Briceno}, {Brooks}, {Brownstein},
  {Buckley-Geer}, {Burden}, {Burtin}, {Busca}, {Cahn}, {Cai}, {Cardiel-Sas},
  {Carlberg}, {Carton}, {Casas}, {Castander}, {Cervantes-Cota}, {Claybaugh},
  {Close}, {Coker}, {Cole}, {Comparat}, {Cooper}, {Cousinou}, {Crocce}, {Cuby},
  {Cunningham}, {Davis}, {Dawson}, {de la Macorra}, {De Vicente}, {Delubac},
  {Derwent}, {Dey}, {Dhungana}, {Ding}, {Doel}, {Duan}, {Ealet}, {Edelstein},
  {Eftekharzadeh}, {Eisenstein}, {Elliott}, {Escoffier}, {Evatt}, {Fagrelius},
  {Fan}, {Fanning}, {Farahi}, {Farihi}, {Favole}, {Feng}, {Fernandez},
  {Findlay}, {Finkbeiner}, {Fitzpatrick}, {Flaugher}, {Flender}, {Font-Ribera},
  {Forero-Romero}, {Fosalba}, {Frenk}, {Fumagalli}, {Gaensicke}, {Gallo},
  {Garcia-Bellido}, {Gaztanaga}, {Pietro Gentile Fusillo}, {Gerard},
  {Gershkovich}, {Giannantonio}, {Gillet}, {Gonzalez-de-Rivera},
  {Gonzalez-Perez}, {Gott}, {Graur}, {Gutierrez}, {Guy}, {Habib}, {Heetderks},
  {Heetderks}, {Heitmann}, {Hellwing}, {Herrera}, {Ho}, {Holland}, {Honscheid},
  {Huff}, {Hutchinson}, {Huterer}, {Hwang}, {Illa Laguna}, {Ishikawa},
  {Jacobs}, {Jeffrey}, {Jelinsky}, {Jennings}, {Jiang}, {Jimenez}, {Johnson},
  {Joyce}, {Jullo}, {Juneau}, {Kama}, {Karcher}, {Karkar}, {Kehoe}, {Kennamer},
  {Kent}, {Kilbinger}, {Kim}, {Kirkby}, {Kisner}, {Kitanidis}, {Kneib},
  {Koposov}, {Kovacs}, {Koyama}, {Kremin}, {Kron}, {Kronig}, {Kueter-Young},
  {Lacey}, {Lafever}, {Lahav}, {Lambert}, {Lampton}, {Landriau}, {Lang},
  {Lauer}, {Le Goff}, {Le Guillou}, {Le Van Suu}, {Lee}, {Lee}, {Leitner},
  {Lesser}, {Levi}, {L'Huillier}, {Li}, {Liang}, {Lin}, {Linder}, {Loebman},
  {Luki{\'c}}, {Ma}, {MacCrann}, {Magneville}, {Makarem}, {Manera}, {Manser},
  {Marshall}, {Martini}, {Massey}, {Matheson}, {McCauley}, {McDonald},
  {McGreer}, {Meisner}, {Metcalfe}, {Miller}, {Miquel}, {Moustakas}, {Myers},
  {Naik}, {Newman}, {Nichol}, {Nicola}, {Nicolati da Costa}, {Nie}, {Niz},
  {Norberg}, {Nord}, {Norman}, {Nugent}, {O'Brien}, {Oh}, {Olsen}, {Padilla},
  {Padmanabhan}, {Padmanabhan}, {Palanque-Delabrouille}, {Palmese},
  {Pappalardo}, {P{\^a}ris}, {Park}, {Patej}, {Peacock}, {Peiris}, {Peng},
  {Percival}, {Perruchot}, {Pieri}, {Pogge}, {Pollack}, {Poppett}, {Prada},
  {Prakash}, {Probst}, {Rabinowitz}, {Raichoor}, {Ree}, {Refregier}, {Regal},
  {Reid}, {Reil}, {Rezaie}, {Rockosi}, {Roe}, {Ronayette}, {Roodman}, {Ross},
  {Ross}, {Rossi}, {Rozo}, {Ruhlmann-Kleider}, {Rykoff}, {Sabiu}, {Samushia},
  {Sanchez}, {Sanchez}, {Schlegel}, {Schneider}, {Schubnell}, {Secroun},
  {Seljak}, {Seo}, {Serrano}, {Shafieloo}, {Shan}, {Sharples}, {Sholl},
  {Shourt}, {Silber}, {Silva}, {Sirk}, {Slosar}, {Smith}, {Smoot}, {Som},
  {Song}, {Sprayberry}, {Staten}, {Stefanik}, {Tarle}, {Sien Tie}, {Tinker},
  {Tojeiro}, {Valdes}, {Valenzuela}, {Valluri}, {Vargas-Magana}, {Verde},
  {Walker}, {Wang}, {Wang}, {Weaver}, {Weaverdyck}, {Wechsler}, {Weinberg},
  {White}, {Yang}, {Yeche}, {Zhang}, {Zhao}, {Zheng}, {Zhou}, {Zhou}, {Zhu},
  {Zou}, \& {Zu}}]{DESICollaboration2016a}
{DESI Collaboration}, {Aghamousa}, A., {Aguilar}, J., {et~al.}
  2016{\natexlab{b}}, arXiv e-prints, arXiv:1611.00037,
  \dodoi{10.48550/arXiv.1611.00037}

\bibitem[{Dewancker {et~al.}(2016)Dewancker, McCourt, \& Clark}]{Dewancker2015}
Dewancker, I., McCourt, M., \& Clark, S. 2016, Bayesian Optimization for
  Machine Learning : A Practical Guidebook.
\newblock \doarXiv{1612.04858}

\bibitem[{{Dierckx}(1993)}]{Dierckx1993}
{Dierckx}, P. 1993, {Curve and surface fitting with splines},
  \dodoi{10.1093/oso/9780198534419.001.0001}

\bibitem[{Ding {et~al.}(2022)Ding, Shi, Wu, Jones, Yan, Li, Gao, Chen, Zhang,
  Liu, Yan, \& Xie}]{Ding_2022}
Ding, M.-Y., Shi, J.-R., Wu, Y., {et~al.} 2022, The Astrophysical Journal
  Supplement Series, 260, 45, \dodoi{10.3847/1538-4365/ac6754}

\bibitem[{{Dotter}(2016)}]{2016ApJS..222....8D}
{Dotter}, A. 2016, \apjs, 222, 8, \dodoi{10.3847/0067-0049/222/1/8}

\bibitem[{Du {et~al.}(2012)Du, Luo, Zhang, Wu, \& Wang}]{du2012comparison}
Du, B., Luo, A., Zhang, J., Wu, Y., \& Wang, F. 2012, in Software and
  Cyberinfrastructure for Astronomy II, Vol. 8451, SPIE, 1043--1051,
  \dodoi{10.1117/12.925970}

\bibitem[{Du {et~al.}(2021)Du, Luo, Zhang, Kong, Guo, Li, Zuo, Wang, Chen, \&
  Zhao}]{Du_2021}
Du, B., Luo, A.-L., Zhang, S., {et~al.} 2021, Research in Astronomy and
  Astrophysics, 21, 202, \dodoi{10.1088/1674-4527/21/8/202}

\bibitem[{Dupree {et~al.}(2016)Dupree, Avrett, \& Kurucz}]{Dupree_2016}
Dupree, A.~K., Avrett, E.~H., \& Kurucz, R.~L. 2016, The Astrophysical Journal,
  821, L7, \dodoi{10.3847/2041-8205/821/1/l7}

\bibitem[{{Fu} {et~al.}(2022){Fu}, {Bragaglia}, {Liu}, {Zhang}, {Xu}, {Wang},
  {Zhang}, {Zhong}, {Chang}, {Li}, {Chen}, {Chen}, {Wang}, {Gjergo}, {Wang},
  {Yue}, \& {Zhang}}]{fu2022lamost}
{Fu}, X., {Bragaglia}, A., {Liu}, C., {et~al.} 2022, \aap, 668, A4,
  \dodoi{10.1051/0004-6361/202243590}

\bibitem[{{Gaia Collaboration} {et~al.}(2023){Gaia Collaboration}, {Vallenari},
  {Brown}, {Prusti}, {de Bruijne}, {Arenou}, {Babusiaux}, {Biermann},
  {Creevey}, {Ducourant}, {Evans}, {Eyer}, {Guerra}, {Hutton}, {Jordi},
  {Klioner}, {Lammers}, {Lindegren}, {Luri}, {Mignard}, {Panem}, {Pourbaix},
  {Randich}, {Sartoretti}, {Soubiran}, {Tanga}, {Walton}, {Bailer-Jones},
  {Bastian}, {Drimmel}, {Jansen}, {Katz}, {Lattanzi}, {van Leeuwen}, {Bakker},
  {Cacciari}, {Casta{\~n}eda}, {De Angeli}, {Fabricius}, {Fouesneau},
  {Fr{\'e}mat}, {Galluccio}, {Guerrier}, {Heiter}, {Masana}, {Messineo},
  {Mowlavi}, {Nicolas}, {Nienartowicz}, {Pailler}, {Panuzzo}, {Riclet}, {Roux},
  {Seabroke}, {Sordo}, {Th{\'e}venin}, {Gracia-Abril}, {Portell}, {Teyssier},
  {Altmann}, {Andrae}, {Audard}, {Bellas-Velidis}, {Benson}, {Berthier},
  {Blomme}, {Burgess}, {Busonero}, {Busso}, {C{\'a}novas}, {Carry}, {Cellino},
  {Cheek}, {Clementini}, {Damerdji}, {Davidson}, {de Teodoro}, {Nu{\~n}ez
  Campos}, {Delchambre}, {Dell'Oro}, {Esquej}, {Fern{\'a}ndez-Hern{\'a}ndez},
  {Fraile}, {Garabato}, {Garc{\'\i}a-Lario}, {Gosset}, {Haigron}, {Halbwachs},
  {Hambly}, {Harrison}, {Hern{\'a}ndez}, {Hestroffer}, {Hodgkin}, {Holl},
  {Jan{\ss}en}, {Jevardat de Fombelle}, {Jordan}, {Krone-Martins}, {Lanzafame},
  {L{\"o}ffler}, {Marchal}, {Marrese}, {Moitinho}, {Muinonen}, {Osborne},
  {Pancino}, {Pauwels}, {Recio-Blanco}, {Reyl{\'e}}, {Riello}, {Rimoldini},
  {Roegiers}, {Rybizki}, {Sarro}, {Siopis}, {Smith}, {Sozzetti}, {Utrilla},
  {van Leeuwen}, {Abbas}, {{\'A}brah{\'a}m}, {Abreu Aramburu}, {Aerts},
  {Aguado}, {Ajaj}, {Aldea-Montero}, {Altavilla}, {{\'A}lvarez}, {Alves},
  {Anders}, {Anderson}, {Anglada Varela}, {Antoja}, {Baines}, {Baker},
  {Balaguer-N{\'u}{\~n}ez}, {Balbinot}, {Balog}, {Barache}, {Barbato},
  {Barros}, {Barstow}, {Bartolom{\'e}}, {Bassilana}, {Bauchet}, {Becciani},
  {Bellazzini}, {Berihuete}, {Bernet}, {Bertone}, {Bianchi}, {Binnenfeld},
  {Blanco-Cuaresma}, {Blazere}, {Boch}, {Bombrun}, {Bossini}, {Bouquillon},
  {Bragaglia}, {Bramante}, {Breedt}, {Bressan}, {Brouillet}, {Brugaletta},
  {Bucciarelli}, {Burlacu}, {Butkevich}, {Buzzi}, {Caffau}, {Cancelliere},
  {Cantat-Gaudin}, {Carballo}, {Carlucci}, {Carnerero}, {Carrasco},
  {Casamiquela}, {Castellani}, {Castro-Ginard}, {Chaoul}, {Charlot}, {Chemin},
  {Chiaramida}, {Chiavassa}, {Chornay}, {Comoretto}, {Contursi}, {Cooper},
  {Cornez}, {Cowell}, {Crifo}, {Cropper}, {Crosta}, {Crowley}, {Dafonte},
  {Dapergolas}, {David}, {David}, {de Laverny}, {De Luise}, {De March}, {De
  Ridder}, {de Souza}, {de Torres}, {del Peloso}, {del Pozo}, {Delbo},
  {Delgado}, {Delisle}, {Demouchy}, {Dharmawardena}, {Di Matteo}, {Diakite},
  {Diener}, {Distefano}, {Dolding}, {Edvardsson}, {Enke}, {Fabre}, {Fabrizio},
  {Faigler}, {Fedorets}, {Fernique}, {Fienga}, {Figueras}, {Fournier},
  {Fouron}, {Fragkoudi}, {Gai}, {Garcia-Gutierrez}, {Garcia-Reinaldos},
  {Garc{\'\i}a-Torres}, {Garofalo}, {Gavel}, {Gavras}, {Gerlach}, {Geyer},
  {Giacobbe}, {Gilmore}, {Girona}, {Giuffrida}, {Gomel}, {Gomez},
  {Gonz{\'a}lez-N{\'u}{\~n}ez}, {Gonz{\'a}lez-Santamar{\'\i}a},
  {Gonz{\'a}lez-Vidal}, {Granvik}, {Guillout}, {Guiraud},
  {Guti{\'e}rrez-S{\'a}nchez}, {Guy}, {Hatzidimitriou}, {Hauser}, {Haywood},
  {Helmer}, {Helmi}, {Sarmiento}, {Hidalgo}, {Hilger}, {H{\l}adczuk}, {Hobbs},
  {Holland}, {Huckle}, {Jardine}, {Jasniewicz}, {Jean-Antoine Piccolo},
  {Jim{\'e}nez-Arranz}, {Jorissen}, {Juaristi Campillo}, {Julbe}, {Karbevska},
  {Kervella}, {Khanna}, {Kontizas}, {Kordopatis}, {Korn}, {K{\'o}sp{\'a}l},
  {Kostrzewa-Rutkowska}, {Kruszy{\'n}ska}, {Kun}, {Laizeau}, {Lambert},
  {Lanza}, {Lasne}, {Le Campion}, {Lebreton}, {Lebzelter}, {Leccia}, {Leclerc},
  {Lecoeur-Taibi}, {Liao}, {Licata}, {Lindstr{\o}m}, {Lister}, {Livanou},
  {Lobel}, {Lorca}, {Loup}, {Madrero Pardo}, {Magdaleno Romeo}, {Managau},
  {Mann}, {Manteiga}, {Marchant}, {Marconi}, {Marcos}, {Marcos Santos},
  {Mar{\'\i}n Pina}, {Marinoni}, {Marocco}, {Marshall}, {Martin Polo},
  {Mart{\'\i}n-Fleitas}, {Marton}, {Mary}, {Masip}, {Massari},
  {Mastrobuono-Battisti}, {Mazeh}, {McMillan}, {Messina}, {Michalik}, {Millar},
  {Mints}, {Molina}, {Molinaro}, {Moln{\'a}r}, {Monari}, {Mongui{\'o}},
  {Montegriffo}, {Montero}, {Mor}, {Mora}, {Morbidelli}, {Morel}, {Morris},
  {Muraveva}, {Murphy}, {Musella}, {Nagy}, {Noval}, {Oca{\~n}a}, {Ogden},
  {Ordenovic}, {Osinde}, {Pagani}, {Pagano}, {Palaversa}, {Palicio},
  {Pallas-Quintela}, {Panahi}, {Payne-Wardenaar}, {Pe{\~n}alosa Esteller},
  {Penttil{\"a}}, {Pichon}, {Piersimoni}, {Pineau}, {Plachy}, {Plum}, {Poggio},
  {Pr{\v{s}}a}, {Pulone}, {Racero}, {Ragaini}, {Rainer}, {Raiteri}, {Rambaux},
  {Ramos}, {Ramos-Lerate}, {Re Fiorentin}, {Regibo}, {Richards}, {Rios Diaz},
  {Ripepi}, {Riva}, {Rix}, {Rixon}, {Robichon}, {Robin}, {Robin}, {Roelens},
  {Rogues}, {Rohrbasser}, {Romero-G{\'o}mez}, {Rowell}, {Royer}, {Ruz Mieres},
  {Rybicki}, {Sadowski}, {S{\'a}ez N{\'u}{\~n}ez}, {Sagrist{\`a} Sell{\'e}s},
  {Sahlmann}, {Salguero}, {Samaras}, {Sanchez Gimenez}, {Sanna},
  {Santove{\~n}a}, {Sarasso}, {Schultheis}, {Sciacca}, {Segol}, {Segovia},
  {S{\'e}gransan}, {Semeux}, {Shahaf}, {Siddiqui}, {Siebert}, {Siltala},
  {Silvelo}, {Slezak}, {Slezak}, {Smart}, {Snaith}, {Solano}, {Solitro},
  {Souami}, {Souchay}, {Spagna}, {Spina}, {Spoto}, {Steele},
  {Steidelm{\"u}ller}, {Stephenson}, {S{\"u}veges}, {Surdej}, {Szabados},
  {Szegedi-Elek}, {Taris}, {Taylor}, {Teixeira}, {Tolomei}, {Tonello}, {Torra},
  {Torra}, {Torralba Elipe}, {Trabucchi}, {Tsounis}, {Turon}, {Ulla}, {Unger},
  {Vaillant}, {van Dillen}, {van Reeven}, {Vanel}, {Vecchiato}, {Viala},
  {Vicente}, {Voutsinas}, {Weiler}, {Wevers}, {Wyrzykowski}, {Yoldas}, {Yvard},
  {Zhao}, {Zorec}, {Zucker}, \& {Zwitter}}]{GaiaCollaboration2023}
{Gaia Collaboration}, {Vallenari}, A., {Brown}, A.~G.~A., {et~al.} 2023, \aap,
  674, A1, \dodoi{10.1051/0004-6361/202243940}

\bibitem[{{Garc{\'\i}a P{\'e}rez} {et~al.}(2021){Garc{\'\i}a P{\'e}rez},
  {S{\'a}nchez-Bl{\'a}zquez}, {Vazdekis}, {Allende Prieto}, {Milone}, {Sansom},
  {Gorgas}, {Falc{\'o}n-Barroso}, {Mart{\'\i}n Navarro}, \&
  {Cacho}}]{garcia2021extension}
{Garc{\'\i}a P{\'e}rez}, A.~E., {S{\'a}nchez-Bl{\'a}zquez}, P., {Vazdekis}, A.,
  {et~al.} 2021, \mnras, 505, 4496, \dodoi{10.1093/mnras/stab076}

\bibitem[{Gardner {et~al.}(2014)Gardner, Kusner, Xu, Weinberger, \&
  Cunningham}]{pmlrv32gardner14}
Gardner, J.~R., Kusner, M.~J., Xu, Z.~E., Weinberger, K.~Q., \& Cunningham,
  J.~P. 2014, in ICML, Vol. 2014, 937--945.
\newblock \url{https://proceedings.mlr.press/v32/gardner14.html}

\bibitem[{Gong {et~al.}(2019)Gong, Liu, Cao, Chen, Fan, Li, Li, Li, Zhang, \&
  Zhan}]{Gong_2019}
Gong, Y., Liu, X., Cao, Y., {et~al.} 2019, The Astrophysical Journal, 883, 203,
  \dodoi{10.3847/1538-4357/ab391e}

\bibitem[{Green {et~al.}(2019)Green, Schlafly, Zucker, Speagle, \&
  Finkbeiner}]{Green_2019}
Green, G.~M., Schlafly, E., Zucker, C., Speagle, J.~S., \& Finkbeiner, D. 2019,
  The Astrophysical Journal, 887, 93, \dodoi{10.3847/1538-4357/ab5362}

\bibitem[{Gunn {et~al.}(2006)Gunn, Siegmund, Mannery, Owen, Hull, Leger, Carey,
  Knapp, York, Boroski, Kent, Lupton, Rockosi, Evans, Waddell, Anderson, Annis,
  Barentine, Bartoszek, Bastian, Bracker, Brewington, Briegel, Brinkmann,
  Brown, Carr, Czarapata, Drennan, Dombeck, Federwitz, Gillespie, Gonzales,
  Hansen, Harvanek, Hayes, Jordan, Kinney, Klaene, Kleinman, Kron, Kresinski,
  Lee, Limmongkol, Lindenmeyer, Long, Loomis, McGehee, Mantsch, Neilsen,
  Neswold, Newman, Nitta, Peoples, Pier, Prieto, Prosapio, Rivetta, Schneider,
  Snedden, \& Wang}]{Gunn_2006}
Gunn, J.~E., Siegmund, W.~A., Mannery, E.~J., {et~al.} 2006, The Astronomical
  Journal, 131, 2332, \dodoi{10.1086/500975}

\bibitem[{Gustafsson {et~al.}(2008)Gustafsson, Edvardsson, Eriksson,
  J{\o}rgensen, Nordlund, \& Plez}]{gustafsson2008grid}
Gustafsson, B., Edvardsson, B., Eriksson, K., {et~al.} 2008, Astronomy \&
  Astrophysics, 486, 951, \dodoi{10.1051/0004-6361:200809724}

\bibitem[{Harris {et~al.}(2020)Harris, Millman, Van Der~Walt, Gommers,
  Virtanen, Cournapeau, Wieser, Taylor, Berg, Smith,
  {et~al.}}]{harris2020array}
Harris, C.~R., Millman, K.~J., Van Der~Walt, S.~J., {et~al.} 2020, Nature, 585,
  357, \dodoi{10.1038/s41586-020-2649-2}

\bibitem[{Hill {et~al.}(2021)Hill, Thomas, Maraston, Yan, Neumann, Lundgren,
  Lazarz, Chen, Cappellari, Holtzman, Imig, Cunha, Stringfellow, Bizyaev, Law,
  Stassun, Drory, Merrifield, \& Beers}]{Hill_2021}
Hill, L., Thomas, D., Maraston, C., {et~al.} 2021, Monthly Notices of the Royal
  Astronomical Society, 509, 4308, \dodoi{10.1093/mnras/stab3263}

\bibitem[{Hill {et~al.}(2022)Hill, Thomas, Maraston, Yan, Lazarz, Chen,
  Stringfellow, Cappellari, Holtzman, Imig, Bizyaev, Law, Stassun, \&
  Drory}]{Hill_2022}
---. 2022, Monthly Notices of the Royal Astronomical Society, 517, 4275,
  \dodoi{10.1093/mnras/stac2992}

\bibitem[{Hobbs {et~al.}(2008)Hobbs, York, Snow, Oka, Thorburn, Bishof,
  Friedman, McCall, Rachford, Sonnentrucker, {et~al.}}]{hobbs2008catalog}
Hobbs, L., York, D., Snow, T., {et~al.} 2008, The Astrophysical Journal, 680,
  1256, \dodoi{10.1086/587930}

\bibitem[{{Hourihane} {et~al.}(2023){Hourihane}, {Fran{\c{c}}ois}, {Worley},
  {Magrini}, {Gonneau}, {Casey}, {Gilmore}, {Randich}, {Sacco}, {Recio-Blanco},
  {Korn}, {Allende Prieto}, {Smiljanic}, {Blomme}, {Bragaglia}, {Walton}, {Van
  Eck}, {Bensby}, {Lanzafame}, {Frasca}, {Franciosini}, {Damiani}, {Lind},
  {Bergemann}, {Bonifacio}, {Hill}, {Lobel}, {Montes}, {Feuillet},
  {Tautvai{\v{s}}ien{\.{e}}}, {Guiglion}, {Tabernero}, {Gonz{\'a}lez
  Hern{\'a}ndez}, {Gebran}, {Van der Swaelmen}, {Mikolaitis}, {Daflon},
  {Merle}, {Morel}, {Lewis}, {Gonz{\'a}lez Solares}, {Murphy}, {Jeffries},
  {Jackson}, {Feltzing}, {Prusti}, {Carraro}, {Biazzo}, {Prisinzano},
  {Jofr{\'e}}, {Zaggia}, {Drazdauskas}, {Stonkut{\'e}}, {Marfil},
  {Jim{\'e}nez-Esteban}, {Mahy}, {Guti{\'e}rrez Albarr{\'a}n}, {Berlanas},
  {Santos}, {Morbidelli}, {Spina}, \&
  {Minkevi{\v{c}}i{\={u}}t{\.{e}}}}]{hourihane2023gaia}
{Hourihane}, A., {Fran{\c{c}}ois}, P., {Worley}, C.~C., {et~al.} 2023, \aap,
  676, A129, \dodoi{10.1051/0004-6361/202345910}

\bibitem[{Hunter(2007)}]{Hunter:2007}
Hunter, J.~D. 2007, Computing in Science \& Engineering, 9, 90,
  \dodoi{10.5281/zenodo.6982547}

\bibitem[{Imig {et~al.}(2022)Imig, Holtzman, Yan, Lazarz, Chen, Hill, Thomas,
  Maraston, Prescott, Stringfellow, Bizyaev, Beaton, \& Drory}]{Imig_2022}
Imig, J., Holtzman, J.~A., Yan, R., {et~al.} 2022, The Astronomical Journal,
  163, 56, \dodoi{10.3847/1538-3881/ac3ca7}

\bibitem[{Ivezi{\'c} {et~al.}(2008)Ivezi{\'c}, Sesar, Juri{\'c}, Bond,
  Dalcanton, Rockosi, Yanny, Newberg, Beers, Prieto,
  {et~al.}}]{ivezic2008milky}
Ivezi{\'c}, {\v{Z}}., Sesar, B., Juri{\'c}, M., {et~al.} 2008, The
  Astrophysical Journal, 684, 287, \dodoi{10.1086/589678}

\bibitem[{Jin {et~al.}(2023{\natexlab{a}})Jin, Wu, Fu, Yao, Ai, Feng, He, Ma,
  Pang, Zhu, {et~al.}}]{jin2023large}
Jin, J.-J., Wu, X.-B., Fu, Y., {et~al.} 2023{\natexlab{a}}, The Astrophysical
  Journal Supplement Series, 265, 25, \dodoi{10.3847/1538-4365/acaf89}

\bibitem[{Jin {et~al.}(2023{\natexlab{b}})Jin, Trager, Dalton, Aguerri, Drew,
  Falcón-Barroso, Gänsicke, Hill, Iovino, Pieri, Poggianti, Smith, Vallenari,
  Abrams, Aguado, Antoja, Aragón-Salamanca, Ascasibar, Babusiaux, Balcells,
  Barrena, Battaglia, Belokurov, Bensby, Bonifacio, Bragaglia, Carrasco,
  Carrera, Cornwell, Domínguez-Palmero, Duncan, Famaey, Fariña, Gonzalez,
  Guest, Hatch, Hess, Hoskin, Irwin, Knapen, Koposov, Kuchner, Laigle, Lewis,
  Longhetti, Lucatello, Méndez-Abreu, Mercurio, Molaeinezhad, Monguió,
  Morrison, Murphy, Peralta de Arriba, Pérez, Pérez-Ràfols, Picó, Raddi,
  Romero-Gómez, Royer, Siebert, Seabroke, Som, Terrett, Thomas, Wesson,
  Worley, Alfaro, Allende Prieto, Alonso-Santiago, Amos, Ashley,
  Balaguer-Núñez, Balbinot, Bellazzini, Benn, Berlanas, Bernard, Best,
  Bettoni, Bianco, Bishop, Blomqvist, Boeche, Bolzonella, Bonoli, Bosma,
  Britavskiy, Busarello, Caffau, Cantat-Gaudin, Castro-Ginard, Couto,
  Carbajo-Hijarrubia, Carter, Casamiquela, Conrado, Corcho-Caballero,
  Costantin, Deason, de Burgos, De Grandi, Di Matteo, Domínguez-Gómez,
  Dorda, Drake, Dutta, Erkal, Feltzing, Ferré-Mateu, Feuillet, Figueras,
  Fossati, Franciosini, Frasca, Fumagalli, Gallazzi, García-Benito,
  Gentile Fusillo, Gebran, Gilbert, Gledhill, González Delgado, Greimel,
  Guarcello, Guerra, Gullieuszik, Haines, Hardcastle, Harris, Haywood, Helmi,
  Hernandez, Herrero, Hughes, Iršič, Jablonka, Jarvis, Jordi, Kondapally,
  Kordopatis, Krogager, La Barbera, Lam, Larsen, Lemasle, Lewis, Lhomé, Lind,
  Lodi, Longobardi, Lonoce, Magrini, Maíz Apellániz, Marchal, Marco, Martin,
  Matsuno, Maurogordato, Merluzzi, Miralda-Escudé, Molinari, Monari, Morelli,
  Mottram, Naylor, Negueruela, Oñorbe, Pancino, Peirani, Peletier, Pozzetti,
  Rainer, Ramos, Read, Rossi, Röttgering, Rubiño-Martín, Sabater, San Juan,
  Sanna, Schallig, Schiavon, Schultheis, Serra, Shimwell, Simón-Díaz, Smith,
  Sordo, Sorini, Soubiran, Starkenburg, Steele, Stott, Stuik, Tolstoy, Tortora,
  Tsantaki, Van der Swaelmen, van Weeren, Vergani, Verheijen, Verro, Vink,
  Vioque, Walcher, Walton, Wegg, Weijmans, Williams, Wilson, Wright,
  Xylakis-Dornbusch, Youakim, Zibetti, \& Zurita}]{Jin2023}
Jin, S., Trager, S.~C., Dalton, G.~B., {et~al.} 2023{\natexlab{b}}, Monthly
  Notices of the Royal Astronomical Society, 530, 2688,
  \dodoi{10.1093/mnras/stad557}

\bibitem[{Jolliffe \& Cadima(2016)}]{Jolliffe2016}
Jolliffe, I.~T., \& Cadima, J. 2016, Philosophical Transactions of the Royal
  Society A: Mathematical, Physical and Engineering Sciences, 374, 20150202,
  \dodoi{10.1098/rsta.2015.0202}

\bibitem[{{J{\"o}nsson} {et~al.}(2020){J{\"o}nsson}, {Holtzman}, {Allende
  Prieto}, {Cunha}, {Garc{\'\i}a-Hern{\'a}ndez}, {Hasselquist}, {Masseron},
  {Osorio}, {Shetrone}, {Smith}, {Stringfellow}, {Bizyaev}, {Edvardsson},
  {Majewski}, {M{\'e}sz{\'a}ros}, {Souto}, {Zamora}, {Beaton}, {Bovy}, {Donor},
  {Pinsonneault}, {Poovelil}, \& {Sobeck}}]{2020AJ....160..120J}
{J{\"o}nsson}, H., {Holtzman}, J.~A., {Allende Prieto}, C., {et~al.} 2020, \aj,
  160, 120, \dodoi{10.3847/1538-3881/aba592}

\bibitem[{Juri{\'c} {et~al.}(2008)Juri{\'c}, Ivezi{\'c}, Brooks, Lupton,
  Schlegel, Finkbeiner, Padmanabhan, Bond, Sesar, Rockosi,
  {et~al.}}]{juric2008milky}
Juri{\'c}, M., Ivezi{\'c}, {\v{Z}}., Brooks, A., {et~al.} 2008, The
  Astrophysical Journal, 673, 864, \dodoi{10.1086/523619}

\bibitem[{Kurucz(1979)}]{Kurucz_1979}
Kurucz, R.~L. 1979, The Astrophysical Journal Supplement Series, 40, 1,
  \dodoi{10.1086/190589}

\bibitem[{Lazarz {et~al.}(2022)Lazarz, Yan, Wilhelm, Chen, Hill, Holtzman,
  Imig, Maraston, M{\'e}sz{\'a}ros, Stringfellow, {et~al.}}]{lazarz2022sdss}
Lazarz, D., Yan, R., Wilhelm, R., {et~al.} 2022, Astronomy \& Astrophysics,
  668, A21, \dodoi{10.1051/0004-6361/202243701}

\bibitem[{Lee {et~al.}(2011)Lee, Beers, Prieto, Lai, Rockosi, Morrison,
  Johnson, An, Sivarani, \& Yanny}]{lee2011segue}
Lee, Y.~S., Beers, T.~C., Prieto, C.~A., {et~al.} 2011, The Astronomical
  Journal, 141, 90, \dodoi{10.1088/0004-6256/141/3/90}

\bibitem[{Li {et~al.}(2021)Li, Liu, Zhang, Tian, Qiu, \& Tian}]{Li_2021}
Li, J., Liu, C., Zhang, B., {et~al.} 2021, The Astrophysical Journal Supplement
  Series, 253, 45, \dodoi{10.3847/1538-4365/abe1c1}

\bibitem[{Lu {et~al.}(2021)Lu, Luo, Wang, Qin, Wang, Chen, Du, Zuo, Hou, Chen,
  {et~al.}}]{lu2021study}
Lu, Y., Luo, A.-L., Wang, L.-L., {et~al.} 2021, Astronomy and Computing, 36,
  100485, \dodoi{10.1016/j.ascom.2021.100485}

\bibitem[{Luo {et~al.}(2012)Luo, Zhang, Zhao, Zhao, Cui, Li, Chu, Shi, Wang,
  Zhang, {et~al.}}]{luo2012data}
Luo, A.-L., Zhang, H.-T., Zhao, Y.-H., {et~al.} 2012, Research in Astronomy and
  Astrophysics, 12, 1243, \dodoi{10.1088/1674-4527/12/9/004}

\bibitem[{{Luo} {et~al.}(2015){Luo}, {Zhao}, {Zhao}, {Deng}, {Liu}, {Jing},
  {Wang}, {Zhang}, {Shi}, {Cui}, {Chu}, {Li}, {Bai}, {Wu}, {Cai}, {Cao}, {Cao},
  {Carlin}, {Chen}, {Chen}, {Chen}, {Chen}, {Chen}, {Chen}, {Chen},
  {Christlieb}, {Chu}, {Cui}, {Dong}, {Du}, {Fan}, {Feng}, {Fu}, {Gao}, {Gong},
  {Gu}, {Guo}, {Han}, {He}, {Hou}, {Hou}, {Hou}, {Hu}, {Hu}, {Hu}, {Huo},
  {Jia}, {Jiang}, {Jiang}, {Jiang}, {Jin}, {Kong}, {Kong}, {Lei}, {Li}, {Li},
  {Li}, {Li}, {Li}, {Li}, {Li}, {Li}, {Li}, {Li}, {Li}, {Li}, {Liang}, {Lin},
  {Liu}, {Liu}, {Liu}, {Liu}, {Lu}, {Luo}, {Mao}, {Newberg}, {Ni}, {Qi}, {Qi},
  {Shen}, {Shi}, {Song}, {Song}, {Su}, {Su}, {Tang}, {Tao}, {Tian}, {Wang},
  {Wang}, {Wang}, {Wang}, {Wang}, {Wang}, {Wang}, {Wang}, {Wang}, {Wang},
  {Wang}, {Wang}, {Wang}, {Wang}, {Wang}, {Wang}, {Wang}, {Wang}, {Wang},
  {Wang}, {Wei}, {Wei}, {Wu}, {Wu}, {Wu}, {Wu}, {Xing}, {Xu}, {Xu}, {Xu},
  {Yan}, {Yang}, {Yang}, {Yang}, {Yang}, {Yao}, {Yu}, {Yuan}, {Yuan}, {Yuan},
  {Yuan}, {Zhai}, {Zhang}, {Zhang}, {Zhang}, {Zhang}, {Zhang}, {Zhang},
  {Zhang}, {Zhang}, {Zhao}, {Zhou}, {Zhou}, {Zhu}, {Zhu}, {Zou}, \&
  {Zuo}}]{luo2015first}
{Luo}, A.~L., {Zhao}, Y.-H., {Zhao}, G., {et~al.} 2015, Research in Astronomy
  and Astrophysics, 15, 1095, \dodoi{10.1088/1674-4527/15/8/002}

\bibitem[{Majewski {et~al.}(2017)Majewski, Schiavon, Frinchaboy, Prieto,
  Barkhouser, Bizyaev, Blank, Brunner, Burton, Carrera, \&
  et~al.}]{Majewski2017}
Majewski, S.~R., Schiavon, R.~P., Frinchaboy, P.~M., {et~al.} 2017, The
  Astronomical Journal, 154, 94, \dodoi{10.3847/1538-3881/aa784d}

\bibitem[{Nguyen {et~al.}(2022)Nguyen, Costa, Girardi, Volpato, Bressan, Chen,
  Marigo, Fu, \& Goudfrooij}]{nguyen2022parsec}
Nguyen, C., Costa, G., Girardi, L., {et~al.} 2022, Astronomy \& Astrophysics,
  665, A126, \dodoi{10.1051/0004-6361/202244166}

\bibitem[{Passegger {et~al.}(2016)Passegger, Wende-von Berg, \&
  Reiners}]{passegger2016fundamental}
Passegger, V.~M., Wende-von Berg, S., \& Reiners, A. 2016, Astronomy \&
  Astrophysics, 587, A19, \dodoi{10.1051/0004-6361/201322261}

\bibitem[{P{\'e}rez {et~al.}(2016)P{\'e}rez, Prieto, Holtzman, Shetrone,
  M{\'e}sz{\'a}ros, Bizyaev, Carrera, Cunha, Garc{\'\i}a-Hern{\'a}ndez,
  Johnson, {et~al.}}]{perez2016aspcap}
P{\'e}rez, A. E.~G., Prieto, C.~A., Holtzman, J.~A., {et~al.} 2016, The
  Astronomical Journal, 151, 144, \dodoi{10.3847/0004-6256/151/6/144}

\bibitem[{{Prugniel} {et~al.}(2007){Prugniel}, {Soubiran}, {Koleva}, \& {Le
  Borgne}}]{prugniel2007new}
{Prugniel}, P., {Soubiran}, C., {Koleva}, M., \& {Le Borgne}, D. 2007, arXiv
  e-prints, astro, \dodoi{10.48550/arXiv.astro-ph/0703658}

\bibitem[{Qiu {et~al.}(2023)Qiu, Tian, Li, Liu, Long, Shi, Yang, \&
  Zhang}]{Qiu_2023}
Qiu, D., Tian, H., Li, J., {et~al.} 2023, Research in Astronomy and
  Astrophysics, 23, 055008, \dodoi{10.1088/1674-4527/acc153}

\bibitem[{Rains {et~al.}(2024)Rains, Nordlander, Monty, Casey, Rojas-Ayala,
  Žerjal, Ireland, Casagrande, \& McKenzie}]{Rains_2024}
Rains, A.~D., Nordlander, T., Monty, S., {et~al.} 2024, Monthly Notices of the
  Royal Astronomical Society, 529, 3171, \dodoi{10.1093/mnras/stae560}

\bibitem[{Rajpurohit {et~al.}(2014)Rajpurohit, Reyl{\'e}, Allard, Scholz,
  Homeier, Schultheis, \& Bayo}]{rajpurohit2014high}
Rajpurohit, A., Reyl{\'e}, C., Allard, F., {et~al.} 2014, Astronomy \&
  Astrophysics, 564, A90, \dodoi{10.1051/0004-6361/201322881}

\bibitem[{Rasmussen \& Williams(2005)}]{rasmussen2006gaussian}
Rasmussen, C.~E., \& Williams, C. K.~I. 2005, {Gaussian Processes for Machine
  Learning} (The MIT Press), \dodoi{10.7551/mitpress/3206.001.0001}

\bibitem[{{Reback} {et~al.}(2020){Reback}, {McKinney}, {Jbrockmendel}, {Van Den
  Bossche}, {Augspurger}, {Cloud}, {Gfyoung}, {Sinhrks}, {Klein}, {Hawkins},
  {Roeschke}, {Tratner}, {She}, {Ayd}, {Petersen}, {MomIsBestFriend}, {Garcia},
  {Schendel}, {Hayden}, {Jancauskas}, {Battiston}, {Saxton}, {Seabold},
  {Alimcmaster1}, {Chris-B1}, {H-Vetinari}, {Hoyer}, {Dong}, {Overmeire}, \&
  {Winkel}}]{reback2020pandas}
{Reback}, J., {McKinney}, W., {Jbrockmendel}, {et~al.} 2020,
  {pandas-dev/pandas: Pandas 1.0.5}, v1.0.5,  Zenodo,
  \dodoi{10.5281/zenodo.3898987}

\bibitem[{Rui {et~al.}(2019)Rui, Luo, Shuo, Wen, Bing, Yihan, Kefei, Jianjun,
  Fang, Li, {et~al.}}]{rui2019analysis}
Rui, W., Luo, A.-L., Shuo, Z., {et~al.} 2019, Publications of the Astronomical
  Society of the Pacific, 131, 024505, \dodoi{10.1088/1538-3873/aaf25f}

\bibitem[{Salaris {et~al.}(2004)Salaris, Weiss, \& Percival}]{salaris2004age}
Salaris, M., Weiss, A., \& Percival, S.~M. 2004, Astronomy \& Astrophysics,
  414, 163, \dodoi{10.1051/0004-6361:20031578}

\bibitem[{S{\'a}nchez-Bl{\'a}zquez {et~al.}(2006)S{\'a}nchez-Bl{\'a}zquez,
  Peletier, Jim{\'e}nez-Vicente, Cardiel, Cenarro, Falcon-Barroso, Gorgas,
  Selam, \& Vazdekis}]{sanchez2006medium}
S{\'a}nchez-Bl{\'a}zquez, P., Peletier, R., Jim{\'e}nez-Vicente, J., {et~al.}
  2006, Monthly Notices of the Royal Astronomical Society, 371, 703,
  \dodoi{10.1111/j.1365-2966.2006.10699.x}

\bibitem[{Santana {et~al.}(2021)Santana, Beaton, Covey, O’Connell,
  Longa-Pe{\~n}a, Cohen, Fern{\'a}ndez-Trincado, Hayes, Zasowski, Sobeck,
  {et~al.}}]{santana2021final}
Santana, F.~A., Beaton, R.~L., Covey, K.~R., {et~al.} 2021, The Astronomical
  Journal, 162, 303, \dodoi{10.3847/1538-3881/ac2cbc}

\bibitem[{Santoni {et~al.}(2024)Santoni, Raponi, De~Leone, \&
  Doerr}]{santoni2024comparison}
Santoni, M.~L., Raponi, E., De~Leone, R., \& Doerr, C. 2024, ACM Transactions
  on Evolutionary Learning, \dodoi{10.1145/3670683}

\bibitem[{Sbordone {et~al.}(2014)Sbordone, Caffau, Bonifacio, \&
  Duffau}]{sbordone2014mygisfos}
Sbordone, L., Caffau, E., Bonifacio, P., \& Duffau, S. 2014, Astronomy \&
  Astrophysics, 564, A109, \dodoi{10.1051/0004-6361/201322430}

\bibitem[{{Sharma} {et~al.}(2016){Sharma}, {Prugniel}, \&
  {Singh}}]{sharma2016new}
{Sharma}, K., {Prugniel}, P., \& {Singh}, H.~P. 2016, \aap, 585, A64,
  \dodoi{10.1051/0004-6361/201526111}

\bibitem[{Smith {et~al.}(2021)Smith, Bizyaev, Cunha, Shetrone, Souto, Prieto,
  Masseron, M{\'e}sz{\'a}ros, J{\"o}nsson, Hasselquist,
  {et~al.}}]{smith2021apogee}
Smith, V.~V., Bizyaev, D., Cunha, K., {et~al.} 2021, The Astronomical Journal,
  161, 254, \dodoi{10.3847/1538-3881/abefdc}

\bibitem[{Soubiran {et~al.}(1997)Soubiran, Friel, Ralite, \&
  Fran{\c{c}}ois}]{soubiran1997catalogue}
Soubiran, C., Friel, E., Ralite, N., \& Fran{\c{c}}ois, P. 1997, Astronomy and
  Astrophysics Supplement Series, 124, 299, \dodoi{10.1051/aas:1997194}

\bibitem[{Soubiran {et~al.}(2016)Soubiran, Le~Campion, Brouillet, \&
  Chemin}]{soubiran2016pastel}
Soubiran, C., Le~Campion, J.-F., Brouillet, N., \& Chemin, L. 2016, Astronomy
  \& Astrophysics, 591, A118, \dodoi{10.1051/0004-6361/201628497}

\bibitem[{{Soubiran} {et~al.}(2020){Soubiran}, {Le Campion}, {Brouillet}, \&
  {Chemin}}]{soubiran2020vizier}
{Soubiran}, C., {Le Campion}, J.~F., {Brouillet}, N., \& {Chemin}, L. 2020,
  {VizieR Online Data Catalog: The PASTEL catalogue (Soubiran+, 2016-)}, VizieR
  On-line Data Catalog: B/pastel. Originally published in:
  2016A\&A...591A.118S.
\newblock \url{https://ui.adsabs.harvard.edu/abs/2020yCat....102029S}

\bibitem[{Soubiran {et~al.}(2010)Soubiran, Le~Campion, De~Strobel, \&
  Caillo}]{soubiran2010pastel}
Soubiran, C., Le~Campion, J.-F., De~Strobel, G.~C., \& Caillo, A. 2010,
  Astronomy \& Astrophysics, 515, A111, \dodoi{10.1051/0004-6361/201014247}

\bibitem[{Soubiran \& Ralite(2001)}]{soubiran2001catalogue}
Soubiran, C., \& Ralite, N. 2001, Astronomy \& Astrophysics, 373, 159,
  \dodoi{10.1051/0004-6361:20010525}

\bibitem[{Stander \& Craig(2002)}]{Stander2002}
Stander, N., \& Craig, K. 2002, International Journal for Computer-Aided
  Engineering and Software (Eng. Comput.), 19,
  \dodoi{10.1108/02644400210430190}

\bibitem[{{Steinmetz} {et~al.}(2006){Steinmetz}, {Zwitter}, {Siebert},
  {Watson}, {Freeman}, {Munari}, {Campbell}, {Williams}, {Seabroke}, {Wyse},
  {Parker}, {Bienaym{\'e}}, {Roeser}, {Gibson}, {Gilmore}, {Grebel}, {Helmi},
  {Navarro}, {Burton}, {Cass}, {Dawe}, {Fiegert}, {Hartley}, {Russell},
  {Saunders}, {Enke}, {Bailin}, {Binney}, {Bland-Hawthorn}, {Boeche}, {Dehnen},
  {Eisenstein}, {Evans}, {Fiorucci}, {Fulbright}, {Gerhard}, {Jauregi}, {Kelz},
  {Mijovi{\'c}}, {Minchev}, {Parmentier}, {Pe{\~n}arrubia}, {Quillen}, {Read},
  {Ruchti}, {Scholz}, {Siviero}, {Smith}, {Sordo}, {Veltz}, {Vidrih}, {von
  Berlepsch}, {Boyle}, \& {Schilbach}}]{2006AJ....132.1645S}
{Steinmetz}, M., {Zwitter}, T., {Siebert}, A., {et~al.} 2006, \aj, 132, 1645,
  \dodoi{10.1086/506564}

\bibitem[{{Sung} \& {Bessell}(1999)}]{sung1999ubvi}
{Sung}, H., \& {Bessell}, M.~S. 1999, \mnras, 306, 361,
  \dodoi{10.1046/j.1365-8711.1999.02522.x}

\bibitem[{Tabernero {et~al.}(2022)Tabernero, Marfil, Montes, \&
  Hern{\'a}ndez}]{tabernero2022steparsyn}
Tabernero, H., Marfil, E., Montes, D., \& Hern{\'a}ndez, J.~G. 2022, Astronomy
  \& Astrophysics, 657, A66, \dodoi{10.1051/0004-6361/202141763}

\bibitem[{{Taylor}(2024)}]{2024arXiv240101156T}
{Taylor}, M. 2024, arXiv e-prints, arXiv:2401.01156,
  \dodoi{10.48550/arXiv.2401.01156}

\bibitem[{{Taylor}(2005)}]{2005ASPC..347...29T}
{Taylor}, M.~B. 2005, in Astronomical Society of the Pacific Conference Series,
  Vol. 347, Astronomical Data Analysis Software and Systems XIV, ed.
  P.~{Shopbell}, M.~{Britton}, \& R.~{Ebert}, 29.
\newblock \url{https://ui.adsabs.harvard.edu/abs/2005ASPC..347...29T}

\bibitem[{Ting {et~al.}(2016)Ting, Conroy, \& Rix}]{ting2016accelerated}
Ting, Y.-S., Conroy, C., \& Rix, H.-W. 2016, The Astrophysical Journal, 826,
  83, \dodoi{10.3847/0004-637x/826/1/83}

\bibitem[{Ting {et~al.}(2019)Ting, Conroy, Rix, \& Cargile}]{ting2019payne}
Ting, Y.-S., Conroy, C., Rix, H.-W., \& Cargile, P. 2019, The Astrophysical
  Journal, 879, 69, \dodoi{10.3847/1538-4357/ab2331}

\bibitem[{Virtanen {et~al.}(2020)Virtanen, Gommers, Oliphant, Haberland, Reddy,
  Cournapeau, Burovski, Peterson, Weckesser, Bright, van~der Walt, Brett,
  Wilson, Millman, Mayorov, Nelson, Jones, Kern, Larson, Carey, Polat, Feng,
  Moore, VanderPlas, Laxalde, Perktold, Cimrman, Henriksen, Quintero, Harris,
  Archibald, Ribeiro, Pedregosa, van Mulbregt, Vijaykumar, Bardelli, Rothberg,
  Hilboll, Kloeckner, Scopatz, Lee, Rokem, Woods, Fulton, Masson, Häggström,
  Fitzgerald, Nicholson, Hagen, Pasechnik, Olivetti, Martin, Wieser, Silva,
  Lenders, Wilhelm, Young, Price, Ingold, Allen, Lee, Audren, Probst, Dietrich,
  Silterra, Webber, Slavič, Nothman, Buchner, Kulick, Schönberger,
  de~Miranda~Cardoso, Reimer, Harrington, Rodríguez, Nunez-Iglesias,
  Kuczynski, Tritz, Thoma, Newville, Kümmerer, Bolingbroke, Tartre, Pak,
  Smith, Nowaczyk, Shebanov, Pavlyk, Brodtkorb, Lee, McGibbon, Feldbauer,
  Lewis, Tygier, Sievert, Vigna, Peterson, More, Pudlik, Oshima, Pingel,
  Robitaille, Spura, Jones, Cera, Leslie, Zito, Krauss, Upadhyay, Halchenko, \&
  Vázquez-Baeza}]{Virtanen_2020}
Virtanen, P., Gommers, R., Oliphant, T.~E., {et~al.} 2020, Nature Methods, 17,
  261, \dodoi{10.1038/s41592-019-0686-2}

\bibitem[{Wang {et~al.}(2023)Wang, Luo, Zhang, Ting, \& O’Briain}]{Wang_2023}
Wang, R., Luo, A.-L., Zhang, S., Ting, Y.-S., \& O’Briain, T. 2023, The
  Astrophysical Journal Supplement Series, 266, 40,
  \dodoi{10.3847/1538-4365/acce36}

\bibitem[{Wang {et~al.}(2016)Wang, Hutter, Zoghi, Matheson, \&
  De~Feitas}]{wang2016bayesian}
Wang, Z., Hutter, F., Zoghi, M., Matheson, D., \& De~Feitas, N. 2016, Journal
  of Artificial Intelligence Research, 55, 361, \dodoi{10.1613/jair.4806}

\bibitem[{Wei {et~al.}(2013)Wei, Luo, Li, Pan, Tu, Jiang, Kong, Shi, Yi, Wang,
  {et~al.}}]{wei2013mining}
Wei, P., Luo, A., Li, Y., {et~al.} 2013, Monthly Notices of the Royal
  Astronomical Society, 431, 1800, \dodoi{10.1093/mnras/stt298}

\bibitem[{Wu {et~al.}(2011)Wu, Luo, Li, Shi, Prugniel, Liang, Zhao, Zhang, Bai,
  Wei, {et~al.}}]{wu2011automatic}
Wu, Y., Luo, A.-L., Li, H.-N., {et~al.} 2011, Research in Astronomy and
  Astrophysics, 11, 924, \dodoi{10.1088/1674-4527/11/8/006}

\bibitem[{Xiang {et~al.}(2015)Xiang, Liu, Yuan, Huang, Huo, Zhang, Chen, Zhang,
  Sun, Wang, {et~al.}}]{xiang2015lamost}
Xiang, M., Liu, X., Yuan, H., {et~al.} 2015, Monthly Notices of the Royal
  Astronomical Society, 448, 822, \dodoi{10.1093/mnras/stu2692}

\bibitem[{Xiang {et~al.}(2019)Xiang, Ting, Rix, Sandford, Buder, Lind, Liu,
  Shi, \& Zhang}]{xiang2019abundance}
Xiang, M., Ting, Y.-S., Rix, H.-W., {et~al.} 2019, The Astrophysical Journal
  Supplement Series, 245, 34, \dodoi{10.3847/1538-4365/ab5364}

\bibitem[{Xiang {et~al.}(2022)Xiang, Rix, Ting, Kudritzki, Conroy, Zari, Shi,
  Przybilla, Ramirez-Tannus, Tkachenko, {et~al.}}]{xiang2022stellar}
Xiang, M., Rix, H.-W., Ting, Y.-S., {et~al.} 2022, Astronomy \& Astrophysics,
  662, A66, \dodoi{10.1051/0004-6361/202141570}

\bibitem[{{Xiang} {et~al.}(2017{\natexlab{a}}){Xiang}, {Liu}, {Shi}, {Yuan},
  {Huang}, {Luo}, {Zhang}, {Zhao}, {Zhang}, {Ren}, {Chen}, {Wang}, {Li}, {Huo},
  {Zhang}, {Wang}, {Zhang}, {Hou}, \& {Wang}}]{Xiang_2016}
{Xiang}, M.~S., {Liu}, X.~W., {Shi}, J.~R., {et~al.} 2017{\natexlab{a}},
  \mnras, 464, 3657, \dodoi{10.1093/mnras/stw2523}

\bibitem[{{Xiang} {et~al.}(2017{\natexlab{b}}){Xiang}, {Liu}, {Yuan}, {Huo},
  {Huang}, {Wang}, {Chen}, {Ren}, {Zhang}, {Tian}, {Yang}, {Shi}, {Zhao}, {Li},
  {Zhao}, {Cui}, {Li}, {Hou}, {Zhang}, {Zhang}, {Wang}, {Wu}, {Cao}, {Yan},
  {Yan}, {Luo}, {Zhang}, {Bai}, {Yuan}, {Dong}, {Lei}, \& {Li}}]{Xiang_2017}
{Xiang}, M.~S., {Liu}, X.~W., {Yuan}, H.~B., {et~al.} 2017{\natexlab{b}},
  \mnras, 467, 1890, \dodoi{10.1093/mnras/stx129}

\bibitem[{Yan {et~al.}(2022)Yan, Li, Wang, Zong, Yuan, Xiang, Huang, Xie, Dong,
  Yuan, {et~al.}}]{yan2022overview}
Yan, H., Li, H., Wang, S., {et~al.} 2022, The Innovation,
  \dodoi{10.1016/j.xinn.2022.100224}

\bibitem[{Yan {et~al.}(2019)Yan, Chen, Lazarz, Bizyaev, Maraston, Stringfellow,
  McCarthy, Meneses-Goytia, Law, Thomas, Barroso, Sánchez-Gallego, Schlafly,
  Zheng, Argudo-Fernández, Beaton, Beers, Bershady, Blanton, Brownstein,
  Bundy, Chambers, Cherinka, Lee, Drory, Galbany, Holtzman, Imig, Kaiser,
  Kinemuchi, Liu, Luo, Magnier, Majewski, Nair, Oravetz, Oravetz, Pan, Sobeck,
  Stassun, Talbot, Tremonti, Waters, Weijmans, Wilhelm, Zasowski, Zhao, \&
  Zhao}]{Yan_2019}
Yan, R., Chen, Y., Lazarz, D., {et~al.} 2019, The Astrophysical Journal, 883,
  175, \dodoi{10.3847/1538-4357/ab3ebc}

\bibitem[{{Yanny} {et~al.}(2009){Yanny}, {Rockosi}, {Newberg}, {Knapp},
  {Adelman-McCarthy}, {Alcorn}, {Allam}, {Allende Prieto}, {An}, {Anderson},
  {Anderson}, {Bailer-Jones}, {Bastian}, {Beers}, {Bell}, {Belokurov},
  {Bizyaev}, {Blythe}, {Bochanski}, {Boroski}, {Brinchmann}, {Brinkmann},
  {Brewington}, {Carey}, {Cudworth}, {Evans}, {Evans}, {Gates}, {G{\"a}nsicke},
  {Gillespie}, {Gilmore}, {Nebot Gomez-Moran}, {Grebel}, {Greenwell}, {Gunn},
  {Jordan}, {Jordan}, {Harding}, {Harris}, {Hendry}, {Holder}, {Ivans},
  {Ivezi{\v{c}}}, {Jester}, {Johnson}, {Kent}, {Kleinman}, {Kniazev},
  {Krzesinski}, {Kron}, {Kuropatkin}, {Lebedeva}, {Lee}, {French Leger},
  {L{\'e}pine}, {Levine}, {Lin}, {Long}, {Loomis}, {Lupton}, {Malanushenko},
  {Malanushenko}, {Margon}, {Martinez-Delgado}, {McGehee}, {Monet}, {Morrison},
  {Munn}, {Neilsen}, {Nitta}, {Norris}, {Oravetz}, {Owen}, {Padmanabhan},
  {Pan}, {Peterson}, {Pier}, {Platson}, {Re Fiorentin}, {Richards}, {Rix},
  {Schlegel}, {Schneider}, {Schreiber}, {Schwope}, {Sibley}, {Simmons},
  {Snedden}, {Allyn Smith}, {Stark}, {Stauffer}, {Steinmetz}, {Stoughton},
  {SubbaRao}, {Szalay}, {Szkody}, {Thakar}, {Sivarani}, {Tucker}, {Uomoto},
  {Vanden Berk}, {Vidrih}, {Wadadekar}, {Watters}, {Wilhelm}, {Wyse}, {Yarger},
  \& {Zucker}}]{2009AJ....137.4377Y}
{Yanny}, B., {Rockosi}, C., {Newberg}, H.~J., {et~al.} 2009, \aj, 137, 4377,
  \dodoi{10.1088/0004-6256/137/5/4377}

\bibitem[{Zasowski {et~al.}(2013)Zasowski, Johnson, Frinchaboy, Majewski,
  Nidever, Pinto, Girardi, Andrews, Chojnowski, Cudworth,
  {et~al.}}]{zasowski2013target}
Zasowski, G., Johnson, J.~A., Frinchaboy, P., {et~al.} 2013, The Astronomical
  Journal, 146, 81, \dodoi{10.1088/0004-6256/146/4/81}

\bibitem[{Zasowski {et~al.}(2017)Zasowski, Cohen, Chojnowski, Santana, Oelkers,
  Andrews, Beaton, Bender, Bird, Bovy, {et~al.}}]{zasowski2017target}
Zasowski, G., Cohen, R., Chojnowski, S.~D., {et~al.} 2017, The Astronomical
  Journal, 154, 198, \dodoi{10.3847/1538-3881/aa8df9}

\bibitem[{Zhao {et~al.}(2012)Zhao, Zhao, Chu, Jing, \& Deng}]{zhao2012lamost}
Zhao, G., Zhao, Y.-H., Chu, Y.-Q., Jing, Y.-P., \& Deng, L.-C. 2012, Research
  in Astronomy and Astrophysics, 12, 723, \dodoi{10.1088/1674-4527/12/7/002}

\end{thebibliography}
\end{document}